%% file: PhD-WD.tex
\documentclass[11pt,twoside,subeqns]{book}
\include{PhD-layout}

\include{PhD-commands}

\begin{document}

\include{PhD-titlepage}
\tableofcontents
\include{PhD-introduction1}

\include{PhD-spectral3}

\include{PhD-vacuum2}

\include{PhD-Conclusions}

\include{PhD-HR}
\include{PhD-Preliminaries_app}

\include{PhD-L2_app}

\include{PhD-L1_app}

\include{PhD-Nnat_app1}

\include{PhD-Cnat_app}

\addcontentsline{toc}{chapter}{Bibliography}
\bibliographystyle{amsalpha}
\bibliography{PhD-bibliography}
\include{PhD-Conventions}

\include{PhD-Acknowledgements}
\end{document}

%% file: PhD-layout.tex
\usepackage[ansinew]{inputenc}
\usepackage{ngerman,a4}
\usepackage{amssymb}   
\usepackage{amsmath}
\usepackage{amsthm,mathrsfs}
\usepackage{amscd}  

\usepackage{graphicx}
\usepackage{psfrag}
\usepackage{epsfig}
\usepackage{setspace}
\usepackage{fancyheadings}
\usepackage[english]{babel} 
\usepackage[setpagesize=false]{hyperref}
\usepackage{url}

\newtheorem{theorem}{Theorem}[section]
\newtheorem{proposition}[theorem]{Proposition}
\newtheorem{definition}[theorem]{Definition}
\newtheorem{lemma}[theorem]{Lemma}
\newtheorem{app-lemma}{Lemma}[subsection]
\newtheorem{corollary}[theorem]{Corollary}

\numberwithin{equation}{section}

%% file: PhD-commands.tex
\usepackage[ansinew]{inputenc}
\usepackage{ngerman,a4}
\usepackage{amssymb}   
\usepackage{amsmath}
\usepackage{amsthm,mathrsfs}
\usepackage{amscd}  
\usepackage{longtable}

\newcommand{\tGad}{\Ga_{N,\de}^{(+)}}
\newcommand{\tThe}{\The^{(+)}}
\newcommand{\Nh}{\overset{{\sswedge}}{\N} }
\newcommand{\Nc}{\overset{{\ssvee}}{\N}}
\newcommand{\Thec}{\overset{{\ssvee}}{\The}}
\newcommand{\Theh}{\overset{\sswedge}{\The}}
\newcommand{\PP}{\mathcal{P}}
\newcommand{\oms}{\om_0}

\newcommand{\vpc}{\vpi^{\prime}}  
\newcommand{\vph}{\vpi^{\prime\prime}} 

\newcommand{\tP}{\tilde{\PP}}
\newcommand{\GadN}{\Gamma_{N,\de_N}}
\newcommand{\vpi}{\hat{\om}}
\newcommand{\vr}{\vec{r}}
\newcommand{\Aa}[1]{A_{#1}(f_{#1,T})}
\newcommand{\slim}{\te{s -}\lim}
\newcommand{\Pm}{P_{[m]}}
\newcommand{\hilm}{\hil^{[m]}}
\newcommand{\At}[1]{A_{#1}(t_{#1},\vx_{#1})}
\newcommand{\Ah}[1]{\widehat{A}_{#1}(\widehat{f}_{{#1},T}^{\hat{s}})}
\newcommand{\As}[1]{A_{#1}(f_{#1,T}^s)}
\newcommand{\cf}{\overset{\,\,\ssvee}{f}}
\newcommand{\hf}{ \overset{\,\,\sswedge}{f} }
\newcommand{\cznn}{C_0^{\infty}(\mathbb{R}^4)}
\newcommand{\Ahhn}[1]{\widehat{A}_{#1}(\widehat{f}_{{#1}})}
\newcommand{\An}[1]{A_{#1}(f_{#1})}
\newcommand{\Ahh}[1]{\widehat{A}_{#1}(\widehat{f}_{{#1},T})}
\newcommand{\og}{\omega(\vep)}
\newcommand{\ft}{\tilde{f}}
\newcommand{\cz}{C_0^{\infty}}
\newcommand{\f}{f}
\newcommand{\en}{\mathring{e}}
\newcommand{\tn}{\mathring{t}}
\newcommand{\T}{\mathring{T}}
\newcommand{\BacR}{(\mfa_{\scc}(\mco(r))^{\times N})^*}

\newcommand{\BaR}{(\mfa(\mco(r))^{\times N})^*}
\newcommand{\mcor}{\mco(r)}
\newcommand{\muh}{\ov{\mu}_a}
\newcommand{\muc}{\ov{\mu}_b}
\newcommand{\btF}{ {\tilde{F}}}
\newcommand{\chka}{\overset{\ssvee}{\ka}}
\newcommand{\haka}{\overset{\sswedge}{\ka}}
\newcommand{\hh}{h}

\newcommand{\mfar}{\mathring{\mfa}}
\newcommand{\alpn}{\al^+}
\newcommand{\Lpmring}{\mathring{\mathcal{L}}^{\pm}_r}
\newcommand{\Lpring}{\mathring{\mathcal{L}}^{+}_r}
\newcommand{\Lmring}{\mathring{\mathcal{L}}^{-}_r}
\newcommand{\Lring}{\mathring{\mathcal{L}}_r}
\newcommand{\Lringz}{\mathring{\mathcal{L}}_{r_0}}
\newcommand{\rzero}{r}
\newcommand{\lil}{n}
\newcommand{\lal}{l}
\newcommand{\hka}{\hat{\ka}}
\newcommand{\uu}{\underline{u}}
\newcommand{\uw}{\underline{w}}
\newcommand{\ub}{\underline{b}}
\newcommand{\balb}{\ov{\al}^{\prime\prime}}
\newcommand{\balp}{\ov{\al}^{\prime}}
\newcommand{\bal}{\ov{\al}}
\newcommand{\Tr}{\te{Tr}}
\newcommand{\ee}{\te{(e)}}
\newcommand{\ei}{\iota_{\te{(e)}} }
\newcommand{\efun}{\un{\fun}}
\newcommand{\eU}{\un{U}}
\newcommand{\uA}{\underline{A}}
\newcommand{\eumfa}{\underline{\mfa}^{\te{(e)} }}
\newcommand{\heumfa}{\underline{\hat\mfa}^{\te{(e)} }}
\newcommand{\alb}{\beta}
\newcommand{\epi}{\pi_{\te{(e)}} }
\newcommand{\epim}{\pi_{\te{(e)}}^{-1} }
\newcommand{\eal}{\underline{\al}}  
\newcommand{\ehil}{\hil^{\te{(e)} }}
\newcommand{\emfa}{\mfa^{\te{(e)} } }
\newcommand{\dmfa}{\mfa^{\te{(d)} } }

\newcommand{\mupmi}{\al^{\pm}}
\newcommand{\hti}{\tilde{h}}
\newcommand{\bti}{\tilde{b}}
\newcommand{\MM}{K}
\newcommand{\cS}{\overset{\ssvee}{S}}
\newcommand{\bsi}{\bar{\si}}
\newcommand{\Sett}{\mathcal{M}}
\newcommand{\Set}{\mathcal{S}}
\newcommand{\F}{F}
\newcommand{\mumi}{\al^{-}}
\newcommand{\mupi}{\al^{+}}
\newcommand{\DF}{D_F}
\newcommand{\uPiEc}{\un{\Pi}_E^{\scc}}
\newcommand{\PiEc}{\Pi_E^{\scc}}

\newcommand{\rz}{r}

\newcommand{\alr}{\gamma}
\newcommand{\tr}{\rho}
\newcommand{\chitr}{\chi(\mco_{\tr})}
\newcommand{\chitrz}{\chi_{\vx}(\mco_{\tr})}
\newcommand{\omar}{\om^{-\alr}}

\newcommand{\cEErs}{\sup_{\om(\vep)\leq E}|\tih(\vep)|^{-2}}

\newcommand{\cEE}{\sup_{\om(\vep)\leq E}|\tilh(\vep)|^{-1}}
\newcommand{\wFz}{\tilde{F}_{\vx}}
\newcommand{\y}{\vec{y}}

\newcommand{\ube}{ \hat{\be}_{\su{\\ \!\!\!\!\!\! \mbox{\boldmath $_\gets$} }} }  
 
\newcommand{\ual}{ \hat{\al}_{\su{\\ \!\!\!\!\!\! \mbox{\boldmath $_\to$} }} }
\newcommand{\uall}{ \hat{\al}_{\su{\\ \!\!\!\!\!\! \mbox{\boldmath $_\gets$} }} }
\newcommand{\ualli}{ \hat{\al}_{\su{i \!\!\!\!\!\!\!\!\! \mbox{\boldmath $_\gets$} }} }
\newcommand{\uali}{ \hat{\al}_{\su{i \!\!\!\!\!\!\!\!\! \mbox{\boldmath $_\to$} }} }
\newcommand{\unu}{\underline{\nu}}
\newcommand{\obe}{\hat{\beta}}
\newcommand{\oal}{\hat{\alpha}}
\newcommand{\Fm}{f^-}
\newcommand{\Fp}{f^+}

\newcommand{\almb}{\al^{\prime\prime-}}
\newcommand{\almp}{\al^{\prime-}}
\newcommand{\alm}{\al^{-}}
\newcommand{\alpb}{\al^{\prime\prime+}}
\newcommand{\alpp}{\al^{\prime+}}

\newcommand{\Gadd}{\Gamma_{M,\de}}
\newcommand{\Pizz}{\Pi^{\prime\prime}}
\newcommand{\Piz}{\Pi^{\prime}}
\newcommand{\BBB}{(\mfar(\mcor)^{\times M})^*}
\newcommand{\Epm}{\Lpm}
\newcommand{\uvx}{\underline{\vx}}

\newcommand{\del}{\gamma}

\newcommand{\SiEh}{\Pi_E}

\newcommand{\ufun}{u}
\newcommand{\khat}{\hat{k}}
\newcommand{\lattice}{\mathbb{J}}
\newcommand{\nor}{|\!|\!|}
\newcommand{\cc}{\scc}
\newcommand{\xx}{\vx}
\newcommand{\hQ}{\hat{Q}}
\newcommand{\he}{\hat{e}}
\newcommand{\too}{\longrightarrow_{\su{\\ \!\!\!\!\!\!\!\!\!\!\!\!\!\! \K\to\infty}}}
\newcommand{\K}{n}
\newcommand{\cA}{\mathring{A}}
\newcommand{\hXi}{\Xi}
\newcommand{\hPi}{\widehat{\Pi}}
\newcommand{\BB}{(\mfa(\mco)^{\times N-1})^* }
\newcommand{\R}{\textrm{Re}}
\newcommand{\I}{\textrm{Im}}
\newcommand{\hB}{B}
\newcommand{\cB}{\mathring{B}}
\newcommand{\vpp}{\theta}
\newcommand{\alp}{\gamma}
\newcommand{\AN}{A_1\times\cdots\times A_N}
\newcommand{\bx}{\underline{\vx}}
\newcommand{\Bac}{(\mfa_{\scc}(\mco)^{\times N})^*}
\newcommand{\Ba}{(\mfa(\mco)^{\times N})^*}
\newcommand{\wh}{\widehat}
\newcommand{\whXi}{\Theta} 
\newcommand{\Nnat}{N_{\bnat}}
\newcommand{\bsharp}{\boldsymbol{\sharp}}
\newcommand{\bflat}{\boldsymbol{\flat}}
\newcommand{\bnat}{\boldsymbol{\natural}}
\newcommand{\Ns}{N_{\bsharp}}
\newcommand{\Cs}{C_{\bsharp}}
\newcommand{\fin}{\mathcal{F}} 
\newcommand{\Cnat}{C_{\bnat}}
\newcommand{\Csq}{C_{\bflat} } 
\newcommand{\BE}{D_B}
\newcommand{\V}{V}

\newcommand{\rad}{r}
\newcommand{\Ppr}{P_{(p,\rad)} }

\newcommand{\vxb}{\underline{\vx}}
\newcommand{\lam}{\la}
\newcommand{\e}{\hat{e}}
\newcommand{\N}{\mathcal{N}}
\newcommand{\tN}{\wt{\mathcal{N}}}
\newcommand{\vxz}{\underline{z}}

\newcommand{\VV}{K}
\newcommand{\KK}{O} 
\newcommand{\ovom}{\om_0}
\newcommand{\A}{L^{(2)}}
\newcommand{\B}{L^{(1)}}
\newcommand{\CC}{\mfa^{(1)}}

\newcommand{\LL}{\mathcal{L}_r}
\newcommand{\Bs}{\B_{\bsharp}}
\newcommand{\nin}{\noindent}
\newcommand{\hmfa}{\hat{\mfa}}
\newcommand{\te}{\textrm}
\newcommand{\PC}{\mathfrak{P}}
\newcommand{\mfc}{\mathfrak{C}}
\newcommand{\mfl}{\mathfrak{L}}
\newcommand{\Sp}{\textrm{Sp}}
\newcommand{\spa}{\textrm{sp}}
\newcommand{\sic}{\textrm{sc}}
\renewcommand{\iff}{\Leftrightarrow}
\newcommand{\hdmfa}{\hat{\mfa}^{\textrm{(d)}}} %
\newcommand{\hemfa}{\hat{\mfa}^{\textrm{(e)}}} %

\newcommand{\ph}{\phantom}
\newcommand{\Full}{\widehat{\hil}}
\newcommand{\hF}{\widehat{F}}

\renewcommand{\Re}{\textrm{Re}}

\newcommand{\M}{M_E}
\newcommand{\m}{k}
\newcommand{\un}{\underline}

\newcommand{\ssvee}{{\scriptscriptstyle\vee}}
\newcommand{\sswedge}{{\scriptscriptstyle\wedge}}

\newcommand{\thec}{\overset{{\ssvee}}{\Theta}_{E,1}}
\newcommand{\theh}{\overset{\sswedge}{\Theta}_{E,1}}
\newcommand{\thet}{\Theta_{E,2}}
\newcommand{\theo}{\Theta_{E,1}}
\newcommand{\tih}{\tilde{h}_{\rzero}}

\newcommand{\wil}{:\!}
\newcommand{\wir}{\!:}
\newcommand{\phip}{\phi_+}
\newcommand{\phim}{\phi_-}
\newcommand{\phipm}{\phi_{\pm}}
\newcommand{\tphip}{\wt{\phi}_+}

\newcommand{\phii}{\phi}

\newcommand{\tphii}{\wt{\phii}}
\newcommand{\field}{\phi}

\newcommand{\Span}{\textrm{Span}}
\newcommand{\pc}{\mathrm{pc}}
\newcommand{\ord}{\mathrm{ord}}
\newcommand{\Ord}{\mathrm{Ord}}

\newcommand{\FH}{\mathrm{FH}}

\newcommand{\ac}{\mathrm{ac}}
\newcommand{\spp}{\mathrm{pp}}
\newcommand{\scc}{\mathrm{c}}

\newcommand{\enorm}{\sup_{\om(\vep)\leq E}|\tih(\vep)|^{-2}}
\newcommand{\tilh}{\tilde{h}}

\newcommand{\pa}{\partial}

\newcommand{\Q}{Q}

\newcommand{\wt}{\widetilde}

\newcommand{\tg}{\widetilde{g}}

\newcommand{\teps}{\tilde{\eps}}

\newcommand{\tF}{\tilde{F}}

\newcommand{\Gpm}{G^{\pm}}

\newcommand{\h}{\half}
\newcommand{\chir}{\chi(\mco_r)}

\newcommand{\czn}{C_0^{\infty}(\real^s)}
\newcommand{\tf}{\tilde{f}}

\newcommand{\dsp}{d^sp_1\ldots d^sp_n}
\newcommand{\ds}{d^sp_2\ldots d^sp_n}
\newcommand{\pp}{\vep_2,\ldots,\vep_n}
\newcommand{\tmn}{\tau_{\mub,\nub}}
\newcommand{\p}{*}

\newcommand{\x}{x_1,\ldots, x_N}

\newcommand{\pin}{\fr{1}{p}}

\newcommand{\fpm}{f^{\pm}}

\newcommand{\omp}{\om^{\h}}
\newcommand{\omm}{\om^{-\h}}
\newcommand{\ompm}{\om^{\pm\h}}

\newcommand{\om}{\omega}

\newcommand{\norm}{|\!|\!|\cdot |\!|\!|}
\newcommand{\lin}{\mathcal{L}}

\newcommand{\vy}{\vec{y}}

\newcommand{\PiE}{\Pi_E}
\newcommand{\si}{\sigma}
\newcommand{\al}{\alpha}
\newcommand{\be}{\beta}
\newcommand{\La}{\Lambda}
\newcommand{\la}{\lambda}
\newcommand{\vp}{\varphi}
\newcommand{\Ga}{\Gamma}
\newcommand{\ga}{\gamma}
\newcommand{\eps}{\varepsilon}
\newcommand{\De}{\Delta}
\newcommand{\de}{\delta}
\newcommand{\The}{\Sigma}
\newcommand{\TEp}{T_{E,+}}
\newcommand{\TEm}{T_{E,-}}
\newcommand{\TEpm}{T_{E,\pm}}

\newcommand{\Tbpm}{T_{h,\pm}}
\newcommand{\Tbp}{T_{h,+}}
\newcommand{\Tbm}{T_{h,-}}
\newcommand{\Thpm}{T_{h,\pm}}
\newcommand{\Thp}{T_{h,+}}
\newcommand{\Thm}{T_{h,-}}
\newcommand{\Lp}{\mathcal{L}^{+}_r}
\newcommand{\Lm}{\mathcal{L}^{-}_r}
\newcommand{\Lpm}{\mathcal{L}^{\pm}_r}

\newcommand{\integer}{\mathbb{Z}}

\newcommand{\nat}{\mathbb{N}}
\newcommand{\spt}{\mathbb{R}^4}
\newcommand{\real}{\mathbb{R}}
\newcommand{\complex}{\mathbb{C}}
\newcommand{\mup}{\mu^+}
\newcommand{\mum}{\mu^-}
\newcommand{\nup}{\nu^+}
\newcommand{\num}{\nu^-}
\newcommand{\mupm}{\mu^\pm}
\newcommand{\nupm}{\nu^\pm}
\newcommand{\mub}{\overline{\mu}}
\newcommand{\nub}{\overline{\nu}}
\newcommand{\nuh}{\nub_a}
\newcommand{\nuc}{\nub_b}
\newcommand{\Bnh}{a(\LL e)^{\nuh}} 
\newcommand{\Bnhs}{a^*(\LL e)^{\nuh}}
\newcommand{\Bnc}{a(\LL e)^{\nuc}}
\newcommand{\Bncs}{a^*(\LL e)^{\nuc}}
\newcommand{\Bm}{a(\LL e)^{\mub}}
\newcommand{\Bms}{a^*(\LL e)^{\mub}}
\newcommand{\Bn}{a(\LL e)^{\nub}}
\newcommand{\Bns}{a^*(\LL e)^{\nub}}
\newcommand{\Smn}{S_{\mub,\nub}}
\newcommand{\ka}{\kappa}
\newcommand{\trace}{\mathcal{T}}
\newcommand{\traceE}{\trace_E}

\newcommand{\traceEB}{\trace_{E,1}}
\newcommand{\traceEBP}{\trace_{E,1}^+}
\newcommand{\traceEP}{\trace_{E}^+}

\newcommand{\tracB}{\trace_{E,1}}
\newcommand{\vep}{\vec{p}}
\newcommand{\veP}{\vec{P}}
\newcommand{\veq}{\vec{q}}

\newcommand{\vu}{\vec{u}}
\newcommand{\vz}{\vec{z}}

\newcommand{\vx}{\vec{x}}
\newcommand{\ctau}{\overset{\ssvee}{\tau}}
\newcommand{\htau}{\overset{\sswedge}{\tau}}

\newcommand{\compsup}{\complex^N_{\sup}}

\newcommand{\ov}{\overline}
\newcommand{\funr}{\fun_{\textrm{Re}}}
\newcommand{\funi}{\fun_{\textrm{Im}}}
\newcommand{\fun}{\vp}

\newcommand{\umu}{\underline{\mu}}

\newcommand{\su}{\substack}

\newcommand{\hil}{\mathcal{H}}
\newcommand{\mfa}{\mathfrak{A}}
\newcommand{\mco}{\mathcal{O}}
\newcommand{\cpo}{P_{+}^{\uparrow}}
\newcommand{\cone}{\overline{V}_+}

\newcommand{\lpo}{L_{+}^{\uparrow}}
\newcommand{\supp}{\textrm{supp}}
\newcommand{\fr}[2]{\frac{#1}{#2}}
\newcommand{\ot}{\otimes}

\newcommand{\non}{\nonumber}

\newcommand{\Gad}{\Gamma_{N,\de}}
\newcommand{\vac}{\Omega}

\newcommand{\half}{\fr{1}{2}}
\newcommand{\lan}{\langle}
\newcommand{\ran}{\rangle}

\def\proof{\noindent{\bf Proof. }}
\def\qed{$\Box$\medskip}

\newcommand{\beq}{\begin{equation}}
\newcommand{\eeq}{\end{equation}}
\newcommand{\beqa}{\begin{eqnarray}}
\newcommand{\eeqa}{\end{eqnarray}}
\newcommand{\ben}{\begin{arabicenumerate}}
\newcommand{\een}{\end{arabicenumerate}}
\newcommand{\bex}{\begin{example}}
\newcommand{\eex}{\end{example}}
\newcommand{\ber}{\begin{remark}}
\newcommand{\eer}{\end{remark}}
\newcommand{\bec}{\begin{corollary}}
\newcommand{\eec}{\end{corollary}}
\newcommand{\bep}{\begin{proposition}}
\newcommand{\eep}{\end{proposition}}
\newcommand{\becr}{\begin{criterion}}
\newcommand{\eecr}{\end{criterion}}

\def\bel{\begin{lemma}}
\def\eel{\end{lemma}}
\def\bet{\begin{theorem}}
\def\eet{\end{theorem}}
\def\bed{\begin{definition}}
\def\eed{\end{definition}}

%% file: PhD-titlepage.tex
\thispagestyle{empty}

\begin{center}

\vspace*{3em}
\vfill

{\Huge \bf 
Spectral Theory of  Automorphism Groups and Particle Structures in Quantum Field Theory
}
\vfill
{\Large 
Dissertation \\
zur Erlangung des Doktorgrades \\
der Mathematisch-Naturwissenschaftlichen Fakult\"aten\\
der Georg-August-Universit\"at zu G\"ottingen

\vfill
  vorgelegt von \\
Wojciech Jan Dybalski\\
aus Warszawa

\vfill
 
G\"ottingen, 2008

\vfill

}
\end{center}

\clearpage  

\thispagestyle{empty}
\mbox{ } \vfill \mbox{ } \\
\noindent
D 7  \\
Referent: Prof.\ Dr.\ D.\ Buchholz \\
Korreferent: Prof.\ Dr.\ K. Fredenhagen \\
Tag der m\"undlichen Pr\"ufung: 15.12.2008

%
%

%% file: PhD-introduction1.tex
\chapter{Introduction}\label{chapter:introduction}

The understanding of quantum theories in terms of particles has been
a fundamental issue for more than four decades. While in the framework
of non-relativistic quantum mechanics this problem has been settled for
a large class of physically relevant models, the situation is
less clear in relativistic quantum field theory (QFT). There emerge new 
phenomena in this setting, like creation of charged particles in collisions or appearance of clouds of massless excitations accompanying a charged particle ('the infraparticle problem'). Their description goes beyond the well understood setting of groups of unitaries acting on a Hilbert space, 
where the notions of spectral measure and measure classes provide the natural conceptual basis for the formulation and 
resolution of the problem of asymptotic completeness. The  language  of groups of automorphisms acting on the algebra of observables, adequate in the relativistic setting, lacks such detailed spectral concepts.

The present work introduces a decomposition of the algebra of observables into spectral subspaces,
whose elements differ in their behavior under translations in space. First,
an ergodic theorem for translation automorphisms is established in theories with physical vacuum structure.
It allows for a natural definition of the pure-point and continuous subspaces, what opens the
door to more detailed spectral analysis: 
Apart from the counterpart of the absolutely continuous subspace, familiar from quantum mechanics, there appears
a new feature - the \emph{point-continuous} subspace - which carries information about the infrared structure
of a theory. It formally belongs to the continuous part, but it is finite dimensional in a large class of models.
In particular, it is shown that this subspace is trivial in all theories complying with a condition proposed in this work.
This new criterion, which identifies a novel class of particle detectors in the algebra of observables, entails
the existence of particles if the theory admits a stress-energy tensor. These results allow for a fresh look at the
problem of asymptotic completeness in quantum field theory from the model-independent perspective.

Formulation of natural assumptions for the ergodic theorem mentioned above 
is the subject of the second part of this work. For this purpose two new phase space conditions
are proposed, inspired by the physical behavior of coincidence measurements and
by the concept of additivity of energy over isolated subregions, respectively.
Among the consequences of these criteria, we obtain the uniqueness of the energetically
accessible vacuum states and convergence of physical states to these vacua under large timelike
translations (i.e. relaxation to the vacuum). \newpage

\section{Particle Content in Quantum Mechanics. Spectrum of Hamiltonian} \label{QM}
In order to emphasize the relevance of detailed spectral concepts to the problem of particle interpretation,
let us consider briefly the familiar case of two-body quantum mechanical scattering:
Here the central object is the (relative motion) Hamiltonian $H=H_0+V$
which consists of the free part $H_0$ and the interaction potential $V$. 
The Hilbert space $\hil$ can be decomposed into the orthogonal subspaces
\beq
\hil=\hil_{\spp}\oplus\hil_{\ac}\oplus\hil_{\sic},\label{hdecomposition}
\eeq
corresponding to the decomposition of the spectral measure of $H$ into the pure-point, absolutely 
continuous and singular-continuous parts. By comparing the free dynamics to the interacting dynamics
at asymptotic times, one obtains, under suitable assumptions on $V$, the wave operators
\beq
W^{\pm}=\te{s-}\!\!\!\!\lim_{t\to\pm\infty}e^{iHt}e^{-iH_0t}.
\eeq
One says that the theory has a complete particle interpretation,
if $\textrm{Ran}\, W^{\pm}=\hil_{\ac}$ and $\hil_{\sic}=\{0\}$. Under these conditions  every state 
in the Hilbert space can be uniquely decomposed into the bound states and the scattering states of the Hamiltonian. The first
proof of asymptotic completeness in the above framework, for a certain class of short-range potentials, is due to Enss \cite{En78}. The argument was later generalized by Sigal and Soffer \cite{SiSo87}, Graf \cite{Gr90} and Derezi\'nski
\cite{De93} to  many body systems interacting with short-range or long-range forces and it forms today a part
of standard textbook material~\cite{DG}.

\section{Wigner's Particle Concept and its Limitations}\label{Wigner}
To understand the limitations of the above approach to the problem of asymptotic completeness, let us 
now describe its implementation in QFT. The general framework, based on the algebra of observables
$\mfa$, is explained in detail in Section~\ref{AQFT}. For the purpose of the present discussion we assume in addition
that the action of the whole Poincar\'e group $\cpo$ is unitarily implemented by a strongly continuous representation $U$
acting on the Hilbert space $\hil$ and that there exists a unique (up to phase) vacuum vector $\vac\in\hil$, invariant under the action of $U$. Since the joint spectrum of the generators of translations $(H,\vec{P})$ is contained in the closed forward light-cone $\cone$,  one can define the mass operator $M=\sqrt{H^2-\vec{P}^2}$.

Following the standard procedure of Wigner \cite{Wi39}, one  unravels the particle content of a theory as follows: First, one finds all the (non-simple) eigenvalues $m$ of the mass operator $M$.
Next, one decomposes the corresponding eigenspaces into subspaces $\hil^{[m,s]}$ which carry irreducible sub-representations of $U$ characterized by a mass $m$ and a spin $s$. Finally, one forms the direct sum of all such subspaces $\hil_{\spa}:=\bigoplus \hil^{[m,s]}$ which is called the single-particle subspace of the theory. 
Scattering theory for massive Wigner particles is well understood by the work of Lehmann, Symanzik and Zimmermann 
\cite{LSZ55} on the one hand and by Haag, Ruelle, Hepp and Herbst \cite{Ha58, Ru62, He65, Her71} on the other.
The situation is less clear in theories with long range forces. There collision theory is under complete control only for massless particles,  by the work of Buchholz \cite{Bu75, Bu77}.
The case of massive particles accompanied by massless excitations was treated by the present author in \cite{Dy05} under
a stability assumption introduced by Herbst \cite{Her71}. The argument  is presented in Appendix~\ref{Haag-Ruelle-Appendix} of this Thesis.

The approach of Haag and Ruelle resembles in many respects the quantum-mechanical setting of the previous section:
One compares the interacting dynamics governed by the Hamiltonian $H$ to the dynamics
of free field theory of mass $m$, for any (non-simple) eigenvalue of the mass operator. There follows the existence of wave operators $W^{\pm}$ which are isometries from
the Fock space over the single-particle space $\Ga(\hil_{\spa})$ to the physical Hilbert space $\hil$
\beq
W^{\pm}:\Ga(\hil_{\spa})\to\hil.
\eeq
Thereby, for every configuration of incoming or outgoing particles
we can find the corresponding vector in  $\hil$. If the wave-operators are invertible, i.e. every vector in the
physical Hilbert space can be interpreted in terms of configurations of incoming and outgoing particles, then
we say that the theory is asymptotically complete in the Wigner sense.

The only known class of interacting theories which satisfy this property are the 
two-dimensional models with factorizing $S$-matrices recently constructed by  Lechner \cite{Le08}.
In particular, no asymptotically complete model exhibiting particle production is known to date.
In the thoroughly studied  $\la(\phi^4)_2$ theory  only states of energy smaller than $3m-\eps$ \cite{GJS73,SZ76}
or from the interval $[3m+\eps,4m-\eps]$ \cite{CD82}, where $\eps\to 0$ with $\la\to 0$, have been
shown to have particle interpretation. More importantly,
the above variant of asymptotic completeness is bound to fail in many physically
relevant situations: Suppose that a pair of charged particles is produced in a collision of neutral particles.
Since the masses of the charged particles are not visible in the energy-momentum spectrum of the vacuum sector, the
vector $\Psi\in\hil$, which corresponds to this process, cannot be interpreted in terms of the Fock space $\Ga(\hil_{\spa})$.
In the case of massive particles and (string-)localized charges one could try to avoid this difficulty by adjoining the charged sectors, determining the masses of charged particles from the extended energy-momentum spectrum and studying 
suitably redefined wave operators \cite{BF82}. This strategy fails, however, for  electrically charged particles, whose
masses are not eigenvalues of $M$, even in the charged sector, due to the presence of  Gauss' Law \cite{Bu86}. In this case the Wigner concept of a particle does not apply and the approach of Haag and Ruelle is invalidated from the very beginning. The presence
of such \emph{infra-particles} \cite{Sch63} is the main motivation for the search for a more general definition of a particle.

\section{Beyond Wigner's Particle Concept. Arveson Spectrum}\label{Arveson}

The first attempt at an intrinsic characterization of theories describing
particles was taken by Haag and Swieca \cite{HS65}. These authors proposed
a physically motivated phase space condition which should hold in all theories
with reasonable particle interpretation. In fact, a large class of models are known to-date to satisfy this criterion \cite{BDL90}
and a general argument,  which corroborates the heuristic reasoning of Haag and Swieca,
 was found by Bros \cite{Br03} in the two-particle situation.
Moreover,  it was shown by Enss \cite{En75} that
the Wigner definition of a (massive) particle is equivalent to its geometric characterization 
as a state which is singly-localized at all times in theories satisfying this phase space condition.

However, as this phase space criterion holds also 
in some models which do not have particle interpretation,
it  is too weak to meet the original goal of Haag and Swieca. 
A number of more stringent restrictions on the phase space
structure of a theory, formulated in terms of compactness and nuclearity conditions, can be found in the 
existing literature \cite{BP90}. They proved very useful in the structural analysis of quantum field theories \cite{BWi86,BJ89,Bu96,Bo05.1,Bo05.2} and in the construction of interacting models \cite{Le08},
but have not offered, as yet, many new insights  into the particle aspects of the theory. Recently a novel phase space condition,
related to additivity of energy over isolated subregions, was proposed by the present author \cite{Dy08.1,Dy08.2}. Among 
other physical consequences, it entails relaxation of any state $\om$ of bounded energy to the vacuum state under large timelike 
translations, i.e.
\beq
\lim_{t\to\infty}\om\big(\al_t (C)\big)=(\vac|C\vac),\quad\quad  C\in\mfa, \label{relaxation}
\eeq
where $\al_{t,\vx}( \, \cdot \, )=U(t,\vx) \, \cdot \, U(t,\vx)^*$ are the translation automorphisms.
The argument, which is given in Chapter~\ref{chapter-vacuum} of the present work, does not require the assumption of asymptotic completeness in the Wigner sense which was used by Araki and Haag \cite{AH67} 
in the first proof of relaxation to the vacuum. 
While relation~(\ref{relaxation}) does not carry information about the particle content of a theory, we recall that in \cite{AH67} 
the authors derived an asymptotic expansion of the function 
$t\to\om\big(\al_t (C)\big)$ as $t\to\infty$  and  demonstrated that the higher-order terms are directly 
related to the asymptotic particle density. A large  part of this analysis was generalized beyond the
framework of Wigner particles by Buchholz, Porrmann and Stein \cite{BPS91}  
leading to a generalized concept of a particle, encompassing also the case of infra-particles.
The remaining part of this section is devoted to a brief sketch and discussion of these developments.
(See \cite{MS85, Sp04, CFP07,He07} for other approaches to the infrared problem).

In order to compensate for the dispersive effects, encoded in equation (\ref{relaxation}),
one paves the whole space with observables and sums up the results. This amounts to studying
the time evolution of the integrals $\int d^sx\, \om(\al_{t,\vx}(C))$  which, however,  make sense only for suitably chosen  $C\in\mfa$.
In order to introduce a class of admissible particle detectors, we need the mathematical concept 
of the Arveson spectrum  \cite{Ar82, Pe}. Postponing the formal definition to Section~\ref{AQFT},
we only recall here that the (local) Arveson spectrum of an element $B\in\mfa$ w.r.t. the group of translation automorphisms $\real^{s+1}\ni x\to\al_x$, denoted by $\Sp_{B}(\al_{\real^{s+1}})$, coincides with the 
energy-momentum transfer of the observable~$B$. In fact, let $\De\to P(\De)$ be
the spectral measure of the energy-momentum operators $(H,\vec{P})$, defined on Borel sets $\De\subset\real^{s+1}$.
Then there holds
\beq
BP(\De)\hil\subset P(\De+\Sp_{B}(\al_{\real^{s+1}}))\hil.
\eeq
In view of this relation we say that an operator $B\in\mfa$ is energy-decreasing if $\Sp_{B}(\al_{\real^{s+1}})\cap\cone=\emptyset$.
It is a crucial result due to Buchholz \cite{Bu90} that for any $B\in\mfa$ which is energy-decreasing 
and almost local\footnote{See Section \ref{AQFT} for the definition of this concept.}, 
and any compact set $\De\subset\real^{s+1}$, 
one can define the integrals
\beq
P(\De)\int d^sx \ \al_{\vx}(B^*B)\, P(\De)  \label{Kintegral}
\eeq
as bounded operators on $\hil$. Having compensated for the dispersive effects,
one extracts information about the particle content of the theory:
For any state $\om$ of bounded energy one studies the behavior of the following expressions as $t\to\infty$
\beq
\si^{(t)}_{\om}(B^*B)=\int d^sx\, \om\big(\al_{t,\vx}(B^*B)\big).
\eeq
It was shown by Porrmann \cite{Po04.1, Po04.2} that the resulting \emph{asymptotic functionals}~$\si^{(+)}_{\om}$, defined on a certain algebra of admissible observables,
can be decomposed into  pure functionals~$\si^{(+)}_{\la}$, the so called pure particle weights. More precisely, for
any $\om$ there exists a measure $d\mu$ on the space of labels $\la$ s.t.
\beq
\si^{(+)}_{\om}=\int d\mu(\la)\si^{(+)}_{\la}. \label{pure} 
\eeq
To each label $\la$ there corresponds a sharp four-momentum $p_{\la}$ and a label
$\ga_{\la}$ which carries information about the internal degrees of freedom, like spin and
charge. Therefore 
we can interpret the weights $\si^{(+)}_{\la}$ as plane wave configurations of the 
particles appearing in the theory. A general algorithm for computation of collision cross-sections of these particles, which does not rely on the existence of charged fields, was developed in \cite{BPS91} and tested in asymptotically complete theories of Wigner particles by Stein \cite{St89}. Also the fundamental problem of the existence of non-trivial asymptotic functionals, which rephrases the question posed by Haag and Swieca \cite{HS65}, has been settled to date only in this restrictive framework.

Two heuristic arguments of more general nature, addressing the question of existence of particles, were proposed by 
Buchholz: The first one combines phase space properties of a theory and the time-slice axiom \cite{Bu87}. The second  relies on the existence of the stress-energy tensor \cite{Bu94}. In Section~\ref{triviality-of-Apc} of this Thesis  we present a rigorous proof of the existence of non-trivial asymptotic functionals which is based on this latter idea and does not rely directly on the Wigner concept of a particle.

Another important issue is convergence of the asymptotic functional approximants $\si^{(t)}_{\om}$ as $t\to\infty$. Again,
a proof is known only in the context of theories which are asymptotically complete in the Wigner sense \cite{AH67}.
We do not present a solution in this work, but a promising strategy is discussed in Chapter~\ref{chapter:conclusions}.
The simpler problem of relaxation to the vacuum, (cf. relation~(\ref{relaxation}) above), which is settled in this Thesis under physically meaningful conditions, should provide a guidance towards a more general proof of convergence of $\si^{(t)}_{\om}$.

Very little is known about asymptotic completeness in this general framework. A possible
formulation was proposed in \cite{Bu94}: It should be possible to determine the energy and momentum
of any physical state $\om$, knowing its particle content from relation (\ref{pure}). More precisely, there should hold
\beq
\om(P)=\int d\mu(\la) p_{\la}, \label{asymptotic-completeness-relation}
\eeq
where the four-momenta $p_{\la}$ label the pure particle weights $\si^{(+)}_{\la}$ and a similar relation
should hold for other conserved quantities which characterize particles, like spin, charges etc.
It is plausible that relation (\ref{asymptotic-completeness-relation}) holds in models admitting a 
stress-energy tensor, but we are
far from an actual proof. Additional assumptions, which may be useful to settle this issue, are 
discussed in Section~\ref{triviality-of-Apc}.

It is evident from the above discussion that the problem of particle interpretation in quantum field
theory is rather poorly understood in comparison to the quantum-mechanical case considered in Section~\ref{QM}.
We see the origin of this disproportion in the absence of adequate mathematical structures on the side of QFT:
Here the natural language for the description of particle aspects 
is that of the group of translation automorphisms $\real^{s+1}\ni x\to\al_{x}$ acting on the $C^*$-algebra
$\mfa$. The Hamiltonian, central for the quantum-mechanical scattering,
is replaced by the  generator of  time translation automorphisms, whose spectrum coincides with the (global) Arveson spectrum of $\al_t$ \cite{Ev76}, defined by formula~(\ref{global-Arveson-spectrum}) below. 
After the spectral theory of automorphisms  was systematized  by Arveson \cite{Ar82},
it became clear that several important notions, familiar from the quantum-mechanical setting, do not have counterparts in this more general context. These include the concept of spectral measure and measure classes consisting of pure-point, Lebesgue absolutely continuous and singular continuous parts, and the corresponding
decomposition~(\ref{hdecomposition}) of the Hilbert space into spectral subspaces.
As we have seen in Section \ref{QM}, these notions are crucial for
the formulation and resolution of the problem of asymptotic completeness in quantum 
mechanics. It is therefore quite certain that the lack of counterparts on the side of automorphism groups impedes the study of  the particle aspects in QFT.
Various steps towards the development
of such more detailed theory of the Arveson spectrum, which can be found in the literature, are discussed
in the next section.

\section{Detailed Theory of Arveson Spectrum in Literature}\label{Literature}
For a group of unitaries $\real^s\ni\vx\to U(\vx)$, acting on a Hilbert space, there holds $U(\vx)\Psi=\int e^{-i\vep\vx} dP(\vep)\Psi$, 
where $\Psi\in\hil$ and $dP$ is the spectral measure. Hence the natural counterpart of the spectral measure in the Arveson theory
are the Fourier transforms of the functions $\real^s\ni\vx\to\al_{\vx}(A)$, where $A\in\mfa$. As the support of the resulting distribution $\vep\to\wt{A}(\vep)$ coincides with the Arveson spectrum of $A$, a more detailed spectral theory should describe also its regularity properties.

Several results in this direction can be found in the existing literature. 
On the physics side, the distributions $\vep\to\om(\wt{A}(\vep))$, where $A$ is a local operator and $\om$ a state of bounded energy, were studied by Buchholz \cite{Bu90}. It was shown that in any local, relativistic QFT they  coincide with square-integrable functions apart from a possible singularity at zero.
This result, stated precisely in Theorem~\ref{theorem-2mollifier} below,  was instrumental for the development of the Haag-Ruelle scattering theory in the presence of massless particles \cite{Dy05}, treated in Appendix~\ref{Haag-Ruelle-Appendix} of this work. It also provides a basis for the spectral decomposition of the algebra of local observables, which we construct in 
Chapter~\ref{chapter:spectral}.

Such Fourier coefficients $\wt{A}(\vep)$  appear also
as a tool in the mathematical literature related to the Rieffel project of extending the notions of proper action and orbit space from the setting of group actions on locally compact spaces to the context of $C^*$-dynamical systems $(\mfa, \real^s,\al)$ \cite{Ri90, Me01}. A number of properties of $\wt{A}(\vep)$, familiar from classical harmonic analysis, like the Fourier inversion formula, have been proven by Exel in \cite{Ex99,Ex00}. Unfortunately, we cannot use these results here as they rely on the assumption of unconditional integrability, which requires that the net
\beq
\big\{\int_K d^sx \ \al_{\vx}(A)\big\}_{K},
\eeq
indexed by compact subsets $K\subset\real^s$, converges in the norm topology of $\mfa$ as $K\nearrow \real^s$. This is
not quite compatible with QFT, where we interpret $A\in\mfa$ as a particle detector, since for any compact region $K$ 
one can prepare a physical state which gives large measurement responses in the complement of $K$. In fact, the
integral~(\ref{Kintegral}) was only shown to converge in the strong operator topology of $B(\hil)$ \cite{Bu90}.

The above two approaches are distinguished by the fact that they exploit the algebraic structure of $\mfa$.
The more general framework  of groups of isometries $\real\ni t\to \al_t$ acting on a Banach space $\mfa$ attracted 
more attention. This direction of research relies on the equality of the (global) Arveson spectrum, given by
\beq
\Sp\,\al_{\real}=\ov{\bigcup_{A\in\mfa}\Sp_{A}\al_{\real} }, \label{global-Arveson-spectrum}
\eeq
and the operator theoretic spectrum of the generator $D=\fr{1}{i}\fr{d}{dt}\al_t|_{t=0}$ which was established  by Evans \cite{Ev76}
and independently by Longo \cite{Lo77}. Related results, known as the spectral mapping theorems, were obtained for the global Arveson spectrum \cite{Co73,DLZ81,Ar82,Ne92} and for the local one \cite{Ne98,LN}, (see \cite{Ar82,LN} for reviews).
As for the more detailed spectral theory, the point spectrum of  $D$ is the best understood one. 
A thorough analysis of the
discrete spectrum, including the mini-max principle, existence and uniqueness theorems
for the ground state and the Rayleigh-Ritz technique, was performed by Jorgensen \cite{Jo82}. Inspired by the Stone formula \cite{RS1}, this  author proposes a suitable limit of resolvents of the generator $D$ to be the counterpart for the spectral measure. This technique has applications also beyond the point spectrum \cite{Jo92}:  Using classical
results  of Fourier analysis, Jorgensen derives integrability and decay properties  of the functions $t\to\om(\al_tA)$, where $A\in \mfa$, $\om\in \mfa^*$, from regularity assumptions on these resolvents. However, such assumptions do not have a clear physical meaning and are difficult to verify in the case of translation automorphisms in QFT, so we do not pursue this approach here.

The subject of Cauchy problems in the theory of differential equations \cite{AB97} includes an interesting line of developments which bears some similarity to the present work: Assume that the spectrum of $D$ is
countable. Then, under some additional conditions,  the eigenvectors of $D$  span a norm dense subspace in $\mfa$ 
\cite{Ba78, Hu99}. In this case the representation $\al$ is called almost periodic, as all its orbits $t\to\al_t(A)$, $A\in\mfa$, belong to the class $AP(\real,\mfa)$ of almost periodic functions with values in $\mfa$. More abstractly,
$u\in AP(\real,\mfa)$ if and only if the set $S(u)=\{ \,u(\cdot+t)\,|\, t\in\real \,\}$  of the translates of $u$
is relatively norm compact in the Banach space $BUC(\real,\mfa)$ of bounded, uniformly continuous  functions
\cite{AB97,Ph93}. Replacing in the above definitions the real line $\real$ with the positive half-line $\real_+$, we obtain the set $S(u)_+$ and the class $AAP(\real_+,\mfa)$ of asymptotically almost periodic  functions which is of interest
in the theory of $C_0$-semigroups $\real_+\ni t\to\al_t$ acting on Banach spaces.
This class has the decomposition \cite{RS90, AB99, BPh90, AB88}
\beq
AAP(\real_+,\mfa)=AP(\real,\mfa)|_{\real_+}\oplus C_0(\real_+,\mfa), \label{AAP}
\eeq
where  $C_0(\real_+,\mfa)$ is the space of continuous, functions which tend to zero in norm as $t\to\infty$.
A larger class $W(\real_+,\mfa)$ of Eberlain weakly almost periodic functions is characterized by the condition that the set $S(u)_+$ is relatively \emph{weakly} compact
\cite{RS90, RS92}. The weak counterpart of $C_0(\real_+,\mfa)$ is the space $W_0(\real_+,\mfa)$ of such functions $u$ from $W(\real_+,\mfa)$, that zero belongs to the weak closure of $S(u)_+$. Again, we have a decomposition \cite{RS90}
\beq
W(\real_+,\mfa)=AP(\real,\mfa)|_{\real_+}\oplus W_0(\real_+,\mfa). \label{W}
\eeq 
All the spaces introduced above  consist of orbits of the  semigroup of translations given by 
$(\al_t u)(t^\prime)=u(t+t^\prime)$ acting on $BUC(\real,\mfa)$. Therefore, relations (\ref{AAP}), (\ref{W}) can 
be interpreted as spectral decompositions of the corresponding generator $D$.
However, in contrast to the Hilbert space case (cf. relation~(\ref{hdecomposition})), these 
decompositions do not involve the whole underlying space $BUC(\real,\mfa)$, but only cover  some small neighborhoods
of the pure-point  part $AP(\real,\mfa)|_{\real_+}$. Moreover, they consist only of the pure-point subspace and its complement which can be understood as  a candidate for the continuous subspace. There does not seem to appear any further decomposition of these continuous subspaces in the literature. Finally, and most importantly, this approach is appropriate for the study of semigroups of operators  rather than groups of isometries we are interested in here. In particular, the space $C_0(\real_+,\mfa)$ does not contain orbits of any group of isometries.

\section{Overview of this Work}

In Chapter~\ref{chapter:introduction}  we gave an overview of scattering theory in quantum mechanics and quantum field theory placing emphasis on the role of spectral analysis. 
The remaining part of this chapter treats 
the framework of Algebraic Quantum Field Theory as well as definitions and results which are particularly useful
in the later part of this Thesis.

In Chapter~\ref{chapter:spectral}  we motivate and develop a detailed spectral theory of space translation automorphisms 
$\real^s\ni\vx\to\al_{\vx}$ acting on the algebra of local observables $\hmfa$. We obtain the decomposition
\beq
\hmfa=\hmfa_{\spp}\oplus\hmfa_{\pc}\oplus\hmfa_{\ac}. 
\eeq
Apart from the pure-point and absolutely
continuous parts, familiar from the Hilbert space context, there appears a new subspace
which we call point-continuous. It formally belongs to the continuous part of the spectrum, but it is finite
dimensional in a large family of models, complying with Condition~$\A$ stated in Section~\ref{space-translations-QFT}. This subspace carries the information about the infrared behavior of the theory which can be assessed with the help of a new quantity which we call the \emph{infrared order} of an operator. 
We provide examples of theories with non-trivial and trivial point-continuous subspaces and compute the infrared orders
of their elements. Triviality of the point-continuous subspace is expected to hold in purely
massive theories and we formulate a general condition~$\B$, stated in Section~\ref{triviality-of-Apc}, which entails this spectral property. A quantitative variant of this new criterion guarantees the existence of non-trivial asymptotic functionals (describing particles) in theories admitting a stress-energy tensor.

In Chapter~\ref{chapter-vacuum} we study the uniqueness of the decomposition of $\hmfa$ into the pure-point and continuous parts. A variant of ergodic theorem for translation automorphisms in QFT, stated as Theorem~\ref{QFT-ergodic-theorem} below, reduces this issue to the problem of uniqueness of the energetically accessible vacuum state. In Section~\ref{vacuum-structure} we introduce a phase space condition~$\Csq$, of algebraic nature, which entails this property and has a number of other physically interesting consequences: The  vacuum state is pure and appears, in particular,  as  a limit of physical states under large spacelike or timelike translations in Lorentz covariant theories. Moreover, it is approximated by states of increasingly sharp energy-momentum values, in accordance with the uncertainty principle. This new condition has a clear physical interpretation in terms of coincidence arrangements of local observables, but its consistency with the basic postulates has been verified only in the realm of massive theories. 
Therefore, in Section~\ref{Condition-Cnat-and-additivity} we introduce another phase space condition~$\Cnat$, involving only the Banach space structure of $\mfa$, which  
can be verified both in massive and massless models. 
We motivate this condition by the physical principle of additivity of energy over isolated regions and show that it has all the physical consequences mentioned above (apart from purity of the vacuum). We also demonstrate that it can be derived from an auxiliary nuclearity condition~$\Nnat$, introduced in Section~\ref{additivity-of-energy} which is better suited for the study of 
concrete examples.

In Chapter~\ref{chapter:conclusions}  we summarize our results  and consider some open problems from a new perspective. In particular, we discuss in detail the problem of convergence of the asymptotic functional approximants.

The main part of this Thesis is accompanied by six appendices: Appendix~\ref{Haag-Ruelle-Appendix}, which develops 
the Haag-Ruelle scattering theory in the presence of massless particles,  complements
our discussion of collision theory in Section~\ref{Wigner}. Appendix~\ref{Preliminaries} summarizes 
the known results
on the phase space structure of scalar free field theory. It provides the basis for the material presented
in the remaining appendices. In Appendices~\ref{Condition-L2} and \ref{Condition-L1} we establish Conditions~$\A$
and $\B$, respectively, in models of non-interacting particles. Appendices~\ref{Condition-Nnat} and \ref{Condition-Csq}
are devoted to verification of Conditions~$\Nnat$ and $\Csq$ in such models. Notably, Conditions~$\A$ and $\Nnat$
hold both in the massive and the massless case.

The first and last section of Chapter~\ref{chapter-vacuum} was published in \cite{Dy08.1}. 
Appendix~\ref{Haag-Ruelle-Appendix} essentially coincides with \cite{Dy05}. The argument from Appendix~\ref{Condition-Nnat}
appeared (for the massless case) in \cite{Dy08.2}.

\section{Technical Background}\label{AQFT}
We adopt here the standard Haag-Kastler framework of Algebraic Quantum Field Theory \cite{Ha,Ar}. Let $\hil$\label{symbol-hil} be
an infinitely dimensional Hilbert  space and let $\mfa=\{\,\mfa(\mco)\subset B(\hil)\,|\, \mco\subset\real^{s+1} \}$ 
\label{symbol-loc}
be a net of local von Neumann algebras attached to open bounded regions $\mco$\label{symbol-mco} of the Minkowski spacetime. We denote
by $\hmfa=\bigcup_{\mco\subset\real^{s+1}}\mfa(\mco)$ \label{symbol-hmfa}
the $*$-algebra
of local observables and its norm closure $\mfa$ \label{symbol-mfa} 
(denoted by the same symbol as the net) is called the global $C^*$-algebra of observables. Finally, let $\al$ \label{symbol-al}
be a representation of the Poincar\'e group $\cpo=\real^{s+1}\rtimes L_+^{\uparrow}$\label{symbol-cpo} in the group of automorphisms of $\mfa$ s.t. for any $A\in\hmfa$ the function $\cpo\ni(x,\La)\to\al_{(x,\La)}(A)$ is continuous in the strong operator topology of $B(\hil)$\label{symbol-Bhil}.
The triple $(\mfa,\al,\hil)$ is called an (algebraic) quantum field theory if it satisfies the following conditions:

\begin{enumerate}
\item[1.] \bf Isotony: \rm If $\mco_1\subset\mco_2$ then $\mfa(\mco_1)\subset \mfa(\mco_2)$.
\item[2.] \bf Locality: \rm If $\mco_1$ and $\mco_2$ are spacelike separated, then 
$\mfa(\mco_1)\subset\mfa(\mco_2)^\prime$, where the prime denotes the commutant in $B(\hil)$.
\item[3.] \bf Irreducibility: \rm 
The global $C^*$-algebra $\mfa$ acts irreducibly on the Hilbert space $\hil$.
\item[4.] \bf Covariance: \rm 
The representation $\al$ of $\cpo$ acts geometrically on the net i.e.
\beq
\al_{(x,\La)}\mfa(\mco)=\mfa(\La\mco+x).
\eeq
\item[5.] \bf Spectrum condition: \rm There exists a strongly continuous unitary representation of translations
$\real^{s+1}\ni x\to U(x)$\label{symbol-U} acting on the Hilbert space $\hil$ which implements the translation automorphisms
i.e. $\al_{x}( \, \cdot \, )=U(x) \, \cdot \, U(x)^*$. The joint spectrum of the infinitesimal generators of 
translations $(H,P_1,\ldots,P_s)$, corresponding to this unitary representation, is contained in the closed future light
cone $\cone=\{\,(p^0,\vep)\in\real^{s+1}\,|\, p^0\geq |\vep|\,\}$.\label{symbol-H}\label{symbol-veP}
\end{enumerate}
We will occasionally consider sub-theories of a given theory: by a sub-theory of $(\mfa,\al,\hil)$ we mean
a triple $(\mathfrak{B},\al,\hil)$ satisfying the above assumptions and  s.t. $\mathfrak{B}(\mco)\subset\mfa(\mco)$
for any open, bounded region $\mco$. At times we use the  notation $A(x)=\al_x(A)$, $\fun_x=\al^*_x\fun$ for translated observables $A\in\mfa$ and functionals $\fun\in\mfa^*$. For any Banach space $X$, we denote by $X_1$ the unit ball in $X$.
Other definitions and results which are central in the later part of this Thesis are listed below:
\begin{itemize}

\item \bf Vectors of bounded energy. \rm  We denote by  $P_E$\label{symbol-PE} be the spectral projection of the Hamiltonian $H$ on vectors of energy bounded by $E$ and define the dense subspace $\BE=\bigcup_{\su{ \\ \\ \!\!\!\!\!\!\!\!\!\!\!\! E \geq 0 }} P_E\hil$ \label{symbol-BE}
of vectors of bounded energy.

\item \bf Functionals of bounded energy. \rm It follows from the irreducibility property  above that the 
space of normal states on $\mfa$ coincides with $B(\hil)_*$ which can be identified with the space $\trace$ \label{symbol-trace}
of trace-class operators on $\hil$.
We introduce the subspace $\traceE=P_E\trace P_E$ \label{symbol-traceE} of normal functionals of restricted
energy. We denote the cone of positive functionals from $\traceE$ by $\traceE^+$ \label{symbol-traceE+} and the subset of states
from $\traceE^+$  by  $S_E$. \label{symbol-SE} The states from $\mfa^*$ which belong to the weak$^*$ closure of
$\traceEB$ for some $E\geq 0$ will be called the energetically accessible states.

It is a well known fact that any normal, self-adjoint functional on a von Neumann algebra can be
expressed as a difference of two normal, positive functionals which are mutually orthogonal \cite{Sa}.
It follows that any $\fun\in\traceEB$ can be decomposed as 
\beq
\fun=\funr^+-\funr^-+i(\funi^+-\funi^-), \label{decomp}
\eeq
where $\funr^\pm$, $\funi^\pm$ are elements of $\traceEB^+$.

\item \bf Almost local observables. \rm  A typical region of spacetime to be used in the sequel is the double cone 
$\mco(r)=\{\,(x^0,\vx)\in\real^{s+1}\,|\, |x^0|+|\vx|<r\,\}$, $r>0$,  \label{symbol-mco(r)} whose
base is the $s$-dimensional ball $\mco_r=\{\,\vx\in\real^s\,|\, |\vx|<r\,\}$ \label{symbol-mcor} of radius $r$ centered at the origin. We say that $A\in\mfa$ is almost local, if there exists a net of local operators $\{\, A_r\in\mfa(\mco(r))\,|\, r>0\,\}$
such that
\beq
\lim_{r\to\infty} r^k\|A-A_r\|=0
\eeq
for any $k\in\nat_0$.

\item\bf Arveson spectrum w.r.t. spacetime translations. \rm Given an observable $B\in\mfa$, its (local) Arveson spectrum w.r.t. the group of translation automorphisms $\real^{s+1}\ni x\to\al_x$ is denoted by $\Sp_{B}(\al_{\real^{s+1}})$ \label{symbol-Arveson} and defined as the support of the operator-valued distribution
\beq
\wt{B}(p)=\fr{1}{(2\pi)^{\fr{s+1}{2}}}\int d^{s+1}\!x \ e^{-ip\cdot x}\al_{x}(B). \label{Fourier}
\eeq 
We say that $B$ is \emph{energy-decreasing} if $\Sp_{B}(\al_{\real^{s+1}})\cap\cone=\emptyset$. There holds the following
important result:
\bet\cite{Bu90}\label{harmonic} Let $B\in\mfa$ be energy-decreasing and almost local.  Let $X\subset\real^{s+1}$
be some subspace spanned by spacelike vectors and $dx$ be a translationally invariant measure on $X$.
Then, for any compact subset $K\subset X$ and any $E\geq 0$, there holds the bound
\beq
\|P_E\int_K dx\, \al_{x}(B^*B) P_E\|\leq c_E,
\eeq
where the constant $c_E$ is independent of $K$.
\eet
The statement holds in particular for $X=\real^s$ or $X=\{\,\la\e\,|\, \la\in\real\,\}$, where $\e$ is some
spacelike unit vector.


\item \bf Arveson spectrum w.r.t. space translations. \rm Given an observable $B\in\mfa$, its  Arveson spectrum w.r.t. the group of space translation automorphisms $\real^{s}\ni \vx\to\al_{\vx}$ is denoted by $\Sp_{A}(\al_{\real^{s}})$ and defined as the support of the operator-valued distribution
\beq
\wt{A}(\vep)=\fr{1}{(2\pi)^{\fr{s}{2}}}\int d^{s}x \ e^{i\vep\vx}\al_{\vx}(A). \label{Fourier1}
\eeq 
The following result is central for our investigations in Chapter \ref{chapter:spectral}: 
\bet\label{theorem-2mollifier}\cite{Bu90} For any $E\geq 0$, $A\in\hmfa$ and $\eps>0$ there holds the bound
\beq
\sup_{\fun\in \traceEB}\int d^sp\,|\vep|^{s+1+\eps}|\fun(\wt{A}(\vep)|^2<\infty. \label{2mollifier}
\eeq
\eet

\item  \bf Vacuum states. \rm We say that $\om_0\in\mfa^*$ \label{symbol-om0} is a vacuum state if it satisfies the following conditions
\begin{enumerate}
\item[(a)] $\al_{x}^*\om_0=\om_0$ for any $x\in\real^{s+1}$.
\item[(b)] $\om_0$ is locally normal i.e. $\om_0|_{\mfa(\mco)}$ belongs to $\mfa(\mco)_*$ for any open,
bounded region~$\mco$.
\item[(c)] In the GNS-representation of $\mfa$ induced by $\om_0$ there holds the spectrum condition
(given by property~5 above).
\end{enumerate}
We note that the local normality property (b) and the fact that local observables form a norm dense
subspace in $\mfa$ ensure continuity of the functions $\real^{s+1}\ni x\to \om_0(AB(x))$, for any $A,B\in\mfa$. 
This entails the existence of the generators of translations $(H_{\om_0},\veP_{\om_0})$ in the GNS-representation 
induced by $\om_0$.

\end{itemize}

%% file: PhD-spectral3.tex
\chapter{Spectral Decomposition and Existence of Particles}\label{chapter:spectral}

The goal of this chapter is twofold: First, to formulate general conditions on the local net of observables
which imply that the particle content of a theory is non-trivial. More precisely, to 
assure that the limit points as $t\to\infty$ of the integrals
\beq
\int d^sx\,\om(\al_{t,\vx}(C)) \label{paving}
\eeq
are finite and  different from zero for some observables $C\in\mfa$ and physical states $\om$.
Following the heuristic reasoning from \cite{Bu94}, our strategy is to link the quasi-local observables
$C$ to the $(0,0)$-component of the stress-energy tensor $T^{00}$, whose integral over the whole space is a constant of motion.
The fact that $T^{00}$ is a local quantity motivates the second aim of this chapter, namely to understand 
the behavior of local observables under translations in space.

As a first orientation we consider in Section~\ref{QM2} the quantum mechanical case of space translations $\vx\to U(\vx)$ acting unitarily on a Hilbert space. With the help of the ergodic theorem we formulate the decomposition of the Hilbert space
into the respective pure-point and continuous subspaces in a way which facilitates its generalization to the Banach space
setting. Proceeding to the more detailed spectral analysis,
we show that the generic property of the functions $\vx\to(\Phi|U(\vx)\Psi)$, where the wavefunction of $\Psi$ is 
localized in space, is not integrability, required in (\ref{paving}), but the weaker property of square-integrability. 

In Section~\ref{space-translations-QFT} we turn our attention to the case of space translation automorphisms $\vx\to\al_{\vx}$
acting on the algebra of local observables $\hmfa$ in QFT. We show that in theories with decent vacuum structure, as
described by Condition~$\V$ stated below,
there holds a counterpart of the ergodic theorem. It offers a natural  decomposition of
$\hmfa$ into the pure-point and continuous parts. The square-integrability of the functions $\vx\to\om(\al_{\vx}(A))$,
where $A$ is a local operator, provides a  meaningful definition of the absolutely continuous subspace $\hmfa_{\ac}$ for 
space translations in QFT. In the present case, however, square-integrability may fail, due to singularities 
at zero momentum transfer, exhibited by some infrared-sensitive observables. A new concept of the infrared order of an
operator is introduced to quantify this effect. Such operators typically span a 
finite-dimensional subspace which we call the point-continuous subspace $\hmfa_{\pc}$. It does not have a quantum mechanical counterpart and carries information about the infrared structure of a theory. Thus we arrive at the decomposition
\beq
\hmfa=\hmfa_{\spp}\oplus\hmfa_{\pc}\oplus\hmfa_{\ac}
\eeq
which is studied in examples in Subsection~\ref{Examples}. The more technical part of this discussion is given
in Appendix~\ref{Condition-L2}.

With integrals~(\ref{paving}) in our main focus, it is certainly of interest to identify
theories with regular infrared behavior. Therefore, in Section~\ref{triviality-of-Apc}, 
we formulate a condition which assures that the point-continuous subspace is trivial. At the same time it
identifies a class of particle detectors which are sufficiently 
close to the stress energy tensor, to conclude that the particle content is non-trivial.
These results demonstrate the interplay between the spectral aspects and the particle structures in QFT.
(The condition is verified in a model of non-interacting massive particles in Appendix~\ref{Condition-L1}).

\section{Space Translations in Quantum Mechanics}\label{QM2}
In this section we revisit the well known spectral theory of strongly continuous unitary 
representations of translations $\real^s\ni\vx\to U(\vx)$ acting on a Hilbert space $\hil$.
Our goal is to motivate its generalization to the context
of translation automorphisms acting on the algebra of observables, which we undertake
in the next section.

The pure-point subspace is spanned by the joint eigenvectors of the generators of translations~i.e.
\beq
\hil_{\spp}=\Span\{\,\Psi\in\hil \,|\, U(\vx)\Psi=e^{-i\vep\vx}\Psi, \te{ for some } \vep\in\real^s 
\te{ and all } \vx\in\real^s \,\}.
\eeq
The continuous subspace $\hil_{\scc}$ is simply the orthogonal complement of $\hil_{\spp}$
in $\hil$. However, it is convenient for our purposes to have a characterization which does not rely 
directly on the concept of orthogonality, specific to Hilbert spaces. Therefore,
we recall that due to the ergodic theorem \cite{RS1}, the spectral projection $P_{\{\vep\}}$
on a point $\vep$ in the joint spectrum of the generators of translations 
is given by
\beq
P_{\{\vep\}}=\te{s-}\!\!\!\lim_{\VV\nearrow\real^s}\fr{1}{|\VV|}\int_\VV d^sx\,U(\vx)e^{i\vep\vx},
\eeq
where the limit is taken w.r.t. to an increasing net of compact sets $\VV\subset\real^s$ ordered by inclusion.
Now the continuous subspace is determined as follows
\beq
\hil_{\scc}=\bigcap_{\vep\in\real^s}\ker\,P_{\{\vep\}}. \label{QM-continuous}
\eeq
This subspace can  be further decomposed into the absolutely continuous and singular
continuous parts $\hil_{\scc}=\hil_{\ac}\oplus\hil_{\sic}$ and we note the following simple
criterion which implies that a vector belongs to the absolutely continuous subspace.
\bep\label{QM-square} Suppose that the function $\real^s\ni\vx\to(\Psi|U(\vx)\Psi)$ 
is square-integrable for some $\Psi\in\hil$. Then $\Psi\in\hil_{\ac}$.
\eep
\nin This statement follows from the Plancherel theorem and the fact that any square-integrable function
is locally integrable. 

In the next section we  propose a similar square-integrability condition as a characterization
of the 'absolutely continuous' subspace of the algebra of observables. In order to motivate such a restrictive
definition, we consider now briefly a concrete unitary representation of translations $\real^s\ni\vx\to U(\vx)$ 
acting on the Hilbert space  $L^2(\real^s, d^sx)$ as follows
\beq
(U(\vx)\Psi)(\vy)=\Psi(\vy-\vx),\quad \Psi\in L^2(\real^s, d^sx).
\eeq
In this setting a natural analogue of a local observable is a vector $\Psi\in L^2(\real^s, d^sx)$ 
which has (a representative with) a support in an open bounded region. In the next  lemma
we show that these vectors satisfy the  square-integrability condition stated in Proposition~\ref{QM-square}.
\bel\label{QM-L1} Let $\Phi,\Psi\in L^2(\real^s, d^sx)$. Suppose that $\Psi$ has (a representative with) a support in an
open, bounded region $\KK\subset\real^s$. Then there holds
\beq
\int d^sx\, |(\Phi|U(\vx)\Psi)|^2\leq |\KK|\,\|\Phi\|^2\,\|\Psi\|^2, \label{square}
\eeq
where $|\KK|$ is the volume of the region $\KK$.
\eel
\proof Using the Plancherel theorem we obtain
\beq
\int d^sx\,|(\Phi|U(\vx)\Psi)|^2=(2\pi)^s\int d^sp\, |\wt{\Phi}(\vep)|^2\,|\wt{\Psi}(\vep)|^2.
\eeq
The support property of $\Psi$ gives
\beqa
|\wt{\Psi}(\vep)|^2\leq(2\pi)^{-s}\bigg( \int d^sx\,|\Psi(\vx)| \bigg)^2
\leq (2\pi)^{-s}|\KK|\,\|\Psi\|^2,
\eeqa 
what concludes the proof. \qed\\
More importantly, the square-integrability of the functions $\real^s\ni\vx\to (\Phi|U(\vx)\Psi)$, proven in the above lemma, 
cannot be improved to integrability  with any lower power. In fact:
\bel For any $0<k<2$ there exist $\Phi,\Psi\in L^2(\real^s, d^sx)$ s.t. $\Psi$ has (a representative with) a support
in an open, bounded region $\KK$  and
\beq
\int d^sx\, |(\Phi|U(\vx)\Psi)|^k=\infty. \label{kintegral}
\eeq
\eel
\proof  Let $\KK$ be the ball of radius $R>0$ centered at the
origin. Let $\chi\in C_0^{\infty}(\real^s)$ be a positive function, s.t. $\chi(\vx)=\chi(-\vx)$,  supported in this ball. 
We set for some $0<\de<s$
\beqa
\tilde{\Phi}(\vep)&:=&\fr{\tilde{\chi}(\vep) }{ |\vep|^{\fr{s-\de}{2}}},\\
\tilde{\Psi}(\vep)&:=&\tilde{\chi}(\vep). 
\eeqa
Making use of the fact that the Fourier transform of $\vep\to |\vep|^{-\fr{s-\de}{2}}$ equals
$\vx\to c_{\de} |\vx|^{-\fr{s+\de}{2}}>0$, where $c_{\de}=2^{\fr{\de}{2}}\fr{\Ga(\fr{s+\de}{4})}{\Ga(\fr{s-\de}{4}) }$,
we obtain
\beqa
(\Phi|U(\vx)\Psi)&=&(2\pi)^{-\fr{s}{2}}\int d^sy\fr{c_{\de}}{|\vy+\vx|^{\fr{s+\de}{2}}}\int d^sz\,\chi(\vz)\chi(\vz-\vy)\\
&\geq& (2\pi)^{-\fr{s}{2}}\fr{c_{\de}}{(|\vx|+2R)^{\fr{s+\de}{2}}}
\bigg(\int d^sy\,\chi(\vy)\bigg)^2.
\eeqa
Consequently, for $0<k<2$ and sufficiently small $\de>0$, 
integral (\ref{kintegral}) diverges. \qed\\
Summing up, the square-integrability of the functions
\beq
\real^s\ni\vx\to(\Phi|U(\vx)\Psi), \label{transition-amplitude}
\eeq
uniformly in $\Phi\in L^2(\real^s,d^sx)$, $\|\Phi\|\leq 1$, is the best possible generic feature of vectors $\Psi$, 
compactly supported in configuration space.  
Turning our attention to quantum field theory, we recall that such vectors are our analogues of
local observables $A\in\hmfa$. Furthermore,
vectors $\Phi$ correspond to  normal functionals  of bounded energy $\fun\in\traceE$ and the group of unitaries $\real^s\ni\vx\to U(\vx)$ is the 
counterpart of the group of translation automorphisms $\real^s\ni\vx\to\al_{\vx}$. Thereby  the transition amplitudes
(\ref{transition-amplitude}) between $\Phi$ and the translates of $\Psi$ provide a toy model for the expectation 
values of translates of local operators on states of bounded energy:
\beq
\real^s\ni\vx\to\fun(\al_{\vx}(A)).
\eeq
These functions are the main subject of the remaining part of this chapter.
\section{Space Translations in Quantum Field Theory. Spectral Decomposition}\label{space-translations-QFT}

In this section we construct the spectral decomposition
of the $*$-algebra of local operators 
$\hmfa=\bigcup_{\mco\subset\real^{s+1} }\mfa(\mco)$
into subspaces which differ in their behavior under translations in space. We first identify the pure-point
and the continuous subspace in Subsection~\ref{subsection-pp-c}. Next, in Subsection~\ref{subsection-ac-pc}, we decompose the continuous subspace into suitably defined absolutely continuous and point-continuous parts. The latter subspace  is a new feature which does not have a counterpart in the Hilbert space setting considered in the previous section.
Since we abstracted our decomposition from the study of physically relevant examples, discussed in Subsection~\ref{Examples},
we do not expect any counterpart of the singular-continuous subspace to appear. In fact, in the quantum-mechanical
framework outlined in Section~\ref{QM} the singular-continuous subspace is  trivial in models with complete particle
interpretation.
\subsection{Pure-Point and Continuous Subspace}\label{subsection-pp-c}
In order to determine the pure-point spectrum, suppose that $A\in\hmfa$ is an eigenvector of the translation automorphisms $\real^s\ni\vx\to\al_{\vx}$ i.e.
\beq
\al_{\vx}(A)=e^{-i\veq\vx}A,\quad\quad \vx\in\real^s
\eeq
for some $\veq\in\real^s$. Then $A$ belongs to the center of $\mfa$, since locality gives
\beq
\|[A,B]\|=\lim_{|\vx|\to\infty}\|[\al_{\vx}(A),B]\|=0,\quad\quad B\in\mfa.
\eeq
The irreducibility assumption from Section \ref{AQFT} ensures that the center of $\mfa$ consists only of
multiples of the unity. Hence the pure-point subspace is given by
\beq
\hmfa_{\spp}=\{\, \la I\, |\, \la\in\complex\,\}. \label{symbol-pure-point}
\eeq
Since we do not have the concept of orthogonality, it is a priori not clear how to choose the 
complementing continuous subspace. In order to restrict the family of admissible projections $P_{\{0\}}$
on $\hmfa_{\spp}$, we proceed along the lines set in the previous section:  we introduce the 
following family of approximants
\beq
P_{\{0\},\VV}(A)=\fr{1}{|\VV|}\int_{\VV} d^sx\,\al_{\vx}(A),\quad\quad A\in\hmfa,
\eeq
which are defined as weak integrals for any compact subset $\VV\subset\real^s$. (They  belong to $\hmfa$ 
by the von~Neumann bicommutant theorem). In the present setting
we do not have the ergodic theorem at our disposal, hence it is not clear if the 
the above net converges as $\VV\nearrow\real^s$. However, there holds the following simple proposition.
For reader's convenience we include the standard argument.
\bep\label{ergodic-theorem} There exists a net $\{\,\VV_{\be}\subset\real^s\,|\, \be\in\mathbb{I}\,\}$ of compact 
subsets of $\real^s$ s.t. $\VV_{\be}\nearrow\real^s$ and for any $A\in\hmfa$ there exists the limit in the 
weak$^*$ topology of $B(\hil)$
\beq
P_{ \{0\} }(A):=\te{\emph{w}$^*$-}\lim_{\be}\fr{1}{|\VV_{\be}|}\int_{\VV_{\be}} d^sx\,\al_{\vx}(A).
\eeq
Moreover, $P_{ \{0\} }(A)=\ovom(A)I$ for some state $\ovom\in\mfa^*$, which is invariant under spacetime translations
and belongs to the closure of $\traceEB$, for any $E\geq 0$, in the  weak$^*$ topology
of $\mfa^*$.
\eep
\proof For any normal functional $\fun\in\trace$ the function $\real^s\ni\vx\to\al_{\vx}^*\fun$ is
continuous w.r.t. the norm topology in $\trace$. Given any
compact subset $\VV\subset\real^s$, we define the functional $\fun_\VV\in\trace$ as the Bochner integral
\beq
\fun_\VV=\fr{1}{|\VV|}\int_\VV d^sx\,\al_{\vx}^*\fun.
\eeq
Now we fix a state $\om\in\trace$  and obtain, from the Banach-Alaoglu theorem, a net 
$\{\,\VV_{\be}\subset\real^s\,|\, \be\in\mathbb{I}\,\}$ and a state $\ovom\in\mfa^*$ s.t.
\beq
\lim_{\be}\om_{\VV_\be}(A)=\ovom(A),\quad\quad A\in\mfa.
\eeq
By locality, $\{P_{\{0\} ,\VV}(A)\}_{\VV\subset\real^s}$ is a central net, i.e. for any
$B\in\mfa$ there holds
\beq
\lim_{\VV\nearrow\real^s}\|[P_{\{0\} ,\VV}(A),B]\|=0.
\eeq
Therefore, all its limit points w.r.t. the weak$^*$ topology
of $B(\hil)$ are multiples of the identity,  by the assumed irreducibility of $\mfa$. 
 It follows that for any $\fun\in\trace$, $A\in\mfa$
\beq
\lim_{\VV\nearrow\real^s}\big(\fun_\VV(A)-\om_\VV(A)\fun(I)\big)=0. \label{V-invariance}
\eeq
Consequently
\beq
\te{w$^*$-}\lim_{\be}\fr{1}{|\VV_{\be}|}\int_{\VV_{\be}} d^sx\,\al_{\vx}(A)=\ovom(A)I.
\eeq
Moreover, by choosing $\fun=\al^*_y\om$ for some $y\in\real^{s+1}$, we obtain from (\ref{V-invariance})
that $\ovom$ is translationally invariant. Similarly, noting that if $\fun\in\traceEB$ then $\fun_\VV\in\traceEB$,
we obtain that $\ovom$ belongs to the weak$^*$ closure of $\traceEB$ in $\mfa^*$ for any $E\geq 0$. \qed\\
In contrast to the ergodic theorem from the Hilbert space setting, this proposition does not provide us with 
a unique projection on the pure-point subspace. However, it restricts the admissible projections to expressions of the form 
$P_{ \{0\} }(\,\cdot\,)=\ovom(\,\cdot\,)I$ where $\ovom$ is a translationally invariant, 
energetically accessible\footnote{We call a state $\om\in\mfa^*$ energetically accessible if it belongs
to the closure of $\traceEB$ for some $E\geq 0$ in the weak$^*$-topology of $\mfa^*$.}
state. A thorough discussion of such states is the subject of Chapter~\ref{chapter-vacuum}, where 
the general postulates from Section~\ref{AQFT} are supplemented with physically motivated phase space
criteria. It is shown that under the existing Condition~$\Cs$ \cite{BP90} every state $\ovom$ satisfying the
conditions from Proposition~\ref{ergodic-theorem} is a vacuum state, as defined in Section~\ref{AQFT}. More
importantly, we introduce new phase space conditions $\Csq$ and $\Cnat$, the former inspired by  the behavior
of coincidence measurements, the latter motivated by the physical principle of additivity of energy, and show that
each of them entails the uniqueness of the energetically accessible vacuum state. These results provide evidence
to the effect that in a large class of physically relevant models there holds the following condition,
which we adopt as a standing assumption in this chapter.
\begin{enumerate}
\item[] \bf Condition $\V:$ \rm \label{cond-V} A state $\ovom\in\mfa^*$, which is invariant under spacetime translations
and can be approximated by elements from $\traceE$ for some $E\geq 0$ in the  weak$^*$ topology
of $\mfa^*$, is unique and is a vacuum state.
\end{enumerate}
Under this condition the  projection on the pure-point subspace $P_{ \{0\} }(\,\cdot\,)=\ovom(\,\cdot\,)I$ is fixed
by the unique, energetically accessible vacuum state $\om_0$. Thus we obtain from Proposition~\ref{ergodic-theorem}
the following ergodic theorem for translation automorphisms in QFT:
\bet\label{QFT-ergodic-theorem} 
Suppose that Condition~$\V$ holds. Then for any $A\in\hmfa$ there exists the limit in the weak$^*$ topology of $B(\hil)$
\beq
P_{ \{0\} }(A)=\te{\emph{w}$^*$-}\lim_{\VV\nearrow\real^s}\fr{1}{|\VV|}\int_{\VV} d^sx\,\al_{\vx}(A).
\eeq
Moreover, $P_{ \{0\} }(A)=\ovom(A)I$, where $\ovom\in\mfa^*$ is the unique energetically accessible vacuum state.
\eet
\nin Guided by relation~(\ref{QM-continuous}), we define the continuous subspace as $\hmfa_{\scc}=\ker P_{ \{0\} }$ or equivalently
\beqa
\hmfa_{\scc}&=&\{\, A\in\hmfa\, |\, \om_0(A)=0\,\}, \label{symbol-continuous}
\eeqa
where $\om_0$ is the unique energetically accessible vacuum state appearing in Theorem~\ref{QFT-ergodic-theorem}. 
For future convenience we also define $\mfa_{\scc}(\mco)=\{\, A\in\mfa(\mco)\, |\, \om_0(A)=0\,\}$.
\label{symbol-local-continuous}

\subsection{Absolutely Continuous and Point-Continuous Subspace}\label{subsection-ac-pc}

Proceeding to more detailed  analysis of the continuous subspace, we note that if Condition~$\V$ holds, then
for any $A\in\hmfa_{\scc}$, spacelike unit vector $\hat{e}\in\real^{s+1}$, and $E\geq 0$, we get
\beq
\lim_{\la\to\infty}\fun\big(\al_{\la\hat{e}}(A)\big)=0, \quad\quad \fun\in \traceE. 
\eeq
This is justified making use of the fact that $\{\al_{\la\hat{e}}(A)\}_{\la\geq 0}$ is a central net
and proceeding as in the proof of Proposition~\ref{ergodic-theorem}. (See also Proposition~\ref{space1} below).
In view of this property, we can base further decomposition of $\hmfa_{\scc}$ on decay properties of
the functions $\real^s\ni\vx\to\om(\al_{\vx}A)$.
Guided by our discussion in Section~\ref{QM2}, we expect that the best possible  generic
property of such functions should be square-integrability.
Let us now demonstrate that in any quantum field theory there is a non-trivial  subspace of local operators
which are square-integrable in the sense made precise in Proposition~\ref{many-square-integrable} below.
For this purpose we consider the operator-valued distribution
\beq
\wt{A}(\vep)=\fr{1}{(2\pi)^{\fr{s}{2}}}\int d^sp\, e^{i\vep\vx}\al_{\vx}(A), \quad\quad A\in\hmfa. \label{2distribution}
\eeq
If $A\notin\hmfa_{\spp}$, then the support of this distribution (i.e. the Arveson spectrum of $A$ w.r.t. space translations)
coincides with $\real^s$ \cite{Bu90}.
A more detailed spectral theory should determine not only the support, but also regularity properties of the 
distribution~(\ref{2distribution}). As we mentioned in Section~\ref{AQFT}, the first step in this direction was taken by 
Buchholz who has shown that  in any local, relativistic  QFT there holds the following bound 
\beq
\sup_{\fun\in \traceEB}\int d^sp\,|\vep|^{s+1+\eps}|\fun(\wt{A}(\vep))|^2<\infty, \quad\quad A\in\hmfa, \label{2regularity}
\eeq
for any $E\geq 0$, $\eps>0$ \cite{Bu90}. Making use of the fact that for any $f\in S(\real^s)$ there holds
\beq
\wt{A(f)}(\vep)=(2\pi)^{\fr{s}{2}}\tf(\vep)\wt{A}(\vep), \label{FT-convolution}
\eeq
where $A(f)=\int d^sx\,\al_{\vx}(A)f(\vx)$, it is not difficult to find non-trivial operators $A\in\hmfa_{\scc}$ 
which satisfy a stronger estimate. 
\bep\label{many-square-integrable} In any quantum field theory admitting a vacuum state $\om_0\in\mfa^*$, there exist $0 \neq A\in\hmfa$ s.t. $\om_0(A)=0$ and
\beq
\|A\|_{E,2}:=\sup_{\fun\in \traceEB}\bigg(\int d^sp\,|\fun(\wt{A}(\vep))|^2\bigg)^{\h}<\infty. \label{square-integrable}
\eeq
\eep
\proof We pick $A\notin\hmfa_{\spp}$. Since the support of the distribution $\real^s\ni\vep\to\wt{A}(\vep)$ coincides
with $\real^s$ and the space $C_0^{\infty}(\real^s)$ of smooth, compactly supported functions is dense in 
$S(\real^s)$ we can find a function $f\in C_0^{\infty}(\real^s)$ s.t. $A(f)=\wt{A}(\tf)\neq 0$. 
Next, for any $n\in\nat$ we introduce functions $f_n\in C_0^{\infty}(\real^s)$ given by
\beq
\tf_n(\vep)=\tf(\vep)|\vep|^{2n}.
\eeq
Then the operators $A(f_n)\in\hmfa$ are also different from zero. (Otherwise the support of $\vep\to\wt{A}(\vep)$
would have to be contained in $\{0\}$). Setting $4n>s+1$ we obtain from identity~(\ref{FT-convolution}) and 
estimate~(\ref{2regularity}) that $\|A(f_n)\|_{E,2}<\infty$ for any $E\geq 0$.
In order to verify that $\om_0(A(f_n))=0$, we use local normality of the state $\om_0$ to exchange its action
with integration and the fact that $\tf_n(0)=0$. \qed\\
We  call operators $A\in\hmfa_{\scc}$, which satisfy the bound~(\ref{square-integrable}) for any $E\geq 0$,
\emph{square-integrable}. We know from Section~\ref{QM2} that the quantum-mechanical analogue of this property
implies absolute continuity of the spectral measure and cannot be improved to integrability with any smaller
power. Thus we define the absolutely continuous subspace of $\hmfa_{\scc}$ as consisting of all square-integrable
observables
\beq
\hmfa_{\ac}=\{\, A\in\hmfa_{\scc}\, |\, \|A\|_{E,2}<\infty\textrm{ for any } E\geq 0\,\}. \label{def-ac}
\eeq
This subspace is non-trivial in any local, relativistic QFT by Proposition~\ref{many-square-integrable}
and we expect from Lemma~\ref{QM-L1} that it 
is large in a certain sense. Also our discussion of quantum mechanical scattering in Section~\ref{QM} suggests that the absolutely continuous
subspace should exhaust the continuous subspace in physically relevant models. In quantum field theory such situation
prevails in massive models, admitting sufficiently many particle detectors, as we show in 
Theorem~\ref{triviality} below. However,
in Subsection~\ref{Examples} we demonstrate that in a large family of massless theories the subspace $\hmfa_{\ac}$ has non-zero,
but finite co-dimension in $\hmfa_{\scc}$. This case study reveals a class of models in which the 
absolutely continuous subspace has the following characterization:
\begin{enumerate}
\item[] \bf Condition $\A$: \rm \label{cond-L2} There exists a finite family of (possibly unbounded) linear functionals $\tau_1,\ldots,\tau_n$
on $\hmfa$, invariant under translations in space, s.t.
\beq
\hmfa_{\ac}=\ker\om_0\cap\ker\tau_1\cap\ldots\cap\ker\tau_n. \label{intersection-of-kernels}
\eeq
\end{enumerate}
We find it noteworthy that the square-integrability requirement from definition~(\ref{def-ac}) can, in many
cases, be replaced with the linear-algebraic condition~(\ref{intersection-of-kernels}). The problem of
constructing the distinguished family of functionals $\{\tau_j\}_1^n$ in particular examples is treated
in Appendix~\ref{Condition-L2}. Here we remark that in theories complying with Condition~$\A$ the
direct sum complement of $\hmfa_{\ac}$ in $\hmfa_{\scc}$ exists and is finite dimensional. This latter property
makes it similar to the pure-point subspace, although it certainly belongs to $\hmfa_{\scc}$. Therefore, we
propose the term \emph{point-continuous} subspace $\hmfa_{\pc}$ \label{symbol-point-continuous} 
for this direct sum complement. Thus we arrive at the decomposition
\beq
\hmfa=\hmfa_{\spp}\oplus\hmfa_{\pc}\oplus\hmfa_{\ac}.
\eeq 
Of course, the point-continuous subspace is non-unique and we do not introduce any natural choice here.
However, its dimension and the behavior of its elements under translations, which we  study below,
do not depend on the selection which is made at this point.

If the point-continuous subspace is non-trivial,  the continuous subspace contains observables which are not 
square-integrable. It is our next goal to quantify their deviation from square-integrability.
For this purpose we introduce the concept of the \emph{infrared order}  of an observable $A\in\hmfa_{\scc}$:
\beq
\ord(A):=\inf\{\,\be\geq 0 \,|\sup_{\fun\in \traceEB}\int d^sp\, |\vep|^{\be}|\fun(\wt{A}(\vep))|^2<\infty \textrm{ for all } E\geq 0 \,\}.
\label{infrared-order}
\eeq
Moreover, we define the set $\Ord(\hmfa_{\scc})=\{\, \ord(A)\, |\, A\in\hmfa_{\scc}\,\}$.
It follows from  estimate~(\ref{2regularity}), due to Buchholz,  that $\ord(A)\leq s+1$ for any  $A\in\hmfa_{\scc}$ in
any local, relativistic QFT. However, the models studied in Subsection~\ref{Examples} below
are far from saturating this bound. These examples demonstrate that the detailed regularity properties
of the distributions~(\ref{2distribution}), which are captured by the dimension of the point-continuous subspace
and the infrared orders of its elements, can provide a distinction between different theories. In contrast, as we mentioned above, the Arveson spectrum of any $A\in\hmfa_{\scc}$ w.r.t. the space translations $\real^s\ni\vx\to\al_{\vx}$ coincides with $\real^s$. 

\subsection{Instructive Examples}\label{Examples} 
In this subsection we study briefly the spectral decomposition in scalar free field theory and
related models, introduced in Section~\ref{scalar-free-field-theory}. This discussion substantiates the point that the detailed
spectral concepts, introduced in the previous subsection, provide a distinction between different models.

The first pair of examples are massive scalar free field theory and its even part, defined in Section~\ref{scalar-free-field-theory}. As we expect perfectly regular infrared
structure here, these models serve as a reference point for our later discussion. In fact there holds:
\bet\label{massive-results1} Massive scalar free field theory and its even part satisfy Condition~$\A$
and have trivial point-continuous subspace for any dimension of space $s\geq 1$.
\eet
\nin This result can be extracted from our discussion in Section~\ref{triviality-of-Apc} below: 
There we formulate Condition~$\B$ (a) which, roughly speaking, assures that the theory admits sufficiently
many particle detectors. We verify in Appendix~\ref{Condition-L1} that massive scalar free field 
theory and its even part satisfy this criterion. Then Theorem~\ref{triviality} entails the above result.

Let us now turn to the more interesting massless case. To exclude from consideration the unduly 
complicated infrared structure of low dimensional massless models, we set $s\geq 3$. 
Let $(\mfa,\al,\hil)$ be the local net generated by the massless scalar free field
acting on the Fock space $\hil$. Let $(\eumfa,\eal,\ehil)$ be the even part of this theory, that is 
the local net generated by even polynomials of the field  acting on the even part 
$\ehil$ of the Fock space. Finally, we denote by $(\dmfa,\al,\hil)$  the sub-theory
of massless scalar free field theory generated by derivatives of the massless scalar field.
For precise definitions  of these models see Section~\ref{scalar-free-field-theory}. 
Condition~$\V$, which is the starting point of our spectral analysis, is verified in 
these examples in Theorem~\ref{altogether}
and Appendix~\ref{Condition-Nnat}. For the proofs of the three theorems stated below 
consult Appendix~\ref{Condition-L2}.

The following theorem describes the dimensions of the point-continuous subspace and
the infrared orders of its elements in the full massless scalar free field theory.
\bet\label{full-results1} Massless scalar free field theory satisfies Condition~$\A$ for $s\geq 3$.
Moreover, there hold the following statements:
\begin{enumerate}
\item[(a)] If $s=3$, then $\dim\hmfa_{\pc}=2 \te{ or } 3$ and $\Ord(\hmfa_{\scc})=\{0,1,2\}$.
\item[(b)] If $s=4$, then $\dim\hmfa_{\pc}=1 \te{ or } 2$ and $\Ord(\hmfa_{\scc})=\{0,2\}$.
\item[(c)] If $s\geq 5$, then  $\dim\hmfa_{\pc}=1$ and $\Ord(\hmfa_{\scc})=\{0,2\}$.
\end{enumerate}
\eet
\nin Clearly, the situation differs substantially from the massive case mentioned above.
As expected, the infrared structure improves with increasing dimension, in the sense that the dimension of the
point-continuous subspace decreases. However, this subspace remains non-trivial for any $s\geq 3$.
In parts (a) and (b) of the above theorem further investigation is needed to determine the dimension of $\hmfa_{\pc}$
exactly.

The next example which we present is the even part of massless free field theory. The following
result demonstrates that the infrared structure is significantly modified if one restricts attention to
even observables.
\bet\label{even-results1} The even part $(\eumfa,\eal,\ehil)$ of massless scalar free field theory 
satisfies Condition~$\A$ for $s\geq 3$ and there hold the following statements:
\begin{enumerate}
\item[(a)] If $s=3$, then $\dim\heumfa_{\pc}=1$ and $\Ord(\heumfa_{\scc})=\{0,1\}$.
\item[(b)] If $s=4$, then $\dim\heumfa_{\pc}=0 \te{ or } 1$ and $\Ord(\heumfa_{\scc})=\{0\}$.
\item[(c)] If $s\geq 5$, then  $\dim\heumfa_{\pc}=0$.
\end{enumerate}
\eet
\nin Here  the point-continuous subspace is again  non-trivial in physical spacetime, but disappears
in higher dimensions. We note that the maximal infrared order of local observables in this model is strictly
smaller than the corresponding quantity in the full theory for any dimension of space $s\geq 3$. Again,
the precise dimension of the point-continuous subspace in part (b) remains to be determined.

Our last example is the sub-theory of massless free field theory generated by the derivatives of the
field.
\bet\label{derivative-results1} The sub-theory $(\dmfa,\al,\hil)$ of massless scalar free field theory satisfies Condition~$\A$
and has trivial point-continuous subspace for any dimension of space $s\geq 3$.
\eet
\nin This theorem demonstrates that triviality of the point-continuous subspace is not a characteristic
feature of massive models. Such mild infrared structure may occur also in massless theories. 

The case study, presented in this subsection, demonstrates that the detailed spectral
concepts of the point-continuous subspace and the infrared order provide quantitative means to
compare the infrared behavior of different quantum field theory models. For example,
for $s=3$ there holds
\beq
\dim\hmfa_{\pc}>\dim\heumfa_{\pc}>\dim\hdmfa_{\pc}, \label{dimensions-comparison}
\eeq
thus we can meaningfully order the three massless models under study w.r.t. their infrared properties.
Presumably, more detailed spectral analysis is needed to distinguish between massive scalar free field theory
and  $(\dmfa,\al,\hil)$. We expect that allowing for negative values of the infrared order (see formula~(\ref{infrared-order})) could serve this purpose, but computations remain to be done.

\section{Triviality of Point-Continuous Subspace and Existence of Particles}\label{triviality-of-Apc}
In the previous section we constructed a general decomposition of the algebra of observables $\hmfa$ 
into subspaces which differ in their behavior under translations in space. Moreover,  we identified the point-continuous
subspace $\hmfa_{\pc}$ which carries  information about the infrared structure. In this section
we formulate a condition which characterizes a class of theories with $\hmfa_{\pc}=\{0\}$. We show that 
models complying with this assumption and admitting in addition a stress-energy tensor, have non-trivial particle content.

We argued in Section~\ref{QM2} that square-integrability under translations in space should be the 
best possible generic feature of local operators. However, almost-local operators may have much better
integrability properties. In fact, Theorem~\ref{harmonic} due to Buchholz gives us a large class of 
observables $C\in\mfa$ which satisfy
\beq
\|C\|_{E,1}:=\sup_{\fun\in\traceEB}\int d^sx |\fun(C(\vx))|<\infty. \label{symbol-integrable-semi}
\eeq
The observables $C\in\mfa$ which satisfy the above bound for any $E\geq 0$ will be called \emph{integrable}. 
A natural framework for our investigations is based on the subspace of $\mfa$ spanned by all such
integrable operators:
\beq
\CC=\{\, C\in\mfa \,|\,\|C\|_{E,1}<\infty \textrm{ for any $E\geq 0$} \,\}.\label{symbol-integrable-space}
\eeq
Equipped with the family of seminorms 
$\{\,\|\,\cdot\,\|_{E,1}\,|\, E\geq 0\,\}$, the space $\CC$ is
a locally convex Hausdorff space and we call the corresponding topology $T^{(1)}$. (This is established as 
in Section 2.2 of \cite{Po04.1}).
The new Condition~$\B$, stated below, specifies another class of
integrable, almost-local observables which, as we shall see below, can be used to approximate the stress-energy tensor. These observables are of the form $A(g)=\int dt\, \al_t(A)g(t)$, where $A\in\hmfa_{\scc}$ and $g\in S(\real)$ is a suitable time-smearing function.
\begin{enumerate}
\item[] \bf Condition $\B$: \rm \label{cond-L1} There exists such $\mu>0$ that for any $g\in S(\real)$ s.t.
$\supp\,\tg\subset ]-\mu,\mu[$ there hold the assertions:
\begin{enumerate}
\item[(a)] $A(g)\in\mfa^{(1)}$ for any $A\in\hmfa_{\scc}$,
\item[(b)] $\|A(g)\|_{E,1}\leq c_{l}\|R^l A R^l\|$ for any $A\in\mfa_{\scc}(\mco)$, $l\geq 0$,
\end{enumerate}
where $R=(1+H)^{-1}$ and the constant $c_l$ depends also on $\mco$, $E\geq 0$ and $g$.
\end{enumerate}
This condition seems to be a generic property of purely massive theories. We verify it in massive
scalar free field theory in  Appendix~\ref{Condition-L1}. Part (a) is verified also in the even
part of this theory (defined  in  Section~\ref{scalar-free-field-theory}) making use of the fact
that the norm of an (even) local operator $A$ does not change upon its restriction to the even subspace of the Fock
space (cf. formula~(\ref{even-norm-equality})). This property reflects the fact that a state consisting
of an odd number of particles can be approximated in any bounded region of spacetime  by a state consisting
of an even number of particles by shifting one particle to spacelike infinity. We expect that a similar reasoning
can be applied to the dislocalized operators $R^l A R^l$, appearing in part~(b) of the condition, but the computation
remains to be done. We also conjecture that the above criterion can be established in theories of charged, non-interacting
massive particles.

Now we are ready to prove the main spectral result of this section, namely that theories satisfying Condition $\B$ (a) 
have trivial point-continuous subspace. In other words, we show that Condition $\B$ (a) implies Condition $\A$ with 
the functionals $\{\tau_i\}_1^n$ equal to zero.

\bet\label{triviality} Assume that Condition $\B$ (a) holds. Then $\hmfa_{\pc}=\{0\}$.
\eet
\proof It suffices to show that for any $A\in\hmfa_{\scc}$ and any $E>\mu$, where
$\mu$ appeared in Condition~$\B$ (a), there holds $\|A\|_{E,2}<\infty$.
To this end, we choose a function
$f\in S(\real)$ s.t. $\tf=(2\pi)^{-\h}$ on $[-E, E]$ and $\supp\,\tf\subset [-2E,2E]$.
With the help of a  smooth partition of unity we can decompose $f$ as follows:
$f=f_-+f_++f_0$, where $\supp\, \tf_-\subset [-2E,-\mu/2]$, $\supp\, \tf_+\subset [\mu/2, 2E]$,
and $\supp\, \tf_0\subset ]-\mu,\mu[$. Then there holds
\beq
P_EAP_E=P_EA(f)P_E=P_EA(f_-)P_E+P_EA(f_+)P_E+P_EA(f_0)P_E.
\eeq
We note that $A(f_0)$ is square-integrable (and even integrable) by Condition $\B$ (a). To the remaining
terms we can apply Theorem~\ref{harmonic}, since both $A(f_-)$ and $A(f_+)^*$ are almost local and 
energy-decreasing. This latter statement follows from the equality
\beq
\wt{A(f_-)}(p)=(2\pi)^{\h}\tf_-(p^0)\wt{A}(p^0,\vep)
\eeq
which implies that the support of $\wt{A(f_-)}$ does not intersect with the closed future light-cone.
(An analogous argument applies to $A(f_+)^*$). Thus we obtain from Theorem~\ref{harmonic}, for any 
compact subset $K\subset\real^s$,
\beqa
\sup_{\om\in S_E}\int_K d^sx\, |\om(A(f_-)(\vx))|^2
\leq \sup_{\om\in S_E}\int_K d^sx\, \om((A(f_-)^*A(f_-))(\vx))
\leq c_E. 
\eeqa
Here $c_E$ is independent of $K$ hence we can take the limit $K\nearrow\real^s$ and the claim 
follows from  decomposition~(\ref{decomp}) and definition~(\ref{square-integrable}) of the 
seminorms $\|\,\cdot\,\|_{E,2}$. The case of $A(f_+)$ is treated analogously. \qed

Proceeding to the particle aspects of the theory, we define, for any $\fun\in \traceE$,
the net $\{\si^{(t)}_{\fun}\}_{t\in\real_+}$  of functionals on $\CC$ given by
\beq
\si^{(t)}_{\fun}(C)=\int d^sx\, \fun(\al_{t,\vx}(C)). \label{symbol-asymptotic-functional-appr}
\eeq
This net satisfies the uniform bound $|\si^{(t)}_{\fun}(C)|\leq\|\fun\|\,\|C\|_{E,1}$. Therefore, by the Alaoglu-Bourbaki theorem
(see \cite{Ja}, Section 8.5), it has limit points $\si^{(+)}_{\om}$\label{symbol-asymptotic-functional} in the topological dual of $(\mfa^{(1)}, T^{(1)})$ 
which are called the asymptotic functionals. The set of such functionals
\beq
\PC=\{\, \si^{(+)}_{\fun}\, |\, \fun\in \traceE \textrm{ for some } E\geq 0\, \} \label{symbol-particle-content}
\eeq
is called the particle content of a theory. This terminology is justified by the fact that
in asymptotically complete theories of Wigner particles, discussed in Section~\ref{Wigner},  
these functionals are related to the asymptotic particle density.
In fact, for one species of Wigner particles, and $C=B^*B$, where $B$ is almost local and energy-decreasing, 
there holds \cite{AH67, En75} 
\beq
\si^{(+)}_{\fun}(C)=(2\pi)^s\int d^sp\, \fun(a^{*\,\te{out} }(\vep)a^{\te{out}}(\vep))\lan \vep|C|\vep \ran , \label{Araki-Haag}
\eeq
where $a^{*\,\te{out} }(\vep)$, $a^{\te{out}}(\vep)$ are the creation and annihilation operators of the mode
$\vep$ from the Fock space of outgoing states and the kernels $\vep\to\lan\vep|C|\vep\ran$ are smooth, positive functions
which are different from zero for suitably chosen $B$. This guarantees the existence of non-trivial asymptotic functionals $\si^{(+)}_{\fun}$ in this restrictive context. It is now our goal to show that $\PC\neq\{0\}$ not relying directly 
on the Wigner concept of a particle.
Since our argument is based on the existence of the stress-energy tensor, which is postulated
in Condition~$T$ below, we recall the definition and properties of pointlike-localized fields:
We set $R=(1+H)^{-1}$\label{symbol-R} and introduce the space of normal functionals with polynomially damped energy 
\beq
\trace_{\infty}=\bigcap_{l\geq 0}R^l\trace R^l,\label{symbol-trace-infty}
\eeq
where $\trace=B(\hil)_*$. We equip this space with the locally convex topology given by the norms 
$\|\,\cdot\,\|_{l}=\|R^{-l}\,\cdot\, R^{-l}\|$  for  $l\geq 0$. The field content of a theory is 
defined as follows \cite{FH81}
\beq
\Phi_{\FH}=\{\,\phi\in\trace_{\infty}^* \,|\, R^l\phi R^l\in\bigcap_{r>0}\overline{R^{l}\mfa(\mco(r))R^{l}}^w
\textrm{ for some } l\geq 0\,\}. \label{symbol-field-content}
\eeq
There holds the following useful approximation property for pointlike-localized fields which is due to
Bostelmann \cite{Bo05.1}:
For any $\phi\in\Phi_{\FH}$ there exists $l\geq 0$ and a net $A_r\in\mfa(\mco(r))$, $r>0$, s.t.
\beq
\lim_{r\to 0}\|R^l(A_r-\phi)R^l\|=0. \label{Bostelmann2}
\eeq
It follows from Proposition~\ref{ergodic-theorem} that for any $E\geq 0$ one can choose a net $\{\fun_\be\}_{\be\in\mathbb{I}}$ of functionals from $\traceEB$, approximating $\om_0$ in the weak$^*$ topology of $\mfa^*$. Therefore, we obtain 
\beq
|\om_0(A)|\leq \|P_EAP_E\|,\quad\quad A\in\mfa,
\eeq
and the approximation property (\ref{Bostelmann2}) implies that $\om_0$ can be uniquely extended to elements from
$\Phi_{\FH}$. We can therefore introduce the continuous part of this space
\beq
\Phi_{\FH,\scc}=\{\, \phi\in\Phi_{\FH} \,|\, \om_0(\phi)=0\,\}.
\eeq 
The approximation property~(\ref{Bostelmann2}) can now be restated as follows:
For any $\phi\in\Phi_{\FH,\scc}$ there exists $l\geq 0$ and 
$A_r\in\mfa_{\scc}(\mco(r))$ s.t.
\beq
\lim_{r\to 0}\|R^l(A_r-\phi)R^l\|=0. \label{Bostelmann1}
\eeq
Making use of Condition $\B$ (b) we also obtain, for any time-smearing function $g\in S(\real)$
s.t. $\supp\,\tg\subset]-\mu,\mu[$,
\beq
\lim_{r\to 0}\|A_r(g)-\phi(g)\|_{E,1}=0. \label{field_approximation}
\eeq
This implies, in particular, that $\|\phi(g)\|_{E,1}<\infty$ for any $\phi\in\Phi_{\FH,\scc}$, what prepares the ground for 
our next assumption:
\begin{enumerate}
\item[] \bf Condition $T$: \rm \label{cond-T} There exists a field $T^{00}\in\Phi_{\FH,\scc}$\label{symbol-T00} which satisfies
\beq
\int d^sx\,\fun(T^{00}(g)(\vx))=\fun(H),\quad\quad \fun\in \traceE,
\eeq
for any $E\geq 0$ and time-smearing function $g\in S(\real)$ s.t. $\supp\,\tg\subset]-\mu,\mu[$, 
$\tg(0)=(2\pi)^{-\fr{1}{2}}$.
\end{enumerate}
\nin This condition holds, in particular, in massive scalar free field theory and its even part as we show in Theorem~\ref{stress-energy}. With the stress-energy tensor at hand, it is easy to prove that the particle content 
of the theory is non-trivial:
\bet\label{existence-of-particles} Suppose that a theory satisfies Conditions~$\B$ and $T$. Then, for any
$\fun\in\traceE$ s.t. $\fun(H)\neq 0$, the corresponding asymptotic functionals
satisfy $\si^{(+)}_{\fun}\neq 0$.
\eet
\proof We choose $g\in S(\real)$ as in Condition $T$ and $0<\eps\leq\h|\fun(H)|$. Making use of Condition~$\B$~(b) and relation~(\ref{field_approximation}), we can find $C\in\CC$  s.t. $\|T^{00}(g)-C\|_{E,1}\leq\eps$.
Then, exploiting Condition~$T$ and invariance of $H$ under time translations, we obtain 
\beq
|\fun(H)|=|\int d^sx\,\fun(T^{00}(g)(\vx))|\leq\eps+|\int d^sx\,\fun(C(t,\vx))|.
\eeq
Thus we obtain a positive lower bound on the asymptotic functional approximants, defined by (\ref{symbol-asymptotic-functional-appr}),  which is uniform in time. \qed\\
We emphasize that we have proven here more than non-triviality of the particle content - we have verified
that every physical state, which has non-zero average energy, gives rise to a non-trivial asymptotic functional.

The framework, which we considered above, is more general than the one considered in the theory of 
particle weights developed by  Buchholz, Porrmann and Stein \cite{BPS91}. These authors introduced the left ideal
\beq
\mfl=\{\, AB\,|\, A, B\in\mfa, B \te{ is almost local and energy-decreasing }\,\}
\eeq
and proposed the $*$-algebra of particle detectors $\mfc=\mfl^*\mfl$\label{symbol-particle-detectors} as the domain of asymptotic functionals. It follows 
from Theorem~\ref{harmonic}, polarization identity and the bound $B^*AB\leq \|A\|\, B^*B$, valid for any self-adjoint operator $A$, that $\mfc\subset\mfa^{(1)}$. Similarly, $\mfc(g)\subset\mfa^{(1)}$, where $\mfc(g)=\{\,C(g)\, |\, C\in\mfc\,\}$ for some time-smearing function
$g\in S(\real)$. As $\mfc$ may be a proper subspace in $\mfa^{(1)}$, it is not clear if the  asymptotic functionals, which are non-trivial on $\mfa^{(1)}$ by Theorem~\ref{existence-of-particles}, remain non-trivial after restriction to $\mfc$. This gap can be closed with the help of the following  strengthened form of Condition~$\B$.
\begin{enumerate}
\item[] \bf Condition $\Bs$: \rm \label{cond-L1s} There exists such $\mu>0$ that for any $g\in S(\real)$ s.t.
$\supp\,\tg\subset ]-\mu,\mu[$ there hold the assertions:
\begin{enumerate}
\item[(a)] $A(g)\in\ov{\mfc(g)}^{T^{(1)}}$ for any $A\in\hmfa_{\scc}$,
\item[(b)] $\|A(g)\|_{E,1}\leq c_{l}\|R^l A R^l\|$ for any $A\in\mfa_{\scc}(\mco)$, $l\geq 0$,
\end{enumerate}
where the constant $c_l$ depends also on $\mco$, $E\geq 0$ and $g$. 
\end{enumerate}
Making use of the fact that $\real^+\ni E\to \|C\|_{E,1}$ is an increasing function for any fixed
$C\in \mfa^{(1)}$, part (a) of the above criterion can be restated as the requirement that for any $E\geq 0$ and $\eps>0$
there exists $C\in\mfc$ s.t.
\beq
\|A(g)-C(g)\|_{E,1}\leq\eps. \label{spectral-part(a)}
\eeq
We verify in Appendix~\ref{Condition-L1} that  this condition holds in massive scalar free field theory,
hence  it is consistent with the basic postulates. However, it seems to be too specific to hold in the even 
part of this theory (defined in Section~\ref{scalar-free-field-theory}) or in models containing charged particles.
As we mentioned above, in the restrictive framework of theories satisfying this condition, we can solve the problem  
of existence of non-trivial asymptotic functionals on~$\mfc$.
\bet Suppose that a theory satisfies Conditions $\Bs$ and $T$. Then, for any
 $\fun\in\traceE$ s.t. $\fun(H)\neq 0$, the corresponding asymptotic functional
satisfies $\si^{(+)}_{\fun}|_{\mfc}\neq 0$.
\eet
\proof We choose $g\in S(\real)$ as in Condition $T$. With the help of Condition~$\Bs$ and formula~(\ref{field_approximation}) we can find, for any $\eps>0$, a detector $C\in\mfc$ s.t. $\|T^{00}(g)-C(g)\|_{E,1}\leq\eps$. There holds
\beqa
|\fun(H)|&=&|\int d^sx\,\om(T^{00}(g)(\vx))|\leq\eps+|\int d^sx\,\fun(C(g)(t,\vx))|\non\\
&=&\eps+|\int d\tau\, g(\tau)\int d^sx\,\om(C(t+\tau,\vx))|,
\eeqa
where in the last step we made use of the fact that $\|C\|_{E,1}<\infty$ and of the Fubini theorem to change 
the order of integration. Assuming that $\lim_{t\to\infty} \int d^sx\,\fun(C(t,\vx))=0$ and making use of the
Lebesgue dominated convergence theorem, we arrive at a contradiction if $\eps<|\fun(H)|$. \qed\\
We recall, that in the framework of Buchholz, Porrmann and Stein any asymptotic functional 
$\si^{(+)}_{\om}|_{\mfc}$, where $\om\in S_E$, can be decomposed into pure functionals~$\si^{(+)}_{\la}$
on $\mfc$, the so called pure particle weights\footnote{As
a matter of fact, the decomposition has been established only on some subalgebra  of
regular particle detectors in $\mfc$. We refer to \cite{Po04.1} for details.}
\beq
\si^{(+)}_{\om}|_{\mfc}=\int d\mu(\la)\si^{(+)}_{\la}, \label{pure1} 
\eeq
which are labeled by sharp four-momentum $p_{\la}$ and a label $\ga_{\la}$ which carries information about the
internal degrees of freedom. This decomposition generalizes
relation~(\ref{Araki-Haag}), valid in the framework of asymptotically complete Wigner particles. 
A possible formulation of the problem of asymptotic completeness in the general context was proposed by Buchholz in \cite{Bu94}. In particular, it should be possible to determine the energy and momentum of any physical state $\om\in S_E$, knowing its particle content from formula~(\ref{pure1}). That is, there should hold
\beq
\om(P)=\int d\mu(\la) p_{\la},
\eeq
where $P=(H,\vec{P})$  and the measure $d\mu$ is suitably normalized. It was conjectured in \cite{Bu94} that 
this relation  holds in models admitting a stress-energy tensor. We note that Condition~$\Bs$ provides a link between
pointlike-localized fields and the algebra of particle detectors $\mfc$ on which the decomposition~(\ref{pure1}) has been 
established (up to the technical details mentioned in the footnote). Therefore, our criterion should be
useful for establishing this form of asymptotic completeness.


%% file: PhD-vacuum2.tex
\chapter{Uniqueness of Spectral Decomposition and Vacuum Structure}\label{chapter-vacuum}

The decomposition of the algebra of local observables $\hmfa$ into spectral subspaces, performed
in the previous chapter, was based on
Condition~$\V$, stated in Section~\ref{space-translations-QFT}. This criterion guarantees
that a state which is translationally invariant and energetically accessible is unique and
is a vacuum state. Among its consequences is the ergodic theorem~\ref{QFT-ergodic-theorem} which
gives the unique projection $P_{\{0\}}$ on the continuous subspace of the algebra of observables.
It is the goal of the present section to derive Condition~$\V$ and other properties of  
vacuum states from physically motivated phase space conditions.

Physical properties of  vacuum states were a subject of study since the early days 
of algebraic quantum field theory \cite{BHS63,Bo65}. In particular, the problem of convergence of
physical states to the vacuum state under large translations attracted much attention. It was
considered under the assumptions of complete  particle interpretation in the sense of Wigner \cite{AH67},
isolated mass hyperboloid \cite{BF82} and asymptotic abelianess in time \cite{BWa92}. As none of these 
assumptions is expected to hold in all physically relevant models, further investigation of the vacuum structure is warranted.

This subject is revisited  here from the point of view of phase space analysis: We show in 
Section~\ref{existence-of-vacuum} that the well-known
compactness condition~$\Cs$ implies that any physical state which is invariant under translations in 
some spacelike ray is a vacuum state. This applies, in particular, to the state $\om_0$ which appears as
a limit point of space averages in Proposition~\ref{ergodic-theorem}.
In order to prove the uniqueness of the energetically
accessible vacuum states, which entails convergence of the space averages, we first reformulate Condition~$\Cs$ in Section~\ref{coincidence-measurement} so that it describes the behavior of coincidence arrangements of local operators. Next, in Section~\ref{vacuum-structure},  we accompany it with a quantitative refinement, motivated by the fact that physical states are localized in space. We show that the resulting Condition~$\Csq$ entails purity and uniqueness of vacuum states which can be prepared with a finite amount of energy.
There follows Condition~$\V$ and, as a consequence, the ergodic theorem~\ref{QFT-ergodic-theorem}.
In addition, we demonstrate that in Lorentz covariant theories there holds convergence of physical states to the vacuum state under large timelike translations i.e. relaxation to the vacuum given by formula~(\ref{relaxation}).

Since Condition~$\Csq$ has clear physical interpretation and holds in massive scalar free field
theory, as shown in Appendix~\ref{Condition-Csq}, we expect that it is a generic feature of massive theories.
However, its status in the realm of  massless theories
is not clear. Therefore, in Section~\ref{Condition-Cnat-and-additivity} we formulate an alternative phase-space condition~$\Cnat$, inspired by the physical principle of additivity of energy over isolated subregions, and show that it has all the physical consequences mentioned above (apart from purity of the vacuum). It is verified in
Appendix~\ref{Condition-Nnat}  that this criterion holds both in massive and massless theory of scalar non-interacting particles. In order to facilitate this argument, we introduce in Section~\ref{additivity-of-energy} an auxiliary nuclearity condition~$\Nnat$, interesting in its own right, which entails Condition~$\Cnat$.

Some results from Sections~\ref{existence-of-vacuum} and \ref{additivity-of-energy} have been published
by the author in \cite{Dy08.1}.

\section{Condition  $\Cs$ and  Existence of Vacuum States}\label{existence-of-vacuum}
It is well known from quantum mechanics that a system of bounded energy, restricted to a finite volume, should have
 a finite number of degrees of freedom. Following the seminal work of Haag and Swieca \cite{HS65}, this fact has 
been formulated in quantum field theory as compactness and nuclearity requirements on certain linear maps \cite{BWi86,BP90}.

For any $E\geq 0$, $\be>0$ and double cone $\mco$ we consider the maps $\Pi_E:\traceE\to\mfa(\mco)^*$, $\whXi_E:\mfa(\mco)\to B(\hil)$ and $\Xi_{\be}:\mfa(\mco)\to B(\hil)$, given by
\beqa
\Pi_E(\fun)&=&\fun|_{\mfa(\mco)},\quad\quad\quad\quad\quad \fun\in\traceE, \label{symbol-PiE}\\
\whXi_E(A)&=&P_EAP_E,\quad\quad\quad\quad A\in\mfa(\mco),\label{hXi} \\
\Xi_{\be}(A)&=&e^{-\be H}Ae^{-\be H},\quad\quad A\in\mfa(\mco).  \label{hhXi}
\eeqa
It is convenient to adapt here the restrictive definition of compactness
from \cite{BP90}: Let $V$ and $W$ be Banach spaces and let $\lin(V,W)$\label{symbol-lin} denote the space of
linear maps from $V$ to $W$ equipped with the standard norm. Let $\fin(V,W)$\label{symbol-fin} be
the subspace of finite rank mappings. More precisely, any $F\in\fin(V,W)$  is
of the form $F=\sum_{i=1}^n\tau_i\, S_i$, where $\tau_i\in W$ and $S_i\in V^*$.
We say that a map $\Pi\in\lin(V,W)$ is compact if it belongs to the closure 
of $\fin(V,W)$ in the norm topology of $\lin(V,W)$.

It was argued by Fredenhagen and Hertel in some unpublished work, quoted in \cite{BP90}, 
that in physically meaningful theories there should  hold the following condition:
\begin{enumerate}
\item[] \bf Condition $\Cs$: \rm\label{cond-Cs}  The maps $\Pi_E$ are compact
for any  $E\geq 0$ and any double cone $\mco$.
\end{enumerate}
We readily obtain equivalent formulations of this criterion which are also useful. (Similar
argument appears in \cite{BP90}).
\bel\label{compactness-equivalence} We fix a double cone $\mco$. Then the following conditions are equivalent:
\begin{enumerate}
\item[(a)] The maps $\Pi_E$ are compact for any $E\geq 0$.
\item[(b)] The maps $\whXi_E$ are compact for any $E\geq 0$.
\item[(c)] The maps $\Xi_{\be}$ are compact for any $\be>0$.
\end{enumerate}
\eel
\proof The implication $(a)\Rightarrow(b)$ can be shown as follows:
Let $\eps>0$ and $F\in\fin(\traceE,\mfa(\mco)^*)$
be a finite rank map s.t. $\|\Pi_E-F\|\leq\eps$. It has the form
\beq
F =\sum_{i=1}^n\tau_i\, S_i
\eeq
for some $\tau_i\in\mfa(\mco)^*$ and $S_i\in\traceE^*$. 
Every element $S_i$ can be extended, by the Hahn-Banach theorem,
to an element $\wh{S}_i\in B(\hil)$. Then the formula $\wh{F}=\sum_{i=1}^n(P_E\wh{S}_iP_E)\,\tau_i$
defines a map in $\fin(\mfa(\mco),B(\hil))$ which satisfies $\|\whXi_E-\wh{F}\|\leq\eps$.
This is easily verified making use of the fact that for any $\fun\in\traceE$ and $A\in\mfa(\mco)$ there holds
$\Pi_E(\fun)(A)=\fun(\whXi_E(A))$. The implication $(b)\Rightarrow(a)$ can be shown by an analogous
reasoning.

To verify $(b)\Rightarrow(c)$, we note that the spectral theorem gives 
\beq
\|\Xi_{\be}(A)-e^{-\be H}\whXi_E(A)e^{-\be H}\|\leq 2\|A\| e^{-\be E}.
\eeq
Thus, choosing sufficiently large $E$, we can approximate the map $\Xi_{\be}$ by finite rank mappings
up to arbitrary accuracy. The opposite implication follows from the identity $\whXi_E(A)=P_E e^{\be H}\Xi_{\be}(A)e^{\be H}P_E$. \qed\\
Turning to the vacuum structure, we note the following elementary lemma which
ensures local normality of energetically accessible states in theories complying
with Condition~$\Cs$. (Similar argument appears in \cite{GJ70} p. 49).
\bel\label{local-normality} Suppose that Condition~$\Cs$ holds. Let $\om\in\mfa^*$ be an element of 
the weak$^*$ closure of $\traceEB$ for some $E\geq 0$. Then $\om$ is locally normal and can be approximated
by a sequence of elements from $\traceEB$ in the weak$^*$ topology. 
\eel
\proof First, we show that $\om$ is locally normal. Let $\{\fun_{\be}\}_{\be\in\mathbb{I}}$ be a
net of elements of $\traceEB$ approximating $\om$ in the weak$^*$ topology of $\mfa^*$. 
By Condition~$\Cs$, for any open bounded region $\mco$ the set $\{\,\fun_{\be}|_{\mfa(\mco)}\,|\, \be\in\mathbb{I}\}$
is compact in the norm topology of $\mfa(\mco)^*$. We can therefore choose a subsequence $\{\fun_{G(n)}\}_{n\in\nat}$,
where $G:\nat\to\mathbb{I}$, approximating $\om$ in this topology. Since a norm limit of a sequence of
normal functionals is normal, we conclude that $\om$ is a normal state upon restriction to any local algebra.
Let $\mco(m)$ denote the double cone of radius $m$.
We note that by choosing subsequences $\{\fun_{G_m(n)}\}_{n\in\nat}$, converging in norm to $\om$ on $\mfa(\mco(m))$,
for any $m\in\nat$, in such a way that $\{\fun_{G_{m+1}(n)}\}_{n\in\nat}$ is a subsequence of $\{\fun_{G_{m}(n)}\}_{n\in\nat}$,
we obtain a diagonal sequence $\fun_n=\fun_{G_{n}(n)}$ which converges to $\om$ in the weak$^*$ topology
of $\mfa^*$, replacing the original net.  \qed\\
We recall from Section~\ref{AQFT} that a vacuum state 
is a state on $\mfa$ which is translationally invariant, locally normal and s.t. the spectrum condition
holds in its GNS-representation. In a theory satisfying Condition~$\Cs$ there holds the following
simple characterization of energetically accessible vacuum states.
\bet\label{disconnected1} Suppose that Condition $\Cs$ holds. Let a state $\om\in\mfa^*$ be an element of 
the weak* closure of $\traceEB$ for some $E\geq 0$ which is invariant under translations along some 
spacelike ray. Then $\om$ is a vacuum state.
\eet
\proof 
We pick any $A\in\mfa(\mco)$, a test function $f\in S(\real^{s+1})$ s.t. $\supp\,\tilde{f}\cap\cone=\emptyset$
and define the energy-decreasing operator $A(f)=\int A(x) f(x) d^{s+1}x$.
Next, we parametrize the ray from the statement of the theorem as $\{ \ \la \e \ | \ \la\in\real \ \}$, 
where $\e\in\real^{s+1}$ is some spacelike unit vector, choose  a compact subset $K\subset\real$ and estimate
\beqa
\om(A(f)^*A(f))|K|&=&\int_{K}d\la \ \om\big( (A(f)^*A(f))(\la \e) \big)\non\\
&=&\lim_{n\to\infty}\fun_n\bigg(\int_{K}d\la \ (A(f)^*A(f))(\la \e) \bigg)\non\\
&\leq& \|P_{E}\int_{K} d\la \ (A(f)^*A(f))(\la \e) \ P_{E}\|. \label{harmonic2}
\eeqa
In the first step we exploited invariance of the state $\om$ under translations along the spacelike ray. In
the second step we made use of local normality of this state, (or of the dominated convergence theorem), in 
order to exchange its action with  integration. Approximating $\om$ by a sequence of functionals $\fun_n\in\traceEB$,
we arrived at the last expression. (Local normality of $\om$ and the existence of an approximating sequence follow from
Lemma~\ref{local-normality}). Now we can apply Theorem~\ref{harmonic} 
to conclude that the last expression on the r.h.s. of (\ref{harmonic2}) is bounded uniformly in $K$.
As $|K|$ can be made arbitrarily large, it follows that
\beq
\om(A(f)^*A(f))=0 \label{zero1}
\eeq
for any $A\in\mfa(\mco)$ and $f$ as defined above. Since equality~(\ref{zero1}) extends to any $A\in\mfa$,
we can proceed with the proof that $\om$ is a vacuum state similarly as in the proof of Theorem~4.5 of \cite{Ar}: 
First, one has to show that the functions 
\beq
\real^{s+1}\ni x\to\om(A^*B(x)) \label{strongcont2}
\eeq
are continuous for any $A, B\in \mfa$. For local operators
$A$, $B$ this follows from local normality of the state $\om$, ensured by Lemma~\ref{local-normality},
and the fact that the functions $\real^{s+1}\ni x\to B(x)$ are continuous in the strong operator topology of $B(\hil)$.
Since local operators are norm dense in $\mfa$, we obtain continuity for all  $A, B\in \mfa$.
In particular, $h(x)=\om(B(x))$, $B\in\mfa$, is continuous and
bounded. Let us now choose any function $f\in S(\real^{s+1})$ such that $\tf$ is compactly supported
and $0\notin\supp\,\tf$. Then it can be decomposed into a sum
$f=f_1+f_2$ s.t. $\supp\,\tf_1\cap\cone=\emptyset$ and 
$\supp\,\tilde{\bar{f}}_2\cap\cone=\emptyset$. Consequently, by the Cauchy-Schwarz inequality
and relation~(\ref{zero1})
\beq
\int h(x)f(x)d^{s+1}x=\om\big(B(f_1)\big)+\overline{\om\big(B^*(\bar{f}_2)\big)}=0,
\eeq
showing that $\supp\,\tilde{h}\subset\{0\}$. This implies that $h(x)=$ const,
what entails translational invariance of the state $\om$.
It follows that translations are unitarily implemented 
in the GNS-representation $(\hil_{\om},\pi_{\om},\vac_{\om})$ induced by $\om$.
The resulting unitary representation $\real^{s+1}\ni x\to U_{\om}(x)$, acting on $\hil_{\om}$,
is strongly continuous by continuity of the functions~(\ref{strongcont2}). To verify the spectral condition, 
we choose $f\in S(\real^{s+1})$, s.t. $\supp\tilde{f}\cap\cone=\emptyset$, $A, B\in\mfa$  and calculate
\beqa
|\int(\pi_{\om}(A)\vac_{\om}|U_{\om}(x)\pi_{\om}(B)\vac)f(x)d^{s+1}x|^2=|\om(AB(f))|^2 & &\non\\
\leq\om(A^*A)\,\om(B(f)^*B(f))=0, & &
\eeqa
where we made use again of identity (\ref{zero1}). \qed\\
Making use of the above theorem, we can construct vacuum states with the help of the method of
'large translations' envisaged first in \cite{BHS63}. The case of spacelike translations is
well known \cite{BP90}, although our proof, based on Theorem~\ref{disconnected1}, appears to be new.
\bep\label{space1} Suppose that Condition~$\Cs$ holds and let $\e\in\real^{s+1}$ be a spacelike unit vector. 
Then there exists 
a net $\{\,\la_{\be}\in\real\,|\, \be\in\mathbb{I}\,\}$ s.t. $\la_{\be}\to\infty$ and a vacuum state
$\om_0\in\mfa^*$ s.t. for any $A\in\mfa$ there holds
\beq
\te{\emph{w}$^*$-}\lim_{\be}A(\la_{\be}\e)=\om_0(A)I,\label{A-spacelike-convergence}
\eeq
where the limit is taken in the weak$^*$ topology of $B(\hil)$.
\eep
\proof Noting that $\{A(\la_{\be}\e)\}_{\be\in\mathbb{I}}$ is a central net in $\mfa$, and
proceeding identically as in the proof of Proposition~\ref{ergodic-theorem}, one obtains
that relation~(\ref{A-spacelike-convergence}) holds for some translationally invariant state
$\om_0$ which belongs to the closure of $\traceE$, for any $E\geq 0$, in the  weak$^*$ topology
of $\mfa^*$. Now making use of Theorem~\ref{disconnected1} we conclude that $\om_0$ is a vacuum state. \qed\\
Now we turn to the more interesting problem of convergence of physical states to the vacuum state
under large timelike translations.
Here we cannot exploit locality directly, but instead we rely on Lorentz transformations. We need
the following regularity assumption on their action $\lpo\ni \La\to\al_{\La}$ on the global algebra
$\mfa$. 
\begin{enumerate}
\item[] \bf Condition $R$: \rm \label{cond-R} Let 
$\om\in\trace$. Then, for any open bounded region $\mco$ and any
timelike unit vector $\e$ there holds
\beq
\sup_{\la\in\real_+}\sup_{A\in\mfa(\mco+\la\e)_1 }|\al^*_\Lambda\om(A)-\om(A)|\underset{\La\to I}{\longrightarrow} 0.
\eeq
\end{enumerate} 
This condition is satisfied, in particular, if the Lorentz transformations are unitarily implemented.
The following result is based on the observation due to Buchholz that the timelike limit points
of physical states are invariant under translations in some spacelike hyperplane.
\bep\label{time1} Suppose that Conditions $\Cs$ and $R$ hold. Let $\om_0$ be a weak$^*$ limit 
point as $\la\to\infty$ of the net $\{\al_{\la \e}^*\om\}_{\la\in\real_+}$ of states on $\mfa$,
where $\e\in\real^{s+1}$ is a timelike unit vector and $\om$ is a state from $\traceE$ for some $E\geq 0$.
Then $\om_0$ is a vacuum state.
\eep
\proof In view of Theorem \ref{disconnected1} it suffices to show that $\om_0$ is invariant 
under translations in the spacelike hyperplane $\{\e^\perp\}=\{  x\in\real^{s+1} \ | \ \e\cdot x =0  \}$, 
where  dot denotes the Minkowski scalar product.

Choose $x\in\{\e^\perp\}$, $x\neq 0$. Then there exists a Lorentz transformation $\La$ and 
$y^0,y^1\in\real\backslash\{0\}$ s.t.
$\La \e=y^0\e_0$, $\La x = y^1\e_1$, where $\e_{\mu}$, $\mu\in\{0,1,\ldots,s\}$ form the canonical basis in $\real^{s+1}$. We set $v=\fr{y^1}{y^0}$ and introduce the family of Lorentz transformations $\La_t=\La^{-1}\tilde{\La}_t\La$, where $\tilde{\La}_t$ denotes the boost in the direction of $\e_1$ with rapidity $\textrm{arsinh}(\fr{v}{t})$.
By the composition law of the Poincar\'e group the above transformations composed with
translations in timelike direction give also rise to spacelike translations
\beq
(0,\La_\lam)(\lam \e,I)(0,\La_\lam^{-1})=(\lam\La_\lam \e,I),\quad
\lam\La_\lam \e=\lam\sqrt{1+\big(v/\lam\big)^2}\e+x.
\eeq
We make use of this fact in the following estimate: 
\beqa
|\al_{\lam \e}^*\om(A)-\al_{\lam \e}^*\om(A(x))|&\leq&
|\om(\al_{\lam \e}A)-\om(\al_{\La_{\lam}}\al_{\lam \e}\al_{\La_{\lam}^{-1}}A)|
\nonumber\\
&+&|\al_{\lam\La_\lam \e}^*\om(A)-\al_{\lam \e}^*\om(A(x))|,
\label{translate1}
\eeqa
where $A\in\mfa(\mco)$. The first term on the r.h.s. of (\ref{translate1}) satisfies the bound
\beqa
& &|\om(\al_{\lam \e}A)-\om\big(\al_{\La_{\lam} }\al_{\lam \e}\al_{\La^{-1}_{\lam} }A\big)|\non\\
& &\phantom{44}\leq|\al_{\lam \e}^*\om(A-\al_{\La^{-1}_{\lam}}A)|+|(\om-\al_{\La_{\lam}}^*\om)
(\al_{\lam \e}\al_{\La_{\lam }^{-1}}A)|\non\\
& &\phantom{44}\leq \|P_{E}(A-\al_{\La_{\lam }^{-1}}A)P_{E}\|+
\sup_{s\in\real_+}\sup_{B\in\mfa(\widetilde{\mco}+s\e)_1}|\om(B)-\al_{\La_{\lam }}^*\om(B)| \, \|A\|,\label{translates}
\eeqa
where $\widetilde{\mco}$ is a slightly larger region than $\mco$.
Making use of Condition $R$ and of the fact that $\La_{\lam}\to I$ for $\lam\to\infty$,
we obtain that the last term above tends to zero in this limit. In view of our general
continuity assumption on the group of automorphisms $\al$ from Section~\ref{AQFT}, the expression
$(A-\al_{\La_{\lam }^{-1}}A)$ tends to zero with $\lam\to\infty$ in the 
strong operator topology. By compactness of the maps $\whXi_E$, the first term on the r.h.s.
of (\ref{translates}) tends to zero as well.
The second term on the r.h.s. of (\ref{translate1}) converges to zero by compactness of the maps $\whXi_E$
and the following bound:
\beqa
& &|\al_{\lam\La_\lam \e}^*\om(A)-\al_{\lam \e}^*\om(A(x))|=
|\om\big(A\big(\lam\sqrt{1+\big(v/\lam\big)^2 }\e+x\big)-A(\lam \e+x)\big)|\non\\
& &\phantom{44444444444444} \leq\|P_{E}\big(A\big(\big\{\sqrt{1+\big(v/\lam\big)^2}+1\big\}^{-1}(v^2/\lam)  \e\big)-A\big)P_{E}\|.
\eeqa  
Thus we have demonstrated that  $\om_0(A)=\om_0(A(x))$ for any local operator $A$.
This result extends by continuity to any $A\in\mfa$. \qed\\ 
Summing up, we have shown that there exist vacuum states in any theory satisfying Condition~$\Cs$.
These vacuum states can be constructed by means of large spacelike or timelike translations of
physical states. (It is an interesting open question if this result holds also for lightlike directions).
However, the uniqueness of the energetically accessible vacuum state does not seem to follow
from this criterion, since the concept of compactness is compatible with many limit points.
Our next task is to find a quantitative variant of Condition~$\Cs$ which entails also this property.
To this end, in the next section we cast this criterion in a form appropriate for a description of
coincidence measurements. Then, in Section~\ref{vacuum-structure}, we propose a strengthened, quantitative
form of Condition~$\Cs$. It accounts for the fact that if the number of separated observables, with vanishing
vacuum observables, is larger than the number of localization centers forming the state under study, 
then the result of the coincidence measurement should be zero.

\section{Condition $\Cs$: Coincidence Measurement Formulation}\label{coincidence-measurement}
To formulate a notion of compactness which is adequate for a description of coincidence arrangements of 
detectors, we need to extend our framework: Let $\Ga$ be a set and let $\lin(V\times\Ga,W)$\label{symbol-lin-times} be the space of 
maps $\Pi$ from $V\times\Ga$ to $W$, linear in the first argument, which are bounded in the norm
\beq
\|\Pi\|=\sup_{\su{v\in V_1\\ x\in\Ga} }\|\Pi(v,x)\|. 
\eeq
The subspace of finite rank maps $\fin(V\times\Ga,W)$\label{symbol-fin-times} contains
all the maps of the form $F=\sum_{i=1}^n\tau_i\,S_i$, where $\tau_i\in W$, $S_i\in\lin(V\times\Ga,\complex)$.
We say that a map $\Pi\in\lin(V\times\Ga,W)$ is compact, if it belongs to the closure of $\fin(V\times\Ga,W)$
in the norm topology of $\lin(V\times\Ga,W)$.

As we are going to consider coincidence measurements, we choose as the target space $W$ the
Banach space $\Ba$\label{symbol-Ba}  of $N$-linear forms on $\mfa(\mco)$ equipped  with  the norm
\beq
\|\psi\|=\sup_{\su{ A_i\in\mfa(\mco)_{1} \\ i\in\{ 1,\ldots, N \}  } }|\psi(A_1\times\cdots\times A_N)|. \label{strangenorm}
\eeq
In order to control the minimal distance  between the regions in which the measurements are performed, we define the set of admissible translates of the region~$\mco$
\beq
\Gad=\{\,\bx=(\vx_1,\ldots, \vx_N)\in\real^{Ns} \, | \,  \forall_{t\in]-\de,\de[, i\neq j} \ \mco+\vx_i\sim \mco+\vx_j+t\hat{e}_0\,\},
\label{symbol-Gad}
\eeq
where the symbol $\sim$ indicates spacelike separation and $\hat{e}_0$ is the unit vector in the time direction.
For any $\bx\in\Gad$ and $\fun\in\traceE$ we introduce the following  elements of $\Ba$
\beq
\fun_{\bx}(\AN)=\fun(A_1(\vx_1)\ldots A_N(\vx_N)), \label{forms}
\eeq
and consider the maps $\Pi_{E,N,\de}\in\lin(\traceE\times\Gad,\Ba)$  given by
\beq
\Pi_{E,N,\de}(\fun,\bx)=\fun_{\bx}. \label{symbol-PiENd}
\eeq
The following theorem provides a reformulation of Condition~$\Cs$ in terms of these maps.
\bet\label{equivalence} A theory satisfies Condition~$\Cs$ if and only if the maps $\Pi_{E,N,\de}$ are compact 
for any $E\geq 0$, $N\in\nat$, $\de>0$ and  double cone $\mco$.
\eet
\nin The 'if' part of the statement holds due to the identity $\Pi_{E,1,\de}=\Pi_{E}$.
The opposite implication is more interesting. It says that the restriction imposed by
Condition~$\Cs$ on the number of states which can be distinguished by measurements with
singly-localized detectors limits also the number of states which can be discriminated 
by coincidence arrangements of such detectors. We start our analysis from the observation 
that the spatial distance between the detectors suppresses the energy transfer between them. 
The proof of the following lemma relies on methods from \cite{BY87}. 
\bel\label{mollifiers} Let $\de>0$, $\be>0$. Define the function $g:\,\, ]-\pi,\pi]\to \complex$
as follows
\beq
g(\vpp)=\fr{\be}{\pi}\ln|\cot\fr{\vpp+\alp}{2}\cot\fr{\vpp-\alp}{2}|,
\eeq
where $\alp=2\arctan e^{-\fr{\pi\de}{2\be}}$. Then, for any pair of bounded operators
$A$, $B$, satisfying $[A(t),B]=0$ for $|t|<\de$, and any functional $\fun\in e^{-\be H}\trace e^{-\be H}$
there holds the identity
\beq
\fun(AB)=\fun([A, \cB_\be]_+)+\fun(Ae^{-\be H}\hB_\be e^{\be H})+
\fun(e^{\be H}\hB_\be e^{-\be H} A),
\label{claim}
\eeq
where $[\,\cdot\, , \,\cdot\, ]_+$ denotes the anti-commutator and we made use of the fact that 
$\fun(e^{\be H}\,\,\cdot\,\,)$, $\fun(\,\,\cdot\,\, e^{\be H})$ are elements of $\trace$. 
Here $\cB_\be$ and  $\hB_\be$ are elements of $B(\hil)$ given by the (weak) integrals
\beqa
\cB_\be&=&\fr{1}{2\pi}\int_{0}^{\alp}d\vpp \ B(g(\vpp))+\fr{1}{2\pi}\int_{\pi-\alp}^{\pi}d\vpp \ B(g(\vpp)),
\label{Bb}\\
\hB_\be&=&\fr{1}{2\pi}\int_\alp^{\pi-\alp}d\vpp \ B(g(\vpp)),\label{Bb1}
\eeqa
where $B(g(\vpp))=e^{ig(\vpp)H}Be^{-ig(\vpp)H}$.
\eel
\proof It suffices to prove the statement for functionals of the form $\fun(\,\cdot\,)=(\Psi_1|\,\cdot\,\Psi_2)$,
where $\Psi_1$ and $\Psi_2$ are vectors from the domain of $e^{\be H}$.
For $\de>0$ and $\be>0$ we define the set 
\beq
G_{\be,\de}=\{ \, z\in \complex \, | \, |\I z|<\be \, \} \backslash \{ \, z \, | \, \I z=0, |\R z|\geq\de \,\}
\eeq
and introduce the following function, analytic on $G_{\be,\de}$ and continuous at its
boundary
\beqa
h(z)=\left\{ \begin{array}{ll}
(\Psi_1|Ae^{izH}Be^{-izH}\Psi_2) &\textrm{ for } 0<\I z<\be \\
(\Psi_1|e^{izH}Be^{-izH}A\Psi_2) &\textrm{ for } -\be<\I z< 0 \\
(\Psi_1|AB(z)\Psi_2)=(\Psi_1|B(z)A\Psi_2) &\textrm{ for } \I z=0 \textrm{ and } |\R z|<\delta.
\end{array} \right.
\eeqa
We make use of the following conformal mapping from the unit
disc  $\{\, w \, | \, |w|<1 \,\}$ to $G_{\be,\de}$ \cite{BY87}
\beq
z(w)=\fr{\be}{\pi}\big\{\ln\fr{1+we^{i\alp}}{1-we^{i\alp}}-\ln\fr{1-we^{-i\alp}}{1+we^{i\alp}}\big\}.
\eeq
Setting $w=re^{i\vpp}$, $0<r<1$, we obtain from the Cauchy formula
\beq
h(0)=\fr{1}{2\pi}\int_0^{2\pi}d\vpp \ h\big(z(re^{i\vpp})\big). \label{Cauchy-vacuum}
\eeq
Since $h(z)$ satisfies the following bound on the closure of $G_{\be,\de}$
\beq
|h(z)|\leq \|A\| \, \|B\| \, \|e^{\be H}\Psi_1\| \, \|e^{\be H}\Psi_2\|,
\eeq
we can, by the dominated convergence theorem, extend the path of
integration in (\ref{Cauchy-vacuum}) to the circle $r=1$. In this limit we have \cite{BY87}
\beqa
\R\, z(e^{i\vpp})&=& g(\vpp),\\
\I\, z(e^{i\vpp})&=&\left\{ \begin{array}{ll}
0 & \textrm{ if } |\vpp|<\alp \textrm{ or } \pi-\vpp<\alp \textrm{ or } \pi+\vpp<\alp \\
\be & \textrm{ if } \alp<\vpp <\pi-\alp \\
-\be & \textrm{ if } \alp<-\vpp <\pi-\alp.
\end{array} \right.
\eeqa
Consequently, we obtain from (\ref{Cauchy-vacuum})
\beqa
& &(\Psi_1|AB\Psi_2)\non\\
&=&\fr{1}{2\pi}\int_{0}^{\alp}d\vpp\,(\Psi_1|[A,B(g(\vpp))]_+\Psi_2)+\fr{1}{2\pi}\int_{\pi-\alp}^{\pi}d\vpp\, (\Psi_1|[A,B(g(\vpp))]_+\Psi_2)\non\\
&+&\fr{1}{2\pi}\int_\alp^{\pi-\alp}d\vpp\,
\bigg((\Psi_1|Ae^{-\be H}B(g(\vpp))e^{\be H}\Psi_2)+(e^{\be H}\Psi_1|B(g(\vpp))e^{-\be H}A\Psi_2)\bigg),\ \ \ \ \
\eeqa
what concludes the proof. \qed\\
From compactness of the maps $\Xi_{\be}$, given by (\ref{hhXi}), there also follows that the mappings
$\hXi_{\be_1,\be_2}:\mfa(\mco)\to B(\hil)$, defined as
\beq
\hXi_{\be_1,\be_2}(A)=e^{-\be_1 H}A_{\be_2}e^{-\be_1 H}, \label{hXi1}
\eeq
are compact for any $\be_1,\be_2>0$ and any double cone $\mco$. Here $A_{\be_2}$ is defined as
in (\ref{Bb1}). After this preparation we are ready to complete the proof of Theorem~\ref{equivalence}.\\
\bf Proof of Theorem \ref{equivalence}:\rm \\
For any $\be>0$ we introduce the auxiliary maps $\hPi_{\be,N,\de}\in\lin(\trace\times\Gad,\Ba)$  given by
\beq
\hPi_{\be,N,\de}(\fun,\vxb)(A_1\times\cdots\times A_N)=\fun(e^{-(N+\h)\be H}A_1(\vx_1)\ldots A_N(\vx_N)
e^{-(N+\h)\be H}).
\eeq
They are related to the maps $\Pi_{E,N,\de}\in\lin(\traceE\times\Gad,\Ba)$ by the following identity,
valid for any $\fun\in\traceE$
\beq
\Pi_{E,N,\de}(\fun,\vxb)=\hPi_{\be,N,\de}(e^{(N+\h)\be H}\fun e^{(N+\h)\be H},\vxb). \label{PihPi}
\eeq 
In order to prove compactness of the maps $\Pi_{E,N,\de}$, it suffices to verify that
the family of mappings $\{\hPi_{\be,N,\de}\}_{\be>0}$ 
is \emph{asymptotically compact} in the following sense: There exists a family of finite rank maps 
$\hF_{\be,N,\de}\in\fin(\trace\times\Gad,\Ba)$ s.t.
\beq
\lim_{\be\to 0}\|\hPi_{\be,N,\de}-\hF_{\be,N,\de}\|=0. \label{AC}
\eeq
If this property holds, then, by identity (\ref{PihPi}), the maps $\Pi_{E,N,\de}$ can be approximated
in norm as $\be\to 0$ by the finite rank maps $F_{\be,N,\de}\in\lin(\traceE\times\Gad,\Ba)$ defined as
\beq
F_{\be,N,\de}(\fun,\vxb)=\hF_{\be,N,\de}(e^{(N+\h)\be H}\fun e^{(N+\h)\be H},\vxb).
\eeq
We establish  property~(\ref{AC}) by induction in $N$:
For $N=1$ the statement follows from compactness of the map $\hXi_{\fr{3}{2}\be}$ given by~(\ref{hXi}).
Next, we assume that the family  $\{\hPi_{\be,N-1,\de}\}_{\be>0}$ is asymptotically compact and prove that
$\{\hPi_{\be,N,\de}\}_{\be>0}$ also has this property. For this purpose we pick $\fun\in\trace_1$, 
$A_1,\ldots, A_N\in\mfa(\mco)_1$ and $\vxb\in\Gad$. Then $A_1(\vx_1)\ldots A_{N-1}(\vx_{N-1})$ and $A_N(\vx_N)$ satisfy the assumptions of Lemma~\ref{mollifiers} and  we obtain
\beqa
& &\hPi_{\be,N,\de}(\fun,\vxb)(A_1\times\cdots\times A_N)\non\\
& &=\fun(e^{-(N+\h)\be H}[A_1(\vx_1)\ldots A_{N-1}(\vx_{N-1}), \cA_{N,N\be}(\vx_N)]_+ e^{-(N+\h)\be H})\non\\
& &+\hPi_{\be,N-1,\de}\big(\{\hXi_{\h\be,N\be}(A_N)(\vx_N)\,\fun\, e^{-\h\be H}\},\vx_1,\ldots,\vx_{N-1}\big)
(A_1\times\cdots\times A_{N-1})\non\\
& &+\hPi_{\be,N-1,\de}\big(\{ e^{-\h\be H}\,\fun\,\hXi_{\h\be,N\be}(A_N)(\vx_N)\},\vx_1,\ldots,\vx_{N-1}\big)
(A_1\times\cdots\times A_{N-1}),\,\,\,\,\,\,\, \label{inductive-step}
\eeqa
where $[\,\cdot,\cdot\,]_+$ denotes the anti-commutator.
The first term on the r.h.s. of (\ref{inductive-step}) satisfies 
\beq
|\fun(e^{-(N+\h)\be H}[A_1(\vx_1)\ldots A_{N-1}(\vx_{N-1}), \cA_{N,N\be}(\vx_N)]_+ e^{-(N+\h)\be H})|\leq \fr{2\alp(\be)}{\pi},
\label{rest1}
\eeq
where we made use of  definition (\ref{Bb}). We recall from the statement
of Lemma~\ref{mollifiers} that $\alp(\be)\to 0$ with $\be\to 0$. To treat the remaining terms we make use of the induction hypothesis: It assures that there exist finite rank mappings $\hF_{\be,N-1,\de}\in\fin(\trace\times\Ga_{N-1,\de},\BB)$ s.t.
\beq
\lim_{\be\to 0}\|\hPi_{\be,N-1,\de}-\hF_{\be,N-1,\de}\|=0. \label{decay1}
\eeq
Next, making use of compactness of the maps $\hXi_{\h\be,N\be}\in\lin(\mfa(\mco),B(\hil))$, given by (\ref{hXi1}), we can find
a family of finite rank mappings $F_{\be}\in\fin(\mfa(\mco),B(\hil))$ s.t.
\beq
\|\hXi_{\h\be,N\be}-F_{\be}\|\leq \fr{\be}{1+\|\hF_{\be,N-1,\de}\| }, \label{decay2}
\eeq
for any $\be>0$. Now the second term on the r.h.s. of (\ref{inductive-step}) can be rewritten as follows
\beqa
& &\hPi_{\be,N-1,\de}(\{\hXi_{\h\be,N\be}(A_N)(\vx_N)\,\fun\, e^{-\h\be H}\},\vx_1,\ldots,\vx_{N-1})\non\\
& &\ph{4444}=(\hPi_{\be,N-1,\de}-\hF_{\be,N-1,\de})\big(\{\hXi_{\h\be,N\be}(A_N)(\vx_N)\,\fun\, e^{-\h\be H}\},\vx_1,\ldots,\vx_{N-1}\big)\non\\
& &\ph{4444}+\hF_{\be,N-1,\de}\big(\{(\hXi_{\h\be,N\be}(A_N)-F_{\be}(A_N))(\vx_N)\,\fun\, e^{-\h\be H}\},\vx_1,\ldots,\vx_{N-1}\big)\non\\
& &\ph{4444}+\hF_{\be,N-1,\de}\big(\{F_{\be}(A_N)(\vx_N)\,\fun\, e^{-\h\be H}\},\vx_1,\ldots,\vx_{N-1}\big).\label{inductive-step1}
\eeqa
We obtain from relations (\ref{decay1}) and (\ref{decay2}) that the first two terms on the r.h.s. 
of (\ref{inductive-step1}) tend to zero with $\be\to 0$ in the norm topology of $\BB$,
uniformly in  $\fun\in\trace_1$, $A_N\in\mfa(\mco)_1$ and $\vxb\in\Gad$.
The last term on the r.h.s. of (\ref{inductive-step1}) coincides with the finite rank  map 
$\hF_{\be,N,\de}^{(1)}\in\fin(\trace\times\Gad,\Ba)$, given by
\beqa
& &\hF_{\be,N,\de}^{(1)}(\fun,\vx)(A_1\times\cdots\times A_N)\non\\
& &\ph{44}=\hF_{\be,N-1,\de}(\{F_{\be}(A_N)(\vx_N)\,\fun\, e^{-\h\be H}\},\vx_1,\ldots,\vx_{N-1})(A_1\times\cdots\times A_{N-1}).\,\,\,\,\,
\eeqa
The last term on the r.h.s. of (\ref{inductive-step}) can be analogously approximated by the maps
$\hF_{\be,N,\de}^{(2)}\in\fin(\trace\times\Gad,\Ba)$ defined as
\beqa
& &\hF_{\be,N,\de}^{(2)}(\fun,\vx)(A_1\times\cdots\times A_N)\non\\
& &\ph{44}=\hF_{\be,N-1,\de}(\{e^{-\h\be H}\,\fun\,F_{\be}(A_N)(\vx_N)\} ,\vx_1,\ldots,\vx_{N-1})(A_1\times\cdots\times A_{N-1}).\,\,\,\,\,
\eeqa
Summing up, we obtain from (\ref{inductive-step}) and  (\ref{rest1}) that 
\beq
\lim_{\be\to 0}\|\hPi_{\be,N,\de}-\hF_{\be,N,\de}^{(1)}-\hF_{\be,N,\de}^{(2)}\|=0,
\eeq
what concludes the inductive argument and the proof of Theorem \ref{equivalence}. \qed
\section{Condition $\Csq$ and Uniqueness of Vacuum}\label{vacuum-structure}
Theorem~\ref{equivalence} opens the possibility to encode the physically expected behavior of 
coincidence arrangement of detectors into the phase space structure of a theory. For this
purpose we first calibrate the observables appropriately: Adopting Condition $\Cs$, from 
Proposition~\ref{ergodic-theorem} and
Theorem~\ref{disconnected1} we obtain a projection $P_{\{0\}}(\,\cdot\,)=\om_0(\,\cdot\,)I$ on the
pure-point subspace $\hmfa_{\spp}$, where $\om_0$ is some energetically accessible vacuum state. As in Section~\ref{space-translations-QFT},
we define the continuous subspace $\hmfa_{\scc}$ as the kernel of $P_{\{0\}} $ and introduce, for
any open, bounded region $\mco$, the local continuous subspace
\beq
\mfa_{\scc}(\mco)=\{\, A\in\mfa(\mco)\,|\, \om_0(A)=0\,\},
\eeq
whose elements will be called the local detectors.
We introduce the Banach space $\Bac$ of $N$-linear forms on $\mfa_{\scc}(\mco)$
equipped with the norm
\beq
\|\psi\|=\sup_{\su{ A_i\in\mfa_{\scc}(\mco)_{1} \\ i\in\{ 1,\ldots, N \}  } }|\psi(A_1\times\cdots\times A_N)|. \label{strangenorm1}
\eeq
We also define the maps $\Pi_{E,N,\de}^{\scc}\in\lin(\traceE\times\Gad,\Bac)$ given by
\beq
\Pi_{E,N,\de}^{\scc}(\fun,\vxb)=\Pi_{E,N,\de}(\fun,\vxb)|_{\mfa_{\scc}(\mco)^{\times N}}. \label{symbol-PiENdc}
\eeq
With the help of Theorem~\ref{equivalence} we can reformulate Condition~$\Cs$ in terms of these maps.
\bet\label{sharp-nat} A theory satisfies Condition~$\Cs$ if and only if there exists an energetically
accessible  vacuum state $\om_0$ and the corresponding maps $\Pi_{E,N,\de}^{\scc}$ are compact for 
any $E\geq 0$, $N\in\nat$, $\de>0$ and any double cone $\mco$.
\eet
\proof The 'only if' part follows immediately from Theorem~\ref{equivalence}. The opposite
implication can be shown as follows: From compactness of the map $\Pi_{E,1,\de}^{\scc}$
we obtain, for any $\eps>0$, a finite rank map $F\in\fin(\traceE\times\real^s,\mfa_{\scc}(\mco)^*)$
s.t.
\beq
\sup_{\su{(\fun,\vx)\in\traceEB\times\real^s \\  A\in\mfa_{\scc}(\mco)_1  }} |\Pi_{E,1,\de}^{\scc}(\fun,\vx)(A)-F(\fun,\vx)(A)|
\leq\eps.
\eeq
Noting that $\Pi_{E,1,\de}^{\scc}(\fun,\vx)=\Pi_E(\fun_{\vx})|_{\mfa_{\scc}(\mco)}$ and making use
of the fact that $\h(A-\om_0(A)I)\in\mfa_{\scc}(\mco)_1$ for any $A\in\mfa(\mco)$, we obtain
\beq
\sup_{\su{\fun\in\traceEB \\ A\in\mfa(\mco)_1}} 
|\Pi_{E}(\fun)(A)-\fun(I)\om_0(A)-F(\fun,0)(A-\om_0(A)I)|\leq 2\eps.
\eeq
Thus we can approximate the maps $\Pi_{E}$ in norm with finite rank mappings i.e.
these maps are compact. \qed\\
In order to strengthen Condition $\Cs$,
we note that any functional from $\traceE$ should consist of a finite number of distinct localization centers. 
Indeed, in a theory describing  particles of mass $m>0$ the maximal number of such centers $N_0(E)$ is given, 
essentially, by $\fr{E}{m}$. If the number of detectors $N$ is larger than $N_0(E)$, then at least one of them
should give no response (since its vacuum expectation value vanishes) and the result of the entire 
coincidence measurement should be zero.

In order to formulate this observation mathematically, we adapt the concept of the 
$\eps$-content \label{epsilon-content} to the present framework: Let $V$, $W$ be Banach spaces and let 
$\Ga$ be some set. Then the $\eps$-content of a map $\Pi\in\lin(V\times\Ga,W)$
is the maximal natural number $\N(\eps)$ for which there exist elements 
$(v_1,x_1),\ldots,(v_{\N(\eps)},x_{\N(\eps)})\in V_1\times\Ga$ s.t.
$\|\Pi(v_i,x_i)-\Pi(v_j,x_j)\|>\eps$ for $i\neq j$. Clearly, this quantity is finite
if the map $\Pi$ is compact.

The $\eps$-content $\N(\eps)_{E,N,\de}^{\scc}$ of the map $\Pi_{E,N,\de}^{\scc}$ gives the number of
distinct measurement results, up to experimental accuracy $\eps$, which can be obtained 
in the coincidence measurement described above. Therefore
we expect that $\N(\eps)_{E,N,\de}^{\scc}=1$ if $N$ is sufficiently large and the local detectors 
are far apart, arriving at the following strengthened, quantitative variant of Condition~$\Cs$:
\begin{enumerate}
\item[ ] \bf Condition \rm $\Csq$ \label{cond-Csq}
\item[(a)] There exists an energetically accessible vacuum state $\om_0$ and the corresponding maps $\Pi_{E,N,\de}^{\scc}$ 
are compact for any $E\geq 0$, $N\in\nat$, $\de>0$ and double cone $\mco$.
\item[(b)] For any $E\geq 0$ there exists  $N_0(E)\in\nat$ s.t. for any $N>N_0(E)$  the $\eps$-contents
$\N(\eps)^{\scc}_{E,N,\de}$ of the maps $\Pi_{E,N,\de}^{\scc}$ satisfy 
\beq
\lim_{\de\to\infty}\N(\eps)_{E,N,\de}^{\scc}=1\quad\textrm{ for any }\quad \eps>0.
\eeq
\end{enumerate}
We verify in Appendix~\ref{Condition-Csq} that this condition holds in massive scalar free field theory and
its even part.

In the remaining part of this section 
we show that the vacuum state $\om_0$, which entered into
the formulation of Condition~$\Csq$, is pure and that it is the only energetically accessible vacuum
state. Then, from Propositions~\ref{space1} and \ref{time1} there follows convergence of physical states to $\om_0$ 
under large spacelike and timelike translations. Moreover, we demonstrate that this vacuum state appears as a 
limit of states of increasingly sharp energy-momentum values. To achieve these goals it is convenient to
reformulate Condition~$\Csq$ (b) in terms of norms of the maps $\Pi_{E,N,\de}$. This is accomplished
in the following lemma.
\bel\label{content} Let $W$ be a Banach space. Let $\{V_{\de}\}_{\de>0}$ and  $\{\Ga_{\de}\}_{\de>0}$ 
be a family of Banach spaces and sets, respectively, ordered by inclusion i.e. $V_{\de_1}\subset V_{\de_2}$ and $\Ga_{\de_1}\subset\Ga_{\de_2}$ for $\de_1\geq\de_2$. We consider a family
$\{\Pi_{\de}\}_{\de>0}$  of compact maps in $\lin(V_\de\times\Ga_\de,W)$ and their respective $\eps$-contents
$\N(\eps)_{\de}$. Then there holds $\lim_{\de\to\infty}\N(\eps)_{\de}=1$ for any $\eps>0$ if and only if $\lim_{\de\to\infty}\|\Pi_{\de}\|=0$.
\eel
\proof First, suppose that $\lim_{\de\to\infty}\N(\eps)_{\de}=1$ for any $\eps>0$. Since
the $\eps$-content takes only integer values, for any $\eps>0$ we can choose $\de_{\eps}$
s.t. $\N(\eps)_{\de}=1$ for $\de\geq\de_{\eps}$. Then, by definition of the $\eps$-content,
there holds for any $\de\geq\de_{\eps}$
\beq
\|\Pi_{\de}\|\leq\eps,
\eeq
what entails $\lim_{\de\to\infty}\|\Pi_{\de}\|=0$.

To prove the opposite implication we proceed by contradiction: We recall that the $\eps$-content
of a compact map is finite for any $\eps>0$. Next, we note that for any fixed $\eps>0$ 
the function  $\de\to\N(\eps)_{\de}$ is decreasing and bounded from below by one, so there exists $\lim_{\de\to\infty}\N(\eps)_{\de}$. Suppose that this limit is strictly  larger than one. Then, 
by definition of the $\eps$-content, there exist nets $(\fun_1^{(\de)},\vx_1^{(\de)})$ 
and $(\fun_2^{(\de)},\vx_2^{(\de)})$ in $V_{\de,1}\times\Ga_{\de}$ s.t.
\beq
\|\Pi_{\de}(\fun_1^{(\de)},\vx_1^{(\de)})-\Pi_{\de}(\fun_2^{(\de)},\vx_2^{(\de)})\|>\eps
\eeq
for any $\de>0$. This inequality, however, contradicts the assumption that the norms
of the maps $\Pi_{\de}$ tend to zero with $\de\to\infty$. \qed\\
With the help of the  above  fact we restate  Condition $\Csq$ (b) as follows: For any $E\geq 0$ there exists
such natural number $N_0(E)$ that for any $N>N_0(E)$ there holds 
\beq
\lim_{\de\to\infty}\|\Pi_{E,N,\de}^{\scc}\|=0.
\eeq
We rewrite the above relation more explicitly 
\beq
\lim_{\de\to\infty}\sup_{\su{A_i\in\mfa_{\scc}(\mco)_{1} \\ i\in\{1,\ldots, N\} \\ \vxb\in\Gad \ } }
\|P_E A_1(\vx_1)\ldots A_N(\vx_N) P_E\|=0 \label{norms1}
\eeq 
and use it to prove the following lemma which is the main technical result of this section.
\bel\label{key} Suppose that Condition $\Csq$ holds.  Then, for any $E\geq 0$, double cone $\mco$, 
and a sequence $\{\de(\K)\}_1^{\infty}$ s.t. $\de(\K)\too\infty$, 
the following assertions hold true:
\begin{enumerate}
\item[(a)] For any family of points $\{\vx_i^{(\K)}\}_1^{\K}\in\Ga_{\K,\de(\K)}$ there holds 
\beq
\lim_{\K\to\infty}\sup_{\su{\fun\in\traceEB \\ A\in\mfa_{\scc}(\mco)_{1}}} \bigg|\fr{1}{\K}\sum_{i=1}^{\K}\fun(A(\vx_i^{(\K)}))\bigg|= 0.\label{average}
\eeq
\item[(b)] For any unit vector $\he\in\real^s$, sequence $\{\la^{(\K)} \}_1^{\infty}\in\real$ and a family 
of  points $\{\vx_i^{(\K)}\}_1^{\K}$,  s.t.
$\{\vx_i^{(\K)}\}_1^{\K}\cup\{\vx_i^{(\K)}+\la^{(\K)}\he\}_1^{\K}\in \Ga_{2\K,\de(\K)}$,
there holds
\beq
\lim_{\K\to\infty}\sup_{\su{\fun\in\traceEB \\ A,B\in\mfa_{\scc}(\mco)_{1}}} \bigg|\fr{1}{\K}\sum_{i=1}^{\K}\fun(A(\vx_i^{(\K)}) B(\vx_i^{(\K)}+\la^{(\K)}\he\ ))\bigg|=0.\label{average2}
\eeq
\end{enumerate}
\eel
\proof 
In view of  decomposition (\ref{decomp}) we can, without loss of generality, substitute $\traceEBP$  for $\traceEB$
in relations (\ref{average}) and (\ref{average2}). Similarly, we can replace $\mfa_{\scc}(\mco)_1$ with the
subset of its self-adjoint elements.

We choose some positive functional $\fun\in\traceEBP$, pick  $m\in\nat$ s.t.
$N=2^m$ is sufficiently large to ensure that (\ref{norms1}) holds.
To prove (a) we  define the operators
$Q_{\K}=\fr{1}{\K}\sum_{i=1}^{\K}A(\vx_i^{(\K)})$, $n\in\nat$, where $A\in\mfa_{\scc}(\mco)$ is self-adjoint,
assume that $\K\geq N$ and compute
\beqa
|\fun(Q_{\K})|^{N} &\leq& \fun(Q_{\K}^{N})=\fr{1}{\K^{N}}\sum_{i_1\ldots i_{N}}
\fun(A(\vx_{i_1}^{(\K)})\ldots A(\vx_{i_{N}}^{(\K)}))\phantom{4444444444}\non\\
&=&\fr{1}{\K^{N}}\sum_{\su{i_1\ldots i_{N} \\ \forall_{k\neq l} i_k\neq i_l  }}\fun(A(\vx_{i_1}^{(\K)})\ldots A(\vx_{i_{N}}^{(\K)}))
\non\\
&+& \fr{1}{\K^{N}}\sum_{\su{i_1\ldots i_{N} \\ \exists_{k\neq l} s.t. i_k=i_l}}
\fun(A(\vx_{i_1}^{(\K)})\ldots A(\vx_{i_{N}}^{(\K)}))\non\\
&\leq& 
\fr{1}{\K^{N}}\!\!\!\!\sum_{\su{i_1\ldots i_{N} \\ \forall_{k\neq l} i_k\neq i_l}}\!\!\!\|P_E A(\vx_{i_1}^{(\K)})\ldots A(\vx_{i_{N}}^{(\K)}) P_E\|+\fr{1}{\K^{N}}\!\!\!\!\sum_{\su{i_1\ldots i_{N} \\ \exists_{k\neq l} s.t. i_k=i_l}}\!\!\! \|A\|^{N}.
\eeqa
In the first step above we applied the Cauchy-Schwarz inequality and in the third step we extracted from the resulting
sum the terms in which all the operators are mutually spacelike separated. Clearly, there are 
${\K \choose N}N!\leq \K^N$ such terms. Therefore, the remainder (the last sum above) 
consists of
\beq
\K^N-{\K \choose N}N!\leq c_N \K^{N-1}
\eeq
terms, where $c_N$ is some constant independent of $\K$. There follows the estimate
\beq
|\fun(Q_{\K})|^{N}\leq 
\sup_{\su{A\in\mfa_{\scc}(\mco)_{1} \\ (\vx_1,\ldots,\vx_N)\in\Ga_{N,\de(\K)} } } \|P_E A(\vx_1)\ldots A(\vx_N) P_E\|+ \fr{c_N}{\K}\|A\|^{N}, \label{norms}
\eeq
whose r.h.s. tends to zero with $\K\to\infty$ by (\ref{norms1}), uniformly in $\fun\in\traceEBP$,  what concludes
the proof of (\ref{average}).

\nin In order to prove (b) we proceed similarly: Let $\hQ_{\K}=\fr{1}{\K}\sum_{i=1}^{\K}A(\vx_i^{(\K)})B(\vx_i^{(\K)}+\la^{(\K)}\he)$,
where $A, B\in\mfa_{\scc}(\mco)$ are self-adjoint. Then,  for $\K\geq N$, we obtain
\beqa
& &|\fun(\hQ_{\K})|^{N}\leq \fun(\hQ_{\K}^{N})\non\\
&=&\fr{1}{\K^{N}}\sum_{i_1\ldots i_{N}}\fun(A(\vx_{i_1}^{(\K)})B(\vx_{i_1}^{(\K)}+\la^{(\K)}\he)\ldots A(\vx_{i_{N}}^{(\K)})B(\vx_{i_{N}}^{(\K)}+\la^{(\K)}\he)) \non\\
&=&\fr{1}{\K^{N}}\sum_{\su{i_1\ldots i_{N}\\ \forall_{k\neq l} i_k\neq i_l  }}
\fun(A(\vx_{i_1}^{(\K)})B(\vx_{i_1}^{(\K)}+\la^{(\K)}\he)\ldots A(\vx_{i_{N}}^{(\K)})B(\vx_{i_{N}}^{(\K)}+\la^{(\K)}\he))\non\\
&+&\fr{1}{\K^{N}}\sum_{\su{i_1\ldots i_{N}\\ \exists_{k\neq l} s.t. i_k=i_l}} \fun(A(\vx_{i_1}^{(\K)})B(\vx_{i_1}^{(\K)}+\la^{(\K)}\he)\ldots A(\vx_{i_{N}}^{(\K)})B(\vx_{i_{N}}^{(\K)}+\la^{(\K)}\he))\non\\
&\leq& 
\fr{1}{\K^{N}}\sum_{\su{i_1\ldots i_{N} \\ \forall_{k\neq l} i_k\neq i_l}}
\|P_E A(\vx_{i_1}^{(\K)})B(\vx_{i_1}^{(\K)}+\la^{(\K)}\he)\ldots A(\vx_{i_{N}}^{(\K)})B(\vx_{i_N}^{(\K)}+\la^{(\K)}\he)P_E\|\non\\
&+&\fr{1}{\K^{N}}\sum_{\su{i_1\ldots i_{N}\\ \exists_{k\neq l} s.t. i_k=i_l}} (\|A\|\, \|B\|)^{N}.
\eeqa
By the same reasoning as in case (a) we obtain the estimate
\beqa
|\fun(\hQ_{\K})|^{N} &\leq& \sup_{\su{A\in\mfa_{\scc}(\mco)_{1} \\ (\vx_1,\ldots,\vx_{2N})\in\Ga_{2N,\de(\K)} } } 
\|P_E A(\vx_1)B(\vx_2)\ldots A(\vx_{2N-1})B(\vx_{2N}) P_E\|\non\\
&+&\fr{c_N}{\K} (\|A\|\,\|B\|)^{N}.
\eeqa
By taking the limit $\K\to\infty$ we conclude the proof of (\ref{average2}). \qed\\
Now we are ready to prove that the vacuum state $\om_0$ is pure and unique in the energetically connected component of the
state space.
\bet\label{main} Suppose that Condition $\Csq$ is satisfied. Then there hold the following assertions:
\begin{enumerate}
\item[(a)] Let $\om\in\mfa^*$ be a state in the weak$^*$-closure of $\traceEB$ for some $E\geq 0$ which is invariant
under translations in some spacelike ray. Then $\om=\om_0$. In particular, there holds Condition~$\V$ stated in 
Section~\ref{space-translations-QFT}.
\item[(b)]  $\om_0$ is a pure state.
\end{enumerate}
\eet
\proof 
(a) By Theorem~\ref{disconnected1}, $\om$ is a vacuum state, in particular it is translationally
invariant. Let $\{\fun_{\be}\}_{\be\in I}$ be a net of functionals from $\traceEB$ approximating $\om$ in the weak$^*$
topology  and  let $A\in\mfa_{\cc}(\mco)$ i.e. $\om_0(A)=0$. We choose  families
of points $\{\xx_i\}_1^{\K}$ in $\real^s$ s.t. $\{\xx_i\}_1^{\K}\in\Ga_{K,\de(\K)}$ for some 
sequence $\{\de(\K)\}_1^{\infty}$ which diverges to infinity with $\K\to\infty$. We note the
following relation
\beqa
|\om(A)|&=&|\fr{1}{\K}\sum_{i=1}^{\K}\om(A(\xx_{i}^{(\K)}))|=\lim_{\be}|\fr{1}{\K}
\sum_{i=1}^{\K}\fun_{\be}(A(\xx_{i}^{(\K)}))|\non\\
        &\leq&\sup_{\fun\in\traceEB}|\fr{1}{\K}\sum_{i=1}^{\K}\fun(A(\xx_{i}^{(\K)}))|_{\su{ \\ \\ \ph{4} \K\to\infty }}
\!\!\!\!\!\!\!\!\!\!\!\!\!\!\longrightarrow 0,
\eeqa
where in the first step we made use of the fact that the state $\om$ is invariant under translations
in  space and in the last step we made use of Lemma \ref{key} (a). Since
local algebras are norm dense in the global algebra $\mfa$, we conclude that $\ker\om_0\subset\ker\om$
and therefore the two states are equal.\\
(b) In order to show purity of $\om_0$, it suffices to verify that for any $A, B\in\mfa_{\cc}(\mco)$, some
unit vector $\he\in\real^s$ and some sequence of real numbers $\{\la^{(\K)}\}_1^{\infty}$ s.t. 
$\la^{(\K)}\too\infty$
there holds
\beq
\lim_{\K\to\infty}\om_0(AB(\la^{(\K)}\he))=0. \label{clustering}
\eeq
To this end, we recall that the vacuum state $\om_0$ is energetically accessible i.e. we can pick a net $\{\fun_{\be}\}_{\be\in I}$ of functionals from $\traceEB$, approximating $\om_0$ in the weak$^*$
topology. Next, we choose families of points $\{\xx_i^{(\K)}\}_1^{\K}$ as in  part (b) of Lemma \ref{key} and compute
\beqa
|\om_0(AB(\la^{(\K)}\he))|&=&|\fr{1}{\K}\sum_{i=1}^{\K}\om_0(A(\xx_i^{(\K)})B(\xx_i^{(\K)}+\la^{(\K)}\he))|\non\\
&=&\lim_{\be}|\fr{1}{\K}\sum_{i=1}^{\K}\fun_{\be}(A(\xx_i^{(\K)})B(\xx_i^{(\K)}+\la^{(\K)}\he))|\non\\
&\leq& \sup_{\fun\in\traceEB}|\fr{1}{\K}\sum_{i=1}^{\K}\fun(A(\xx_i^{(\K)})B(\xx_i^{(\K)}+\la^{(\K)}\he))|_{\su{ \\ \\ \ph{4} \K\to\infty }}
\!\!\!\!\!\!\!\!\!\!\!\!\!\!\longrightarrow 0,
\eeqa
what proves relation (\ref{clustering}). \qed\\
From part (a) of the above theorem and from Propositions~\ref{space1} and \ref{time1} we readily
obtain convergence of states of bounded energy to the vacuum state $\om_0$ under large spacelike 
or timelike translations.
\bec\label{relaxation-to-vacuum} Let Condition $\Csq$  be satisfied. Then, for any state $\om\in\traceE$, $E\geq 0$, and 
a spacelike  unit vector $\he\in\real^{s+1}$, there holds
\beq
\lim_{\la\to\infty}\om_{\la\he}(A)=\om_0(A) \textrm{ for } A\in\mfa.
\eeq
If, in addition, there holds Condition $R$, then the above relation is also true for any timelike unit vector $\he$.
\eec
To conclude this survey of applications of Condition $\Csq$, let us mention 
another physically meaningful procedure for preparation of vacuum states: It is to construct functionals with 
increasingly sharp values of energy and momentum and exploit the uncertainty principle.
Let $\Ppr$\label{symbol-Ppr} be the spectral projection corresponding to the ball of radius $r$ centered around some point $p$ in the energy-momentum spectrum and $\trace_{(p,r)}=\Ppr\trace\Ppr$. \label{symbol-traceppr} Then, in a theory satisfying Condition~$\Csq$,  any sequence of states $\om_r\in\trace_{(p,r)}$ converges, uniformly on local algebras, to the vacuum state $\om_0$ as $r\to 0$, since this is the only energetically accessible state which is completely dislocalized in spacetime. This fact is reflected in the following proposition:
\bep\label{shrinking} Suppose that Condition $\Csq$ is satisfied. Then, for any $p\in\cone$ and
double cone $\mco$, there holds
\beq
\lim_{r\to 0}\sup_{\su{\fun\in\trace_{(p,r),1} \\ A\in\mfa(\mco)_{1}} }|\fun(A)-\fun(I)\om_0(A)|=0.
\eeq
\eep
\proof 
We pick $A\in B(\hil)$, $\fun\in\trace_{(p,r)}$, $\xx\in\real^{s}$ and estimate the deviation of this functional 
from translational invariance
\beqa
|\fun(A)-\al^*_{\xx}\fun(A)|&=&|\fun(\Ppr A \Ppr)-\fun(\Ppr e^{i(\vec{P}-\vep)\xx}A e^{-i(\vec{P}-\vep)\xx}\Ppr)|\non\\
&=&|\fun(\Ppr e^{i(\vec{P}-\vep)\xx}A(1-e^{-i(\vec{P}-\vep)\xx})\Ppr)\non\\ 
&+&\fun(\Ppr(1-e^{i(\vec{P}-\vep)\xx})A\Ppr)|\leq 2\|\fun\|\,\|A\|\,|\xx|\,\rad, \label{spectral}
\eeqa
where in the last step we applied the spectral theorem. Consequently, for any
$\xx_1,\ldots, \xx_{\K}\in\real^{s}$ and $A\in\mfa(\mco)$ there holds
\beqa
|\fun(A)-\fun(I)\om_0(A)|\leq\bigg|\fr{1}{\K}\sum_{k=1}^{\K}[\fun(A)-\fun(A(\xx_k))]\bigg|+
\bigg|\fr{1}{\K}\sum_{k=1}^{\K}\fun(A(\xx_k)-\om_0(A)I)\bigg|& &\non\\
\leq \|A\| \sup_{k\in\{1,\ldots,N\}} \|\fun-\al_{\xx_k}^*\fun\|\,+2\|A\|\sup_{B\in\mfa_{\cc}(\mco)_{1}}|\fr{1}{\K}\sum_{k=1}^{\K}\fun(B(\xx_k))|.
& &\eeqa
Applying to the first term on the r.h.s. estimate (\ref{spectral}) and to the second term
Lemma~\ref{key}~(a) we obtain the statement of the proposition. \qed
\section{Condition $\Cnat$: Additivity of Energy}\label{Condition-Cnat-and-additivity}
In the previous section we introduced the phase space condition $\Csq$, inspired by the physically
expected behavior of coincidence arrangements of local detectors. We showed that this condition implies the
uniqueness and purity of the energetically accessible vacuum state as well as various approximation
procedures for this state. However, its status in massless theories is not clear. Since such theories play
an important role in our study of spectral theory of automorphism groups in Chapter~\ref{chapter:spectral}, 
we think it is worthwhile to fill this gap. For this purpose we introduce a different phase space 
condition~$\Cnat$ stated below. We note that in contrast to Condition~$\Csq$ the vacuum state does not enter 
into the formulation of the present condition. Thus Condition~$\Cnat$ is a property of the local net and not of a particular vacuum state,
what is certainly an advantage. This criterion is motivated by a heuristic argument based on additivity
of energy over isolated subregions and it is shown that it has all the consequences mentioned in Section~\ref{vacuum-structure} (except for purity of the vacuum). In Section~\ref{additivity-of-energy} and Appendix~\ref{Condition-Nnat} we verify that this condition holds in massive and massless scalar free field theory. In Chapter~\ref{chapter:conclusions} we discuss its  generalizations
which may be useful for the problem of convergence of asymptotic functional approximants.

The concept of additivity of energy does not have an unambiguous meaning in the general framework of 
local relativistic quantum field theory and we rely here on the following formulation: We introduce the 
family of maps $\The_{E,N,\de}\in\lin(\traceE\times\Gad,\mfa(\mco)^*\ot\compsup)$, 
given by
\beq
\The_{E,N,\de}(\fun,\vxb)=\big(\PiE(\fun_{\xx_1}),\ldots,\PiE(\fun_{\xx_N}) \big), \label{theta}
\eeq
where $\compsup$\label{symbol-compsup} denotes the space $\complex^N$ with the norm $\|\vxz\|=\sup_{k\in\{1,\ldots,N\}}|z_k|$.
We note that $\mfa(\mco)^*\ot\compsup$, equipped with the norm, 
\beq
\|(\fun_1,\ldots,\fun_N)\|=\sup_{k\in \{1,\ldots,N\}}\|\fun_k\| \label{compsup-topology}
\eeq
is a Banach space.
It is clear that the map $\The_{E,N,\de}$ is compact in theories satisfying Condition~$\Cs$. We claim that a mild (polynomial) growth of the $\eps$-contents
$\N(\eps)_{E,N,\de}$ of these maps with $N$, (when $\de$ tends to infinity), is a signature of additivity of energy over isolated subregions. In order to justify this formulation we provide a heuristic argument: 

The physical meaning of the maps $\The_{E,N,\de}$ is most easily elucidated if we consider their restrictions
to $S_E\times\Gad$. Thus we are interested in the $\eps$-contents $\tN(\eps)_{E,N,\de}$ of the 
sets\footnote{The $\eps$-content $\tN(\eps)$ of some set $S$ in a Banach space is the maximal number
of elements $\om_1,\ldots,\om_{\N(\eps)}\in S$ s.t. $\|\om_i-\om_j\|>\eps$ for $i\neq j$.\label{symbol-eps-content-set}}
\beq
S_{E,N,\de}(\mco)=\{\, (\PiE(\om_{\xx_1}),\ldots,\PiE(\om_{\xx_N})) \in \mfa(\mco)^*\ot\compsup\,|\, \om\in S_E,\,
\vxb\in\Gad\,\}.\label{symbol-SENde}
\eeq
Given a state $\om\in S_E$, we denote by $E_k$ the 'local energy content' of the restricted state $\om|_{\mfa(\mco+x_k)}$. Additivity of energy should then imply that $E_1+\cdots+E_N\leq E$ for large spacelike distance $\de$ between the regions $\mco+\vx_1,\ldots,\mco+\vx_N$, where $\vxb\in\Gad$. This suggests that to calculate
$\tN(\eps)_{E,N,\de}$ one should count all the families of states
$(\om_1,\ldots,\om_N)$, $\om_k\in S_{E_k}$, $E_1+\cdots+E_N\leq E$, which can be distinguished, up to
accuracy $\eps$, by measurements in $\mco+\vx_1,\ldots,\mco+\vx_N$. Relying on this heuristic reasoning we write
\beqa
\tN(\eps)_{E,N,\de}=\#\{\, (n_1,\ldots, n_N)\in\nat^{N}\, |\,
n_1\leq \tN(\eps)_{E_1},\ldots, n_N \leq\tN(\eps)_{E_N}, \non\\  
\textrm{ for some } E_1,\ldots,E_N\geq 0\textrm{ s.t. }  E_1+\cdots+E_N\leq E \,\}, \label{heuristic1}
\eeqa
where we made use of the fact that  the number of states from $S_{E_k}$ which can be discriminated, up to $\eps$, by observables localized in the region $\mco+x_k$ is equal to the $\eps$-content $\tN(\eps)_{E_k}$ of the set
\beq
S_{E_k}(\mco+\vx_k)=\{\, \om|_{\mfa(\mco+\vx_k)} \,|\, \om\in S_{E_k}\,\}. \label{set-SEm}
\eeq
Anticipating that $\tN(\eps)_{E_k}$ tends to one for small $E_k$, we may assume 
that
\beq
\tN(\eps)_{E_k}\leq 1+c_0(\eps, E)E_k \label{ff1}
\eeq
for $E_k\leq E$. This bound obviously holds in theories with lower mass gap, satisfying Condition~$\Cs$, where $S_E$
contains at most the vacuum state $(\vac|\cdot\vac)$ for sufficiently small $E$. (If the vacuum vector  $\vac$ exists, it must
be unique (up to phase) by the irreducibility assumption from Section~\ref{AQFT}. See e.g. Theorem~4.6 of \cite{Ar}).
We also expect that estimate~(\ref{ff1}) holds in massless theories, where de Broglie wavelengths of states
from $S_{E_k}(\mco+\vx_k)$ are much larger than the extent of the region $\mco+\vx_k$
if $E_k$ is sufficiently small. Thus the states should be indistinguishable by measurements in this
region, up to the experimental accuracy $\eps$. (The existence of  massless theories satisfying
the bound~(\ref{ff1}) is indicated in the last part of this section).
From the heuristic formula (\ref{heuristic1}) and the bound (\ref{ff1}) we obtain the estimate which grows only polynomially with $N$
\beqa
\tN(\eps)_{E,N,\de}\leq\#\{\, (n_1,\ldots, n_N)\in\nat^{N}\, |\, n_1+\cdots+n_N\leq N+c_0(\eps,E)E\, \}\non\\
\phantom{444444444444444444444444444444444444444444444444444}\leq (N+1)^{c_0(\eps,E)E},
\eeqa
where the last inequality can be verified by induction in $N$. Omitting the key condition $E_1+\cdots+E_N\leq E$ in (\ref{heuristic1}) and setting  $E_k=E$ instead, one would arrive at an exponential growth of $\tN(\eps)_{E,N,\de}$ as a function of $N$. Thus the moderate (polynomial) increase of this quantity with regard to $N$ is in fact a clear-cut signature of additivity of energy over isolated subsystems. 
We encode this fact into the following strengthened variant of Condition~$\Cs$.
\begin{enumerate}
\item[ ] \bf Condition $\Cnat$:\rm  \label{cond-Cnat}
\item[(a)] The maps $\The_{E,N,\de}$ are compact for any $E\geq 0$, $N\in\nat$, $\de>0$ and double cone~$\mco$.
\item[(b)] The $\eps$-contents $\N(\eps)_{E,N,\de}$ of the maps $\The_{E,N,\de}$ satisfy, for any $\eps>0$, 
\beq
\lim_{\de\to\infty}\N(\eps)_{E,N,\de}\leq (N+1)^{c(E,\eps)}, \label{addenergy}
\eeq
for some constant $c(E,\eps)$ independent of $N$.
\end{enumerate}
The remaining part of this section is devoted to the proof that Condition~$\Cnat$ has all
the consequences, pertaining to the vacuum structure, which were discussed in Section~\ref{vacuum-structure}
(apart from purity of the vacuum). For this purpose it suffices to show that the present condition
implies relation~(\ref{average}) for some specific sequence $\{\de_n\}_{n\in\nat}$ and some energetically 
accessible vacuum state $\om_0$. This goal will be
accomplished in the two lemmas below. Since no
distinguished vacuum state enters into Condition~$\Cnat$, we have to prepare such a state first:
We fix a  unit vector $\e$ in space direction and obtain from Proposition~\ref{space1} a net of real numbers 
$\{\la_\be\}_{\be\in\mathbb{I}}$ s.t. $\la_\be\to\infty$  and a vacuum state $\om_0$ s.t. for any $A\in\mfa$
\beq
\te{w$^*$-}\lim_{\be}A(\la_{\be}\e)=\om_0(A)I.
\eeq
We will call the triple $\{\, \e, \,\{\la_{\be}\}_{\be\in\mathbb{I}},\, \om_0\}$ a spacelike asymptotic vacuum state.

In the subsequent discussion we keep $E\geq 0$ and a double cone $\mco$ fixed. Moreover, for any $\om\in S_E$ and $\vxb\in\Gad$
we denote by $\om^{\vxb}$ the element of $S_{E,N,\de}(\mco)$ given by
\beq
\om^{\vxb}(A)=(\om(A(\xx_1)),\ldots,\om(A(\xx_N))),\quad\quad A\in\mfa(\mco).\label{symbol-omvxb}
\eeq
Furthermore, $\ov{S}_{E,N,\de}(\mco)$ denotes the closure of $S_{E,N,\de}(\mco)$ 
in $\mfa(\mco)^*\ot\compsup$ in the topology given by the norm~(\ref{compsup-topology}).
The following simple lemma summarizes the essential properties of the sets $\ov{S}_{E,N,\de}(\mco)$. 
\bel\label{properties} Assume that Condition $\Cs$ is satisfied. Let $\vpi=(\vpi_1,\ldots,\vpi_N)\in \ov{S}_{E,N,\de}(\mco)$,
let $\{\, \e, \,\{\la_{\be}\}_{\be\in\mathbb{I}},\, \om_0\}$ be a spacelike asymptotic vacuum state 
and $\PP_N$ the group of permutations of an $N$-element set. Then:
\begin{enumerate}
\item[(a)] $\vpi_\pi:=(\vpi_{\pi(1)},\ldots,\vpi_{\pi(N)})\in \ov{S}_{E,N,\de}(\mco)$, for
any $\pi\in \PP_N$.
\item[(b)] $\vpc:=(\vpi_1,\ldots,\vpi_{N-1})\in \ov{S}_{E,N-1,\de}(\mco)$.
\item[(c)] $\vph:=(\vpi_1,\ldots,\vpi_N,\underbrace{\oms,\ldots,\oms}_L)\in \ov{S}_{E,N+L,\de}(\mco)$.
\end{enumerate}
\eel
\proof To prove part (a), we first define the action of the group of
permutations on the sets $\Gad$. Given $\vxb=(\xx_1,\ldots, \xx_N)\in\Gad$ and
$\pi\in \PP_N$,
we set $\vxb_{\pi}=(\xx_{\pi(1)},\ldots, \xx_{\pi(N)})$
which is again an element of $\Gad$. This induces an action
of permutations on the sets $S_{E,N,\de}(\mco)$ in the obvious manner:
Given $\om^{\vxb}\in S_{E,N,\de}(\mco)$ we define $(\om^{\vxb})_{\pi}=\om^{\vxb_{\pi}}$.
Consequently, all the sets  $S_{E,N,\de}(\mco)$, $\de>0$ are invariant under the permutations
of the entries of their elements. This property carries over to their closures:
In fact given $\vpi\in\ov{S}_{E,N,\de}(\mco)$, there exists for every $\eps>0$ some 
$\om^{\vxb}\in S_{E,N,\de}(\mco)$ s.t. $\|\om^{\vxb}-\vpi\|\leq\eps$. Then $\|\om^{\vxb_\pi}-\vpi_{\pi}\|\leq\eps$.

Part (b) of the lemma follows directly from the definition of the sets $S_{E,N,\de}(\mco)$ and their closures.
In order to prove part (c) we pick again $\om^{\vxb}\in S_{E,N,\de}(\mco)$ s.t. $\|\om^{\vxb}-\vpi\|\leq\eps$. 
According to Lemma~\ref{local-normality} there exists a subsequence $\{\la_{n}\}_{n\in\nat}$ of 
$\{\la_{\be}\}_{\be\in\mathbb{I}}$ s.t. 
\beq
\lim_{n\to\infty}\om_{\la_n\e}(A)=\om_0(A),\quad\quad A\in\mfa. \label{spacelike-vacuum-convergence}
\eeq
We choose its subsequences  $\la^{(1)}_{n},\ldots,\la^{(L)}_{n}$ 
s.t. for $k\neq l$ there holds $|\la^{(k)}_{n}-\la^{(l)}_{n}|\to\infty$ when $n\to\infty$. 
Consequently,
\beq
\vph_n:=(\om_{\xx_1},\ldots,\om_{\xx_N}, \om_{\xx_1+\la^{(1)}_n \e}
,\ldots,\om_{\xx_1+\la^{(L)}_{n} \e} ) \in S_{E,N+L,\de}
\eeq
for sufficiently large $n$, where $\e$ is the unit vector in a space direction  which entered into the construction of 
the state $\oms$. It follows immediately from (\ref{spacelike-vacuum-convergence})  and Condition~$\Cs$ that
\beq
\lim_{n\to\infty}\|\om_{\xx_1+\la^{(k)}_{n}\e}-\oms\|_{\mfa(\mco)}=0
\eeq
for $k\in\{1,\ldots, L\}$. Consequently, $\lim_{n\to\infty}\|\vph_n-\vph\|\leq\eps$, what concludes the proof. \qed\\
The next lemma demonstrates that a state of bounded energy can deviate only locally 
from a vacuum state.
\bel\label{fluctuations} Suppose that Condition~$\Cnat$ holds. Let $\om_0$ be the vacuum state which appears
in Lemma~\ref{properties}. Then there exists a sequence of positive numbers $\{\de_N\}_{N\in\nat}$ s.t. 
$\de_N\nearrow\infty$ and
\beq
\sup_{\vpi\in \ov{S}_{E,N,\de_N}(\mco)}\fr{1}{N}\sum_{k=1}^N\|\vpi_k-\oms\|_{\mfa(\mco)}\to 0
\textrm{ for } N\to\infty.
\label{mean}
\eeq
\eel
\proof First, making use of Condition~$\Cnat$ and the diagonal trick, we can find a sequence $\de_N\nearrow\infty$
s.t. for any $\eps>0$  the $\eps$-contents $\tN(\eps)_{E,N,\de_N}$ of the sets $S_{E, N,\de_N}(\mco)$ satisfy
\beq
\tN(\eps)_{E,N,\de_N}\leq 2(N+1)^{c(\eps,E)}, \label{Cnat-restated}
\eeq
if $N$ is sufficiently large.
Next, we fix some $\eps>0$, $0<q<1$ and show that for any 
$\vpi=(\vpi_1,\ldots,\vpi_N)\in \ov{S}_{E, N,\de_N}(\mco)$,
the inequality
\beq
\|\vpi_k-\oms\|_{\mfa(\mco)}>\eps\label{deviations}
\eeq
holds for less than $[N^q]$ entries if $N$ is sufficiently large.
In fact, suppose the opposite is true i.e. that for any $N_0\in\nat$ there exists $N>N_0$ and an element 
$\vpi\in \ov{S}_{E,N,\de_N}(\mco)$ s.t. the bound (\ref{deviations}) holds for $[N^q]$
entries or more. Making use of Lemma~\ref{properties} (a) we 
can assume that (\ref{deviations}) is satisfied for $k\in\{1,\ldots, [N^q]\}$ and  proceed
to the element
\beq
\vph:=(\vpi_1,\ldots,\vpi_{[N^q]},\underbrace{\oms,\ldots,\oms}_{N-[N^q]})\in\ov{S}_{E,N,\de_N}(\mco).
\eeq
By permuting the entries of the above expression we obtain a family of elements $\vph_{\pi}\in\ov{S}_{E,N,\de_N}(\mco)$ s.t. 
\beq
\|\vph_{\pi_1}-\vph_{\pi_2}\|>\eps\label{distance}
\eeq
at least for $\pi_1,\pi_2\in \tP_N:=\PP_N/(\PP_{[N^q]}\times \PP_{N-[N^q]})$, $\pi_1\neq\pi_2$.
The cardinality of $\tP_N$ satisfies the bound
\beq
\#\tP_N=\fr{N!}{[N^q]!(N-[N^q])!}\geq\bigg(\fr{N-[N^q]+1}{[N^q]}\bigg)^{[N^q]}\geq 2^{N^q-1}, \label{cardinality}
\eeq
where the last inequality holds for $N$ sufficiently large. It follows from
formulas~(\ref{distance}), (\ref{cardinality}) that the $\eps$-contents of the sets 
$\ov{S}_{E,N,\de_N}(\mco)$ grow with $N$ faster than any polynomial. Since the 
$\eps$-content of a set and its closure coincide, we arrive at
a contradiction with relation~(\ref{Cnat-restated}).

With the above information at hand it is easy to estimate the mean, appearing in the statement 
of the lemma. In fact, for sufficiently large $N$ we obtain
\beq
\sup_{\vpi\in \ov{S}_{E,N,\de_N}(\mco)}\fr{1}{N}\sum_{k=1}^N\|\vpi_k-\oms\|_{\mfa(\mco)}\leq 
2\fr{N^q}{N}+\fr{N-[N^q]}{N}\eps\leq\fr{2}{N^{1-q}}+\eps.
\eeq
Since $\eps>0$ is arbitrary, the desired result follows. \qed\\
It is an immediate consequence of the above lemma and of decomposition~(\ref{decomp}) 
that for any  $\vxb^{(N)}=(x_1^{(N)},\ldots, x_N^{(N)})\in\GadN$ there holds
\beq
\lim_{N\to\infty}\sup_{\su{\fun\in \traceEB \\ A\in\mfa(\mco)_1 }}\fr{1}{N}\sum_{k=1}^N |\fun(A(x_k^{(N)}))-\fun(I)\om_0(A)|=0.
\label{mean1}
\eeq
This statement coincides with relation~(\ref{average}) from Lemma~\ref{key} (for the special sequence $\{\de_N\}_{N\in\nat}$
introduced in Lemma~\ref{fluctuations}). Since Condition~$\Csq$ was used only via this relation
in the proofs of Theorem~\ref{main}~(a), Corollary~\ref{relaxation-to-vacuum} and 
Proposition~\ref{shrinking},
these results still hold after replacing Condition~$\Csq$ with Condition~$\Cnat$ in their assumptions.
Thus we arrive at the following theorem.
\bet\label{altogether} Suppose that Condition~$\Cnat$ is satisfied and let $\om_0$ be any vacuum state 
in the weak$^*$-closure of $\traceEB$ for some $E\geq 0$.
Then, for any $E\geq 0$, there hold the following assertions:
\begin{enumerate}
\item[(a)] Let $\om\in\mfa^*$ be a state in the weak$^*$-closure of $\traceEB$ which is invariant
under translations in some spacelike ray. Then $\om=\om_0$. In particular there holds Condition~$\V$ stated in 
Section~\ref{space-translations-QFT}.
\item[(b)] For any spacelike  unit vector $\he\in\real^{s+1}$ and $\om\in S_E$ there holds
\beq
\lim_{\la\to\infty}\om_{\la\he}(A)=\om_0(A) \textrm{ for } A\in\mfa.
\eeq
If, in addition, Condition $R$, stated in Section~\ref{existence-of-vacuum}, is satisfied, then the above 
relation is also true for any timelike unit vectors $\he$.
\item[(c)] For any $p\in\cone$ and double cone $\mco$, there holds
\beq
\lim_{r\to 0}\sup_{\su{\fun\in\trace_{(p,r),1} \\ A\in\mfa(\mco)_{1}} }|\fun(A)-\fun(I)\om_0(A)|=0.
\eeq
\end{enumerate}
\eet
\nin Part (c) of this theorem allows us to show that estimate~(\ref{ff1}), which we used in 
our heuristic discussion, holds in all theories complying with Condition~$\Cnat$. Since the
$\eps$-contents $\tN(\eps)_E$ of the sets $S_{E}(\mco)$, given by (\ref{set-SEm}), take only integer 
values, it suffices to show that 
\beq
\lim_{E\downarrow 0}\tN(\eps)_E=1 \label{eps-content-energy}
\eeq
for any $\eps>0$. If this relation did not hold, then, by definition of the $\eps$-content,
we could find nets $\{\om_1^{E}\}_{E>0}$, $\{\om_2^{E}\}_{E>0}$ s.t. 
$\|\om_1^{E}-\om_2^{E}\|_{\mfa(\mco)}>\eps$ for any $E>0$. But choosing $p=0$ in 
Theorem~\ref{altogether} (c), we arrive at a contradiction:
\beq
\|\om_1^{E}-\om_2^{E}\|_{\mfa(\mco)} \leq 2\sup_{\om\in S_E}\|\om-\om_0\|_{\mfa(\mco)}
\underset{E\downarrow 0}{\to}0.
\eeq
As we show in the next section and in Appendix~\ref{Condition-Nnat}, the set of theories
complying with Condition~$\Cnat$ contains also massless models, where relation~(\ref{eps-content-energy})
is a non-trivial statement about the infrared structure. (See the discussion after formula~(\ref{ff1})).

\section{Condition $\Nnat$ implies Condition $\Cnat$} \label{additivity-of-energy}

In the previous section we introduced the phase space condition $\Cnat$ and  studied
it physical consequences. Although this criterion was motivated by the firm physical principle
of additivity of energy over isolated subregions, its consistency with the general postulates
from Section~\ref{AQFT} remains to be verified by establishing it in a model.
Since the $\eps$-contents of the maps $\The_{E,N,\de}$ are difficult to control directly in 
concrete theories, we proceed as follows: We introduce a nuclearity condition~$\Nnat$, stated
below, which is verified  by a relatively straightforward computation
in massive and massless scalar free field theory in Appendix~\ref{Condition-Nnat}.
In this section we show that this new condition, which is interesting in its own right, implies Condition~$\Cnat$.

To begin with, let us recall the concept of nuclearity in the form which is suitable for our investigations:
Let $V$, $W$ be Banach spaces 
and $\norm$ be a norm on the space $\lin(V,W)$ of linear maps from $V$ to $W$. We say that a map $\Pi: V\to W$ is $p$-nuclear w.r.t. the norm $\norm$ if there exists a decomposition $\Pi(v)=\sum_n\Pi_n(v)$ into rank-one maps, convergent for any $v\in V$ in the norm topology in $W$, s.t. $\nu:=(\sum_n\nor\Pi_n\nor^p)^{\fr{1}{p}}<\infty$. The $p$-norm $\nor\Pi\nor_{p}$\label{symbol-pnorm} of this map is the smallest such $\nu$ over the set of all admissible decompositions. This concept was first used for the description of phase space in
quantum field theory in the work of Buchholz and Wichmann \cite{BWi86}. Our starting point is a stronger nuclearity condition,  introduced by Buchholz and Porrmann in \cite{BP90}:
\begin{enumerate}
\item[] \bf Condition $\Ns$: \rm \label{cond-Ns} The maps $\Pi_E$ are $p$-nuclear w.r.t. the standard norm on $\lin(\traceE,\mfa(\mco)^*)$
for any $0<p\leq 1$, $E\geq 0$ and  double cone $\mco$.
\end{enumerate}
For verification of this condition in models, which we present, following \cite{Bo00}, in Appendix~\ref{Preliminaries},  it is convenient to have an equivalent formulation in terms of the maps $\whXi_E$. It is stated in 
the following simple lemma. A proof, up to obvious modifications, can be found in \cite{BP90}.
(See also Lemma~\ref{compactness-equivalence} above for a similar argument).
\bel\label{equivalent-Nsharp} We fix a double cone $\mco$ and $0<p\leq 1$. Then the following conditions 
are equivalent:
\begin{enumerate}
\item[(a)] The maps $\Pi_E$ are $p$-nuclear for any $E\geq 0$.
\item[(b)] The maps $\whXi_E$ are $p$-nuclear for any $E\geq 0$.
\end{enumerate}
\eel
\nin Clearly, Condition $\Ns$ is stronger than  Condition $\Cs$. Thus in theories complying with this
nuclearity criterion 
we can introduce the local continuous
subspace $\mfa_{\scc}(\mco)=\{\, A\in\mfa(\mco)\,|\, \om_0(A)=0\,\}$, fixed by some energetically accessible 
vacuum state $\om_0$, similarly as in the first part of Section~\ref{vacuum-structure}.
Moreover, we define the maps $\Pi_E^{\scc}\in\lin(\traceE,\mfa_{\scc}(\mco)^*)$
given by
\beq
\Pi_E^{\scc}(\fun)=\fun|_{\mfa_{\scc}(\mco)}, \label{symbol-PiEc}
\eeq
which satisfy the following identity, valid for any $\fun\in\traceE$ and $A\in\mfa(\mco)$,
\beq
\Pi_E(\fun)(A)=\PiEc(\fun)(A-\om_0(A)I)+\om_0(A)\fun(I).
\eeq
One easily obtains from the above equality that Condition~$\Ns$ holds if and only if there exists an 
energetically accessible vacuum state and the corresponding maps $\PiEc$ are $p$-nuclear for any $0<p\leq 1$. However,
we note that this condition  is still somewhat conservative, since it does not take into account
the fact that for any $\fun\in\traceE$ the restricted functionals $\fun_{\vx}|_{\mfa_{\scc}(\mco)}$ should be arbitrarily close to zero apart from translations varying in some compact subset of $\real^{s}$, depending on $\fun$. It seems therefore desirable to introduce a family of norms on $\lin(\traceE,X)$, where $X$ is some Banach space, given for any $N\in\nat$ and $\x\in\real^{s+1}$ by
\beq
\|\Pi\|_{\x}=\sup_{\fun\in\traceEB}\bigg(\sum_{k=1}^N\|\Pi(\al_{x_k}^\p\fun)\|^2\bigg)^{\half},\quad \Pi\in \lin(\traceE,X) \label{Nnorm}
\eeq
and the corresponding family of $p$-norms $\|\Pi\|_{p,\x}$.\label{symbol-pnormx}
It is easily seen that if $\PiEc$ is $p$-nuclear w.r.t. the standard norm on $\lin(\traceE, \mfa_{\scc}(\mco))$, then 
it is also $p$-nuclear with respect to the norms~(\ref{Nnorm}) and vice versa. Important additional information 
is contained in the dependence of the nuclear $p$-norms on $N$. The assumption, which is consistent with the
basic postulates 
and suitable for the purpose of deriving Condition~$\Cnat$,
is the following: 
\begin{enumerate}
\item[] \bf Condition $\Nnat$: \rm \label{cond-Nnat}
\begin{enumerate}
\item[(a)] There exists an energetically accessible vacuum state $\om_0$ and the corresponding 
maps $\PiE^{\scc}$ are $p$-nuclear w.r.t. the norms $\|\cdot~\|_{\x}$
for any $N\in\nat$, $\x\in\real^{s+1}$,  $0<p\leq 1$, $E\geq 0$ and double cone $\mco$.
\item[(b)] The nuclear $p$-norms of the maps $\PiE^{\scc}$ satisfy
\beq
\limsup\|\PiE^{\scc}\|_{p,\x}\leq c_{p,E}, \label{natbound}
\eeq
where $c_{p,E}$ is independent of $N$ and the limit is taken for configurations $\x$, where all $x_i-x_j$, $i\neq j$, 
tend to spacelike infinity.
\end{enumerate}
\end{enumerate}
We verify this criterion in massive and massless free field theory, as well as in their even parts, in Appendix~\ref{Condition-Nnat}.
It is clear that if this condition holds in a given theory then it holds also in its sub-theories. In particular it is
satisfied by the sub-theory of massless free field theory generated by derivatives of the field. 
(See Corollary~\ref{Nnat-even-derivatives}).

We note as an aside that the quantitative refinement~(\ref{natbound}) implies
that for any $E\geq 0$, double cone $\mco$ and sequence $\{\de(\K)\}_1^{\infty}$ s.t. $\de(\K)\too\infty$
there holds
\beqa
\sup_{\su{\fun\in\traceEB \\ A\in\mfa_{\scc}(\mco)_{1}}} \fr{1}{\K}\sum_{i=1}^{\K}|\fun(A(\vx_i^{(\K)}))|
\leq\fr{1}{\sqrt{n}}\|\Pi_E^{\scc}\|_{x_1^{(n)},\ldots,x_n^{(n)} }\underset{n\to\infty}{\to}0,
\eeqa
where $\{x_k^{(n)}\}_1^n\in\Ga_{\K,\de(\K)}$. This formula coincides with relation~(\ref{mean1}),
so we immediately conclude that Condition $\Nnat$ has all the implications listed in  
Theorem~\ref{altogether}. This approach to the study of the vacuum structure
was taken in \cite{Dy08.1}. However, in the present work we derived properties of the vacuum states
from Conditions~$\Cnat$ and $\Csq$ whose physical bases are more solid. Thus
Condition~$\Nnat$ serves here only as an auxiliary step in the proof that Condition~$\Cnat$ is consistent
with the basic postulates. Our goal in this section is to prove the following theorem:
\bet \label{Nregions} In any quantum field theory Condition~$\Nnat$ implies Condition~$\Cnat$.
\eet
\nin Proceeding towards the proof of this theorem, we first decompose the map 
$\The_{E,N,\de}\in\lin(\traceE\times\Gad,\mfa(\mco)^*\ot\compsup)$, 
given by (\ref{theta}), as follows
\beqa
& &\The_{E,N,\de}=\Theh_{E,N,\de}+\Thec_{E,N,\de}, \label{The-decomposition}\\
& &\Theh_{E,N,\de}(\fun,\vxb)=(\Pi_E(\fun_{\xx_1})-\fun(I)\om_0,\ldots,\Pi_E(\fun_{\xx_1})-\fun(I)\om_0),\\
& &\Thec_{E,N,\de}(\fun,\vxb)=\fun(I)\om_0\ot(1,\ldots, 1).
\eeqa
Since $\Thec_{E,N,\de}$ is a rank-one map, its $\eps$-content $\Nc(\eps)_{E,N,\de}$ satisfies the bound
\beq
\Nc(\eps)_{E,N,\de}\leq 1+\fr{8^2\|\Thec_{E,N,\de}\|^2}{\eps^2}=1+\fr{8^2}{\eps^2} \label{rank-one-eps-cont}
\eeq
which is independent of $N$. In order to estimate the $\eps$-content $\Nh(\eps)_{E,N,\de}$ of the map
$\Theh_{E,N,\de}$, we introduce the auxiliary mapping $\The^{\scc}_{E,N,\de}\in \lin(\traceE\times\Gad,\mfa_{\scc}(\mco)^*\ot\compsup)$ given~by
\beq
\The^{\scc}_{E,N,\de}(\fun,\vxb)=(\PiEc(\fun_{\xx_1}),\ldots,\PiEc(\fun_{\xx_1})).\label{map-thescc}
\eeq
We note that for any $(\fun_1,\vxb_1), (\fun_2,\vxb_2)\in \traceE\times\Gad$ there holds
\beq
\|\Thec_{E,N,\de}(\fun_1,\vxb_1)-\Thec_{E,N,\de}(\fun_2,\vxb_2)\|\leq 2\|\The^{\scc}_{E,N,\de}(\fun_1,\vxb_1)-\The^{\scc}_{E,N,\de}(\fun_2,\vxb_2)\|.
\eeq
Thus, by definition of the $\eps$-content, we obtain $\Nh(\eps)_{E,N,\de}\leq \N_{\scc}(\eps/2)_{E,N,\de}$,
where $\N_{\scc}(\eps)_{E,N,\de}$ is the $\eps$-content of the map $\The^{\scc}_{E,N,\de}$. We will control
this quantity with the help of Condition~$\Nnat$ and the following key lemma.
\begin{lemma}
\label{key-key} Let $V$ and $W$ be Banach spaces and $\Ga$ be a set. Let $S_k\in \lin(V\times\Ga,\complex)$ for 
$k\in\{1,\ldots,N\}$ and $\tau\in W$ be s.t. $\|\tau\|=1$. Then the  $\eps$-content of the map
$\The\in\lin(V\times\Ga, W\ot\compsup)$ given by
\beq
\The(v,x)=\tau\, (S_1(v,x),\ldots, S_N(v,x)),\quad (v,x)\in V\times\Ga, \label{standard}
\eeq
satisfies the bound
\beq
\N(\eps)_{\The}\leq (4eN)^{\fr{2^7\pi \|\The\|_2^2}{\eps^2}}, \label{standard1}
\eeq
where $\|\The\|_2=\sup_{(v,x)\in V_1\times\Ga}(\sum_{k=1}^N|S_k(v,x)|^2)^{\fr{1}{2}}$.
\end{lemma}
\proof Fix $\eps>0$ and let $\lattice_0=\{ (n_1+in_2)\eps \ | \ n_1,n_2\in\integer\}$. For each
$k\in\{1,\ldots, N\}$ and $\ufun=(v,x)\in V_1\times\Ga$ we choose $J_k(\ufun)\in \lattice_0$ so that
$|S_k(\ufun)-J_k(\ufun)|\leq \sqrt{2}\eps$ and $|J_k(\ufun)|\leq |S_k(\ufun)|$. Define
the set $\lattice=\{J_1(\ufun),\ldots,J_N(\ufun) \ | \ \ufun\in V_1\times\Ga \}$ of all $N$-tuples 
appearing in this way. We claim that $\#\lattice\geq \N(4\eps)_{\The}$. In fact, assume that there  
are $\ufun_1,\ldots,\ufun_K\in V_1\times\Ga$, $K>\#\lattice$, s.t. for $i\neq j$ there holds
\beq
4\eps<\|\The(\ufun_i)-\The(\ufun_j)\|
=\sup_{k\in\{1,\ldots, N\}} |S_k(\ufun_i)-S_k(\ufun_j)|.
\eeq
Then there exists such $\khat$, depending on $(i,j)$, that $4\eps<|S_{\khat}(\ufun_i)-S_{\khat}(\ufun_j)|$.
Consequently, by a $3\eps$-argument
\beq
|J_{\khat}(\ufun_i)-J_{\khat}(\ufun_j)|
\geq|S_{\khat}(\ufun_i)-S_{\khat}(\ufun_j)|-2\sqrt{2}\eps>\eps,
\eeq
which shows that there are at least $K$ different elements of $\lattice$ in contradiction
to our assumption.

In order to estimate the cardinality of the set $\lattice$, we define
$M=\big[\fr{\|\The\|_2^2}{\eps^2}\big]$, assume for the moment that $0<M\leq 2N$ and denote by 
$V_M(R)\leq e^{2\pi R^2}$ the volume of the $M$-dimensional ball of radius $R$. Then
\beqa
\#\lattice\leq\sum_{\su{n_1,\ldots,n_{2N}\in\integer \\ n_1^2+\cdots+n_{2N}^2\leq M} } 1 \leq
\binom{2N}{M} 2^M V_{M}(2\sqrt{M})\leq (4Ne)^{8\pi M}. \label{ball}
\eeqa
Here we noted that each admissible combination of integers $n_1,\ldots, n_{2N}$ contains at most $M$ 
non-zero entries. Thus to estimate the above sum we picked $M$ out of $2N$ indices and considered the points 
$(n_{i_1},\ldots, n_{i_M})\in\integer^M$ which  belong to the  $M$-dimensional ball of radius $\sqrt{M}$.
Each such point is a vertex of a unit cube which fits into a ball of radius $2\sqrt{M}$ (since $\sqrt{M}$
is the length of the diagonal of the cube). As in $M$ dimensions a cube has $2^M$ vertices, there
can be no more than $2^M V_{M}(2\sqrt{M})$ points $(n_{i_1},\ldots,n_{i_M})\in\integer^M$ satisfying the 
restriction $n_{i_1}^2+\cdots+n_{i_M}^2\leq M$. In the case $M\geq 2N$ a more stringent bound
(uniform in $N$) can be established by a similar reasoning. For $M=0$ there obviously holds $\#\lattice=1$.
\qed\\
Next, by a straightforward combination of Lemma~2.3 and Lemma~2.4 from \cite{BD95}, we obtain
the following useful lemma, which says, essentially, that the $\eps$-content of a sum of
maps is equal to the product of their respective $\eps$-contents.
\bel\label{sum-eps-content} Let $V$, $W$ be Banach spaces, let $\{\The_n\}_{n\in\nat}$ be a family of compact maps in $\lin(V\times\Ga, W)$ 
and $\N(\eps)_n$ be the corresponding $\eps$-contents. Suppose that $\sum_n \|\The_n\|<\infty$. Then the 
$\eps$-content $\N(\eps)$ of the (compact) map $\The:=\sum_n\The_n$ satisfies, for any sequence of positive numbers $\{\eps_n\}_{n\in\nat}$ s.t. $\sum_n\eps_n\leq \eps/4$, the following bound
\beq
\N(\eps)\leq \liminf_{k\to\infty}\prod_{n=1}^k\N(\eps_n)_n. \label{product1}
\eeq
\eel
\nin After this preparation we are ready to prove Theorem~\ref{Nregions}. The argument relies
on techniques from the proof of Proposition~2.5 (ii) in \cite{BD95}. \\
\nin\bf Proof of Theorem \ref{Nregions}. \rm  Fix $0<p<\fr{2}{3}$. Then Condition $\Nnat$ provides, for any $\del>0$, 
a decomposition of the map $\PiEc$ into rank-one mappings $\Pi_n(\,\cdot\,)=\tau_n\,S_n(\,\cdot\,)$, where
$\tau_n\in\mfa_{\scc}(\mco)^*$ and $S_n\in\traceE^*$,
s.t. 
\beq
(\sum_{n=1}^{\infty}\|\Pi_n\|^p_{\x})^{\fr{1}{p}}\leq (1+\del)\|\PiEc\|_{p,\x}. \label{summability-of-pimaps}
\eeq
Assuming that the norms $\|\Pi_n\|_{\x}$ are given in  descending order with $n$, we obtain the bound
\beq
\|\Pi_n\|_{\x}\leq \fr{(1+\del)\|\PiEc\|_{p,\x}}{n^{1/p}}. \label{xxnorm}
\eeq 
Similarly, we can decompose the map $\The_{E,N,\de}^{\scc}$, given by (\ref{map-thescc}), 
into a sum of maps $\The_n\in\lin(\traceE\times\Gad,\mfa_{\scc}(\mco)^*\ot\compsup )$ of the form
\beqa
\The_n(\fun,\vxb)=\big(\Pi_n(\fun_{\vx_1}),\ldots,\Pi_n(\fun_{\vx_N}) \big)
            =\tau_n\big(S_n(\fun_{\vx_1}),\ldots,S_n(\fun_{\vx_N})\big).
\eeqa
Now we apply Lemma~\ref{key-key} with $\tau=\tau_n/\|\tau_n\|$ and 
$S_k\in\lin(\traceE\times\Gad,\complex)$, $k\in\{1,\ldots,N\}$ given by 
\beq
S_k(\fun,\vxb)=\|\tau_n\|\,S_n(\fun_{\vx_k}).
\eeq
From estimate~(\ref{xxnorm}) we obtain
\beqa
\|\The_n\|_2&=&\sup_{(\fun,\vxb)\in\traceEB\times\Gad}(\sum_{k=1}^N\|\tau_n\|^2|S_n(\fun_{\vx_k})|^2)^{\fr{1}{2}}
\non\\
&=&\sup_{\vxb\in\Gad}\|\Pi_n\|_{\x}\leq\sup_{\vxb\in\Gad}\fr{(1+\del)\|\PiEc\|_{p,\x}}{n^{1/p}},
\eeqa
where $x_k=(0,\vx_k)$, $k\in\{1,\ldots,N\}$.
Substituting this inequality to the bound~(\ref{standard1}) we get 
\beq
\N(\eps)_n\leq (4eN)^{\fr{2^7\pi(1+\del)^2\big(\sup_{\vxb\in\Gad}\|\SiEh\|_{p,\x}^2\big)}{\eps^2 n^{2/p}}}. \label{factor}
\eeq
Since $\|\The_n\|=\|\Pi_n\|$, it follows from (\ref{summability-of-pimaps}) that the summability assumption of 
Lemma~\ref{sum-eps-content} is satisfied.
We choose $\eps_n=\fr{\eps}{4} \fr{n^{-2/(3p) } }{\sum_{n_1=1}^\infty n_1^{-2/(3p)}}$, 
make use of the bounds (\ref{factor}) and (\ref{product1}), and take the infinum w.r.t. $\del>0$. There follows
\beq
\N_{\scc}(\eps)_{E,N,\de}\leq (4eN)^{ \fr{2^{11}\pi \big(\sup_{\vxb\in\Gad} 
\|\SiEh\|_{p,\x}^2\big)}{\eps^2}(\sum_{n=1}^\infty n^{-2/(3p)} )^3 }.\label{Nscceps}
\eeq
From decomposition~(\ref{The-decomposition}), estimate~(\ref{rank-one-eps-cont}) and Lemma~\ref{sum-eps-content} we 
obtain the bound on the $\eps$-content of the map $\The_{E,N,\de}$
\beq
\N(\eps)_{E,N,\de}\leq \N_{\scc}(\eps/8)_{E,N,\de}\Nc_{\scc}(\eps/8)_{E,N,\de}
\leq \N_{\scc}(\eps/8)_{E,N,\de}\big(1+\fr{8^4}{\eps^2}\big).
\eeq
The l.h.s. of this inequality is a  decreasing and bounded from below function of $\de$, so we can
take the limit $\de\to\infty$. With the help of formula~(\ref{Nscceps}) and Condition~$\Nnat$ we obtain the 
bound~(\ref{addenergy}) in the statement of Condition~$\Cnat$. \qed

%% file: PhD-Conclusions.tex
\chapter{Conclusions and Outlook}\label{chapter:conclusions}

In this work we have developed a detailed spectral theory of translation
automorphisms in quantum field theory, pursuing the programme initiated in \cite{Bu90}.
Motivated by the spectral properties of unitary representations of translations
acting on a Hilbert space, we proposed a decomposition of the algebra of local
observables $\hmfa$ into subspaces which differ in their behavior under translations
in space. Our investigation has led to new insights into the infrared structure
of quantum field theories, their particle content and properties of the vacuum
states.

First, we identified the counterpart of the pure-point subspace.
In order to find the natural projection  $P_{\{0\}}$ on this  subspace in the 
absence of orthogonality,  we have established a variant of the ergodic theorem for the space
translation automorphisms. It relies on physically motivated phase space conditions,
whose consistency with the basic postulates has been verified by explicit computations
in models of non-interacting particles. We have shown that these criteria have, in
addition, a number of consequences pertaining to the vacuum structure, including the uniqueness
of the energetically accessible vacuum state~$\om_0$ and relaxation of physical states to $\om_0$ 
under large timelike translations. Moreover, this vacuum state is simply related to the above 
mentioned projection: $P_{\{0\}}(\,\cdot\,)=\om_0(\,\cdot\,)I$. The continuous subspace has been defined as 
the kernel of this projection, or equivalently $\hmfa_{\scc}=\ker\om_0$.

Proceeding to a more detailed spectral analysis, we 
took the quantum-mechanical framework as a guide: We noted that square-integrability of the
transition amplitude between a vector and its translate implies that this vector belongs to the
absolutely continuous subspace. We convinced ourselves that this property always holds for 'local' vectors
whose wavefunctions are compactly supported in configuration space, which serve as analogues of
local operators. 
More importantly, transition amplitudes between an arbitrary vector and translates of a 'local' 
one are also square-integrable, but not necessarily integrable with any smaller power. Summing up, 
square-integrability of the transition amplitudes on the one hand implies absolute continuity 
of the spectrum and on the other hand is the strongest decay property to be expected from 'local' vectors.
Therefore, in quantum field theory we have defined the absolutely continuous subspace $\hmfa_{\ac}$ as 
consisting of square-integrable elements i.e. $A\in\hmfa_{\scc}$ satisfying $\|A\|_{E,2}<\infty$ for any $E\geq 0$.
We have shown that in a number of cases this subspace can be expressed as the intersection of kernels
of a finite family of linear functionals i.e. $\hmfa_{\ac}=\ker\om_0\cap\ker\tau_1\cap\ldots\cap\ker\tau_n$.
Then its direct sum complement is finite dimensional, akin to the pure-point part, what motivates the term
'point-continuous subspace' $\hmfa_{\pc}$. We emphasize that this subspace does not have a quantum mechanical 
counterpart - it carries information about the infrared structure which is specific 
to quantum field theories. If $\hmfa_{\pc}$ is non-trivial, then $\hmfa_{\scc}$ contains
some elements $A$ which are not square-integrable. Their deviation from square-integrability can be quantified
with the help of a new concept:  the 'infrared order' $\ord(A)$ of the observable~$A$. It captures the 
regularity properties of the distribution $\wt{A}(\vep)$ which, as we argued in Section~\ref{Literature}, is
a natural analogue of the spectral measure for the Arveson spectrum. Since the infrared orders of observables from 
$\hmfa_{\scc}$ can also be computed with the help of the functionals $\{\tau_j\}_1^n$ (see the proof of Theorem~\ref{full-results1} in Section~\ref{HA-in-models-section}), a model-independent construction of such functionals would constitute a major progress in our understanding of the spectral theory of automorphism groups. Taking our study of non-interacting examples as a guide, decent
phase space properties and sufficiently rich field content of a theory should be relevant to the study of 
this problem.

As explained in Section~\ref{Arveson},  to unravel particle aspects of a theory, one needs integrable
observables i.e. such $C\in\mfa$ that $\|C\|_{E,1}<\infty$ for any $E\geq 0$. In view of the above discussion we conjecture that
there do not exist non-zero local operators which satisfy this property. A class of almost local, integrable observables
was found in \cite{Bu90}. However, apart from the case of Wigner particles, it was not known
under what conditions the particle content of a theory is non-trivial. We have proposed a criterion, 
suitable for a class of massive theories, which introduces a new family of particle detectors. 
With these novel integrable observables one can approximate pointlike localized fields in the topology 
generated by the seminorms $\|\,\cdot\,\|_{E,1}$.
Assuming in addition the existence of the stress-energy tensor, we have obtained non-triviality 
of the particle content, 
substantiating the strategy put forward in \cite{Bu94}. We expect that the conditions mentioned above, or their strengthened
variants, have other interesting consequences, perhaps even some weak form of asymptotic completeness, 
as discussed in the last part of Section~\ref{triviality-of-Apc}. It would therefore be desirable to find similar criteria
in the realm of massless theories. There we have acquired thorough understanding of square-integrability properties
of observables, but integrable detectors, which could approximate the stress-energy tensor, have not been found as yet.

Another interesting problem is  convergence of the asymptotic functional approximants $\{\si^{(t)}_{\fun}\}_{t\in\real_+}$,
defined by formula (\ref{symbol-asymptotic-functional-appr}), as $t\to\infty$.
This property is certainly expected on physical grounds, since results of 
particle physics experiments stabilize for sufficiently large times. Although we do not know any examples,
where this property would fail, a proof relying only on the basic postulates of QFT is out of sight at the moment. On the
other hand, our study of the problem of convergence to the vacuum in Chapter~\ref{chapter-vacuum} suggests an approach
based on phase space conditions. Let us outline briefly a possible strategy: First, we note that Condition~$\Cnat$,
stated in Section~\ref{Condition-Cnat-and-additivity}, provides a promising starting point as it does not require any
a priori knowledge of the limiting state. This is a great advantage in the present context, since the
asymptotic functionals $\si^{(+)}_{\fun}$, in contrast to asymptotic vacuum states, depend in a complicated manner on the initial functional~$\fun$. In fact, they should describe all the possible asymptotic particle configurations, so no uniqueness result is expected here. To reformulate Condition~$\Cnat$ so that it describes the results of particle measurements, separated
by large time intervals, we define 
\beq
\tGad=\{\,\un{t}:=(t_1,\ldots,t_N)\in\real^N_+\,|\, |t_i-t_j|\geq\de \te{ for }i\neq j\,\}.
\eeq
Next, we adopt Condition~$\B$ stated in Section~\ref{triviality-of-Apc},  and note that 
$\mfa_{\scc}(\mco)\ni A\to \si^{(t)}_{\fun}(A(g))$ is a bounded linear functional  for any
$t\in\real_+$,  $\fun\in\traceE$ and a suitable time-smearing function $g\in S(\real)$.
It is  convenient to shift the smearing on the functional $\fun$ i.e. to 
write $\si^{(t)}_{\fun_g}(A)$ instead of $\si^{(t)}_{\fun}(A(g))$.
Now we  introduce the maps $\tThe_{E,N,\de}: \traceE\times\tGad\to\mfa_{\scc}(\mco)^*\ot\compsup$ given by
\beq
\tThe_{E,N,\de}(\fun,\un{t})=(\si^{(t_1)}_{\fun_g},\ldots,\si^{(t_N)}_{\fun_g}).
\eeq
The limits of $\{\si^{(t)}_{\fun_g} \}_{t\in\real_+}$ as $t\to\infty$ are directly related to the asymptotic functionals
which constitute the particle content of a theory. Thus in physical terms the ranges of these maps consist of collections 
of data (summed over time-slices) obtained in particle physics experiments. In a typical experiment, after a finite number  of rescattering events 
there emerges a stable, asymptotic configuration. (See \cite{Ha}, Section VI.2.3). As a first orientation,
let us make the 'uniform dispersion' assumption  which says that the number of such events,
which can be detected up to some accuracy $\eps$, is bounded by some constant $c(\eps, E)$, independent of the
initial functional $\fun\in\traceEB$.
Hence we obtain that the number of distinguishable collections of data should not exceed 
${N\choose c(\eps,E)}\leq N^{c(\eps,E)}$. Therefore, we impose on the maps $\tThe_{E,N,\de}$ the following
variant of Condition~$\Cnat$:
\begin{enumerate}
\item[ ] \bf Tentative Condition \rm $\Cnat^{(+)}$:
\item[(a)] The maps $\tThe_{E,N,\de}$ are compact for any $E\geq 0$, $N\in\nat$, $\de>0$ and double cone 
$\mco\subset\real^{s+1}$.
\item[(b)] The $\eps$-contents $\N(\eps)_{E,N,\de}^{(+)}$ of the maps $\tThe_{E,N,\de}$ satisfy, for any $\eps>0$, 
\beq
\lim_{\de\to\infty}\N(\eps)_{E,N,\de}^{(+)}\leq (N+1)^{c(E,\eps)}, \label{spread-wave}
\eeq
for some constant $c(E,\eps)$ independent of $N$.
\end{enumerate}
Under this condition  we immediately obtain
convergence of the asymptotic functional approximants on the class of detectors introduced
in Condition~$\B$:  We fix $\fun\in\traceEB$. By compactness of the map $\tThe_{E,1,\de}$, 
there exist limit points $\si^{(i,+)}_{\fun_g}$ of the net $\{\si^{(t)}_{\fun_g}\}_{t\in\real_+}$ as $t\to\infty$, which can be
approximated by sequences $\{\si^{(t_{i,n})}_{\fun_g}\}_{n\in\nat}$ in the norm topology of $\mfa_{\scc}(\mco)^*$.
Suppose that there are two different limit points $\si^{(1,+)}_{\fun_g}$ and $\si^{(2,+)}_{\fun_g}$ i.e. there holds
$|\si^{(1,+)}_{\fun_g}(A)-\si^{(2,+)}_{\fun_g}(A)|>\eps$ for some $A\in\mfa_{\scc}(\mco)_1$ and $\eps>0$. By choosing
suitable subsequences of their approximating sequences, one can easily show that for any $i_1,\ldots,i_N\in\{1,2\}$ 
the elements
\beq
(\si^{(i_1,+)}_{\fun_g},\ldots,\si^{(i_N,+)}_{\fun_g})\in\mfa_{\scc}(\mco)^*\ot\compsup
\eeq
belong to the closures of the ranges of the maps $\tThe_{E,N,\de}$ for any $\de>0$. Since there are $2^N$
such elements and their mutual norm distances are larger than~$\eps$, the bound~(\ref{spread-wave}) is violated.
We conclude that all the limit points $\si^{(i,+)}_{\fun_g}$ are equal. Thus, for any $\fun\in\traceE$, we obtain
the unique functional $\si^{(+)}_{\fun_g}=\lim_{t\to\infty}\si^{(t)}_{\fun_g}$ describing the asymptotic  
particle configuration. Moreover, by part (a) of the above assumption, the set $\{\, \si_{\fun_g}^{(+)} \,|\, \fun\in\traceEB\,\}$
is compact in the norm topology of $\mfa_{\scc}(\mco)^*$.

Unfortunately, we do not have a complete verification argument for the above tentative condition. Preliminary
computations in massive scalar free field theory confirm the qualitative part~(a). On the
other hand, the quantitative refinement~(b) seems to be satisfied only after replacing the  space $\mfa_{\scc}(\mco)$ 
with a smaller, but infinitely dimensional subspace $\mfa_{\scc}(\mco)^{(+)}$ in the definition of the maps $\tThe_{E,N,\de}$.
This indicates that our 'uniform dispersion' picture can be maintained only after excluding some 'oversensitive' detectors.
In this particular model it can be shown that the subspace $\mfa_{\scc}(\mco)^{(+)}$ is sufficiently large to 
ensure non-triviality of the asymptotic functionals,
but it is not clear how to formulate this requirement in general terms. Interestingly,
$\mfa_{\scc}(\mco)^{(+)}$ can be expressed as the intersection of kernels of a (countable) family of functionals,
akin to the spectral subspaces discussed above. Better understanding of the origin of this subspace may,
on the one hand, shed light on the question of convergence of the asymptotic functional approximants, on the other
hand open the door to  deeper understanding of the spectral theory of automorphism groups.

%% file: PhD-HR.tex
\appendix
\chapter{Haag-Ruelle Scattering Theory in Presence of Massless Particles} \label{Haag-Ruelle-Appendix}

In this appendix, which complements our discussion in Section~\ref{Wigner},
we construct a scattering theory of stable, massive particles without assuming mass gaps.
This extension of the Haag-Ruelle theory is based on advances in the harmonic analysis of local operators \cite{Bu90}
restated in Theorem~\ref{theorem-2mollifier} of the present Thesis. Our construction is restricted to theories 
complying with a regularity property introduced by Herbst, stated as Condition~$A^\prime$
below. The appendix concludes with a brief discussion of the status of this assumption.
The analysis presented here has been published in \cite{Dy05}.

\section{Introduction}

It is the aim of this appendix to prove that the Haag-Ruelle collision theory can be 
extended to stable massive particles, obeying a sharp dispersion law, in the presence of massless excitations.
Thus we do not touch upon the infraparticle problem \cite{Sch63,Bu86}, mentioned in Section~\ref{Wigner},
but our arguments are applicable, for example, to electrically neutral, stable particles such as atoms in quantum electrodynamics.
Before we enter into this discussion we briefly outline our notation, state our assumptions and comment
on previous approaches to this problem.

Similarly as in Section~\ref{Wigner}, we adopt here a more restrictive framework than the 
one expounded in Section~\ref{AQFT}.
First, we restrict attention to the physical case of dimension of space $s=3$. Next, we assume that the Poincar\'e 
transformations are unitarily implemented i.e. there exists a
continuous unitary representation $\cpo\ni(x,\La)\to U(x,\La)$ acting on the
Hilbert space $\hil$ s.t. $\al_{(x,\La)}(\,\cdot\,)=U(x,\La)\,\cdot\, U(x,\La)^*$.
It is also required that there exists in $\hil$ a unique (up to a phase) vacuum vector $\vac$
which is invariant under the action of $U$. Finally, we suppose that the point spectrum of the mass operator
$M=\sqrt{H^2-\vec{P}^2}$ consists, apart from $0$, of a unique eigenvalue $m>0$. We also postulate that the 
representation $U$, restricted to the corresponding spectral subspace $\hilm$, coincides with the irreducible representation of the 
Poincar\'e group of mass $m$ and spin $0$. In other words, we consider a single species of massive,
spinless particles. 
In the pioneering work of Haag \cite{Ha58} and Ruelle \cite{Ru62} these 
general postulates were amended by two additional requirements:
\begin{enumerate}
\item[] \bf Condition $A$: \rm \label{cond-A} The time-dependent operators $A(f_T)=\int A(x)f_T(x) d^4x$, 
constructed from $A(x)=U(x)AU(x)^{-1}$, $A\in \mfa(\mco)$ and suitably chosen sequences 
of functions $f_T\in S(\spt)$, satisfy $A(f_T)\vac\neq 0$, $A(f_T)\vac\in\hilm$ and $\fr{d}{dT}A(f_T)\vac=0$. 
\item[] \bf Condition $M$: \rm \label{cond-M} The vacuum is isolated from the rest of the energy-momentum spectrum.
 \end{enumerate}
Both of these conditions are ensured if the mass $m$ is an isolated eigenvalue of the mass operator $M$. 
On the other hand, if the mass
of the particle in question is an embedded eigenvalue, then it seems difficult to
meet the requirement $A$. It was, however, noticed  
by Herbst \cite{Her71} that, in fact, it is only needed in the proof that 
$\slim_{T\to\infty}A(f_T)\vac$ is a non-zero vector in $\hilm$ and 
$\|\fr{d}{dT}A(f_T)\vac\|$ is an integrable function of $T$. 
We summarize here Herbst's analysis since it will be the starting point of our 
considerations: The operators $A(f_T)$ are constructed in a slightly different manner 
than in the work of Haag and Ruelle: 
First, a local operator $A$ is smeared in space with a regular solution of the Klein-Gordon equation 
$f(t,\vx)=\fr{1}{(2\pi)^{3/2}} \int e^{-i\og t+i\vep\vx}\ft(\vep)d^3p $, (where $\ft\in\czn$,
$\og=\sqrt{\vep^2+m^2}$):
\beq
A_t(f)=\int A(t,\vx)f(t,\vx)d^3x. \label{fTdef1}
\eeq
Next, to construct the time averaging function, we choose $s(T)=T^{\nu}$, 
$0<\nu<1$ and a positive function $h\in S(\mathbb{R})$ such that its Fourier transform satisfies $\hti\in\cz (\mathbb{R})$, $\hti(0)=(2\pi)^{-2}$. Then we set $h_T(t)=\fr{1}{s(T)}h(\fr{t-T}{s(T)})$ for $T\geq 1$ and define \cite{Her71,Bu77}:
\beq
A(f_T)=\int h_T(t) A_t(f) dt. \label{fTdef2}
\eeq 
It is clear from the formulas (\ref{fTdef1}) and (\ref{fTdef2}) that $f_T(x)=h_T(x^0)f(x^0,\vec{x})$.  
Its Fourier transform $\ft_T$ has a compact support which approaches a compact subset of the
mass hyperboloid as $T\to\infty$. In view of this fact we will refer to $A(f_T)$ as creation operators
and to $A(f_T)^*$ as annihilation operators. This terminology is also supported by the following
simple calculation:
\beq
\slim_{T\to\infty}A(f_T)\vac=\Pm A(f)\vac,
\eeq
where $A(f)=A_{t=0}(f)$ and $\Pm$ is the projection on the single-particle space $\hilm$. The integrability condition requires the following assumption:
\begin{enumerate}
\item[] \bf Condition $A^\prime$: \rm \label{cond-Ap} There exist operators $A\in\mfa(\mco)$ such that $\Pm A\vac\neq 0$ and 
for every $\delta\geq 0$
\beq
\|P(m^2-\delta\leq p^2\leq m^2+\delta)(1-\Pm)A\vac\|\leq c\,\delta^\eps,
\label{Regularity1}
\eeq
where $P(\,\cdot\,)$ is the spectral measure of the energy-momentum operators and  $c,\eps>0$. We refer to such operators as 'regular'.
\end{enumerate}
For regular operators there holds the bound
\beq
\|\fr{d}{dT}A(f_T)\vac\|\leq\fr{c}{s(T)^{1+\eps}} \label{derivative}
\eeq
which implies integrability if $\nu>\fr{1}{1+\eps}$. Now we are ready to state the main 
result of Herbst; we restrict attention to the outgoing asymptotic states $\Psi^+$, since 
the case of incoming states is completely analogous.
\bet\cite{Her71}\label{Herbsttheorem}  Suppose that the theory respects Conditions $M$ and $A^\prime$.
Then, for regular operators $A_i$, $i\in\{1,\ldots, n\}$, there exists the limit
\beq
\Psi^{+}=\slim_{T\to\infty} A_1(f_{1T})\ldots A_n(f_{nT})\vac
\eeq
and it depends only on the single-particle states $\Pm A_i(f_i)\vac$.
Moreover, given two states $\Psi^{+}$ and $\hat\Psi^{+}$, constructed as above using creation
operators $\Aa{i}$ and $\Ahh{i}$, $i\in\{1,\ldots,n\}$, their scalar product can be calculated as follows
\beq 
(\Psi^{+}|\widehat\Psi^{+})=
\sum_{\si\in S_n}(\vac|\An{1}^*\Pm \Ahhn{\si_{1}}\vac)\ldots(\vac|\An{n}^*\Pm \Ahhn{\si_{n}}\vac).
\eeq
Here the sum is over all permutations of an n-element set.
\eet
\nin It was, however, anticipated already by Ruelle \cite{Ru62} that in a purely massive theory 
Condition~$A$ can be replaced by the following, physically meaningful, stability requirement:
 \begin{enumerate}
 \item[] \bf Condition $S$: \rm \label{cond-S} In a theory satisfying M a particle can only be stable if, in its 
 superselection sector, its mass is separated from the rest of the spectrum by a
 lower and upper mass gap.
 \end{enumerate}
This condition is also stated in Herbst's work \cite{Her71}, but he expects that scattering
theory should be a necessary tool to study the superselection structure.
Subsequent analysis by  Buchholz and Fredenhagen \cite{BF82} clarified this issue:
There exist interpolating fields which connect the vacuum with the sector of the
given particle. Although they are, in general, localized in spacelike cones, they
can be used to construct a collision theory. Thereby there exists a prominent alternative to
the approach of Herbst in the realm of massive theories.  

It is the purpose of our investigations to extend Herbst's result to the situation where 
massless particles are present, that is  Conditions $M$ and $S$ do not hold.
A model physical example of a system with a sharp mass immersed in a spectrum
of massless particles is the hydrogen atom in its ground state from the point of view 
of quantum electrodynamics.
Although the approach of Herbst seems perfectly adequate to study such situations, the 
original proof of Theorem~\ref{Herbsttheorem} does not work because of the slow, 
quadratic decay of the correlation functions. In order to overcome this difficulty,
we apply the  bounds on creation operators obtained by  Buchholz \cite{Bu90}.
Namely, if $\Delta$ is any compact subset of the energy-momentum spectrum and 
$\ft$ vanishes sufficiently fast at zero then: 
\beqa
\|A(f_T)P(\Delta)\|\leq c,  \label{bound1}\\
\|A(f_T)^*P(\Delta)\|\leq c, \label{bound2}
\eeqa
where the constant $c$ does not depend on time. In fact, these bounds easily follow from
Theorem~\ref{theorem-2mollifier}: First, we note that by the Cauchy-Schwarz inequality
there holds for any $\eps>0$, $E\geq 0$
\beq
\|P_E A_t(f) P_E\|\leq \sup_{\fun\in\traceEB}\big(\int d^3p\,|\vep|^{4+\eps}|\fun(\wt{A}(\vep))|^2\big)^\h\big(\int d^3p\,\fr{1}{|\vep|^{4+\eps}}|\tf(\vep)|^2\big)^\h. \label{PEAtPE}
\eeq
In particular, the r.h.s. is independent of $t$. Next, making use of the fact that the
Fourier transform of $f_T$ has a compact support, we can find for any compact set $\Delta\subset \real^4$
another compact subset $\Delta^\prime$ of the energy-momentum spectrum  and $E\geq 0$ s.t.
\beq
\|A(f_T)P(\Delta)\|=\|P(\Delta^\prime)A(f_T)P(\Delta)\|\leq \|h\|_1\sup_{t\in\real}\|P_EA_t(f) P_E\|.
\eeq
This bound together with  estimate~(\ref{PEAtPE}) justifies relation~(\ref{bound1}). The case of
$A(f_T)^*$ is treated analogously.

The remaining part of this appendix is organized as follows: In Section~\ref{Asymptotic-states} we prove 
the existence of asymptotic states 
and verify that the limits are independent of the
actual value of the parameter $0<\nu<1$ chosen in the time averages of the operators $A(f_T)$. 
This property allows us to apply in Section~\ref{Fock-structure-AS}  the methods from
the collision theory of massless bosons \cite{Bu77} in order to calculate the scalar  
product of asymptotic states. In the Conclusion we summarize our results and discuss 
the status of Condition~$A^\prime$.

\section{Existence of Asymptotic States}\label{Asymptotic-states}

In order to prove the existence of asymptotic states we need information about 
the time evolution of the operators $A(f_T)$ and their commutators. It is the 
purpose of the two lemmas below to summarize the necessary properties.
Before we enter into these investigations, we recall that the regular solutions 
of the Klein-Gordon equation satisfy the bounds \cite{Ru62}
\beqa
|f_t(\vx)|&\leq& c(1+|t|)^{-3/2}, \label{KG-1}\\
\int d^3x\,|f_t(\vx)|&\leq& c(1+|t|)^{3/2}, \label{KG-2}
\eeqa
where the constant $c$ is independent of $t$ and $\vx$.
\bel \label{norm} Let $A(f_T)^{\#}$ denote $A(f_T)$ or $A(f_T)^*$. Then:
\begin{enumerate}
\item[(a)] $\|A(f_T)^{\#}\|\leq cT^{3/2}.$ 
\item[(b)] $P(\Delta_1)A(f_T)^{\#}P(\Delta_2)=0$ if 
$\Delta_1\cap(\Delta_2\pm\textrm{supp}\ft_T)=\emptyset$. The $(+)$ sign holds for
$A(f_T)$, $(-)$ for $A(f_T)^*$.
\item[(c)] Suppose that the functions $\ft_i$, $i\in\{1,\ldots,n\}$, vanish sufficiently fast at zero. 
Then, for any compact subset $\Delta$ of the energy-momentum spectrum:
 \beq
 \|\Aa{1}^{\#}\ldots \Aa{n}^{\#}P(\Delta)\|\leq c_1. \label{multiplebound}
 \eeq
\end {enumerate}
The constants $c$, $c_1$ do not depend on $T$.
\eel
\proof
\begin{enumerate}
\item[(a)] The statement follows from the estimate
\begin{eqnarray}
\|A(f_T)^{\#}\|&\leq&\|A^{\#}\|\int dt h_T(t)\int d^3x|f(t,\vx)|
\leq c_0\int dt\, h_T(t)(1+|t|)^{3/2}\nonumber\\
&=& c_0\int dt\, h(t)(1+|s(T)t+T|)^{3/2}
\leq c T^{3/2},
\end{eqnarray}
where in the second step we used property~(\ref{KG-1}) of regular solutions 
of the Klein-Gordon equation.
\item[(b)] See, for example, \cite{Ar82}.
\item[(c)] For $n=1$ the assertion follows from (\ref{bound1}) and (\ref{bound2}).
Assuming that (\ref{multiplebound}) is valid for $n-1$ and making use of
part (b) of this lemma we estimate
\beqa
& &\|\Aa{1}^{\#}\ldots \Aa{n}^{\#}P(\Delta)\|\nonumber\\
&=&\|\Aa{1}^{\#}\ldots\Aa{n-1}^{\#}P(\Delta\pm\textrm{supp}\ft_{n,T}) \Aa{n}^{\#}P(\Delta)\|\nonumber\\
&\leq&\|\Aa{1}^{\#}\ldots\Aa{n-1}^{\#}P(\Delta\pm\textrm{supp}\ft_{n,T})\|\|\Aa{n}^{\#}P(\Delta)\|.\quad
\eeqa 
The last expression is bounded by the inductive assumption and the support properties
of functions $\ft_T$. \qed
\end {enumerate} 
Now we turn our attention to the commutators of the operators $A(f_T)$.
It will simplify this discussion to decompose the function $f_T$ into
its compactly supported dominant contribution and a spatially extended,
but rapidly decreasing remainder \cite{BBS01}. To this end, let us define the velocity
support of the function $\ft$ 
\beq
\Ga(\ft) =\{(1,\fr{\vep}{\og})\,\big|\, \vep\in \mathrm{supp}\ft \}.
\eeq
We introduce a function $\chi_{\delta}\in\cznn$ such that $\chi_{\delta}=1$ on $\Ga(\ft)$
and $\chi_{\delta}=0$ in the complement of a slightly larger set $\Ga(\ft)_{\delta}$. 
$\hf_T(x)=f_T(x)\chi_{\delta}(x/T)$ is the asymptotically dominant part of
$f_T$, whereas $\cf_T(x)=f_T(x)(1-\chi_{\delta}(x/T))$ tends rapidly to 
zero with $T\to\infty$ \cite{Her71,He66}. In particular, for
each natural $N$ and some fixed $N_0>4$ there exists a constant $c_N$ such that
\beq
\int |\cf_T(x)|d^4x\leq c_N\fr{s(T)^{N+N_0}}{T^N}. \label{integral}
\eeq
We remark that this bound relies on the slow increase of the function $s(T)$, 
so Condition~$A^\prime$ cannot be eliminated simply by modifying this function. 

As was observed first by Hepp \cite{He65}, particularly strong estimates on commutators 
can be obtained in the case of particles moving with different velocities.
\bel \label{com} Let $A_1(f_{1,T})$, $A_2(f_{2,T})$, $A_3(f_{3,T})$ be defined as above. 
Moreover, let $\ft_1$, $\ft_2$ have disjoint velocity supports. Then, for each natural $N$, there
exists a constant $c_N$ such that:
\begin {enumerate}
\item[(a)] $\|[A_1(f_{1,T}),A_2(f_{2,T})]\|\leq \fr{c_N}{T^N}.$
\item[(b)] $\|[A_1(f_{1,T}),[A_2(f_{2,T}),A_3(f_{3,T})]]\|\leq\fr{c_N}{T^N}.$
\end{enumerate}
The same estimates are valid if some of the operators $A(f_T)$ are replaced
by their adjoints or time derivatives.
\eel
\proof
\begin {enumerate}
\item[(a)]  Making use of the decomposition $f_T=\hf_T+\cf_T$, we obtain 
 \beqa
 [A_1(f_{1,T}),A_2(f_{2,T})]
 =[A_1(\hf_{1,T}),A_2(\hf_{2,T})]+[A_1(\hf_{1,T}),A_2(\cf_{2,T})] \nonumber\\
 +[A_1(\cf_{1,T}),A_2(\hf_{2,T})]+[A_1(\cf_{1,T}),A_2(\cf_{2,T})].
 \eeqa
The first term on the r.h.s. is a commutator of two local operators. For
sufficiently large $T$ their localization regions become spatially separated
because of disjointness of the velocity supports of $\ft_1$ and $\ft_2$. Then the
commutator vanishes by virtue of locality. Each of the remaining terms contains
a factor $A(\cf_T)$ which decreases in norm faster than any power of $T^{-1}$
by  estimate (\ref{integral}). It is multiplied by $A(\hf_T)$ which increases
in norm only as $T^{3/2}$ by Lemma~\ref{norm} (a).

\item[(b)] First, let us suppose that $\ft_3$ and $\ft_2$ have disjoint velocity supports. 
Then $[A_2(f_{2,T}),A_3(f_{3,T})]$ decreases fast in norm as a consequence 
of part (a) of this lemma. Recalling that the norm of $A_1(f_{1,T})$ increases at most as 
$T^{3/2}$ the assertion follows. Now suppose that $\ft_3$ and $\ft_1$ have disjoint velocity supports. 
Then, by application of the Jacobi identity, we arrive at the previous situation.
In the general case we use a smooth partition of unity to decompose $\ft_3$ into a
sum of two functions, each belonging to one of the two special classes studied
above. 
\end{enumerate}
The statement about adjoints is obvious. To justify the claim concerning  
derivatives we note that:
\beq
\fr{d}{dT}A(f_T)=-\fr{1}{s(T)}\bigg\{\fr{ds(T)}{dT}\big(A(f_T)+A(f_{aT})\big)+A(f_{bT})\bigg\}, 
\label{derivative1}
\eeq
where $f_{aT}$ is constructed using $h_a(t)= t\fr{dh(t)}{dt}$ and $f_{bT}$ 
contains $h_b(t)=\fr{dh(t)}{dt}$. Although $h_a$ and $h_b$ do not satisfy 
all the conditions imposed previously on functions $h$, they are elements of 
$S(\mathbb{R})$. This property suffices to prove the decomposition 
$f_T=\hf_T+\cf_T$. \qed\\
Having constructed creation operators and studied their properties, it
will be fairly simple to demonstrate the existence of asymptotic
states. The following theorem uses the original method of Haag \cite{Ha58} 
modified by Araki \cite{Ar}.
\bet\label{Haag}\label{asymptotic-states} 
Suppose that local operators $A_1,\ldots, A_n$ are regular, $\ft_1,\ldots, \ft_n$ have
disjoint velocity supports and vanish sufficiently fast at zero. Moreover, $s(T)=T^{\nu}$,
$\fr{1}{1+\eps}<\nu<1$, where $\eps$ is the exponent appearing in the regularity condition $A^\prime$.
Let us denote $\Psi(T)=A_1(f_{1T})\ldots A_n(f_{nT})\vac$. Then there exists the limit 
$\Psi^+=\slim_{T\to\infty}\Psi(T)$ and it is called the asymptotic state.
\eet
\proof
We verify the Cauchy condition using Cook's method
\beq
\|\Psi(T_2)-\Psi(T_1)\|\leq \int_{T_1}^{T_2}\big\|\fr{d\Psi(T)}{dT}\big\|\, dT.\label{Cauchy}
\eeq
It now has to be checked whether the integrand decays sufficiently
fast when $T\to\infty$. By using the Leibniz rule, and then commuting the derivatives of creation operators 
with the other operators until they act on the vacuum, we arrive at the 
following expression
\beqa
\fr{d\Psi}{dT}&=&\sum_{k=1}^{n} \Aa{1}\ldots \fr{d}{dT}\Aa{k}\ldots \Aa{n}\vac\nonumber\\
&=&\sum_{k=1}^{n}\big\{\sum_{l=k+1}^{n}\Aa{1}\ldots [\fr{d}{dT}\Aa{k},\Aa{l}]\ldots\Aa{n}\vac\nonumber\\
&+&\Aa{1}\ldots\check{k}\ldots\Aa{n}\fr{d}{dT}\Aa{k}\vac\big\},
\eeqa
where $\check{k}$ denotes omission of $A_k(f_{kT})$.
Each term containing commutators vanishes in norm faster than any power of 
$T^{-1}$ since the rapid decay of commutators, proved
in Lemma~\ref{com}~(a), suppresses the polynomial increase of $\|A(f_T)\|$ shown in
Lemma~\ref{norm}~(a). To estimate the 
remaining terms we first note that, by virtue of  formula~(\ref{derivative1})
and Lemma~\ref{norm} (b), the vector $\fr{d}{dT}\Aa{k}\vac$ has a compact spectral support
$\Delta$. Consequently,
\beqa
 &  & \|\Aa{1}\ldots\check{k}\ldots\Aa{n}\fr{d}{dT}\Aa{k}\vac\|  \nonumber\\
 &= & \|\Aa{1}\ldots\check{k}\ldots\Aa{n}P(\Delta)\fr{d}{dT}\Aa{k}\vac\| \nonumber\\
 &\leq& \|\Aa{1}\ldots\check{k}\ldots\Aa{n}P(\Delta)\|\|\fr{d}{dT}\Aa{k}\vac\|
 \leq \fr{c}{s(T)^{1+\eps}},
\eeqa
where in the last step we made use of Lemma~\ref{norm} (c) and  estimate~(\ref{derivative}).
As $\nu(1+\eps)>1$, the integral~(\ref{Cauchy}) tends to zero when $T_1,T_2\to\infty$ and 
the Cauchy condition is satisfied. \qed\\
It is a remarkable feature of asymptotic states with disjoint velocity supports
that already at this stage it is possible to prove that they depend only
on the single-particle states $\Pm A(f)\vac$ rather than on the
specific $A$, $\ft$, $h$, and $s$ that were used to construct them. As the possibility
to relax the increase of functions $s(T)$ is particularly important for us, we 
temporarily introduce the notation $A(f_T^s)$ to distinguish between operators
containing different functions $s(T)$. The following lemma is due to D. Buchholz.
\bel\label{sT} Suppose that the families of operators $\As{1},\ldots, \As{n}$, resp. 
$\Ah{1},\ldots,\Ah{n}$,  satisfy the following conditions:
\begin{enumerate} 
\item[(a)] The functions $\ft_1,\ldots,\ft_n$, resp. $\tilde{\widehat{f}_1},\ldots,\tilde{\widehat{f}_n}$,
have, within each family, disjoint velocity supports and vanish sufficiently fast at zero.
\item[(b)] $\Pm A_i(f_i)\vac=\Pm\widehat{A}_i(\widehat{f_i})\vac$, $i\in\{1,\ldots, n\}$, i.e. the single-particle 
states corresponding to the two families of operators coincide.
\item[(c)] $\Psi^{+}=\slim_{T\to\infty}A_{1}(f_{1,T}^s)\ldots A_{n}(f_{n,T}^s)\vac$
exists.
\end{enumerate}
Then the limit $\widehat\Psi^{+}=\slim_{T\to\infty}\Ah{1}\ldots \Ah{n}\vac$ exists and
coincides with $\Psi^{+}$.
\eel
\proof
We proceed by induction. For $n=1$ the assertion is satisfied by assumption.
Let us assume that it is satisfied for states involving $n-1$ creation operators.
Then the following inequality establishes the strong convergence of the net $\As{1}\Ah{2}\ldots \Ah{n}\vac$
\beqa
\|\As{1}\As{2}\ldots \As{n}\vac-
\As{1}\Ah{2}\ldots \Ah{n}\vac\|\nonumber\\ 
\leq \|\As{1}P(\Delta)\| \|\As{2}\ldots \As{n}\vac-\Ah{2}\ldots \Ah{n}\vac\|,
\eeqa
where $\Delta$ is the spectral support of the product of creation operators acting on the vacuum which  
is compact by Lemma~\ref{norm} (b). The r.h.s. of this expression vanishes in the 
limit of large $T$ as a consequence of  estimate~(\ref{bound1}) and the induction hypothesis. 
By applying the bound on commutators proved in Lemma~\ref{com} (a) and the estimate from 
Lemma~\ref{norm} (a) we verify that also $\Ah{2}\ldots \Ah{n}\As{1}\vac$ converges strongly and has
the same limit. Finally, our claim follows from the estimate
\beqa
&  &  \|\Ah{2}\ldots \Ah{n}\big(\Ah{1}-\As{1}\big)\vac\|\nonumber\\
&\leq& \|\Ah{2}\ldots \Ah{n}P(\Delta_1)\|\|\big(\Ah{1}-\As{1}\big)\vac\|,
\eeqa
where $\Delta_1$ is again a compact spectral support.
The r.h.s. of this inequality tends to zero with $T\to\infty$ by assumption~(b)
and the bound in Lemma~\ref{norm} (c). \qed\\

\section{Fock Structure of Asymptotic States}\label{Fock-structure-AS}

It was instrumental in the original proof of the existence of asymptotic states that $s(T)=T^\nu$, where
$\nu$ was sufficiently close to one. Lemma~\ref{sT} allows us to relax this condition and choose any
$0<\nu<1$. Using this piece of information we will verify the Fock structure of the
scattering states by the following strategy: First, we establish a counterpart of the relation
$a a^*\vac=(\vac|a a^*|\vac)\vac$ satisfied by the ordinary creation and annihilation operators.
Once this equality is proven in the sense of strong limits, we combine it with
the double commutator bound from Lemma~\ref{com} (b) to obtain the factorization of 
the scalar product of the scattering states.

We start from two definitions: $\mco(r)$ is the double cone of radius $r$.
By $\mfa_0$ we denote the weakly dense subalgebra of $\mfa$ consisting of operators 
for which the operator valued functions $ x\to A(x)$ are infinitely often differentiable 
in the norm topology. (We remark that, given any regular operator, we can construct
a regular operator in $\mfa_0$ by smearing it with a smooth function).

Now we proceed along the lines set in the scattering theory of massless bosons.
(See \cite{Bu77} p.169.) The analysis is  based on the following result, due to Araki, Hepp and Ruelle,
on the two-point function of operators from $\mfa_0$ localized in double cones.
\bel\label{AHR}\cite{AHR62}
Let $C_1$ and $C_2$ be local operators in $\mfa_0$, localized in double cones  $\mco(r_1)$,
 $\mco(r_2)$. Then for all $|\vx|\geq 2(r_1+r_2)$
\beq
|(\vac|C_1(\vx)(1-P_0)C_2\vac)|\leq c |\vx|^{-2}(r_1+r_2)^3\{\|C_1^*\vac\|\|\pa_0 C_2\vac\|+
\|C_2^*\vac\|\|\pa_0 C_1\vac\|\},
\eeq
where $P_0=|\vac)(\vac|$  and $c$ is a constant which depends neither on $\vx$
nor on $C_1$, $C_2$ and $r_1$,~$r_2$. 
\eel
\nin We will apply this lemma, 
taking as $C_1$ and $C_2$ the commutators  $[A_i(t_i,\vx_i),A_j(t_j,\vx_j)]$. 
It will be used in the proof of the following statement  that these operators are localized 
in regions of finite volume, as long as the differences $|t_i-t_j|$ are kept small.
\bel\label{AHRB-lemma} Let $A_1,\ldots, A_4\in\mfa_0$ be localized in double cones $\mco(r_1),\ldots,\mco(r_4)$.
We define
\beq
K=(\vac|[A_1(t_1,\vx_1),A_2(t_2,\vx_2)](1-P_0)[\At{3},\At{4}]\vac),
\eeq
where $P_0=|\vac)(\vac|$. Then the following estimate holds
\beq
|K|\leq c\chi(|\vx_1-\vx_2|\leq R) \chi(|\vx_3-\vx_4|\leq R)\cdot\left\{\begin{array}{cc}
 1 & \textrm{if $|\vx_2-\vx_3|\leq 4R$}\\
\fr{R^3}{|\vx_2-\vx_3|^2+R^2} & \textrm{if $|\vx_2-\vx_3|>4R$}
\end{array}\right. 
\eeq
where $R=\sum_{i=1}^4(r_i+|t-t_i|)$, $t=\fr{1}{4}(t_1+t_2+t_3+t_4)$\label{symbol-new-R} and 
$\chi$ are the characteristic functions of the respective sets. The constant $c$ depends neither on 
$t_1,\ldots,t_4$ nor on $r_1,\ldots, r_4$.
\eel
\proof Using translation invariance of the vacuum, we can rewrite $K$ as follows:
\beq
K=(\vac|[A_1(t_1-t,\vr_{12}),A_2(t_2-t,\vr_{21})]U(\vx)(1-P_0)[A_3(t_3-t,\vr_{34}),A_4(t_4-t,\vr_{43})]\vac),
\eeq
where $\vr_{ik}=\fr{1}{2}(\vx_i-\vx_k)$, $\vx=\fr{1}{2}(\vx_1+\vx_2-\vx_3-\vx_4)$.
The commutator $[A_i(t_i-t,\vr_{ik}),A_k(t_k-t,\vr_{ki})]$ is localized in a double cone
of radius $r$ given by
\beq
r=|\vr_{ik}|+r_i+r_k+|t_i-t|+|t_k-t|.
\eeq
But it is zero (owing to locality) if
\beq
2|\vr_{ik}|\geq r_i+r_k+|t_i-t|+|t_k-t|.
\eeq
Consequently, there follows the bound for $r$
\beq
r\leq\fr{3}{2}(r_i+r_k+|t_i-t|+|t_k-t|).
\eeq
Now we can apply Lemma~\ref{AHR} and get for $|\vx|\geq 3\sum_{i=1}^{4}(r_i+|t_i-t|)$
\beqa
|K| &\leq& c|\vx|^{-2}\big(\sum_{i=1}^{4}(r_i+|t_i-t|)\big)^3\nonumber\\
    &\cdot&\{\|[\At1,\At2]^*\vac\|\,\|[\pa_0\At3,\At4]\vac \nonumber\\
    &+&[\At3,\pa_0\At4]\vac\|+(1\leftrightarrow 3,2\leftrightarrow 4)\}.\label{Coulomb} 
\eeqa
The r.h.s. of this expression vanishes, if either of the following two conditions holds
\beqa
|\vx_1-\vx_2|\geq |t_1-t_2|+r_1+r_2,\\ 
|\vx_3-\vx_4|\geq |t_3-t_4|+r_3+r_4.
\eeqa
We may therefore estimate $|\vx|$ as follows:
\beqa
|\vx|&=&\fr{1}{2}\big|2(\vx_2-\vx_3)-(\vx_2-\vx_1)-(\vx_4-\vx_3)\big| \nonumber\\
&\geq& |\vx_2-\vx_3|-\fr{1}{2}(|\vx_2-\vx_1|+|\vx_4-\vx_3|) \nonumber\\
&\geq& |\vx_2-\vx_3|-\fr{1}{2}\sum_{i=1}^{4}(r_i+|t_i-t|)=|\vx_2-\vx_3|-\h R.
\eeqa
If $|\vx_2-\vx_3|> 4R$ then $|\vx|\geq 3R$ 
and it follows from the bound (\ref{Coulomb}) that
\beq
|K|\leq cR^3(|\vx_2-\vx_3|^2+R^2)^{-1} \chi(|\vx_1-\vx_2|\leq R) \chi(|\vx_3-\vx_4|\leq R).
\eeq
When $|\vx_2-\vx_3|\leq 4R$ we estimate trivially
\beqa
|K|&\leq&\|[\At1,\At2]\|\, \|[\At3,\At4]\|\non\\
&\leq& c^\prime \chi(|\vx_1-\vx_2|\leq R) \chi(|\vx_3-\vx_4|\leq R),
\eeqa
what concludes the proof. \qed\\
\nin Now we are ready to prove that a product of a creation and  annihilation operator acting on 
the vacuum reproduces it.
\bep Suppose that $A_1$, $A_2$ are local operators from $\mfa_0$, $s(T)=T^{\nu}$, $\nu<1/8$. Then
\beq
\slim_{T\to\infty}\An{1}^*\An{2}\vac=(\vac|\An{1}^*\Pm\An{2}\vac)\vac. 
\eeq
\eep
\proof
We start by performing the integration of the function $K$ from the preceding lemma 
with the regular Klein-Gordon wave-packets 
and estimating the behavior of the resulting function of $t_1,\ldots, t_4$. We will change the variables
of integration to $\vu_1=\vx_2-\vx_1$, $\vu_2=\vx_2$, $\vu_3=\vx_3-\vx_2$, $\vu_4=\vx_4-\vx_3$.
In the region $|\vx_2-\vx_3|> 4R$ we obtain
\beqa
&  &\int d^3x_1|f_1(t_1,\vx_1)|\ldots \int d^3x_4|f_4(t_4,\vx_4)| |K|\nonumber\\
&\leq& cR^9(1+|t_1|)^{-3/2}(1+|t_4|)^{-3/2}\int d^3u_2\int d^3 u_3
\fr{|f_2(t_2,\vu_2)||f_3(t_3,\vu_3+\vu_2)|}{|\vu_3|^2+R^2}\nonumber\\
&\leq& cR^9(1+|t_1|)^{-3/2}(1+|t_4|)^{-3/2}(1+|t_2|)^{3/2}.\quad\label{integrals}
\eeqa
Here in the first step we applied the bound~(\ref{KG-1}) to $f_1$ and $f_4$. 
In the second step we exploited estimate~(\ref{KG-2}) to control the integral w.r.t. $\vu_3$ and 
we used the inequality  $2\fr{|f_3(\cdot,\cdot)|}{|\vu_3|^2+R^2}\leq 
(\fr{1}{|\vu_3|^2+R^2})^2+|f_3(\cdot,\cdot)|^2$ in order to verify that the integral over $\vu_3$
is bounded in $t_3$. In the region  $|\vx_2-\vx_3|\leq 4R$ our estimate becomes improved by the
factor $(1+|t_3|)^{-3/2}$, so (\ref{integrals}) holds without restrictions.

Since $A(f_T)^*\vac=0$ for sufficiently large $T$, we can estimate:
\beqa
& &\|\Aa{1}^*\Aa{2}\vac-(\vac|\Aa{1}^*\Aa{2}\vac)\vac\|^2\nonumber\qquad\\
&=&(\vac|[\Aa{2}^*,\Aa{1}](1-P_0)[\Aa{1}^*,\Aa{2}]\vac)\nonumber\\
&\leq&\int dt_1 h_T(t_1)\ldots\int dt_4 h_T(t_4)
cR^9(1+|t_1|)^{-3/2}(1+|t_4|)^{-3/2}(1+|t_2|)^{3/2}\nonumber\\
&\leq& C\fr{s(T)^{12}}{T^{3/2}}.
\eeqa
Now the assertion follows from the slow increase of the function $s(T)$. \qed\\
After this preparation it is straightforward to calculate the scalar product of two asymptotic
states. 
\bet\label{scalar-product-as} Suppose that  $\Aa{1},\ldots,\Aa{n}$, resp. $\Ahh1,\ldots,\Ahh{n}$, are two families 
of creation operators constructed using local operators from $\mfa_0$, functions
 $\ft_i$, resp. $\widehat\ft_i$, $i=1,\ldots, n$, vanishing sufficiently fast at zero 
and having, within each family, disjoint velocity supports. Moreover, $s(T)=T^{\nu}$, where  
$0<\nu<1$. Then
\beqa
\lim_{T\to\infty}
(\vac|\Aa{n}^*\ldots\Aa1^*\Ahh1\ldots\Ahh{n}\vac)\nonumber\\
=\sum_{\si\in S_n}(\vac|\An{1}^*\Pm\Ahhn{\si_{1}}\vac)\ldots(\vac|\An{n}^*\Pm\Ahhn{\si_{n}}\vac).
\eeqa
Here the sum is over all permutations of an n-element set.
\eet
\proof
First, we make use of Lemma~\ref{sT} to ensure a sufficiently slow
increase of the function $s(T)$. Next,  
we proceed by induction. For $n=1$ the theorem is trivially true.
Let us assume that it is true for $n-1$ and compute
\beqa
& &(\vac|\Aa{n}^*\ldots\Aa{1}^*\Ahh1\ldots\Ahh{n}\vac)\nonumber\\
&=&\sum_{k=1}^{n}(\vac|\Aa{n}^*\ldots\Aa{2}^*\Ahh1\ldots[\Aa{1}^*,\Ahh{k}]\ldots\Ahh{n}\vac)
\nonumber\\
&=&\sum_{k=1}^{n}\big\{\sum_{l=k+1}^{n}
(\vac|\Aa{n}^*\ldots\Aa{2}^*\Ahh1\ldots\nonumber\\
&\ldots& [[\Aa{1}^*,\Ahh{k}],\Ahh{l}]\ldots\Ahh{n}\vac)
\nonumber\\
&+&(\vac|\Aa{n}^*\ldots\Aa{2}^*\Ahh1\ldots\check{k}\ldots\Ahh{n}\Aa{1}^*\Ahh{k}\vac)\big\}.\quad\quad\quad 
\eeqa
Terms containing double commutators vanish in the limit by Lemma~\ref{com} (b) and Lemma~\ref{norm} (a).
The remaining terms factorize by the preceding proposition and by Lemma~\ref{norm} (b) and (c)
\beqa
& &\lim_{T\to\infty}(\vac|\Aa{n}^*\ldots\Aa2^*\Ahh1\ldots\check{k}\ldots\Ahh{n}\Aa{1}^*\Ahh{k}\vac) 
\nonumber\\
&=&\lim_{T\to\infty}(\vac|\Aa{n}^*\ldots\Aa{2}^*\Ahh1\ldots\check{k}\ldots\Ahh{n}|\vac)\cdot \nonumber\\
& &\ph{44444444444444444444444444444444444444}\cdot(\vac|\An1^*\Pm\Ahhn{k}\vac).
\qquad
\eeqa
This quantity factorizes into two-point functions by the induction 
hypothesis. \qed\\
It is also evident from the proof that the scalar product of two asymptotic
states involving different numbers of operators is zero. Now the  Fock 
structure of asymptotic states and the construction of the wave operators follows by standard 
density arguments: By Condition~$A^\prime$ there exists a regular operator $A\in\mfa(\mco)$ s.t.
$\Pm A\vac\neq 0$. Since the representation $U$ acts irreducibly in $\hilm$, the vectors
$\{\,U(x,\La)\Pm A\vac\,|\, (x,\La)\in\cpo\,\}$ span a dense set in this subspace. Making use
of the fact that a Poincar\'e transformation of a regular operator is regular, we obtain that the
set of vectors 
\beq
\{\, \Pm A(f)\vac\, |\, A\te{-regular}, \tf\in C_0^\infty(\real^3)\te{ vanishes sufficiently fast at $0$ }\}
\label{admissible}
\eeq
is total in $\hilm$. In view of Theorem~\ref{asymptotic-states} the wave operator $W^+$ can be defined on a dense
set in $\Ga(\hilm)$ extending by linearity the following relation
\beq
W^+\big(a^*(\Pm A_1(f_1)\vac)\ldots a^*(\Pm A_n(f_n)\vac)\vac\big)=\slim_{T\to\infty}\Aa{1}\ldots\Aa{n}\vac,
\eeq
where $A$ and $f$ satisfy the conditions stated in (\ref{admissible}). Due to Theorem~\ref{scalar-product-as},
$W^+$ preserves norms, thus it extends to an isometry from $\Ga(\hilm)$ to $\hil$. The wave operator $W^-$
is constructed analogously, making use of the incoming states, and the scattering matrix $S: \Ga(\hilm)\to\Ga(\hilm)$ 
is given by $S=(W^+)^*W^-$.

\section{Conclusion}
We have constructed a scattering theory of massive particles without
the lower and upper mass gap assumptions. The Lorentz covariance of
the construction can be verified by application of standard arguments 
\cite{Ar}. Including fermions would cause no additional difficulty, 
as the fermionic creation operators are bounded uniformly in time \cite{Bu75}.

The only remaining restriction is the regularity assumption $A^\prime$. We note that it
was used only to establish the existence of scattering states - the construction 
of the Fock structure was independent of this property. Moreover, we would like
to point out that it does not seem possible to derive it from general postulates.
In fact, let us consider the generalized free field $\phi$ with the commutator fixed
by the measure $\sigma$:
\beq
[\phi(x),\phi(y)]=\int d\sigma(\lambda)\Delta_{\lambda}(x-y),
\eeq 
where $\Delta_{\lambda}$ is the commutator function of the free field of mass $\sqrt{\lambda}$.  
Suppose that the measure $\sigma$ contains a discrete mass $m$ and in its neighborhood
is defined by the function $F(\lambda)=1/\ln|\lambda-m^2|$. Then it is easy to
find polynomials in the fields smeared with Schwartz class functions which
violate the bound from Condition $A^\prime$. However, the existence of scattering states, constructed with the
help of such polynomials, can easily be verified
using the properties of generalized free fields. These observations indicate that 
Condition $A^\prime$ is only of  technical nature. To relax it one should probably look for a construction 
of asymptotic states which avoids Cook's method - perhaps similarly to the scattering theory
of massless particles \cite{Bu75,Bu77}.

%% file: PhD-Preliminaries_app.tex
\chapter{Scalar Free Field Theory and its Phase Space Structure} \label{Preliminaries}
In this appendix we collect some known results on the phase space structure of scalar free field theory 
in a slightly modified form, suitable for our purposes. They provide a basis for the proofs that the new
Conditions~$\A$, $\B$, $\Nnat$ and $\Csq$, introduced in the main body of this work, hold in the 
model of scalar, non-interacting particles. These arguments are given in the subsequent appendices.
For Conditions~$\B$ and $\Csq$ only the massive case will  be considered. 

Verification of phase space conditions in models is an integral part of phase
space analysis in QFT as it proves consistency of the introduced criteria with the basic postulates of quantum field
theory. This issue was treated already in the seminal paper of Haag and Swieca \cite{HS65}, who verified 
their compactness condition in massive scalar free field theory. Serious technical improvements,
including the reduction of the problem  to a single-particle question, appeared in the 
work of Buchholz and Wichmann \cite{BWi86}, who noted the importance of nuclearity.
The massless case was included for the first time in \cite{BJa87}.
Conditions~$\Cs$ and $\Ns$, stated in Sections~\ref{existence-of-vacuum} and \ref{additivity-of-energy}, 
respectively, were first verified in \cite{BP90}. A particularly flexible and explicit formulation of 
the subject was given by Bostelmann in \cite{Bo00} and our presentation relies primarily on this work.

The  goal is to construct an expansion of the map $\whXi_E: \mfa(\mco(r))\to B(\hil)$, given by $\whXi_E(A)=P_EAP_E$,
into rank-one mappings. More precisely, we are looking for functionals $\tau_i\in\mfa(\mco(r))^*$ and 
operators $S_i\in B(\hil)$, $i\in\nat$, s.t. 
\beqa
\whXi_E(A)=\sum_i\tau_i(A)S_i,\quad\quad A\in\mfa(\mco(r)), \label{expansion-of-whXi}\\
\sum_{i}\|\tau_i\|^p\,\|S_i\|^p<\infty,\quad\quad 0<p\leq 1. \label{nuclearity-of-whXi}
\eeqa
Thus, in view of Lemma~\ref{equivalent-Nsharp},  we corroborate the well known fact that Condition~$\Ns$ holds in 
scalar free field theory \cite{BP90}. Moreover, we establish  properties of the functionals $\{\tau_i\}_1^\infty$ and $\{S_i\}_1^{\infty}$ which will be needed to verify the new conditions introduced in this Thesis.

This appendix is organized as follows: In Section \ref{multiindex-notation} we explain our multiindex notation.
Section~\ref{scalar-free-field-theory} introduces scalar free field theory, its even part and
its sub-theory generated by derivatives of the field.
In Section \ref{special-functionals} we construct the functionals $\{\tau_i\}_1^\infty$  and estimate
their norms. Apart from the material familiar from Section~7.2.B of \cite{Bo00}, we establish certain energy
bounds on these functionals which we  need to verify Condition~$\B$. In Section~\ref{single-particle-expansions}
two expansions of the single-particle wavefunctions are developed, following Sections 7.2.2 and 7.2.3 of \cite{Bo00}.
They give rise to an expansion of the map $\whXi_E$ into rank-one mappings, which we introduce in Section~\ref{full-expansion}.
It is a variant of the 'mixed expansion' from Section~7.2.6 of \cite{Bo00}. In order
to avoid the unduly complicated infrared structure and phase space properties of low dimensional
massless models, in this appendix we adopt the following:\\
\nin\bf Standing assumption: \rm Unless stated otherwise, all statements concerning 
scalar free field theory hold  either for  $m>0$ and $s\geq 1$ or $m=0$ and $s\geq 3$.

\section{Multiindex Notation} \label{multiindex-notation}

A multiindex is a sequence $\mu=\{\mu(i)\}_1^{\infty}$ of elements from
$\nat_0$ s.t. only a finite number of components is different from zero.
Addition of multiindices is performed component-wise. The length of a multiindex
is given by
\beq
|\mu|=\sum_{i} \mu(i).
\eeq
The factorial of a multiindex is defined as $\mu!=\prod_{i}\mu(i)!$. 
Given a sequence $a=\{a_i\}_{1}^{\infty}$, valued in any set with multiplication,
its multiindex power is defined as
\beq
a^{\mu}=\prod_ia_i^{\mu(i)}.
\eeq
It is convenient to extend the above conventions to pairs of multiindices $\mub=(\mup,\mum)$:
The length of such 2-multiindex is given by $|\mub|=|\mup|+|\mum|$ and the factorial is
defined as $\mub!=\mup!\mum!$. \label{symbol-2multi} Given any sequence of pairs $b=\{b_i^+,b_i^-\}_{1}^{\infty}$,
we define the 2-multiindex power of $b$ as follows
\beq
b^{\mub}=(b^+)^{\mup}(b^-)^{\mum}.
\eeq
Finally, an $n$-index $\ka$ \label{symbol-ka} is a multiindex s.t. $\ka(i)=0$ for $i>n$. All the above conventions
extend naturally to $n$-indices.

We recall that for any sequence of complex numbers $\{t_j\}_1^{\infty}$ and $k\in\nat$ there holds
the multinomial formula
\beq
\big(\sum_{j=1}^n t_j\big)^{k}=\sum_{\mu,|\mu|=k}\fr{|\mu|}{\mu!}t^{\mu}, \label{multiform}
\eeq
where the sum on the r.h.s. extends over all $n$-indices of length $k$. Assuming that the sequence $\{t_j\}_1^{\infty}$
is absolutely summable, we can take the limit $n\to\infty$, obtaining on the r.h.s. the sum over all
multiindices $\mu$ of length $k$.

\section{Scalar Free Field Theory and Related Models}\label{scalar-free-field-theory}
In this section  we recall from  Section~X.7 of \cite{RS2} some basic 
properties of scalar free field theory of mass $m \geq 0$ in $s$ space dimensions.
The single-particle space of this theory is $L^2(\real^s, d^sp)$. On this space there act the multiplication operators $\om(\vep)=\sqrt{|\vep|^2+m^2}$\label{symbol-om(p)} and $p_1, \ldots, p_s$ which are self-adjoint on a suitable dense domain. The  unitary representation of the Poincar\'e group $\cpo\ni(x,\La)\to U_1(x,\La)$, acting on the single-particle space, 
is given by
\beq 
(U_1(x,\La)f)(\vep)=e^{i(\om(\vep)x^0-\vep \vx)}f(\La^{-1}\vep),\quad f\in L^2(\real^s, d^sp),
\eeq
where  $\La^{-1}\vep$ is the spatial part of the four-vector $\La^{-1}(\om(\vep),\vep)$.
The full  Hilbert space $\hil$ of the theory is the symmetric Fock space over
$L^2(\real^s, d^sp)$. By the method of second quantization we obtain the unitary representation
of the Poincar\'e group $U(x,\La)=\Ga(U_1(x,\La))$ which implements the corresponding family
of automorphisms acting on $B(\hil)$
\beq
\al_{(x,\La)}(\,\cdot\,)=U(x,\La)\,\cdot\, U(x,\La)^*.
\eeq
The Hamiltonian $H=d\Ga(\om)$, and the momentum operators $P_i=d\Ga(p_i)$, $i=1,2,\ldots,s$ 
are defined on a suitable domain in $\hil$. The joint spectrum of this family of commuting, self adjoint 
operators is contained in the closed forward light cone. 

The elements of $\hil$ have the form $\Psi=\{\Psi_n\}_{n=0}^\infty$, where
$\Psi_n$ is an $n$-particle vector. We denote by $\DF$\label{symbol-DF} the dense subspace of finite particle vectors, consisting 
of such $\Psi$ that $\Psi_n\neq 0$ only for finitely many $n$. On this subspace we define the annihilation
operator $a(f)$, $f\in L^2(\real^s, d^sp)$, given by the formula
\beq
(a(f)\Psi)_n(\vep_1,\ldots,\vep_n)=\sqrt{n+1}\int d^sp \bar{f}(\vep)\Psi_{n+1}(\vep,\vep_1,\ldots,\vep_n) \label{symbol-a(f)}
\eeq
and its adjoint $a^*(f)$. With the help of these operators we construct the (time zero) canonical fields and momenta
for any $g\in S(\real^s)$
\beqa
\phip(g)&=&\fr{1}{\sqrt{2}}\big(a^*(\om^{-\h}{\tilde{g}})+a(\om^{-\h}\tilde{\bar{g}} )\big),\label{canonical-field}\\
\phim(g)&=&\fr{1}{\sqrt{2}}\big(a^*(i\om^{\h}{\tilde{g}})+a(i\om^{\h}{\tilde{\bar{g}}})\big).\label{canonical-momentum}
\eeqa
The algebra $\mfa(\mco(r))$ of observables localized in the double cone $\mco(r)$, whose base is the $s$-dimensional ball $\mco_{r}$ of radius $r$, is given by
\beq
\mfa(\mco(r))=\{\, e^{i\sqrt{2}\phip(\F^+)+i\sqrt{2}\phim(\F^-)} \, |\, \F^\pm\in D(\mco_{r})_{\real}\, \}^{\prime\prime},
\label{local-algebra}
\eeq
where $D(\mco_{r})_{\real}$ \label{symbol-DOr} is the space of real-valued test functions supported in this ball. Let us mention
an equivalent definition of the local algebra which is more convenient for some purposes: We define
the following (non-closed) subspaces in $L^2(\real^s, d^sp)$
\beq
\Lpmring=\om^{\mp\h}\wt{D}(\mco_r), \label{symbol-Lpmring}
\eeq
where tilde denotes  Fourier transform. We denote by $J$ \label{symbol-J} the complex conjugation in configuration space and set
\beq
\Lring=(1+J)\Lpring+(1-J)\Lmring.
\eeq
Then every $f\in\Lring$ has the form $f=f^++if^-$, where 
\beq
\tf^{\pm}=\om^{\mp\h}\btF^{\pm}, \label{fpm}  
\eeq
for some $\F^{\pm}\in D(\mco_{r})_{\real}$. For any $f\in\Lring$ we define the Weyl operator
\beq
W(f):=e^{i(a^*(f)+a(f))}=e^{i\sqrt{2}\phip(\F^+)+i\sqrt{2}\phim(\F^-)}. \label{Weyl-operator}
\eeq
In view of the second equality and definition~(\ref{local-algebra}) there holds
\beq
\mfa(\mco(r))=\{\, W(f) \, |\, f\in\Lring\, \}^{\prime\prime}.
\eeq
With the help of the translation automorphisms $\al_x$, introduced above, we define local 
algebras attached to double cones centered at any point $x$ of spacetime
\beq
\mfa(\mco(r)+x)=\al_x(\mfa(\mco(r))).
\eeq
Now for any open, bounded region $\mco$ we set 
\beq
\mfa(\mco)=\bigg(\bigcup_{\su{ r,x \\ \mco(r)+x\subset\mco}}\mfa(\mco(r)+x)\bigg)^{\prime\prime},
\eeq
obtaining the local net $\mfa$. The global algebra, denoted by the same symbol, is the $C^*$-inductive 
limit of all such local algebras.
It is well known that the triple $(\mfa,\al,\hil)$ satisfies the postulates 1-5 from Section~\ref{AQFT} \cite{RS2}.

We can immediately construct two related theories: First, we define the local algebra generated by the derivatives of the free field
\beq
\dmfa(\mco(r))=\{\, e^{i\sqrt{2}\phip(\sum_{j=1}^s\pa_{x_j}\F^+_j)+i\sqrt{2}\phim(\F^-)} \, |\, 
\F^+_j,\F^-\in D(\mco_{r})_{\real},\, j\in\{1,\ldots,s\}\,  \}^{\prime\prime}, \label{symbol-dmfa}
\eeq
which is clearly a subalgebra of $\mfa(\mco(r))$. In the massive case one can show that $\dmfa(\mco(r))=\mfa(\mco(r))$
making use of the equation of motion. In the massless case, however, the inclusion is proper. The global $C^*$-algebra,
constructed as in the case of the full theory, is denoted by $\dmfa$ and the theory $(\dmfa,\al,\hil)$ satisfies the 
postulates from Section~\ref{AQFT}.

The second example is the even part of scalar free field theory. Here the local
algebra, attached to the double cone $\mco(r)$, is given by
\beq
\emfa(\mco(r))=\{\, \cos(\sqrt{2}\phip(\F^+)+\sqrt{2}\phim(\F^-)) \, |\, \F^+,\F^-\in D(\mco_{r})_{\real} \}^{\prime\prime}.
\eeq
The corresponding local net $\emfa$, \label{symbol-emfa} constructed as above, gives rise to the theory $(\emfa,\al,\hil)$
which satisfies all the postulates from Section~\ref{AQFT} except for irreducibility. In order to ensure
this latter property, we represent the algebra $\emfa$ on the subspace $\ehil$ \label{symbol-ehil} in $\hil$ spanned by vectors
with even particle numbers. We define $\eumfa=\emfa|_{\ehil}$, \label{symbol-eumfa} $\eU(x,\La)=U(x,\La)|_{\ehil}$ and 
$\eal_{(x,\La)}(\,\cdot\,)=\eU(x,\La)\,\cdot\,\eU(x,\La)^{-1}$. \label{symbol-eal} The resulting
theory $(\eumfa,\eal,\ehil)$ satisfies all the general postulates.

For future convenience we discuss the relation
between the theory $(\emfa,\al,\hil)$, acting on the full Fock space, and the theory $(\eumfa,\eal,\ehil)$ acting on $\ehil$.
First, we define the $C^*$-representation $\epi: \emfa \to \eumfa$ given by
\beq
\epi(A)=\uA:=A|_{\ehil},\quad\quad A\in \emfa. \label{symbol-epi}
\eeq
Due to the Reeh-Schlieder property of the vacuum and the fact that local operators are norm
dense in $\mfa$ we obtain that $\epi$ is a faithful representation.
Thus there holds, by Proposition~2.3.3 of \cite{BR},
\beq
\|\uA\|=\|A\|,\quad\quad A\in \emfa. \label{even-norm-equality}
\eeq
Next, we note that the natural embedding $\ehil\hookrightarrow\hil$ induces the embedding
of the preduals $\ei:\trace^{\te{(e)}}\hookrightarrow\trace$,\label{symbol-ei}  where $\trace^{\te{(e)}}=B(\ehil)_*$. \label{symbol-tracee}
There clearly holds
\beq
\ei(\efun)(A)=\efun(\uA),\quad\quad A\in \hemfa, \efun\in\trace^{\te{(e)}},\label{even-functoriality}
\eeq
and it is easy to see that $\ei$ is an isometry: 
The absolute value of any $\efun\in\trace^{\te{(e)}}$ can be expressed as $|\efun|=\sum_ip_i|\Psi_i)(\Psi_i|$, 
where $p_i>0$ and $\Psi_i\in\ehil$ form an orthonormal system. Completing it to an orthonormal basis in $\hil$,
we obtain
\beq
\|\efun\|=\Tr|\efun|=\Tr|\ei(\efun)|=\|\ei(\efun)\|.
\eeq
Thus, making use of the fact that the embedding $\ehil\hookrightarrow\hil$ preserves the energy of
a vector, we obtain that
\beq
\ei: \traceEB^{\ee}\to\traceEB, \label{even-norm-conservation}
\eeq
what implies, together with relation~(\ref{even-functoriality}), that $\|\uA\|_{E,2}\leq \|A\|_{E,2}$ for 
square-integrable operators $A\in\hemfa$.

Our next goal is to describe the field content of these theories. (Cf. Section~\ref{triviality-of-Apc} for a general
discussion of this concept). For this purpose we 
introduce the domain $D_{S}=\{\,\Psi\in \DF\,|\, \Psi_n\in S(\real^{s\times n})\,\}$ \label{symbol-DS} on which there acts
the annihilation operator of the mode $\vep\in\real^s$
\beq
(a(\vep)\Psi)_n(\vep_1,\ldots,\vep_n)=\sqrt{n+1}\Psi_{n+1}(\vep,\vep_1,\ldots,\vep_n). \label{a(p)}
\eeq
Clearly, on $D_{S}$ there holds the equality
\beq
a(f)=\int d^sp \bar{f}(\vep)a(\vep),\quad\quad f\in S(\real^s).
\eeq
The adjoint $a^*(\vep)$ is only a quadratic form on $D_{S}\times D_{S}$. The Hamiltonian
can be expressed as a quadratic form on this domain by
\beq
H=\int d^sp\,\om(\vep)\,a^*(\vep)a(\vep). \label{Hamiltonian-representation}
\eeq
Similarly, we define the pointlike localized canonical fields and momenta as quadratic forms on
$D_{S}\times D_{S}$
\beqa
\phip(\vx)&=&\fr{1}{(2\pi)^{s/2}}\int \fr{d^s p}{\sqrt{2\om(\vep)}} 
\big(e^{-i\vep\vx}a^*(\vep)+e^{i\vep\vx}a(\vep)\big),\label{field+}\\
\phim(\vx)&=&\fr{i}{(2\pi)^{s/2}}\int d^sp \sqrt{\fr{\om(\vep)}{2}}
\big( e^{-i\vep\vx}a^*(\vep)-e^{i\vep\vx}a(\vep)\big). \label{field-}
\eeqa
This terminology is justified by the fact that on this domain there holds for any $g\in S(\real^s)$
\beqa
\phipm(g)&=&\int d^sx\, \phipm(\vx) g(\vx).
\eeqa
Moreover, for any $s$-index $\ka$, (see Section~\ref{multiindex-notation} for our multiindex notation), we define the derivatives $\pa^{\ka}\phipm:=\pa^{\ka}\phipm(\vx)|_{\vx=0}$ 
and the corresponding Wick monomials 
\beq
\wil\pa^{\ka_{1}^+}\phip\ldots\pa^{\ka_{k^+}^+}\phip\pa^{\ka^-_{1}}\phim\ldots\pa^{\ka^-_{k^-}}\phim\wir \label{symbol-Wick}
\eeq
which are also quadratic forms on $D_S\times D_S$. The Wick powers are defined by the standard 
prescription consisting in shifting all the creation operators to the left disregarding the 
commutators\footnote{We warn the reader that the Wick ordering is not linear on the algebra of (smeared) creation and annihilation
operators.}.

It is a well known fact (see e.g. \cite{Bo00}) that the Wick monomials
can be extended by continuity to bounded functionals on $\trace_{\infty}$ and the field 
content of scalar free field theory is given by
\beqa
\Phi_{\FH}=\Span\{\, \wil\pa^{\ka_{1}^+}\phip\ldots\pa^{\ka_{k^+}^+}\phip\pa^{\ka^-_{1}}\phim\ldots\pa^{\ka^-_{k^-}}\phim\wir\, |\, \ka^\pm_{j^\pm}\in \nat_0^s, j^{\pm}\in\{1,\ldots, k^{\pm}\}, & &\non\\
k^{\pm}\in\nat_{0}\,\} \,\,\,\, \label{FH-in-full-theory}
\eeqa
i.e. it consists of finite linear combinations of the Wick monomials.
Furthermore, it can be extracted from Section 7.4.2. of \cite{Bo00} that the field content of the sub-theory $(\dmfa,\al,\hil)$ generated by  derivatives of the field has the form
\beqa
\Phi^{\te{(d)}}_{\FH}=\Span\{\, \wil\pa^{\ka_{1}^+}\phip\ldots\pa^{\ka_{k^+}^+}\phip\pa^{\ka^-_{1}}\phim\ldots\pa^{\ka^-_{k^-}}\phim\wir\, |\, \ka^\pm_{j^\pm}\in \nat_0^s, \,\, j^{\pm}\in\{1,\ldots, k^{\pm}\}, & &\non\\ 
k^{\pm}\in\nat_0,|\ka^+_{j^+}|>0 \,\}.\quad  & & \label{symbol-Phid}
\eeqa
(As mentioned above, in the massive case one can show that $\Phi^{\te{(d)}}_{\FH}=\Phi_{\FH}$, making use of the
equation of motion. In the massless case, however, $\Phi^{\te{(d)}}_{\FH}$ is a proper subspace of $\Phi_{\FH}$).
Finally, the even part $(\eumfa,\eal,\ehil)$ of scalar free field theory has the following field content
\beqa
\Phi^{\te{(e)}}_{\FH}=\Span\{\, \underline{\wil\pa^{\ka_{1}^+}\phip\ldots\pa^{\ka_{k^+}^+}\phip\pa^{\ka^-_{1}}\phim\ldots\pa^{\ka^-_{k^-}}\phim\wir}\, |\, \ka^\pm_{j^\pm}\in \nat_0^s,\,\, j^{\pm}\in\{1,\ldots, k^{\pm}\}, & &\non\\
k^{\pm}\in\nat_0,\,\, k^++k^-\te{ is even }\},\quad & & \label{symbol-Phie}
\eeqa
where the underlining indicates that the Wick monomials act on the Hilbert space $\ehil$.

To close the present section, we show that Condition~$T$, stated in Section~\ref{triviality-of-Apc}, holds in 
massive scalar free field theory and its even part.
\bet\label{stress-energy} Massive scalar free field theory and its even part satisfy Condition~$T$ 
for any dimension of space $s\geq 1$.
\eet
\proof First, we consider the full massive scalar free field theory.
The $(0,0)$-component of the stress-energy tensor $T^{00}\in\Phi_{\FH}$ is given by
\beq
T^{00}=\h\wil \phim^2 \wir+\h\sum_{j=1}^s\wil (\pa_j \phip)^2 \wir+\h m^2\wil \phip^2\wir.
\eeq
The standard computation, which makes use of representation~(\ref{Hamiltonian-representation}) of the Hamiltonian, gives 
for $\Psi, \Phi\in D_S$
\beq
\int d^sx\,(\Psi|T^{00}(\vx)\Phi)=(\Psi|H\Phi). \label{Too-jest-hamiltonian}
\eeq
Certainly, equality (\ref{Too-jest-hamiltonian}) still holds if 
$T^{00}$ is replaced with $T^{00}(g)$, where the time-smearing function $g$ is specified in 
Condition $T$. It is shown in Appendix~\ref{Condition-L1} that massive scalar free field theory 
satisfies Condition~$\B$ which ensures that  $\|T^{00}(g)\|_{E,1}<\infty$.
(See relation~(\ref{field_approximation})). Due to this bound and the fact that $D_S\cap P_E\hil$
is dense in $P_E\hil$, we obtain for any $\Psi,\Phi\in P_E\hil$
\beq
\int d^sx\,(\Psi|T^{00}(g)(\vx)\Phi)=(\Psi|H\Phi).
\eeq
Making use of the fact that any functional $\fun\in\traceE$ is of the form $\fun(\,\cdot\,)=\sum_j(\Psi_j|\,\cdot\,\Phi_j)$,
where $\Psi_j, \Phi_j\in P_E\hil$ and $\sum_j\|\Psi_j\|\,\|\Phi_j\|<\infty$ \cite{BR}, we conclude the proof for the full theory. 
The even part is treated analogously, exploiting the fact that $\underline{T^{00}}$
is an element of $\Phi^{\te{(e)}}_{\FH}$ and restricting attention to $\Psi,\Phi\in\ehil$. \qed

\section{Special Functionals on Local Algebra} \label{special-functionals}
In this section we construct suitable functionals on the local algebra $\mfa(\mco(r))$, estimate their norms and
establish certain energy bounds. For this last purpose we introduce the Sobolev spaces \cite{RS2} given, 
for any $l\geq 0$, by 
\beq
L^2(\real^s,d^sx)_l=\{\, f\in L^2(\real^s, d^sx)\, |\, \int d^sp (1+|\vep|^2)^l|\tf(\vep)|^2<\infty\, \}.
\label{symbol-Sobolev-space}
\eeq
These  spaces are equipped with the norm 
\beq
\|f\|_{2,l}=(\int d^sp (1+|\vep|^2)^l|\tf(\vep)|^2)^\h.\label{Sobolev-norm}
\eeq
Clearly, $\|\,\cdot\,\|_{2,0}=\|\,\cdot\,\|_{2}$. The goal of this section is to prove the following 
proposition which is a slight generalization of Lemmas~7.8 and 7.9 of \cite{Bo00}.
\bep\label{Bfunctionals} Let $l\geq 0$ and $\{b_i^+\}_1^{\infty}, \{b_i^-\}_1^{\infty}$
be two sequences of $J$-invariant vectors from $L^2(\real^s, d^sx)_l$. Let $\mup,\mum$ be two multiindices, $\fpm\in\Lpmring$ 
and $f=f^++if^-$. Then there exists a normal functional $\si_{\mup,\mum}$ on $B(\hil)$ s.t.
\beqa
\si_{\mup,\mum}(W(f))&=&e^{-\h\|f\|^2_2}\lan b^+| f^+\ran^{\mup} \lan b^-| f^-\ran^{\mum},\\
\|R^{-l}\si_{\mup,\mum}R^{-l}\|&\leq& (2c_l)^{|\mup|+|\mum|}\sqrt{(|\mup|+|\mum|)!}\, \|b^+\|_{2,l}^{\mup}\|b^-\|_{2,l}^{\mum},
\label{funcbound}
\eeqa
where  $R=(1+H)^{-1}$, $c_l=(12+2m^2)^{l/2}$, and $m$ is the mass of the theory. 
Moreover, if $b_i^+=b_i^-$ for any $i\in\nat$ and $\{b_i^+\}_1^\infty$ form an orthonormal system of vectors in $L^2(\real^s, d^sx)$,
then there holds
\beq
\|\si_{\mup,\mum}\|\leq 4^{|\mup|+|\mum|}\sqrt{\mup!\mum!}. \label{orthogonal-bound}
\eeq
\eep
\nin For the proof of this proposition we need two simple lemmas. First, we introduce energy damping
operators which are more convenient than the resolvents $R$ considered above.
\bel\label{Gl} Let $l\geq 0$ and $G_l=\int d^sp\, \log(2+\om(\vep))^l\, a^*(\vep)a (\vep)$. Then there holds
\beq
\|R^{-l}e^{-G_l}\|\leq 1.
\eeq
\eel
\proof Let $n\geq 1$ and $\Psi_n$ be an $n$-particle (Schwartz-class) wavefunction. Then there holds
\beqa
(\Psi_n|R^{-l}e^{-G_l}\Psi_n)&=&\int d^sp_1\ldots d^sp_n|\Psi_n(\vep_1,\ldots,\vep_n)|^2
\fr{\big(1+\om(\vep_1)+\cdots+\om(\vep_n)\big)^l}{(2+\om(\vep_1))^l\ldots\big(2+\om(\vep_n)\big)^l}\non\\
&\leq& \|\Psi_n\|^2, \label{psin}
\eeqa
where we made use of the fact that $R^{-l}e^{-G_l}$ is a positive operator and that
$(x+y)\leq xy$ for any $x\geq 2$, $y\geq 2$. We can thus drop the restriction that $\Psi_n$ is
a Schwartz-class function. Now making use of the fact that any vector $\Psi$ in the Fock space can be expressed as
$\Psi=\sum_{n=0}^{\infty}\Psi_n$, where 
$\Psi_0=c\vac$ for some $c\in\complex$ and $\|\Psi\|^2=\sum_{n=0}^{\infty}\|\Psi_n\|<\infty$, we obtain
from (\ref{psin})
\beq
(\Psi|R^{-l}e^{-G_l}\Psi)\leq \|\Psi\|^2,
\eeq
what concludes the proof. \qed\\
Next, we recall the elementary combinatorial Lemma~7.6 of \cite{Bo00} which gives useful bounds on the
lengths of certain vectors in the Fock space.
\bel\label{vector-lengths} For any family of vectors $b_1\ldots b_n\in L^2(\real^s, d^sx)$ there holds
\beq
\|a(b_1)\ldots a(b_k)a^*(b_{k+1})\ldots a^*(b_{n})\vac\|\leq \sqrt{n!}\|b_1\|\ldots \|b_n\|.\label{vector-lengths1}
\eeq
Moreover, if $\{b_i\}_1^\infty$ form an orthonormal system of vectors in $L^2(\real^s, d^sx)$ 
and $\al$, $\be$ are multiindices, then there holds
\beq
\|a(e)^{\al}a^*(e)^{\be}\vac\|\leq \sqrt{(\al+\be)!}. \label{vector-lengths2}
\eeq
\eel
\nin After this preparation we are in position to prove the main result of this section.\\
\bf Proof of Proposition \ref{Bfunctionals}. \rm This is a slight generalization of
the argument from Appendix~4.2.B of \cite{Bo00}. 
Consider the generating functional
\beqa
F(\uu,\uw)(A)=(\vac| e^{\h a(i\uu\,\ub^+-\uw\,\ub^- )}e^{\h a^*(i\uu\,\ub^+-\uw\,\ub^-)}A e^{-\h a^*(i\uu\,\ub^+-\uw\,\ub^-)}\vac),\label{Fdef}
\eeqa
where $A$ is a bounded operator on the Fock space and $\uu=\{u_i\}_1^{\infty}$, $\uw=\{w_i\}_1^{\infty}$ are
sequences of real numbers with only finite number of entries different from zero. We denote
$\uu\ub^+=\sum_{i}u_ib_i^+$,  $\uw\ub^-=\sum_{i}w_ib_i^-$ and define the exponentials above 
by their Taylor series expansions which converge in view of Lemma~\ref{vector-lengths}. 
In particular, for $A=W(f)$ we obtain 
\beqa
F(\uu,\uw)(W(f))=e^{-\h\|f\|^2_2}e^{\uu\lan \ub^+ | f^+ \ran +\uw\lan \ub^- | f^-\ran}.
\eeqa
Differentiating the above formula, we obtain
\beq
\pa_u^{\mup}\pa_w^{\mum}|_{(\uu,\uw)=(0,0)}F(\uu,\uw)(W(f))=e^{-\h\|f\|^2_2}\lan b^+| f^+\ran^{\mup} \lan b^-| f^-\ran^{\mum}.
\eeq
Therefore, we can define
\beq
\si_{\mup,\mum}(A):=\pa_u^{\mup}\pa_w^{\mum}|_{(\uu,\uw)=(0,0)}F(\uu,\uw)(A).
\eeq
From this definition and formula (\ref{Fdef}) we obtain
\beqa
\si_{\mup,\mum}(A)&=&
\bigg(\h\bigg)^{|\mup|+|\mum|}\!\!\!\!\!\!\!i^{|\mup|+2|\mum|}\!\!\!\!\!\!\sum_{\su{\alpn+\alpp+\alpb=\mup\\ \alm+\almp+\almb=\mum}} 
(-1)^{|\alpn|+|\almb|} 
\fr{\mup!}{\alpn!\alpp!\alpb!}\fr{\mum!}{\alm!\almp!\almb!}\non\\
&\cdot&(\vac|a(b^+)^{\alpn}a(b^-)^{\alm}a^*(b^+)^{\alpp}a^*(b^-)^{\almp}Aa^*(b^+)^{\alpb}a^*(b^-)^{\almb}\vac)\non\\
&=&\bigg(\h\bigg)^{|\mup|+|\mum|}\!\!\!\!\!\!\!i^{|\mup|+2|\mum|}\!\!\!\!\!\!\sum_{\bal+\balp+\balb=\mub} 
(-1)^{|\alpn|+|\almb|}\fr{\mub!}{\bal!\balp!\balb!}\cdot\non\\  
& &\ph{4444444444444444444444444444}\cdot(\vac|a(b)^{\bal}a^*(b)^{\balp}Aa^*(b)^{\balb}\vac).\label{sigma}
\eeqa
Here  $\bal=(\alpn,\alm)$ etc. are 2-multiindices.
Since the operator $G_l$, introduced in Lemma~\ref{Gl}, satisfies $\|R^{-l}e^{-G_l}\|\leq 1$ and
$e^{G_l}a^*(b)e^{-G_l}=a^*((2+\om)^lb)$, there holds
\beqa
& &\sup_{A\in B(\hil)_1}|(\vac|a(b)^{\bal}a^*(b)^{\balp}R^{-l}AR^{-l}a^*(b)^{\balb}\vac)|\non\\
&\leq& \sup_{A\in B(\hil)_1}|(\vac|a((2+\om)^{-l}b)^{\bal}a^*((2+\om)^lb)^{\balp}Aa^*((2+\om)^lb)^{\balb}\vac)|\non\\
&\leq& c_l^{|\mup|+|\mum|}\sqrt{(|\mup|+|\mum|)!}\, \|b^+\|_{2,l}^{\mup}\,\|b^-\|_{2,l}^{\mum}, \label{functional-bound}
\eeqa
where in the last step we made use of estimate~(\ref{vector-lengths1}) and $c_l$ is defined in the statement
of the proposition. It follows from this relation and formula~(\ref{sigma}) that $R^{-l}\si_{\mup,\mum}R^{-l}$
is a bounded functional. Making use of  relation~(\ref{sigma}) and exploiting the following formula, valid for any
$n\in\nat$ 
\beq
\sum_{ \su{a,a^\prime,a^{\prime\prime}\in\nat_0  \\ a+a^\prime+a^{\prime\prime}=n}}\fr{n!}{a!a^\prime!a^{\prime\prime}!}=3^{n},
\label{3-nomial}
\eeq
we prove estimate~(\ref{funcbound}) from the statement of the proposition.
In order to verify relation~(\ref{orthogonal-bound}),  suppose that $b_i^+=b_i^-$ for $i\in\nat$ and $\{b_i^+\}_1^\infty$
form an orthogonal system. Then estimate~(\ref{vector-lengths2}) gives
\beq
\sup_{A\in B(\hil)_1}|(\vac|a(b)^{\bal}a^*(b)^{\balp} A a^*(b)^{\balb}\vac)|\leq \sqrt{(\mup+\mum)!}\leq 
2^{|\mup|+|\mum|}\sqrt{\mup!\mum!}.
\eeq
Substituting this bound to formula~(\ref{sigma}) and making use again of identity~(\ref{3-nomial}), we conclude the proof
of the proposition. \qed\\

\section{Expansions in Single-Particle Space} \label{single-particle-expansions}

In this section we derive two expansions of the functions $\F^{\pm}\in D(\mco_r)$ which enter
into definition (\ref{local-algebra}) of the local algebra. The first expansion relies on the
fact that $\tF^{\pm}$ are analytic functions and thus can be expanded into convergent Taylor series.
The coefficients of these expansions are given in terms of vectors $\{\bti_{\ka_j,r}^\pm\}_1^\infty$ defined
by (\ref{b}) below. We show in Lemma~\ref{b-lemma}, stated below,  that these vectors have finite Sobolev norms.
Thus the corresponding functionals  $\si_{\mup,\mum}$, constructed in Lemma~\ref{Bfunctionals},
satisfy the energy bounds~(\ref{funcbound}).
Due to the rapid growth with  $|\mup|+|\mum|$, this
estimate does not suffice to establish convergence of the sum (\ref{nuclearity-of-whXi}) in the
massless case. In order to exploit the more tame estimate~(\ref{orthogonal-bound}), 
in the second part of this section we construct a suitable orthonormal basis expansion.
\subsection{Taylor Expansion}\label{Taylor}
We introduce  approximate characteristic function $\vep\to\chi_E(\vep)\in\czn$ \label{symbol-chiE} which is equal to one on the set 
$\{\, \vep\in\real^s\, |\, |\om(\vep)|\leq E\}$ and $\vx\to\chir(\vx)\in\czn$ \label{symbol-chir} which is equal to one on the ball $\mco_r$, and define, for any $s$-index $\ka$, functions from $L^2(\real^s, d^sp)$
\beqa
 \bti_{\ka,r}^\pm(\vep)&=&\fr{1}{(2\pi)^{\fr{s}{2}}}\fr{\om(\vep)^{\pm\h}\wt{x^{\ka}\chir}(\vep) }{\ka!},\label{b}\\ 
 \hti_{\ka,E}^\pm(\vep)&=&(-1)^{|\ka|}\om(\vep)^{\mp\h}(ip)^{\ka}\chi_E(\vep). \label{h}
\eeqa
We recall that $\fpm\in\Lpmring$ have the form $\tf^\pm=\om^{\mp\h}\tilde{\F}^\pm$, where $\F^\pm\in D(\mco_r)_{\real}$, 
set  $\tilde{\F}^{\pm}_E:=\chi_E\tilde{\F}^{\pm}$ and note that
the functions $\tilde{\F}^{\pm}$ are analytic. From their Taylor expansions we obtain \cite{Bo00}
\beq
\om^{\mp\h}\tilde{\F}^{\pm}_E=\chi_E \tf^{\pm}=\sum_{\ka} \lan b_{\ka,r}^\pm |f^{\pm} \ran \hti_{\ka,E}^\pm=
\sum_{j=1}^\infty \lan b_{\ka_j,r}^\pm |f^{\pm} \ran \hti_{\ka_j,E}^\pm, \label{hb-expansion}
\eeq
where in the second step we numbered the $s$-indices $\ka$ with some index $j\in\nat$ in such a way that
$\ka_1=0$. There holds the following proposition.
\bep\cite{Bo00}\label{hb-expansion-convergence} Expansion~(\ref{hb-expansion}) converges in $L^2(\real^s,d^sp)$.
\eep
\nin This statement follows from  estimate~(\ref{sp-convergence}) below which relies
on the bounds on the norms of functions (\ref{b}) and (\ref{h}), established in Lemmas~\ref{b-lemma} 
and \ref{h-lemma}, respectively. Actually,
we  derive here more general estimates on the Sobolev norms of these functions, (see definition (\ref{Sobolev-norm})),
which are needed in Lemma~\ref{Bfunctionals} above.
First, we note the following auxiliary fact.
\bel\label{Leibniz} Let $\chi\in C_0^{\infty}(\real^s)$. Then, for any $s$-index $\ka$, $i\in\{1,\ldots,s\}$ and $n\in\nat$,
there holds
\beq
\int d^sx\, |\pa_{x_i}^{n} (x^{\ka}\chi(\vx))|^2\leq (c_{n})^{|\ka|+1}
\eeq
for some  constant $c_n$, independent of $\ka$.
\eel
\proof We note the following identity which follows from the Leibniz rule
\beq
\pa_{x_i}^{\lil} (x^{\ka}\chi)(\vx)=x^{\hka}\sum_{k=0}^{\min(\lil,\ka(i))}\fr{\lil!}{(\lil-k)!k!}
\fr{\ka(i)!}{(\ka(i)-k)!}x_i^{\ka(i)-k}\pa_{x_i}^{\lil-k}\chi(\vx).
\eeq
where the $s$-index $\hka$ is obtained from $\ka$ by setting $\ka(i)=0$.
There easily follows the bound
\beq
\int d^sx\, |\pa_{x_i}^{n} (x^{\ka}\chi(\vx))|^2 \leq 2^n(c_{\lil})^{|\ka|}
\int d^sx\big(\sum_{k=0}^n\, |\pa_{x_i}^{k}\chi(\vx)|\big)^2,
\eeq
for some constant $c_{\lil}$, independent of $\ka$. \qed\\
Now we are ready to prove the required bounds on the Sobolev norms of $b_{\ka}^{\pm}$.
\bel\label{b-lemma} For any $\lal\geq 0$ the functions $b_{\ka,r}^\pm$, defined by (\ref{b}), satisfy the bound 
\beq
\|b_{\ka,r}^{\pm}\|_{2,\lal}\leq \fr{(c_{\lal,r})^{|\ka|+1}}{\ka!},
\eeq
where the constant $c_{\lal,r}$, is independent of $\ka$.
\eel
\proof First, we note that $\|b_{\ka,r}^{+}\|_{2,\lal}\leq (1+m)^{1/4}\|b_{\ka,r}^{-}\|_{2,\lal+\h}$,
so it suffices to consider the $(-)$ case. Clearly $\lal\to\|b_{\ka,r}^{-}\|_{2,\lal}$ is a monotonically
increasing function, so it is enough to establish the bound for $\lal=\lil+\h$, $\lil\in\nat$.
We consider the expression
\beqa
\|\omm\wt{x^{\ka}\chir}\|^2_{2,\lil+\h}=\big(\wt{x^{\ka}\chir}\big|\om(\vep)^{-1}(1+|\vep|^2)^{\lil+\h}\wt{x^{\ka}\chir}\big)& &\non\\
\leq c_{\lil}\bigg(\|\omm\wt{x^{\ka}\chir}\|^2_2+
\big(\wt{x^{\ka}\chir}\big|(\sum_{i=1}^s|p_i|^{2\lil})\wt{x^{\ka}\chir}\big)\bigg),& &\label{twoterms}
\eeqa
where on the r.h.s. above we mean the scalar product in $L^2(\real^s, d^sp)$  and the constant $c_{\lil}$ depends
only on $\lil$ and $s$. In order to study the first term on the r.h.s. above, we
write $\omm=(\omm)_++(\omm)_-$, where $(\omm)_{\pm}(\vep)=\om(\vep)^{-\h}\theta(\pm(|\vep|-1))$.
There clearly holds
\beqa
& &\|(\omm)_+\wt{x^{\ka}\chir}\|_2\leq\|x^{\ka}\chir\|_2\leq \h (c_r)^{|\ka|+1},\\ 
& &\|(\omm)_-\wt{x^{\ka}\chir}\|_2\leq \|\wt{x^{\ka}\chir}\|_{\infty}\|(\omm)_-\|_2\leq \h (c_r)^{|\ka|+1},
\eeqa
for some constant $c_r\geq 0$, independent of $\ka$, and therefore 
\beq
\|\omm\wt{x^{\ka}\chir}\|_2\leq (c_r)^{|\ka|+1}. \label{firstterm}
\eeq
Next, we study the second term on the r.h.s. of (\ref{twoterms}). Making use of Lemma~\ref{Leibniz}, we obtain
\beq
\big(\wt{x^{\ka}\chir}\big||p_i|^{2\lil}\wt{x^{\ka}\chir}\big)=\int d^sx |\pa_{x_i}^{\lil} (x^{\ka}\chir)(\vx)|^2
\leq (c_{\lil,r})^{|\ka|+1},
\eeq
where the constant $c_{\lil,r}$ does not depend on $\ka$. This concludes the proof of the lemma. \qed\\
While the above result holds for any $m\geq 0$, in the next statement, concerning the Sobolev norms
of the functions $\hti_{\ka,E}^{\pm}$, we have to make a distinction between the massive and massless case.
\bel\label{h-lemma} In massive scalar free field theory for $s\geq 1$ the functions $\hti_{\ka,E}^{\pm}$, defined by relation (\ref{h}),  satisfy,
for any $\la\geq 0$, $\be\in\real$,
\beq
\|\om^{-\be}\hti_{\ka,E}^{\pm}\|_{2,\la}\leq  (c_{\la,\be,E})^{|\ka|+1}, \label{h-Sobolev-bound}
\eeq
where the constants $c_{\la,\be,E}$ are independent of $\ka$. In massless scalar free field theory for $s\geq 3$ the bound~(\ref{h-Sobolev-bound}) holds (in particular) in the following two cases:
\begin{enumerate}
\item[(a)] For any  $\be<1$ and $\la=0$.
\item[(b)] For $\be=\mp\h$ and any $\la\geq 0$. (The $\pm$ signs are correlated with these appearing in 
formula~(\ref{h-Sobolev-bound})).
\end{enumerate}
\eel
\proof We define $\chi_E^{\pm}(\vep):=\om(\vep)^{-\be\mp\h}\chi_E(\vep)$. Then
\beq
(\om^{-\be}\hti_{\ka,E}^\pm)(\vep)=(-1)^{|\ka|}(ip)^{\ka}\chi_E^{\pm}(\vep). \label{h-representation}
\eeq
We first consider the case $m=0$. Then the functions $\chi_E^{\pm}$ are square-integrable for $\be<1$ and there holds
for some constant $c_E$, independent of $\ka$, 
\beq
\|\om^{-\be}\hti_{\ka,E}^\pm\|_2\leq c_E^{|\ka|}\|\chi_E^{\pm}\|_2, \label{massless-and-massive}
\eeq
what proves part (a) and part (b) for $\la=0$.
For $m>0$  inequality (\ref{massless-and-massive}) holds for any $\be\in\real$.

In the massive case and the case considered in part~(b),
$\vep\to\chi_E^{\pm}(\vep)$ are smooth, compactly supported functions.
It suffices to take into account $\la\in\nat$, since 
the functions $\la\to\|\om^{-\be}\hti_{\ka,E}^{\pm}\|_{2,\la}$ are monotonically increasing.
By H\"older's inequality applied to the term $(1+|\vx|^2)^\la$, identity~(\ref{h-representation}), 
Lemma~(\ref{Leibniz}) and relation~(\ref{massless-and-massive}) there holds
the following bound
\beqa
\|\om^{-\be}\hti_{\ka,E}^{\pm}\|_{2,\la}^2\leq c_{\la}\bigg(\|\om^{-\be}\hti_{\ka,E}^{\pm}\|_2^2+
\sum_{i=1}^s\int d^sp\,|\pa_{p_i}^{\la}\big(p^{\ka}\chi_E^{\pm}(\vep)\big)|^2\bigg)\leq (c_{\la,\be,E})^{|\ka|+1},
\eeqa
where the constant $c_{\la,\be,E}$ is independent of $\ka$. \qed\\
From Lemmas~\ref{b-lemma} and \ref{h-lemma} we obtain the bound for any combination of $\pm$-signs 
any $0\leq\be\leq 1$ and $0<p\leq 1$
\beq
\sum_{\ka\in\nat^s_0} \|b_{\ka,r}^{\pm}\|^{p}_2 \, \|\om^{-\be}\hti_{\ka,E}^{\pm}\|_2^{p}
\leq \sum_{\ka\in\nat^s_0}\fr{(c_{0,r}\,c_{0,\be,E})^{p(|\ka|+1) } }{(\ka!)^p}<\infty. \label{sp-convergence}
\eeq
Thus we have proven Proposition \ref{hb-expansion-convergence}.
\subsection{Orthonormal Basis Expansion}\label{orthonormal}

Let $\Q_E$ \label{symbol-QE} be a projection on the single-particle space onto  states of energy lower than $E$.
Let $h_{\rzero}\in D(\mco_{\rzero})_{\real}$\label{symbol-hrzero} be s.t. $\tih>0$. We introduce the closed, linear
subspaces $\Lpm=[\om^{\pm\h}\wt{D}(\mco_r)]$ in $L^2(\real^s,d^sp)$\label{symbol-Lpm} and denote the respective projections by the same symbol. We choose $\h\leq\ga<\fr{s-1}{2}$ 
and define operators $\TEpm=\omm \Q_E\Lpm$, $\Thpm=\om^{-\ga}\tih^{1/2}\Lpm$, \label{symbol-Toper} where $\tih$ is the corresponding
multiplication operator in momentum space. We recall, that the $p$-norm of an operator $A$ is given, for any $p>0$, 
by $\|A\|_p=\||A|^p\|_1^{1/p}$,\label{symbol-Hilbert-pnorm} where $\|\,\cdot\,\|_1$ denotes the trace norm. By a slight modification of Lemma 3.5  from \cite{BP90}, one obtains  the following result: 
\bel\label{TEpm-lemma} For any $p>0$ the operators $\TEpm$ and $\Thpm$ are $p$-nuclear i.e. there holds
\beqa
\|\TEpm\|_p&<&\infty,\\
\|\Thpm\|_p&<&\infty.
\eeqa
\eel
\proof In order to show that the operators $\Tbpm$ are $p$-nuclear for any $p>0$, it suffices to demonstrate that their 
adjoints $\Tbpm^*$ are products of an arbitrary number of Hilbert-Schmidt operators.
(The Hilbert-Schmidt property is preserved under the adjoint operation due to the cyclicity of the trace).
We set, as in \cite{BP90}, for $i\in\nat$
\beqa
h_i&=&\om(1+\om^2)^{(i-1)s}\chir(1+\om^2)^{-is}\om^{-1}\\
k_i&=&(1+\om^2)^{(i-1)s}\chir(1+\om^2)^{-is}.
\eeqa
These operators, and also $\omm h_1$, $\omm k_1$, are in the Hilbert-Schmidt class  \cite{BP90}.
For any $n\in\nat$ there hold the identities
\beqa
\Tbp^*&=&\Lp\omm h_1\ldots h_n\om^{-\alr+\h}(1+\om^2)^{ns}\tilh^{1/2},\\
\Tbm^*&=&\Lm\omm k_1\ldots k_n\om^{-\alr+\h}(1+\om^2)^{ns}\tilh^{1/2},
\eeqa
so it suffices to check that $h_n\om^{-\alr+\h}$ and $k_n \om^{-\alr+\h}$ are 
Hilbert-Schmidt. Since the Hilbert-Schmidt norm $\|\,\cdot\,\|_{\te{HS}}$ of an operator is
equal to the $L^2$-norm  of its integral kernel, we obtain
\beqa
\|h_n\om^{-\alr+\h}\|_{\te{HS}}^2
=(2\pi)^{-s}\int d^sp\, d^sq\, \om(\vep)^2(1+\om(\vep)^2)^{2(n-1)s}
|\wt{\chi(\mco_r)}(\vep-\veq)|^2& &\non\\
\cdot\fr{1}{(1+\om(\veq)^2)^{2ns}}\fr{1}{\om(\veq)^{2\alr+1}}.& &
\eeqa
Making use of the following two identities (see eq. formula~(7.2.109) of \cite{Bo00} for the proof of the first 
statement)
\beqa
\fr{1+\om(\vep)^2}{1+\om(\veq)^2}\leq (|\vep-\veq|+1)^2,\\
\om(\vep)^2\leq 2(\om(\vep-\veq)^2+\om(\veq)^2),
\eeqa
we arrive at the bound
\beqa
& &\|h_n\om^{-\alr+\h}\|_{\te{HS}}^2\ph{44444444444444444444444444444444444444444444444444444} \non\\
&\leq& 2(2\pi)^{-s} \int d^sp\,d^sq\, \fr{1}{(1+\om(\veq)^2)^{2s}}\fr{1}{\om(\veq)^{2\alr-1}} 
(|\vep|+1)^{4(n-1)s} 
|\wt{\chi(\mco_r)}(\vep)|^2 \non\\
&+& 2(2\pi)^{-s} \int d^sp\,d^sq\, \fr{1}{(1+\om(\veq)^2)^{2s}}\fr{1}{\om(\veq)^{2\alr+1}} \om(\vep)^2(|\vep|+1)^{4(n-1)s}
|\wt{\chi(\mco_r)}(\vep)|^2.\quad\quad\quad
\eeqa
These integrals are clearly convergent for $0\leq\alr<\fr{s-1}{2}$. The (simpler) case of $k_n \om^{-\alr+\h}$ is
treated analogously. Finally, the $p$-nuclearity of the operators $\TEpm$ follows from the fact that $Q_E\tih^{-1}$
is a bounded operator and there holds $\TEpm=(Q_E\tih^{-1})\Thpm$ for $\alr=\h$. \qed\\
We define the operator $T$ as follows
\beq
T=(|\TEp|^2+|\TEm|^2+|\Thp|^2+|\Thm|^2)^\h. \label{T-definition}
\eeq
Making use of the fact \cite{Ko84} that for any $0<p\leq1$ and any pair of positive operators $A$, $B$ s.t.
$A^p$, $B^p$ are trace-class, there holds $\|(A+B)^p\|_1\leq \|A^p\|_1+\|B^p\|_1$, we get
\beqa
\|T\|_p^p\leq\|\TEp\|_p^p+\|\TEm\|^p_p +\|\Tbp\|^p_p+\|\Tbm\|^p_p \label{lub3}.
\eeqa
Since $T$ commutes with the operator $J$ of complex conjugation in configuration space, it has a $J$-invariant orthonormal basis of eigenvectors $\{e_j\}_1^\infty$\label{symbol-J-basis}
and we denote the corresponding eigenvalues by $\{t_j\}_1^\infty$\label{symbol-T-eigenvectors}. As we will see in the next section, the expansion
\beq
Q_E f^{\pm}=\sum_{j=1}^{\infty}\lan e_j  | f^{\pm} \ran Q_E\Lpm e_j,
\eeq
valid for any $f^{\pm}\in\Lpmring$, has the required convergence properties.
\section{Expansion of $\whXi_E$ into Rank-One Mappings}\label{full-expansion}
In this section we will decompose the map $\whXi_E: \mfa(\mco(r))\to B(\hil)$, given by 
$\whXi_E(A)=P_EAP_E$, as follows
\beq
\whXi_E(A)=\theh(A)+\thec(A)+\thet(A), \quad\quad A\in\mfa(\mco). \label{decomposition}
\eeq
Here $\theh$ is a finite rank map, 
the part $\thec(A)$ collects the terms involving high derivatives of the field and
$\thet(A)$ contains the contributions to $A$ coming from high Wick powers.
In order to construct such a decomposition, we evaluate the map $\whXi_E$ on a Weyl operator
$W(f)$, $f\in\Lring$, given by definition (\ref{Weyl-operator}).
We obtain the following expansion valid in the sense of quadratic forms on $\DF\times\DF$:
\beqa
\Theta_E(W(f))&=&e^{-\h\|f\|^2_2} P_E\wil e^{i\sqrt{2}\phip(\F^+)}e^{i\sqrt{2}\phim(\F^-)}\wir P_E\non\\
&=&e^{-\h\|f\|^2_2}\sum_{\m^+,\m^-\in\nat_0}\fr{(i\sqrt{2})^{\m^++\m^-} }{\m^+!\m^-!}P_E\wil\phip(\F^+)^{\m^+}\phim(\F^-)^{\m^-}\wir P_E,
\eeqa
where $\tf^{\pm}=\om^{\mp\h}\btF^{\pm}$.
Now we introduce natural numbers $\MM^+$, $\MM^-$ 
and the set of indices $\Set=\{(\m^+,\m^-)\in\nat_{0}^{2}\,|\, \m^+\leq \MM^+  \textrm{ and }
\m^-\leq \MM^-\}$.\label{symbol-set} We decompose the above sum into two parts
\beqa
\Theta_E(W(f))&=&\theo(W(f))+\thet(W(f)),\label{symbol-first-decomposition}\non\\
\theo(W(f))&=&e^{-\h\|f\|^2_2}\sum_{\m^+,\m^-\in \Set } 
\fr{(i\sqrt{2})^{\m^++\m^-} }{\m^+!\m^-!}P_E\wil\phip(\F^+)^{\m^+}\phim(\F^-)^{\m^-}\wir P_E,\label{onesum}\\
\thet(W(f))&=&e^{-\h\|f\|^2_2}\sum_{\su{\m^+,\m^-\in \Set^\prime }  }\fr{(i\sqrt{2})^{\m^++\m^-} }{\m^+!\m^-!}
P_E\wil\phip(\F^+)^{\m^+}\phim(\F^-)^{\m^-}\wir P_E,\quad\label{twosums}
\eeqa
where $\Set^\prime$ is the complement of $\Set$ and $\theo$, $\thet$ are understood as 
linear maps from the $*$-algebra $\mfar(\mco(r))$ of finite linear combinations of the Weyl
operators, given by
\beq
\mfar(\mco(r))=\Span\{\, W(f)\,|\, f\in\Lring\,\} \label{symbol-mfar}
\eeq
to quadratic forms on $\DF\times\DF$. In the sequel we will show that $\theo$ and $\thet$ have range in $B(\hil)$. Moreover,
we will extend their domain, by continuity, to the whole local algebra $\mfa(\mco(r))$.
For this purpose we will use the expansions of the functions $\fpm\in\Lpmring$ introduced in Section~\ref{single-particle-expansions}.
\subsection{Expansion of $\theo$}
First, recalling that $\tilde{\F}^{\pm}_E:=\chi_E\tilde{\F}^{\pm}$ and exploiting
the Wick ordering we obtain
\beq
P_E\wil\phip(\F^+)^{\m^+}\phim(\F^-)^{\m^-}\wir P_E=P_E\wil\phip(\F^+_E)^{\m^+}\phim(\F^-_E)^{\m^-}\wir P_E.
\eeq
With the help of the multinomial formula~(\ref{multiform}) and  expansion~(\ref{hb-expansion}) we get the following identity
\beqa
& &P_E\wil\phip(\F^+_E)^{\m^+}\phim(\F^-_E)^{\m^-}\wir P_E\non\\
& &=\sum_{\su{\mupi,\mumi \\ |\mupmi|=\m^{\pm}} }\fr{|\mupi|!||\mumi|!}{\mupi!\mumi!}\lan b_{\ka,r}^+|f^+\ran^{\mupi}
\lan b_{\ka,r}^-| f^-\ran^{\mumi}
P_E\wil(\pa\phip)^{\mupi}(\pa\phim)^{\mumi}\wir P_E,\quad\label{multinomial-hb}
\eeqa
where $\mupmi$ are multiindices and we introduced the short-hand notation 
\beq
(\pa\phipm)^{\mupmi}=\prod_{j=1}^{\infty}(\pa^{\ka_j}\phipm)^{\mupmi(j)}, \label{derivatives-convention}
\eeq
which refers to the numbering of the $s$-indices $\ka$, introduced in formula~(\ref{hb-expansion}).
Identity~(\ref{multinomial-hb}) relies on the fact that the 
vectors $b_{\ka,r}^\pm$ and $f^{\pm}$  are real in configuration space and exploits the relation
\beqa
\bigg(\fr{1}{\sqrt{2}}\bigg)^{|\mupi|+|\mumi|}P_E\wil\big(a^*(\hti_{\ka,E}^+)+a(\hti_{\ka,E}^+)\big)^{\mupi}
\big(a^*(i\hti_{\ka,E}^-)+a(i\hti_{\ka,E}^-)\big)^{\mumi}\wir P_E\non\\
\ph{44444444444444444444}=P_E\wil(\pa\phip)^{\mupi}(\pa\phim)^{\mumi}\wir P_E.
\eeqa
Substituting  expansion~(\ref{multinomial-hb}) to (\ref{onesum}), we obtain
\beqa
\theo(W(f))
&=&\sum_{\su{\mupi,\mumi \\ (|\mupi|,|\mumi|)\in\Set} }\fr{(i\sqrt{2})^{|\mupi|+|\mumi|} }{\mupi!\mumi!}
e^{-\h\|f\|^2_2} \lan b_{\ka,r}^+|f^+\ran^{\mupi}\lan b_{\ka,r}^-| f^-\ran^{\mumi}\non\\
& &\phantom{44444444444444444}\cdot P_E\wil(\pa\phip)^{\mupi}(\pa\phim)^{\mumi}\wir P_E. \label{preexpansion}
\eeqa
Next, we choose some finite subset $\Sett$\label{symbol-sett} of the set of all pairs of multiindices
and decompose the map $\theo$ as follows
\beqa
\theo(W(f))&=&\theh(W(f))+\thec(W(f)), \label{symbol-second-decomposition} \\
\theh(W(f))&=& e^{-\h\|f\|^2_2}\!\!\!\!\!\!\!\sum_{\su{(\mupi,\mumi)\in\Sett \\ (|\mupi|,|\mumi|)\in \Set}}
\fr{(i\sqrt{2})^{|\mupi|+|\mumi|} }{\mupi!\mumi!}  \lan b_{\ka,r}^+|f^+\ran^{\mupi}\lan b_{\ka,r}^-| f^-\ran^{\mumi}\non\\ 
& &\ph{4444444444444444444444444}\cdot P_E\wil(\pa\phip)^{\mupi}(\pa\phim)^{\mumi}\wir P_E, \label{thehdef}\quad\\
\thec(W(f))&=& e^{-\h\|f\|^2_2}\!\!\!\!\!\!\!\sum_{\su{(\mupi,\mumi)\in\Sett^{\prime} \\ (|\mupi|,|\mumi|)\in \Set}}
 \fr{(i\sqrt{2})^{|\mupi|+|\mumi|} }{\mupi!\mumi!} \lan b_{\ka,r}^+|f^+\ran^{\mupi}\lan b_{\ka,r}^-| f^-\ran^{\mumi}\non\\ 
& &\ph{4444444444444444444444444}\cdot P_E\wil(\pa\phip)^{\mupi}(\pa\phim)^{\mumi}\wir P_E. \quad \label{thecdef}
\eeqa
The above expressions can be restated in terms of  suitable normal functionals on $B(\hil)$. 
We denote by $\bsi^{(r)}_{\mupi,\mumi}$\label{symbol-very-spacial-functionals} the functionals from Lemma~\ref{Bfunctionals}, corresponding to the families of vectors
$\{b_{\ka_j,r}^\pm\}_{j\in\nat}$. Next, we define the normal functionals $\htau^{(r)}_{\mupi,\mumi}$ on $B(\hil)$ given by 
\beqa
& &\htau_{\mupi,\mumi}^{(r)}=\fr{(i\sqrt{2})^{|\mupi|+|\mumi|} }{\mupi!\mumi!}\bsi^{(r)}_{\mupi,\mumi}. \label{htau}
\eeqa
Making use of formula (\ref{thehdef}), we can write
\beq
\theh(A)=\sum_{\su{(\mupi,\mumi)\in\Sett \\ (|\mupi|,|\mumi|)\in \Set} }\htau_{\mupi,\mumi}^{(r)}(A)\,
P_E\wil(\pa\phip)^{\mupi}(\pa\phim)^{\mumi}\wir P_E, \label{finiterank0}
\eeq
where $A\in\mfar(\mco(r))$ is any finite linear combination of Weyl operators. The Wick monomials $\wil(\pa\phip)^{\mupi}(\pa\phim)^{\mumi}\wir$ belong to the field content of the theory (See 
definition~(\ref{symbol-field-content}) and relation~(\ref{FH-in-full-theory})), hence 
they are elements of $\trace_{\infty}^*$. It follows that 
$\|P_E\wil(\pa\phip)^{\mupi}(\pa\phim)^{\mumi}\wir P_E\|<\infty$, thus $\theh$ is a finite rank map from
$\mfar(\mco(r))$ to $B(\hil)$. Since the functionals $\htau_{\mupi,\mumi}^{(r)}$ are normal, $\theh$ extends
to a finite rank map from $\mfa(\mco(r))$ to $B(\hil)$.

In order to simplify the expression of the map $\thec$, we note
that for any two families of functions $\{\F^+_j\}_{j\in\nat}$, $\{\F^-_j\}_{j\in\nat}$
from $S(\real^s)_{\real}$ and for any pair of multiindices $(\mupi,\mumi)$ there holds the identity
\beqa
2^{\h |\mupi|} 2^{\h |\mumi|}:\phip(\F^+)^{\mupi}\phim(\F^-)^{\mumi}:\phantom{444444444444444444444444444444444444}\non\\
=\sum_{\su{\mu^{\pm},\nu^{\pm}\\ \mup+\nup=\mupi\\ \mum+\num=\mumi}}
\fr{\mupi!\mumi!}{\mup!\nup!\mum!\num!}
a^*(\tf^+)^{\mup}a^*(i \tf^-)^{\mum}a(\tf^+)^{\nup}a(i \tf^-)^{\num},\quad \label{normal-ordering}
\eeqa
where $\tf^{\pm}=\om^{\mp\h}\btF^\pm$ 
and the equality holds in the sense of quadratic forms on $\DF\times\DF$. Consequently, we obtain 
from (\ref{preexpansion})
\beqa
\thec(W(f))\ph{44444444444444444444444444444444444444444444444444444}\non\\
=\!\!\!\!\!\!\!\!\!\!\sum_{\su{\mub,\nub \\ \mub+\nub\in \Sett^{\prime} \\ (|\mup|+|\nup|,|\mum|+|\num|)\in \Set}}
\!\!\!\!\!\!\!\!\!\!
\fr{i^{|\mup|+|\nup|+2|\mum|}}{\mub!\nub!}e^{-\h\|f\|^2_2} \lan b_{\ka,r}|\,f\ran^{\mub+\nub} 
\, P_E a^*(h_{\ka,E})^{\mub}a(h_{\ka,E})^{\nub}P_E. \label{thec-expansion}
\eeqa
We introduce normal  functionals
$\ctau_{\mub,\nub}^{(r)}$ and quadratic forms $\cS_{\mub,\nub}$ on $\DF\times\DF$ given by
\beqa
\ctau_{\mub,\nub}^{(r)}&:=&\fr{i^{|\mup|+|\nup|+2|\mum|}}{\mub!\nub!}\bsi^{(r)}_{\mub+\nub}, \label{ctau}\\
\cS_{\mub,\nub}&:=&P_E a^*(h_{\ka,E})^{\mub}a(h_{\ka,E})^{\nub}P_E.\label{cS}
\eeqa 
Now the expansion (\ref{thec-expansion}) takes the form
\beq
\thec(A)=\!\!\!\!\!\!\!\!\!\!\sum_{\su{\mub,\nub \\ \mub+\nub\in \Sett^{\prime} \\ (|\mup|+|\nup|,|\mum|+|\num|)\in \Set}}
\!\!\!\!\!\!\!\!\!\!\ctau_{\mub,\nub}^{(r)}(A)\,\cS_{\mub,\nub}. \label{thec-expansion1}
\eeq
In order to study its convergence properties we collect several auxiliary results. From
Proposition~\ref{Bfunctionals} we obtain the following lemma.
\bel\label{ttau-bound-lemma} For any $l\geq 0$, the functionals $\ctau_{\mub,\nub}^{(r)}$, defined by (\ref{ctau}), satisfy the bound
\beq
\|R^{-l}\ctau_{\mub,\nub}^{(r)}R^{-l}\|\leq (2c_l)^{|\mub|+|\nub|}\sqrt{(|\mub|+|\nub|)!}\|b_{\ka,r}\|_{2,l}^{\mub+\nub}, \label{ttau-bound}
\eeq
where  $c_l=(12+2m^2)^{l/2}$, and $m$ is the mass of the theory.
\eel
\nin Next, to study the forms $\cS_{\mub,\nub}$, we recall the so-called energy bounds \cite{BP90}:
\bel\label{energy-bounds-lemma} For any $h_1,\ldots,h_n\in L^2(\real^s, d^sp)$ in the domain of $\omp$ there 
holds the  bound 
\beq
\|a(\omp h_1)\ldots a(\omp h_n)P_E\|
\leq E^{\fr{n}{2}}\| h_1\|_2\ldots \| h_n\|_2. \label{energy-bounds}
\eeq
\eel
\nin Making use of the above result and  Lemma~\ref{h-lemma}, we obtain
that the forms $\cS_{\mub,\nub}$, defined by~(\ref{cS}), are actually elements of $B(\hil)$
and satisfy the bound
\beq
\|\cS_{\mub,\nub}\|\leq E^{\fr{|\mub|+|\nub|}{2}} \|\omm \hti_{\ka,E}\|_2^{\mub+\nub}. \label{tS-bounds}
\eeq 
Now we are ready to prove convergence of  expansion (\ref{thec-expansion1}).
\bep\cite{Bo00}\label{bh-sum-convergence} Let  $\Set=\{(\m^+,\m^-)\in\nat_{0}^{2}\,|\, \m^+\leq \MM^+  \textrm{ and }
\m^-\leq \MM^-\}$ for some $\MM^{\pm}\in \nat_0$. Then, for any $0<p\leq 1$, there holds
\beq
\sum_{\su{\mub,\nub \\ (|\mup|+|\nup|,|\mum|+|\num|)\in \Set}}\!\!\!\!\!\!\!\!\!\!\!\!\!\!\!\!\!\!
\|\ctau_{\mub,\nub}^{(r)}\|^p \,\|\cS_{\mub,\nub}\|^p<\infty.
\eeq
\eep
\proof From estimates (\ref{ttau-bound}), (\ref{tS-bounds}) we obtain
\beqa
\sum_{\su{\mub,\nub \\ (|\mup|+|\nup|,|\mum|+|\num|)\in \Set}}\!\!\!\!\!\!\!\!\!\!\!\!\!\!\!\!\!\!
\|\ctau_{\mub,\nub}^{(r)}\|^p\,\|\cS_{\mub,\nub}\|^p
\leq \ph{4444444444444444444444444444444444444444} & &\non\\
\leq \big(2^{\MM^++\MM^-}\sqrt{(\MM^++\MM^-)!}\big)^p\sum_{\su{\mub,\nub \\ |\mub|+|\nub|\leq \MM^++\MM^-} }
\big(\|b_{\ka,r}\|_2^p\,\|\omm \hti_{\ka,E}\|_2^p\big)^{\mub+\nub}.& &
\eeqa
The sum on the r.h.s. can be estimated as follows
\beqa
\sum_{\su{\mub,\nub \\ |\mub|+|\nub|\leq \MM^++\MM^-} }\!\!\!\!\!\!\!\!\!\!\!\!
\big(\|b_{\ka,r}\|_2^p\,\|\omm \hti_{\ka,E})\|_2^p\big)^{\mub+\nub}\leq
\bigg(\sum_{k^+=0}^{\,\,\,\M^++\M^-}\sum_{\su{ \mup \\ |\mup|=k^+} }
\big(\|b_{\ka,r}^+\|_2^p\,\|\omm \hti_{\ka,E}^+\|_2^p)^{\mup}\bigg)^2& &\non\\
\cdot\bigg(\sum_{k^-=0}^{\,\,\,\M^++\M^-}\sum_{\su{ \mum \\ |\mum|=k^-}}
\big(\|b_{\ka,r}^-\|_2^p\,\|\omm \hti_{\ka,E}^-\|_2^p \big)^{\mum}\bigg)^2.& &\non\\
\eeqa
Making  use of the fact that the
multinomial coefficients are larger or equal to one and of the multinomial formula~(\ref{multiform}), we get
\beq
\sum_{\su{ \mup \\ |\mup|=k^+}}\big(\|b_{\ka,r}^+\|^p_2\,\|\omm \hti_{\ka,E}^+\|^p_2)^{\mup}\leq
\big(\sum_{\ka\in\nat_0^s}\|b_{\ka,r}^+\|_2^p\,\|\omm \hti_{\ka,E}^+\|_2^p)^{k^+}
\eeq
and similarly for the sum w.r.t. $\mum$.
The last expression is finite due to relation (\ref{sp-convergence}). \qed\\
This proves uniform convergence of
expansion (\ref{thec-expansion}). It follows that the map $\thec$, which was defined on 
the norm dense subalgebra $\mfar(\mco(r))$, can be extended to the full local algebra $\mfa(\mco(r))$.

From definition~(\ref{cS}) we obtain that in massive scalar free field theory of mass $m>0$
there holds $\cS_{\mub,\nub}=0$ for $|\mub|>\M$ or $|\nub|>\M$, where $\M=\fr{E}{m}$. Thus, choosing
$\Sett=\emptyset$ and $\MM^+=\MM^-=2[\M]$, we obtain that $\whXi_E=\thec$. Hence 
Proposition~\ref{bh-sum-convergence} and Lemma~\ref{equivalent-Nsharp} give the  known fact \cite{BP90, Bo00}
that Condition~$\Ns$ holds  in massive scalar free field theory.
\bep\label{massive-condition-Ns} In massive scalar free field theory there holds the identity
\beq
\whXi_E(A)=\sum_{\mub,\nub}\ctau_{\mub,\nub}^{(r)}(A)\,\cS_{\mub,\nub},\quad\quad A\in\mfa(\mco(r)),
\eeq
in the sense of norm convergence in $B(\hil)$. Moreover, for any $0<p\leq 1$ there holds the bound
\beq
\sum_{\su{\mub,\nub} }
\|\ctau_{\mub,\nub}^{(r)}\|^p \,\|\cS_{\mub,\nub}\|^p<\infty.
\eeq
\eep
\nin However, the methods of the present subsection do not suffice to verify Condition~$\Ns$
in massless free field theory. This goal is accomplished in the next subsection.
\subsection{Expansion of $\thet$}
Our last task is to complete the construction of the map $\thet$. In subsection (\ref{orthonormal}) we
introduced the $p$-nuclear positive operator $T$ which has a $J$-invariant orthonormal basis of 
eigenvectors $\{e_i\}_1^{\infty}$.
We expand the functions $f^{\pm}\in\Lpmring$  in this basis 
\beq
f^{\pm}=\sum_{j=1}^{\infty}\lan e_j  | f^{\pm} \ran e_j
\eeq
and make use of  the  multinomial formula~(\ref{multiform}), obtaining the following equality valid on $\DF$
\beq
a^{(*)}(f^\pm)^{m^\pm}=\sum_{\mu^\pm,|\mu^\pm|=m^\pm}\fr{m^\pm!}{\mu^\pm!}\lan e| f \ran^{\mu^\pm} 
a^{(*)}(\Lpm e)^{\mu^\pm}. \label{multinomial}
\eeq
Using relation (\ref{normal-ordering}), we obtain from definition~(\ref{twosums}) the following expansion,
understood in the sense of quadratic forms on $\DF\times\DF$
\beqa
\thet(W(f))\phantom{44444444444444444444444444444444444444444444444444444}\non\\
=\sum_{\su{\mub,\nub \\ (|\mup|+|\nup|,|\mum|+|\num|)\in \Set^\prime }  }
\fr{i^{|\mup|+|\nup|+2|\mum|}}{\mub!\nub!}e^{-\h\|f\|^2_2}\lan e | f \ran^{\mub+\nub}
P_Ea^*(\LL e)^{\mub}a(\LL e)^{\nub}P_E.
\eeqa
From the second part of Lemma~\ref{Bfunctionals}
there follows the existence of normal functionals $\tau_{\mub,\nub}$ on $B(\hil)$
which have the property
\beq
\tau_{\mub,\nub}(W(f))=\fr{i^{|\mup|+|\nup|+2|\mum|}}{\mub!\nub!}e^{-\h\|f\|^2_2}\lan e | f \ran^{\mub+\nub}. \label{functionals1}
\eeq
and satisfy the bound 
\beq
\|\tau_{\mub,\nub}\|\leq \fr{4^{|\mub|+|\nub|} }{(\mub!\nub!)^\h} \bigg(\fr{(\mub+\nub)!}{\mub!\nub!}\bigg)^\h\leq 
\fr{2^{\fr{5}{2}(|\mub|+|\nub|)}}{(\mub!\nub!)^\h},\label{tauestimate2}
\eeq
where we used properties of multinomial coefficients. Finally, we define the quadratic forms on $\DF\times\DF$
given by 
\beq
\Smn=P_E a^*(\LL e)^{\mub}a(\LL e)^{\nub} P_E. \label{Smn-def}
\eeq
We note that $a(\LL^\pm e)^{\nub} P_E=a(Q_E\LL^\pm e)^{\nub} P_E$ and $\omm Q_E\LL^\pm e_i=\TEpm e_i$, where 
$\TEpm$ are bounded operators by Lemma~\ref{TEpm-lemma}. Thus we obtain from 
the energy bounds 
(\ref{energy-bounds}) and definition~(\ref{T-definition}) of the operator $T$
\beq
\|\Smn\|\leq E^{\fr{|\mub|+|\nub|}{2}}\|\omm\Q_E \LL e\|_2^{\mub}\|\omm\Q_E\LL e\|_2^{\nub} 
\leq E^{\fr{|\mub|+|\nub|}{2}}t^{\mub} t^{\nub}, \label{Sestimate2}
\eeq 
where $\{t_j\}_1^\infty$ are eigenvalues of $T$. We have arrived at the following expansion, still in the
sense of quadratic forms on $\DF\times\DF$
\beq
\thet(A)=\sum_{\su{\mub,\nub \\ (|\mup|+|\nup|,|\mum|+|\num|)\in \Set^\prime }  }\tau_{\mub,\nub}(A)\,\Smn,
\label{lastexpansion}
\eeq
valid for any $A\in\mfar(\mco(r))$ i.e. for any  finite, linear combination of Weyl operators.
Our task is to  establish convergence of this sum in the
norm topology of $B(\hil)$ and extend this map by continuity to all $A\in\mfa(\mco(r))$.  It suffices to consider the case $\Set=\{0,0\}$ when there holds
\beq
\whXi_E(A)=\om_0(A)P_E+\thet(A),\quad\quad A\in\mfa(\mco).
\eeq
The following proposition verifies the known fact that Condition $\Ns$ holds in scalar free field theory.
\bep\cite{BP90,Bo00}\label{N-sharp} In massive and massless scalar free field theory there holds the identity
\beq
\whXi_E(A)=\sum_{\mub,\nub}\tau_{\mub,\nub}(A)\Smn, \quad\quad A\in\mfa(\mco(r)), 
\eeq
in the sense of norm convergence in $B(\hil)$. Moreover, for any $0<p\leq 1$ there holds the bound
\beq
\sum_{\su{\mub,\nub} }\|\tmn\|^p \, \|\Smn\|^p <\infty.
\eeq
\eep
\proof We note the following estimate
\beqa
\sum_{\mub,\nub}\|\tmn\|^p \, \|\Smn\|^p 
&\leq& \sum_{\mub,\nub} 
\fr{(2^5E)^{\half p (|\mub|+|\nub|)} } {(\mub!)^{\half p }(\nub!)^{\half p}} t^{p\mub}t^{p\nub} \non\\
&\leq& \sum_{m^\pm,n^\pm\in\nat_0}
\fr{(2^{5/2}E^{1/2}\|T\|_p)^{p(m^++m^-+n^++n^-)} }{(m^+!m^-!n^+!n^-!)^{\h p}},\label{scale-noninvariant}
\eeqa
where in the first step we used (\ref{tauestimate2}) and  (\ref{Sestimate2}), in the second step 
we made use of the fact that the multinomial coefficients are larger or equal to one and
of the identity
\beq
\sum_{\mupm, |\mupm|=m^\pm}\fr{m^\pm!}{\mupm!} (t^p)^{\mupm}=(\|T\|_p^p)^{m^\pm},
\eeq 
and similarly for the sums w.r.t. $\nupm$. \qed\\
Thus the map $\thet$, given by (\ref{lastexpansion}), has range in $B(\hil)$ and extends by continuity
to the whole local algebra $\mfa(\mco(r))$. We denote the resulting map by the same symbol.

%% file: PhD-L2_app.tex
\chapter{Verification of Condition $\A$ in Scalar Free Field Theory} \label{Condition-L2} 

The goal of this appendix is to verify that Condition~$\A$, introduced in Section~\ref{space-translations-QFT},
holds in scalar free field theory. In the massive case this fact follows from Condition~$\B$, verified
in Appendix~\ref{Condition-L1}, and Theorem~\ref{triviality}. Thus our main interest in the present appendix is in  massless scalar free field theory, although some results will be stated for general $m\geq 0$ to facilitate their application in
other contexts. Our aim is to prove Theorems~\ref{full-results1}, \ref{even-results1} and \ref{derivative-results1} which 
are at the basis of our discussion in Subsection~\ref{Examples}.

The proofs of the above three statements are given in Section~\ref{HA-in-models-section}. They rely 
on the the auxiliary Theorem~\ref{HA-in-models}, stated below, whose proof is the subject of the
later part of this appendix. In Section~\ref{Part1} we define the functionals $\tau_k$, $k\in\{1,2,3\}$, 
on $\hmfa$ and verify that they have the properties required in the statement
of Theorem~\ref{HA-in-models}. We also show that the map $R^{(2)}$, defined in Theorem~\ref{HA-in-models},
can be expressed in terms of the maps $\thec$ and $\thet$, introduced in Appendix~\ref{Preliminaries}.
In Section~\ref{Part2} we show that the range of $R^{(2)}$
consists of square-integrable operators. The argument is divided into three parts: In Subsection~\ref{KeyLemmas} we prove a variant of Theorem~\ref{harmonic} which is applicable to the present problem. In Subsections~\ref{square-integrability-1} and  
\ref{square-integrability-2} we apply this result to prove square-integrability of the ranges of the maps $\thec$ and $\thet$, respectively.

\section{Proofs of Theorems 2.2.5, 2.2.6  and 2.2.7 
based on Theorem C.1.1} \label{HA-in-models-section} 
Our discussion in this section  is based on the following theorem, whose proof is given in Sections~\ref{Part1} and \ref{Part2}.
\bet\label{HA-in-models} In massless scalar free field theory in $s\geq 3$ dimensional space 
there exist linear functionals $\tau_1$, $\tau_2$, $\tau_3$ on $\hmfa$, invariant under translations 
in space, s.t.  
for any $A\in\hmfa$ the quantity
\beqa
R^{(2)}(A):= A-\om_0(A)I-\tau_1(A)\phip-\tau_2(A)\wil\phip^2\wir-\tau_3(A)\wil\phip^3\wir, \label{R2-in-models}
\eeqa
defined as a quadratic form on the domain $\BE\times\BE$ of vectors of bounded energy,
satisfies $\|R^{(2)}(A)\|_{E,2}<\infty$ for any $E\geq 0$. Moreover, $\om_0, \tau_1, \tau_2, \tau_3$
form a linearly independent family. In addition, there hold the following assertions:
\beqa
& &\hemfa\subset\ker\tau_1\cap\ker\tau_3 \te{ and }\tau_2|_{\hemfa}\neq 0,\label{even-in-kernels}\\
& &\hdmfa\subset\ker\tau_1\cap\ker\tau_2\cap\ker\tau_3. \label{derivatives-in-kernels}
\eeqa
\eet
\nin It is an immediate consequence of this theorem that the theory $(\dmfa,\al,\hil)$, generated by
derivatives of the massless scalar free field, satisfies Condition~$\A$ and has trivial point-continuous
subspace. In fact: \\
\bf Proof of Theorem \ref{derivative-results1}: \rm The statement follows directly from relations 
(\ref{derivatives-in-kernels}) and (\ref{R2-in-models}). \qed

Let us now consider the full massless scalar  free field theory $(\mfa,\al,\hil)$. Theorem~\ref{HA-in-models} 
reduces the analysis of the point-continuous subspace in this model to the study of the three pointlike-localized fields: 
$\phip$, $\wil\phip^2\wir$, $\wil\phip^3\wir$. We will show below that
the concepts of square-integrability and of the infrared order of an operator, defined for observables from $\hmfa_{\scc}$ by relations~(\ref{square-integrable}) and (\ref{infrared-order}), respectively, can be extended to the fields in question.  Moreover, there holds the following proposition, whose proof is given in the later part of this section.
\bep\label{Wick3} In massless scalar free field theory the following statements hold true:
\begin{enumerate}
\item[(a)] If $s\geq 3$, then  $\ord(\phip)=2$.
\item[(b)] If $s=3$, then $\ord(\wil\phip^2\wir)=1$. \\
           If $s=4$, then $\ord(\wil\phip^2\wir)=0$. \\
           If $s\geq 5$, then  $\|\wil\phi^2\wir\|_{E,2}<\infty$ for any $E\geq 0$.
\item[(c)] If $s=3$, then $\ord(\wil\phip^3\wir)=0$. \\
           If $s\geq 3$, then $\|\wil\phi^3\wir\|_{E,2}<\infty$ for any $E\geq 0$.
\end{enumerate}
Part (b) also holds if $\wil\phip^2\wir$ is replaced with  $\un{\wil\phip^2\wir}\in\Phi^{\te{(e)}}_{\FH}$.
\eep
\nin We note that vanishing of the infrared order does not imply that a given operator
is square-integrable. There remains an open question if $\wil\phip^2\wir$ for $s=4$ and $\wil\phip^3\wir$
for $s=3$ have the property of square-integrability. Its resolution would allow one to determine exactly the dimensions
of the point-continuous subspaces in Theorem~\ref{full-results1}~(a) and (b), and Theorem~\ref{even-results1}~(b).

After this preparation we estimate the dimension of the point-continuous subspace in (full) massless
scalar free field theory and compute the infrared orders of its elements. As expected, the
infrared structure improves with increasing dimension, in the sense that the dimension of the
point-continuous subspace decreases. However, this subspace remains non-trivial for any $s\geq 3$.\\
\bf Proof of Theorem~\ref{full-results1}: \rm
We  consider only the case $s=3$ as the remaining cases can be proven analogously.
Since we do not know whether $\wil\phip^3\wir$, whose infrared order is zero by Proposition~\ref{Wick3},
is also square-integrable, we have to consider both possibilities. 
First, we show that if $\|\wil\phip^3\wir\|_{E,2}=\infty$ for some $E\geq 0$, then for any $A\in\hmfa_{\scc}$ 
there hold the following statements:
\begin{enumerate}
\item[(i)] $\tau_1(A)\neq 0\iff \ord(A)=2$.
\item[(ii)] $(\tau_1(A)=0\te{ and }\tau_2(A)\neq 0)\iff\ord(A)=1$.
\item[(iii)] $(\tau_1(A)=0, \tau_2(A)=0\te{ and }\tau_3(A)\neq 0)\iff(\ord(A)=0 \te{ and }
\exists_{E\geq 0}\te{ s.t. } \|A\|_{E,2}=\infty)$.
\item[(iv)] $(\tau_1(A)=0,\tau_2(A)=0\te{ and }\tau_3(A)=0)\iff \forall_{E\geq 0}\,\,\|A\|_{E,2}<\infty$.
\end{enumerate}
To justify  these claims, suppose that $\tau_0(A)=\cdots=\tau_{l-1}(A)=0$ for some $l\in\{1,2,3,4\}$.
($\tau_0:=\om_0$ is understood here). Then relation~(\ref{R2-in-models}) gives 
the following bounds for any $\fun\in\traceEB$ 
\beqa
& &\pm\bigg\{\big(\int d^sp\, |\vep|^{\be}|\fun(\wt{A}(\vep))|^2\big)^\h-
|\tau_l(A)|\big(\int d^sp\, |\vep|^{\be}|\fun(\wt{\wil\phip^{\scriptstyle{l}}\wir}(\vep))|^2\big)^\h\bigg\}\non\\
&\leq&\sum_{k=l+1}^3|\tau_k(A)|\sup_{\fun^\prime\in\traceEB}\big(\int d^sp\, |\vep|^{\be}|\fun^\prime(\wt{\wil\phip^k\wir}(\vep))|^2\big)^\h+
E^\fr{\be}{2}\|R^{(2)}(A)\|_{E,2}.
\eeqa
We note that all terms in this expression are finite for sufficiently large $\be$ by estimate~(\ref{2regularity})
and Proposition~\ref{Wick3}. We will now study their behavior with decreasing $\be$.
By Proposition~\ref{Wick3},  $\ord(\wil\phip^l\wir)>\ord(\wil\phip^k\wir)$, for $k>l$ in the above formula. Thus, by considering
$\be$ in a small neighborhood of $\ord(\wil\phip^l\wir)$ and taking supremum w.r.t. $\fun\in\traceEB$,  we easily
obtain that $\ord(A)=\ord(\wil\phip^l\wir)$ if and only if $\tau_l(A)\neq 0$. Hence, there holds (i) and (ii).
In part (iii) we set $\be=0$ and make use of our assumption that $\|\wil\phip^3\wir\|_{E,2}=\infty$ for some $E\geq 0$. 
In (iv) the implication $(\Rightarrow)$ follows trivially from square-integrability of $R^{(2)}(A)$. The opposite
implication is a consequence of (i), (ii) and (iii).

Thus we have verified that $\Ord(\hmfa_{\scc})=\{0,1,2\}$ and the subspace $\hmfa_{\ac}$, consisting of
square-integrable observables, can be expressed as follows
\beqa
\hmfa_{\ac}=\ker\om_0\cap\ker\tau_1\cap\ker\tau_2\cap\ker\tau_3, \label{intersection-of-kernels1}
\eeqa
Now computation of the dimension of the point-continuous subspace is a simple
exercise in linear algebra: Since the above functionals are linearly independent (by Theorem~\ref{HA-in-models}),
we can find $A_1,A_2,A_3\in\hmfa_{\scc}$ s.t. $\tau_i(A_j)=\de_{i,j}$. For any $A\in\hmfa_{\scc}$ we  obtain
the decomposition
\beq
A=\big(A-A_1\tau_1(A)-A_2\tau_2(A)-A_3\tau_3(A)\big)+A_1\tau_1(A)+A_2\tau_2(A)+A_3\tau_3(A),
\eeq
where the term in bracket belongs to $\hmfa_{\ac}$ due to (\ref{intersection-of-kernels1}). Choosing the 
point-continuous subspace as $\hmfa_{\pc}=\Span\{A_1,A_2,A_3\}$ and noting that $\{A_j\}_1^3$ are linearly
independent we obtain that $\dim\hmfa_{\pc}=3$.

Assuming that for any $E\geq 0$ there holds $\|\wil\phip^3\wir\|_{E,2}<\infty$, we can incorporate the term 
$\tau_3(\,\cdot\,)\wil\phip^3\wir$ to  $R^{(2)}(\,\cdot\,)$ in formula~(\ref{R2-in-models}). Thus, proceeding
analogously as in the previous case, we verify the following facts for any $A\in\hmfa_{\scc}$:
\begin{enumerate}
\item[(i$^\prime$)] $\tau_1(A)\neq 0\iff \ord(A)=2$.
\item[(ii$^\prime$)] $(\tau_1(A)=0\te{ and }\tau_2(A)\neq 0)\iff\ord(A)=1$.
\item[(iii$^\prime$)] $(\tau_1(A)=0\te{ and }\tau_2(A)=0)\iff(\forall_{E\geq 0}\,\, \|A\|_{E,2}\leq\infty)$.
\end{enumerate}
Again, it follows that $\Ord(\hmfa_{\scc})=\{0,1,2\}$. However, the absolutely continuous subspace is now
given by
\beq
\hmfa_{\ac}=\ker\om_0\cap\ker\tau_1\cap\ker\tau_2. \label{intersection-of-kernels2}
\eeq
Hence the  point-continuous subspace is two-dimensional in this case. In view of relations~(\ref{intersection-of-kernels1})
and (\ref{intersection-of-kernels2}) the theory satisfies Condition~$\A$. \qed

The last example which we consider is the even part of massless scalar free field theory. 
Making use of Theorem~\ref{HA-in-models}, we will show that the theory $(\eumfa,\eal,\ehil)$, 
introduced in Section~\ref{scalar-free-field-theory},
satisfies Condition~$\A$ and we will analyze the resulting point-continuous subspace. To this end, we define the
functionals $\un{\om}_0=\om_0\circ\epim$, $\un{\tau}_2=\tau_2\circ\epim$ and set
for any $\uA\in\eumfa$
\beq
R^{(2)}_{\te{(e)}}(\uA):= \uA-\un{\om}_0(\uA)\un{I}-\un{\tau}_2(\uA)\un{\wil\phip^2\wir}\label{even-R2-in-models}
\eeq 
as a quadratic form on states of bounded energy in $\ehil$. Then, due to relation~(\ref{even-functoriality})
and the fact that $\efun(\un{\wil\phip^2\wir})=\ei(\efun)(\wil\phip^2\wir)$ for any $\efun\in\traceE^{\ee}$,
there holds
\beq
\efun\big(R^{(2)}_{\te{(e)}}(\uA)\big)=\ei(\efun)\big(R^{(2)}(\epim(\uA))\big),\quad\quad \uA\in\eumfa,\,\,\, \efun\in\traceE^{\ee}.
\eeq
Finally, making use of relation~(\ref{even-norm-conservation}), we obtain
$\|R^{(2)}_{\te{(e)}}(\uA)\|_{E,2}\leq \|R^{(2)}(\epim(\uA))\|_{E,2}<\infty$, where the last
bound follows from Theorem~\ref{HA-in-models}. With the help of Proposition~\ref{Wick3} we 
obtain the description of the point-continuous subspace in the even part of massless scalar free field theory.
Here the only possible obstruction to square-integrability of observables form $\heumfa_{\scc}$ is the presence 
of the term $\un{\wil\phip^2\wir}$ in relation~(\ref{even-R2-in-models}).\\
\bf Proof of Theorem~\ref{even-results1}: \rm Exploiting relation~(\ref{even-R2-in-models})
and the subsequent discussion, and  proceeding as in the proof of Theorem~\ref{full-results1} above, we obtain the result. \qed

The remaining part of this section is devoted to the proof of Proposition~\ref{Wick3} which was
the main technical input of the above discussion. We note that
any field $\phii\in\Phi_{\FH}$ belongs to $\trace_{\infty}^*$ and therefore
$\fun(\phii(\vx))$ is a bounded, continuous function for any $\fun\in\trace_E$. We are interested 
in the regularity properties of its  Fourier transform $\fun(\tphii(\vep))$ which is a tempered distribution.
To begin with, we prove a simple, technical  lemma.
\bel\label{technical1} Let $0<\be<s$, $\field\in \traceE^*$ and $D\subset\hil$ be a dense domain s.t.
$|\vep|^{\fr{\be}{2}}(\Psi_1|\tphii(\vep)\Psi_2)$ is square integrable, uniformly
in $\Psi_1,\Psi_2\in (P_E\hil\cap D)_1$. Then  $|\vep|^{\fr{\be}{2}}\fun(\tphii(\vep))$ is square
integrable, uniformly in $\fun\in\traceE$, and
\beq
\sup_{\Psi_1,\Psi_2\in(P_E\hil\cap D)_1}\int d^sp|\vep|^{\be}|(\Psi_1|\wt{\phii}(\vep)\Psi_2)|^2=
\sup_{\fun\in\traceEB}\int d^sp|\vep|^{\be}|\fun(\wt{\phii}(\vep))|^2.
\label{c}
\eeq
\eel
\proof By the Cauchy-Schwarz inequality there holds for any $g\in S(\real^s)$
and $\Psi_1,\Psi_2\in(P_E\hil\cap D)$
\beqa
|(\Psi_1|\phii(g)\Psi_2)|\phantom{444444444444444444444444444444444444444444444444444444444444}\non\\
\leq \|\Psi_1\|\,\|\Psi_2\|\sup_{{\Psi'}_1,{\Psi'}_2\in(P_E\hil\cap D)_1}\big(\int d^sp|\vep|^{\be}
|({\Psi'}_1| \tphii(\vep){\Psi'}_2)|^2\big)^{\h}\,
\big(\int d^sp\fr{1}{|\vep|^{\be}}|\tilde{g}(\vep)|^2\big)^\h. \,\,\,\,\,\,\label{Riesz}
\eeqa
The above bound extends, by continuity, to any $\Psi_1,\Psi_2\in P_E\hil$ and we can proceed
as in the proof of Theorem 2.5 from \cite{Bu90}: Let $L^2(\real^s)_{(\be)}$ be the Hilbert 
space of (classes of) functions $h$ on $\real^s$ for which
\beq
\|h\|_{2,(\be)}=\int d^sp\,|\vep|^{-\be} |h(\vep)|^2<\infty.
\eeq
Clearly, the subspace of test functions from $S(\real^s)$ is dense in  $L^2(\real^s)_{(\be)}$.
It follows from relation (\ref{Riesz}) that  for any $\fun\in\traceEB$
\beqa
& &|\fun(\phii(g))|\non\\
& &\leq \|P_E \phii(g) P_E\|\leq \|g\|_{2,(\be)}\sup_{{\Psi'}_1,{\Psi'}_2\in(P_E\hil\cap D)_1}\big(\int d^sp|\vep|^{\be}
|({\Psi'}_1| \tphii(\vep){\Psi'}_2)|^2\big)^{\h}.\quad\quad
\eeqa
The above inequality and the Riesz theorem imply that $\fun(\wt{\phii}(\vep))\in L^2(\real^s)_{(-\be)}$ and
\beq
\sup_{\fun\in\traceEB}\int d^sp|\vep|^{\be}|\fun(\wt{\phii}(\vep))|^2\leq \sup_{{\Psi'}_1,{\Psi'}_2\in(P_E\hil\cap D)_1}\big(\int d^sp|\vep|^{\be}
|({\Psi'}_1| \tphii(\vep){\Psi'}_2)|^2\big)^{\h}.
\eeq
The opposite inequality is trivial, since the supremum on the r.h.s extends over a smaller set. \qed\\
Setting in the above lemma $\phi=\phip$ and $D=D_S$ we obtain a prescription for computation
of the infrared order of the field $\phip$: \\ 
\bf Proof of Proposition \ref{Wick3} (a): \rm We will establish the bound
\beq
\sup_{\fun\in\traceEB}\int d^sp|\vep|^{2}|\fun(\tphip(\vep))|^2<\infty,
\eeq
and show that the power of the mollifier $|\vep|^{2}$ cannot be reduced. By Lemma~\ref{technical1} it suffices to consider $\fun(\,\cdot\,)=(\Psi_1|\, \cdot\, \Psi_2)$ where $\Psi_1,\Psi_2\in D_S\cap P_E\hil$. 
Making use of the fact that 
\beq
\tphip(\vep)=\fr{1}{\sqrt{2|\vep|}}\big(a^*(\vep)+a(-\vep)\big), \label{ft}
\eeq
in the sense of quadratic forms on $D_S\times D_S$, we obtain the estimate
\beqa
& &\int d^sp|\vep|^{2}|(\Psi_1|{\tphip}(\vep)\Psi_2)|^2\non\\
& &\phantom{44444444}\leq \int d^sp |\vep|\big(\|\Psi_1\|^2 (\Psi_2|a^*(\vep)a(\vep)\Psi_2)+
\|\Psi_2\|^2 (\Psi_1|a^*(\vep)a(\vep)\Psi_1)\big)\non\\
& &\phantom{44444444}\leq 2E\|\Psi_1\|^2\,\|\Psi_2\|^2,
\label{Ebound}
\eeqa
where we used representation (\ref{Hamiltonian-representation})  of the Hamiltonian.
To show that the bound for $\phip$ is optimal, we construct a suitable sequence of functionals: We choose a 
positive function $h(\vep)\in\czn$ s.t. 
$\supp\, h(\vep)\subset \{\,\vep\in\real^s\,|\, |\vep|\leq E\}$, $\int |h(\vep)|^2 d^sp=1$
and set  $H_n(\vep)=n^{\fr{s}{2}}h(n\vep)$ for $n\in\nat$. Then the functionals on $B(\hil)$ given by
\beqa
\fun_n^{(1)}(\,\cdot\,)=(H_n^{\otimes (n-1)}|\,\cdot\, H_n^{\otimes n})
\eeqa
clearly belong to $\traceEB$. We note that for any $\eps>0$
\beqa
\int d^sp |\vep|^{2-\eps}|\fun_n^{(1)}(\tphip(\vep))|^2 &=& \fr{1}{2}\int d^sp |\vep|^{1-\eps}
|(H_n^{\otimes n-1}|a(\vep)H_n^{\otimes n})|^2\non\\
&=&\fr{n^\eps}{2}\int d^sp |\vep|^{1-\eps}|h(\vep)|^2.
\eeqa
By taking the limit $n\to\infty$ the claim follows. \qed\\
In order to compute the infrared orders of higher  Wick powers of $\phip$, we need the following lemma.
\bel\label{Energy1} Let $\Psi_1,\Psi_2$ be  normalized vectors from $D_S\cap P_E\hil$. Then
\beq
\int d^sp_1\ldots d^sp_n |\vep_1|^{2}\ldots |\vep_n|^{2} 
|(\Psi_1|\wil\tphip(\vep_1)\ldots\tphip(\vep_n)\wir \Psi_2)|^2
\leq c_{n,s,E},\label{Energy2}
\eeq
for some constant $c_{n,s,E}$ independent of $\Psi_1$, $\Psi_2$. 
\eel
\proof Due to formula (\ref{ft})
it is clear that the expression on the l.h.s. of (\ref{Energy2}) can be bounded by a linear combination of terms of the form
\beqa
& &\int d^sp_1\ldots d^sp_n |\vep_1|\ldots|\vep_n| 
|(\Psi_1|a^*(\vep_1)\ldots a^*(\vep_k)a(\vep_{k+1})\ldots a(\vep_n)\Psi_2)|^2\non\\
& &\leq \int d^sp_1\ldots d^sp_k |\vep_1|\ldots|\vep_k| 
(\Psi_1|a^*(\vep_1)\ldots a^*(\vep_k)a(\vep_{k})\ldots a(\vep_1)\Psi_1)\non\\
& &\cdot\int d^sp_{k+1}\ldots d^sp_n |\vep_{k+1}|\ldots|\vep_n| 
(\Psi_2|a^*(\vep_{k+1})\ldots a^*(\vep_n)a(\vep_n)\ldots a(\vep_{k+1})\Psi_2)
\leq E^n, \non\\
\eeqa
where in the first step we made use of the Cauchy-Schwarz inequality and in the second step of the representation~(\ref{Hamiltonian-representation}) of the Hamiltonian. \qed\\
After this preparation we turn to the Wick powers of the field $\phip$.
\bel\label{Wick2} Let  $n>1$, $s\geq 3$. Then,
for any $\alb\geq 0$  s.t. $\alb>2-(s-2)(n-1)$ there holds the bound
\beq
\sup_{\fun\in\traceEB}\int d^sp |\vep|^\alb |\fun(\wil\tphip^n\wir(\vep))|^2<\infty. \label{mollifier}
\eeq
\eel
\proof We fix $\Psi_1,\Psi_2\in (D_S\cap P_E\hil)_1$ and define the functions
\beqa
& &F_n(\vep)=\int d^sq_1\ldots d^sq_{n-1}|\vep-\veq_1|^{2}|\veq_1-\veq_2|^{2}\ldots |\veq_{n-1}|^{2}\non\\
& &\phantom{44444444444444444}\cdot|(\Psi_1|\wil\tphip(\vep-\veq_1)\tphip(\veq_1-\veq_2)\ldots\tphip(\veq_{n-1})\wir|\Psi_2)|^2,
\eeqa
which, according to Lemma \ref{Energy1}, belong to $L^1(\real^s,d^sp)$  and their $L^1$-norms are uniformly 
bounded in $\Psi_1$, $\Psi_2$ from the above set. We introduce the function 
$\chi_E(\vep)\in \czn$ which is equal to one on the set $\{\, \vep\in\real^s |\, |\vep|\leq E\,\}$ and obtain the following string of inequalities
\beqa
\int d^sp|\vep|^{\alb}|(\Psi_1|\wil\tphip^n\wir(\vep)\Psi_2)|^2
\phantom{4444444444444444444444444444444444444444444444444}& &\non\\
\leq (2\pi)^{-s(n-1)}\int d^sp|\vep|^{\alb}
\bigg|\int d^sq_1\ldots d^sq_{n-1}\fr{\chi_E(\vep-\veq_1)}{|\vep-\veq_1|}
\fr{\chi_E(\veq_1-\veq_2)}{|\veq_1-\veq_2|}\ldots
\fr{\chi_E(\veq_{n-1})}{|\veq_{n-1}|}\ph{4444444} & &\non\\ 
\cdot|\vep-\veq_1|\,|\veq_1-\veq_2|\ldots |\veq_{n-1}|
(\Psi_1|\wil\tphip(\vep-\veq_1)\tphip(\veq_1-\veq_2)\ldots\tphip(\veq_{n-1})\wir|\Psi_2)\bigg|^2\ph{444444} & &\non\\
\leq (2\pi)^{-s(n-1)}n^{\alb+1}\|F_n\|_1 \phantom{444444444444444444444444444444444444444444444444444444}& &\non\\ 
\sup_{\vep_1\in\real^s}
\int d^sq_1\ldots d^sq_{n-1}\fr{|\chi_E(\vep_1-\veq_1)|^2}{|\vep_1-\veq_1|^{2-\alb}} 
\fr{|\chi_E(\veq_1-\veq_2)|^2}{|\veq_1-\veq_2|^{2}} \ldots
\fr{|\chi_E(\veq_{n-1})|^2}{|\veq_{n-1}|^{2}}.\quad\quad\quad & & \label{longg}
\eeqa
Here in the first step we made use of the fact that the Fourier transform of a product is a convolution of 
the Fourier transforms of the factors. We also used the support properties of the wavefunctions corresponding to
$\Psi_1$, $\Psi_2$. In the second step we applied the Cauchy-Schwarz inequality and the bound
$|\vep|^\alb\leq n^{\alb}(|\vep-\veq_1|^\alb+|\veq_1-\veq_2|^\alb\cdots+|\veq_{n-1}|^\alb)$ valid for any $\alb\geq 0$. To show that expression (\ref{longg})  is bounded in the cases considered in the lemma we make use of the Young inequality 
\cite{RS2} which implies that for $f_i\in L^{r_i}(\real^s, d^sp)$, $i\in\{1,\ldots,n\}$, there holds 
\beq
\|f_1*\cdots*f_n\|_r\leq \|f_1\|_{r_1}\ldots\|f_n\|_{r_n},
\eeq
if $1\leq r,r_i\leq\infty$ are s.t. $r_1^{-1}+\cdots+r_n^{-1}=n-1+r^{-1}$. First, we assume that $\alb\geq 2$.
Then, by  choosing $r_1^{-1}=0$, $r_2^{-1}=\cdots=r_n^{-1}=1$, we conclude the proof of estimate~(\ref{mollifier}).
For $0\leq \alb <2$, it follows from our assumptions that there exists $\eps>0$ s.t. $(2-\be)(1+\eps)<\min(s,(s-2)(n-1))$.
We choose $r_1^{-1}=\fr{(2-\alb)(1+\eps)}{s}$,
$r_2^{-1}=\cdots=r_n^{-1}=1-\fr{r_1^{-1}}{(n-1)}$. One easily checks that these parameters satisfy the conditions specified
above and, moreover, $(2-\be)r_1<s$, $2r_i<s$ for $i\in\{2,\ldots,n\}$.   \qed\\ 
\bf Proof of Proposition \ref{Wick3} (b) and (c): \rm Upper bounds from part (b) and part (c) follow from Lemma \ref{Wick2}.
It remains to prove the lower bound in part (b) for $s=3$. Let $H_n$ be defined as in the proof of 
Proposition~\ref{Wick3}~(a) above and let us consider the family of functionals from $\traceEB$ given by
\beq
\fun_n^{(2)}(\,\cdot\,)=(H_n^{\otimes (n-2)}|\,\cdot\, H_n^{\otimes n}).
\eeq
We fix  $\eps>0$ and compute
\beqa
& &\int d^sp |\vep|^{1-\eps}|\fun_n^{(2)}(:\tphip^2:(\vep))|^2\non\\
& &\phantom{444444444}=\fr{(2\pi)^{-s}}{4}\int d^sp |\vep|^{1-\eps}
\bigg|\int d^sq \fr{1}{|\vep-\veq|^\h|\veq|^\h}  (H_n^{\otimes n-2}|a(\vep-\veq)a(\veq)H_n^{\otimes n})\bigg|^2\non\\
& &\phantom{444444444}=\fr{(2\pi)^{-s}}{4}\fr{n^3(n-1)}{n^{s+1}}n^\eps\int d^sp |\vep|^{1-\eps}
\bigg|\int d^sq\fr{h(\vep-\veq)}{|\vep-\veq|^\h}\fr{h(\veq)}{|\veq|^\h}\bigg|^2.
\eeqa
By taking the limit $n\to\infty$ we verify that $\ord(\wil\phip^2\wir)=1$ for $s=3$. 
In the case of $\un{:\phip^2:}\in\Phi_{\FH}^{\te{(e)}}$, the lower bound
is established analogously, restricting attention to even $n$. Similarly, the
upper bound established in Lemma~\ref{Wick2} still holds, since the supremum in relation~(\ref{mollifier}) 
extends now over a smaller set of functionals. \qed\\

\section{Proof of Theorem C.1.1 (I): Functionals $\{\tau_i\}_1^3$}\label{Part1} 

In Section~\ref{full-expansion} we introduced the decomposition of the map $\whXi_E:\mfa(\mco(r))\to B(\hil)$, given by 
$\whXi_E(A)=P_EAP_E$, into three components
\beq
\whXi_E=\theh+\thec+\thet, \label{whXi-decomposition}
\eeq
which are determined by the sets $\Set$ and $\Sett$.
With the statement of Theorem~\ref{HA-in-models} in mind, we 
choose $\MM^+=3$, $\MM^-=0$ in the definition of the set $\Set$ and define the set $\Sett$ as 
consisting of four 2-multiindices
\beq
\Sett=\{\ov{\al}_{0},\ldots,\ov{\al}_{3}\} \label{Sett}
\eeq
which are given by:  $\al^+_{k}(j)=k\de_{j,1}$, $\al^-_{k}(j)=0$ for $k\in\{0,1,2,3\}$, $j\in\nat$.
Then, making use of definition (\ref{finiterank0}) of the map $\theh$, and recalling that our numbering of
the $s$-indices $\{\ka_j\}_1^\infty$, introduced after formula (\ref{hb-expansion}), was chosen so that $\ka_1=0$,
we obtain
\beq
\theh(A)=P_E\big(\om_0(A)I+\htau_{\ov{\al}_{1}}^{(r)}(A)\phi+\htau_{\ov{\al}_{2}}^{(r)} (A)\wil\phi^2\wir+
\htau_{\ov{\al}_{3}}^{(r)}(A)\wil\phi^3\wir\big)P_E, \label{theh-concrete}
\eeq
for any  $A\in\mfa(\mco(r))$. Now we are in position to construct the functionals $\tau_k$, $k\in\{1,2,3\}$,
appearing in Theorem~\ref{HA-in-models}, and verify that they have the required properties.
\bep In massless scalar free field theory for $s\geq 3$ there exist linear functionals $\tau_k$, $k\in\{1,2,3\}$ on $\hmfa$ which satisfy
\beqa
& &\tau_k|_{\mfa(\mco(r))}=\htau_{\ov{\al}_{k}}^{(r)},\label{tau-consisitency}\\
& &\tau_k(A(\vx))=\tau_k(A) \label{tau-invariance}
\eeqa
for any $r>0$, $A\in\hmfa$ and  $\vx\in\real^s$. Moreover, $\om_0,\tau_1,\tau_2,\tau_3$ form a linearly independent 
family. In addition there holds:
\beqa
& &\hemfa\subset\ker\tau_1\cap\ker\tau_3\te{ and }\tau_2|_{\hemfa}\neq 0,\label{even-in-kernels1}\\
& &\hdmfa\subset\ker\tau_1\cap\ker\tau_2\cap\ker\tau_3. \label{derivatives-in-kernels1}
\eeqa
\eep
\proof Let $\ov{\al}_k$, $k\in\{1,2,3\}$ be the multiindices introduced in (\ref{Sett}) above.
By definition~(\ref{htau}), there holds
\beq
\htau_{\ov\al_k}^{(r)}=\fr{(i\sqrt{2})^{|k|} }{k!}\bsi^{(r)}_{\ov{\al}_k}, \label{htauk-def}
\eeq
where $\bsi^{(r)}_{\ov{\al}_k}$ are the functionals introduced in Lemma~\ref{Bfunctionals}, corresponding to the 
families of vectors $\{b_{\ka_j,r}^\pm\}_{j\in\nat}$. Thus for any $f\in\Lringz$ and  $r\geq r_0$ we obtain
\beq
\bsi^{(r)}_{\ov{\al}_k}(W(f))=e^{-\h\|f\|^2_2}\lan b_{0,r}^+| f^+\ran^k, \label{tau-of-one-is-zero}
\eeq
where we made use of our convention that $\ka_1=0$.
We recall that $\tf^{\pm}=\om^{\mp\h}\btF^\pm$, where $\F^\pm\in D(\mco_{r_0})_{\real}$. Moreover, by definition~(\ref{b}),
\beq
\bti_{0,r}^\pm(\vep)=\fr{1}{(2\pi)^{\fr{s}{2}}}\om(\vep)^{\pm\h}\wt{\chir}(\vep),
\eeq
where $\chir=1$ on $\mco_r$. We note that for $r\geq r_0$ and any $\vy\in\real^s$ s.t. $\mco_{r_0}+\vy\subset\mco_r$
there holds
\beqa
\lan b_{0,r}^+| U(\vy)f^+\ran=\fr{1}{(2\pi)^{\fr{s}{2}}} \lan \chir|U(\vy) F^{+}\ran
&=&\fr{1}{(2\pi)^{\fr{s}{2}}}\int\, d^sx\,\chir(\vx) \F^{+}(\vx-\vy)\non\\
&=&\fr{1}{(2\pi)^{\fr{s}{2}}}\int\, d^sx\,\F^{+}(\vx), \label{htauk-int}
\eeqa
i.e the functions $r\to \lan b_{0,r}^+ | U(\vy) f^+ \ran$ are constant for $r\geq r_0$ and independent of
$\vy$ within the above restrictions. Thus for any $r,r^\prime\geq r_0$, 
 $A\in\mfar(\mco(r_0))$ (i.e. $A$ is a finite liner combination of  Weyl operators)
and $\vy$ as specified above, we obtain
\beqa
& &\htau_{\ov{\al}_k}^{(r)}(A)-\htau_{\ov\al}^{(r^{\prime})}(A)=0,\non\\
& &\htau_{\ov{\al}_k}^{(r)}(A)-\htau_{\ov\al}^{(r)}(A(\vy))=0.
\eeqa
Since the functionals above are normal, the identities extend to any $A$ from $\mfa(\mco(r_0))$.
In view  of the above relation and the fact that any $A\in\hmfa$ belongs to $\mfa(\mco(r_0))$
for sufficiently large $r_0$, the following formulas
\beq
\tau_{k}(A)=\lim_{r\to\infty}\htau_{\ov\al_k}^{(r)}(A),\quad\quad A\in\hmfa
\eeq
define  linear functionals on $\hmfa$ which satisfy conditions (\ref{tau-consisitency}) and (\ref{tau-invariance}).
By this former condition, relation~(\ref{htauk-def}) and formula (\ref{htauk-int}) we obtain
\beq
\tau_{k}(W(f))=\htau_{\ov\al_k}^{(r_0)}(W(f))=\fr{(i\sqrt{2})^{|k|} }{k!} 
e^{-\h\|f\|^2_2}
\bigg(\fr{1}{(2\pi)^{\fr{s}{2}}}\int\, d^sx\,\F^{+}(\vx)\bigg)^k, \label{kernels-of-functionals}
\eeq
for $f$, $F^+$  defined as above. The last integral vanishes if $\F^{+}(\vx)=\sum_{j=1}^s\pa_{x_j}\F^{+}_j(\vx)$ for some
$\F^+_j\in D(\mco_{r_0})_{\real}$. From this fact and the strong continuity of $\htau_{\ov\al_k}^{(r_0)}$ 
there follows statement~(\ref{derivatives-in-kernels1}). Next, we note that by formula (\ref{kernels-of-functionals})
\beq
\tau_{k}(W(f)+W(-f))=\htau_{\ov\al_k}^{(r_0)}(W(f)+W(-f))=(1+(-1)^k)\,\, \htau_{\ov\al_k}^{(r_0)}(W(f)). \label{even-verification}
\eeq
Thus for $k\in\{1,3\}$ the functionals $\htau_{\ov\al_k}^{(r_0)}$ are zero on $\emfa(\mco(r_0))$, since
they vanish on a strongly dense subspace of this algebra.
This implies that $\hemfa$ belongs to $\ker\tau_1\cap\ker\tau_3$. Choosing $f$ so that $\int d^sx\, F^+(\vx)\neq 0$,
we obtain from formulas (\ref{kernels-of-functionals}), (\ref{even-verification})  that $\tau_2((W(f)+W(-f))\neq 0$,
what concludes the proof of statement~(\ref{even-in-kernels1}). 

We still have to show that the functionals $\om_0,\tau_1,\tau_2,\tau_3$ are linearly independent. Suppose that
\beq
c_0\,\om_0+c_1\,\tau_1+c_2\,\tau_2+c_3\,\tau_3=0
\eeq
for some complex numbers $c_0,\ldots, c_3$. Evaluating this expression on the unity operator
we obtain from relation~(\ref{tau-of-one-is-zero}) that $c_0=0$. Since there exist $A\in\hemfa$ s.t.
$\tau_2(A)\neq 0$, we get $c_2=0$. It remains to find $B\in\hmfa$ s.t. $\tau_1(B)=0$ and $\tau_3(B)\neq 0$.
Given $f$, $F^+$ as introduced above, with the additional condition that $\int d^sx\, F^+(\vx)\neq 0$, we pick
a function $h\in S(\real)$ s.t.
\beqa
& &\int e^{-\h u^2\|f\|_2^2}uh(u)\,du=0, \label{h-property1}\\
& &\int e^{-\h u^2\|f\|_2^2}u^3h(u)\,du=1.\label{h-property2}
\eeqa
(Existence of such function can be established with the help of the Gram-Schmidt procedure
as in the proof of Lemma~\ref{Smn-B*B}). The weak integral
\beq
B:=\int W(uf)h(u)du
\eeq
defines an element of $\mfa(\mco(r_0))$ by the von Neumann bicommutant theorem. With the help
of relations~(\ref{htauk-def}), (\ref{tau-of-one-is-zero}) and (\ref{htauk-int}) as well as properties
(\ref{h-property1}), (\ref{h-property2}) of the function $h$ we obtain 
that $\tau_1(B)=0$ and $\tau_3(B)\neq 0$. Since $\tau_1$ is non-zero, this concludes the proof
of linear independence of the functionals. \qed\\
The functionals $\tau_k$, $k\in\{1,2,3\}$, constructed in the above proposition, determine
the map $R^{(2)}$ introduced in the statement of Theorem~\ref{HA-in-models}. For any $A\in\mfa(\mco(r))$ we obtain
\beq
R^{(2)}(A)= A-\om_0(A)I-\tau_1(A)\phip-\tau_2(A)\wil\phip^2\wir-\tau_3(A)\wil\phip^3\wir,
\eeq
in the sense of quadratic forms on the domain $\BE\times\BE$ of vectors of bounded energy. In view of the
property~(\ref{tau-consisitency}), formula~(\ref{theh-concrete}) and decomposition~(\ref{whXi-decomposition})
there holds
\beq
P_ER^{(2)}(A)P_E=\thec(A)+\thet(A),\quad\quad A\in\mfa(\mco(r)). \label{R2-relation}
\eeq
Thus to conclude the proof of Theorem~\ref{HA-in-models} we have to verify that the ranges of the maps $\thec$, $\thet$ 
are square-integrable. This  is the subject of  the next section.

\section{Proof of Theorem C.1.1 (II): Square-Integrability of $R^{(2)}$ }\label{Part2} 
In this section we complete the proof of Theorem~\ref{HA-in-models}. After introducing the necessary
technical background in Subsection~\ref{KeyLemmas}, we prove the square-integrability of the ranges of
the maps $\thec$ and $\thet$ in Subsections~\ref{square-integrability-1} and \ref{square-integrability-2}, respectively. 

\subsection{Key Lemma}\label{KeyLemmas}

The main goal of this subsection is to prove Lemma~\ref{master-harmonic} below, which is inspired
by Lemma~2.2 of \cite{Bu90}. To state and prove this result it is convenient to proceed to the full
(non-symmetrized) Fock space $\Full$ on which there act the (prototype) creation operators $b^*(\Psi)$, 
$\Psi\in L^2(\real^s, d^sp)$, given by
\beq
b^*(\Psi)\Phi=\Psi\ot\Phi, \qquad \Phi\in\Full
\eeq
and their adjoints $b(\Psi)$. Upon restriction to the symmetric Fock space $\hil$, the formula $a(\Psi)=\sqrt{\widehat{N}}b(\Psi)$, 
where $\widehat{N}$ is the number operator, gives the standard annihilation operator introduced in Subsection~\ref{scalar-free-field-theory}.
With these definitions at hand we proceed to the main technical result of this appendix.
\bel\label{master-harmonic} Let  $E\geq 0$ and $\hh$ be a Borel function on $\real^s$ which is 
bounded on  $\{\,\vep\in\real^s\,|\, \om(\vep)\leq E\}$.
We denote the operator of multiplication by $\hh$ on $L^2(\real^s,d^sp)$ by the same symbol.
Let $\{g_{1,i}\}_{1}^{\infty}$, $\{g_{2,i}\}_{1}^{\infty}$ be two families
of functions from $L^2(\real^s, d^sp)$ which belong to the domain of $\hh\omp$ and let $\mu$, $\nu$ be
multiindices. Then
\beqa
\sup_{\fun\in\traceEBP}\int_{K}d^sx\, \fun(a^*(\hh\omp g_{1,\vx})^\mu a(\hh\omp g_{1,\vx})^\mu)
\fun(a^*(\hh\omp g_{2,\vx})^\nu a(\hh\omp g_{2,\vx})^\nu)\ph{444444}& &\non\\
\leq (\sup_{\om(\vep)\leq E}|\hh(\vep)|^{2} E)^{|\mu|+|\nu|}
\int_{\De K}d^sy |(\vac|b(g_1)^\mu b^*(g_{1,\vy})^\mu\vac)|\ph{,}& &\non\\
|(\vac|b(g_2)^\nu b^*(g_{2,\vy})^\nu\vac)|,& & \label{master-estimate}
\eeqa
where $K$ is any compact subset of $\real^s$, $\De K=\{\vx-\vy \ | \ \vx,\vy\in K \}$,
$g_{k,i,\vx}=U_1(\vx)g_{k,i}$, $k\in\{1,2\}$, $i\in\nat$.
\eel
\proof Let 
$Q=\int_K d^sx\, b^*(g_{1,\vx})^\mu b(g_{1,\vx})^\mu\ot
b^*(g_{2,\vx})^\nu b(g_{2,\vx})^\nu$ act on $\Full\ot\Full$ and $\Psi_1$, $\Psi_2\in\Full$ be 
$|\mu|$-, resp. $|\nu|$-, particle vectors. For $|\mu|=0$ (resp. $|\nu|=0$) we choose $\Psi_1$ (resp. $\Psi_2$)
to be a multiple of $\vac$. Then $\fun(\,\cdot\,)=(\Psi_1|\, \cdot\, \Psi_1)(\Psi_2|\, \cdot \,\Psi_2)$ is a 
positive functional on $B(\Full)\ot B(\Full)$ and we obtain
\beqa
\fun(Q)^2
\leq\fun(I)\fun(QQ)=\fun(I)\int_{K}d^sx\int_{K}d^sy\, (\Psi_1|b^*(g_{1,\vx})^\mu b(g_{1,\vx})^\mu
b^*(g_{1,\vy})^\mu b(g_{1,\vy})^\mu\Psi_1)& &\non\\
(\Psi_2|b^*(g_{2,\vx})^\nu b(g_{2,\vx})^\nu b^*(g_{2,\vy})^\nu b(g_{2,\vy})^\nu\Psi_2)& &\non\\
\leq\fun(I)\int_{K}d^sx\int_{K}d^sy\,|(\Psi_1|b^*(g_{1,\vx})^\nu\vac)(\vac| b(g_{1,\vx})^\nu b^*(g_{1,\vy})^\nu\vac) (\vac|b(g_{1,\vy})^\nu\Psi_1)|& &\non\\
|(\Psi_2|b^*(g_{2,\vx})^\nu\vac)(\vac|b(g_{2,\vx})^\nu b^*(g_{2,\vy})^\nu\vac)(\vac|b(g_{2,\vy})^\nu\Psi_2)|& &\non\\
\leq\fun(I)\fun(Q)
\int_{\De K}d^sy\, |(\vac|b(g_1)^{\mu}b^*(g_{1,\vy})^{\mu}\vac)|\,|(\vac|b(g_2)^{\nu}b^*(g_{2,\vy})^{\nu}\vac)|.& &\quad\,\,
\eeqa
Where in the last step we made use of the inequality $|f(\vx)f(\vy)|\leq \fr{1}{2}(|f(\vx)|^2+|f(\vy)|^2)$.
Consequently
\beqa
& &\int_K d^sx\, (\Psi_1|b^*(g_{1,\vx})^\mu b(g_{1,\vx})^\mu|\Psi_1)
(\Psi_2|b^*(g_{2,\vx})^\nu b(g_{2,\vx})^\nu\Psi_2)\non\\
& &\ph{444}\leq \|\Psi_1\|^2\|\Psi_2\|^2
\int_{\De K} d^sy\,|(\vac|b(g_1)^{\mu}b^*(g_{1,\vy})^{\mu}\vac)|\, |(\vac|b(g_2)^{\nu}b^*(g_{2,\vy})^{\nu}\vac)|.
\label{ha2}
\eeqa
We pick $n_1,n_2\in\nat$ s.t. $n_1\geq|\mu|$, $n_2\geq|\nu|$.
Given $n_1$- and $n_2$-particle vectors $\Psi_{n_1}, \Psi_{n_2}\in (P_E\hil)_1$, 
whose wavefunctions are Schwartz-class,
we define $|\mu|$-, resp. $|\nu|$-particle vectors, where the remaining arguments are treated 
as parameters
\beqa
& &\Psi_1(\vep_1,\ldots,\vep_{|\mu|})_{\vep_{|\mu|+1},\ldots,\vep_{n_1}}
:=\bar{\hh}(\vep_1)\om(\vep_1)^{\h}\ldots \bar{\hh}(\vep_{|\mu|})\om(\vep_{|\mu|})^{\h}
\Psi_{n_1}(\vep_1,\ldots,\vep_{n_1}),\non\\
& &\Psi_2(\veq_1,\ldots,\veq_{|\nu|})_{\veq_{|\nu|+1},\ldots,\veq_{n_2}}
:=\bar{\hh}(\veq_1)\om(\veq_1)^{\h} \ldots \bar{\hh}(\veq_{|\nu|})\om(\veq_{|\nu|})^{\h}
\Psi_{n_2}(\veq_1,\ldots,\veq_{n_2}).\quad\quad\quad
\eeqa
With these definitions we compute
\beqa
\int_K d^sx\,(\Psi_{n_1}|a^*(\hh\omp g_{1,\vx})^\mu a(\hh\omp g_{1,\vx})^\mu\Psi_{n_1})
(\Psi_{n_2}|a^*(\hh\omp g_{2,\vx})^\nu a(\hh\omp g_{2,\vx})^\nu\Psi_{n_2})\ph{}& &\non\\
=\fr{n_1!}{(n_1-|\mu|)!}\fr{n_2!}{(n_2-|\nu|)!}\int d^sp_{|\mu|+1}\ldots d^sp_{n_1}
\int d^sq_{|\nu|+1}\ldots d^sq_{n_2}& &\non\\
\int_K d^sx\, (\Psi_{1,\vep_{|\mu|+1},\ldots,\vep_{n_1}} |b^*(g_{1,\vx})^\mu b(g_{1,\vx})^\mu\Psi_{1,\vep_{|\mu|+1},\ldots,\vep_{n_1}} )& &\non\\
(\Psi_{2,\veq_{|\nu|+1},\ldots,\veq_{n_2}}|b^*(g_{2,\vx})^\nu b(g_{2,\vx})^\nu\Psi_{2,\veq_{|\nu|+1},\ldots,\veq_{n_2}} )& &\non\\
\leq(\sup_{|\vep|\leq E}|\hh(\vep)|^{2} E)^{|\mu|+|\nu|}
\int_{\De K}d^sy\, |(\vac|b(g_{1} )^\mu b^*(g_{1,\vy})^\mu\vac)|\non\\
|(\vac|b(g_{2})^\nu b^*(g_{2,\vy} )^\nu\vac)|,& &\quad\quad
\label{long}
\eeqa
where in the last step we applied formula (\ref{ha2}) and made use of the following
computation to control the $n_1$ and $n_2$ dependence
\beqa
& & \fr{n_1!}{(n_1-|\mu|)!}
\int d^sp_1\ldots d^sp_{n_1} \om(\vep_1)\ldots \om(\vep_{|\mu|}) |\Psi_{n_1}(\vep_1,\ldots,\vep_{n_1})|^2\non\\
& &\phantom{4444}=|\mu|! \int d^sp_1\ldots d^sp_{n_1}\sum_{i_1<\ldots<i_{|\mu|}} \om(\vep_{i_1})
\ldots \om(\vep_{i_{|\mu|}}) |\Psi_{n_1}(\vep_1,\ldots,\vep_{n_1})|^2\non\\
& &\phantom{4444}\leq \int d^sp_1\ldots d^sp_{n_1}\, \big(\om(\vep_1)+\cdots+\om(\vep_{n_1})\big)^{|\mu|}
|\Psi_{n_1}(\vep_1,\ldots,\vep_{n_1})|^2\leq E^{|\mu|}.\quad\quad
\eeqa 
Here in the first step we used the symmetry of the wavefunction and in the second
step we applied the multinomial formula. By an approximation argument, the bound (\ref{long})
holds also without the restriction that $\Psi_{n_1}$, $\Psi_{n_2}$ have Schwartz-class wavefunctions.

To conclude the argument we need several simple, geometrical observations: 
We note that any $\fun\in\traceEBP$ has the form $\fun(\,\cdot\,)=\sum_{k=1}^\infty p_k(\Psi_k|\,\cdot\, \Psi_k)$, where
$p_k\geq 0$, $\sum_{k=1}^{\infty} p_k\leq 1$, and $\Psi_k\in (P_E\hil)_1$. Let $C_1$, $C_2$ be a pair
of  bounded, positive operators. (The operators $P_Ea^*(\hh\omp g_{1})^\mu a(\hh\omp g_{1})^\mu P_E$,
$P_E a^*(\hh\omp g_{2})^\nu a(\hh\omp g_{2})^\nu P_E$, appearing on the l.h.s. of relation~(\ref{long}) are bounded due to
the energy bounds (\ref{energy-bounds}) and the fact that $\|Q_E\hh\|<\infty$).
There  holds the simple estimate
\beqa
& &\sup_{\fun\in\traceEBP}\int_K d^sx\,\fun(C_1(\vx))\,\fun(C_2(\vx))\non\\
& &\ph{44444444444444}\leq\sup_{\Psi_1,\Psi_2\in(P_E\hil)_1}\int_K d^sx\,(\Psi_1|C_1(\vx)\Psi_1) (\Psi_2|C_2(\vx)\Psi_2).
\eeqa
Finally, we can decompose any $\Psi\in (P_E\hil)_1$ as $\Psi=c_0\vac+\sum_{n=1}^{\infty} c_n\Psi_n$, where $\Psi_n$ are
normalized $n$-particle wavefunctions 
and $|c_0|^2+\sum_{n=1}^{\infty} |c_n|^2\leq 1$. Consequently, for $C_1$ and $C_2$
as above, conserving particle number, we obtain
\beqa
& &\int_K d^sx\,(\Psi_1|C_1(\vx)\Psi_1)\, (\Psi_2|C_2(\vx)\Psi_2)\non\\
& &\ph{44444444444444}\leq \sup_{n_1,n_2}\int_K d^sx(\Psi_{1,n_1}|C_1(\vx)\Psi_{1,n_1}) (\Psi_{2,n_2}|C_2(\vx)\Psi_{2,n_2}).
\eeqa
Now the bound in the statement of the lemma follows from estimate (\ref{long}). \qed\\
In the next section we will use the above lemma to prove the square-integrability of individual terms
appearing in the expansions of the maps $\thec$ and $\thet$. Then the square-integrability of the
ranges of these maps follows from the special case of the Minkowski inequality \cite{Hal} which we state below.
\bel For any family of $f_n\in L^2(\real^s, d^sp)$, $n\in\nat$, whose norms $\|f_n\|_2$ form a 
summable sequence, there holds the bound
\beq
\big(\int d^sp|\sum_{n=1}^\infty f_n(\vep)|^2\big)^\h\leq \sum_{n=1}^{\infty}\big(\int d^sp |f_n(\vep)|^2\big)^\h.
\label{Minkowski}
\eeq
\eel
\nin After this technical preparation  we can proceed with the proof of Theorem~\ref{HA-in-models}.
\subsection{Square-Integrability of $\thec$}  \label{square-integrability-1} 
In this subsection we apply Lemma \ref{master-harmonic} to prove the square-integrability of the range of the map $\thec$.
In order to control the expressions appearing on the r.h.s. of estimate~(\ref{master-estimate}), we need
the following result:
\bel\label{Sobolev-application} Let $\la>\fr{s}{2}$ and suppose that $\tilde{h}\in L^2(\real^s, d^sp)_{\la}$.  Then the function 
$\real^s\ni\vx\to \lan \hti|U(\vx)\hti\ran$ is absolutely integrable and there holds
\beq
\int d^sx\, |\lan \hti|U(\vx)\hti\ran|\leq c_{\la}\,\|\hti\|_{2,\la}^2, 
\eeq
where the constant $c_{\la}$ depends only on $\la$ and $s$. (Here $\lan\,\cdot\, |\, \cdot\,\ran$ denotes the
scalar product in $L^2(\real^s, d^sp)$).
\eel
\proof We pick $\hti\in S(\real^s)$, a compact subset $K\subset\real^s$ and estimate 
\beqa
\int_K d^sx\, \big|\lan \hti|U(\vx)\hti\ran\big|&=&\int_K d^sx\,\big|\int d^sp\, e^{-i\vep\vx} \bar{\tilde{h}}(\vep) \tilde{h}(\vep)\big|\non\\
&\leq& \int d^sx\,\big|\int d^sy\, \bar{h}(\vy) h(\vy+\vx)\big|\leq \big(\int d^sy\, |h(\vy)|\big)^2\non\\
&\leq& \int d^sx\,\fr{1}{(1+|\vx|^2)^\la}\int d^sy\, (1+|\vy|^2)^\la|h(\vy)|^2\non\\
&=& c_{\la}\|\tilde{h}\|_{2,\la}^2,
\eeqa
where in the fourth step we made use of the Cauchy-Schwarz inequality. Since $S(\real^s)$ is dense
in $L^2(\real^s, d^sp)_{\la}$, this estimate extends to any $\hti\in L^2(\real^s, d^sp)_{\la}$. Finally,
as the constant $c_{\la}$ is independent of $K$, we can take the limit $K\nearrow\real^s$.
\qed\\
With our choice of the sets $\Set$ and $\Sett$ (see definition (\ref{Sett})) the map $\thec$, given by
formula~(\ref{thec-expansion1}), has the following form
\beq
\thec(A)=\!\!\!\!\!\!\!\!\!\!\sum_{\su{\mup,\nup \\ \exists_{j>1} \te{ s.t. } \mup(j)+\nup(j)\neq 0 \\ 
|\mup|+|\nup|\leq 3}}
\!\!\!\!\!\!\!\!\!\!\ctau_{\mup,\nup}^{(r)}(A)\,\cS_{\mup,\nup}, \quad\quad A\in\mfa(\mco(r)), \label{thec-expansion2}
\eeq
where we made use of the fact that $\mum=\num=0$, since $K^-=0$. We recall from formula~(\ref{cS}) that
\beq
\cS_{\mup,\nup}=P_E a^*(h^+_{\ka,E})^{\mup}a(h^+_{\ka,E})^{\nup}P_E \label{cS1}
\eeq
and that the numbering $\{\ka_j\}_1^\infty$ of the $s$-indices is chosen so that $\ka_1=0$.
The following lemma establishes the square-integrability of these operators under the restrictions on
the multiindices $\mup,\nup$ given in the sum~(\ref{thec-expansion2}). We will exploit the useful
fact that for any $C\in B(\hil)$ which satisfies $\|C\|_{E,2}<\infty$ there holds
\beq
\|C\|_{E,2}\leq 4\sup_{\fun\in\traceEBP}\big(\int d^sx\,|\fun(C(\vx))|^2\big)^\h. \label{decomp-appl}
\eeq
This bound follows immediately from  decomposition (\ref{decomp}).
\bel Suppose that $\mup(j)+\nup(j)\neq 0$ for some $j>1$. Then in massless scalar free field theory for $s\geq 3$
there holds the bound
\beq
\|\cS_{\mup,\nup}\|_{E,2}\leq (c_E^{|\ka|+1})^{\mup+\nup} \label{cS-L2-bound}
\eeq
for some constant $c_E$ independent of $\ka$.
\eel
\proof We  assume, without restriction, that there exist $j>1$ s.t. $\nup(j)\neq 0$.
Thus we can choose multiindices $\nup_a$, $\nup_b$ in such a way that $\nup=\nup_a+\nup_b$ and
$|\nup_a|=\nup_a(j)=1$. Making use of formula~(\ref{cS1}) we obtain
for any compact subset $K\subset\real^s$
\beqa
& &\sup_{\fun\in\traceEBP}\int_K d^sx\,|\fun(\cS_{\mup,\nup}(\vx))|^2\non\\
& &\leq \sup_{\fun\in\traceEBP}\int_K d^sx\,\fun(a^*(h^+_{\ka,E})^{\mup}a(h^+_{\ka,E})^{\mup})\,
\fun(a^*(h^+_{\ka,E})^{\nup}a(h^+_{\ka,E})^{\nup}) \non\\
& &\leq E^{|\mup|+|\nup_b|}\|\omm h^+_{\ka,E}\|_2^{2(|\mup|+|\nup_b|)}
\sup_{\fun\in\traceEBP}\int_K d^sx\,\fun(a^*(h^+_{\ka_j,E})a(h^+_{\ka_j,E})),\quad \label{CS-plus}
\eeqa
where in the first step we applied the Cauchy-Schwarz inequality and in the second step
we exploited the energy bounds~(\ref{energy-bounds}), Lemma~\ref{h-lemma}, the fact that 
$a(h_{\ka,E})^{\nup}P_E=a(h_{\ka,E})^{\nup_b}P_Ea(h_{\ka,E})^{\nup_a}P_E$ and the 
properties of $\nup_a$. 

We will apply Lemma~\ref{master-harmonic} to the integral on the r.h.s. of relation~(\ref{CS-plus}):
Since $\ka_j\neq 0$, we can write $\ka_j=\haka_j+\chka_j$ in such a way that $|\haka_j|=1$.
Next, we set $\hh(\vep)=\fr{(ip)^{\haka_j}}{\om(\vep)}$ for $\vep\neq 0$ and $\hh(0)=0$. 
There holds $\|\hh\|_{\infty}\leq 1$ and $\hti^+_{\ka_j,E}=\hh\omp(\omp \hti^+_{\chka_j,E})$. 
(We recall that  $\hti_{\ka,E}^+(\vep)=(-1)^{|\ka|}\om(\vep)^{-\h}(ip)^{\ka}\chi_E(\vep)$ by definition~(\ref{h})).
Moreover, by Lemma~\ref{h-lemma} there holds for any $\la\geq 0$
\beqa
& &\|\omp \hti^+_{\ka,E}\|_{2,\la}\leq (c_{\la,-\h,E})^{|\ka|+1}, \label{h-lemma-repeated1}\\ 
& &\|\omm \hti^+_{\ka,E}\|_2\leq (c_{0,\h,E})^{|\ka|+1}, \label{h-lemma-repeated2}
\eeqa
where the constants $c_{\la,-\h,E}$, $c_{0,\h,E}$ are  independent of $\ka$.
We pick some $\la>\fr{s}{2}$ and obtain from Lemmas~\ref{master-harmonic} and \ref{Sobolev-application}
\beqa
\sup_{\fun\in\traceEBP}\int_K d^sx\,\fun(a^*(h^+_{\ka_j,E})a(h^+_{\ka_j,E}))
&\leq& E\int_{\Delta K}d^sx\, |\lan \omp \hti^+_{\chka_j,E}|U(\vx)\omp \hti^+_{\chka_j,E}\ran|\non\\
&\leq& Ec_{\la}\|\omp \hti^+_{\chka_j,E}\|^2_{2,\la}\non\\
&\leq& (c_E)^{|\ka_j|+1}= \big((c_E)^{|\ka|+1}\big)^{\nup_a}.
\eeqa
Here  in the third step we
used the bound~(\ref{h-lemma-repeated1}), absorbed the constants involved into one constant
$c_E$, independent of $\chka_j$, and made use of the fact that $|\ka_j|=|\chka_j|+1$. Substituting
the above bound to relation~(\ref{CS-plus}) and making use of estimates~(\ref{h-lemma-repeated2})
and (\ref{decomp-appl}), we obtain the estimate in the statement of the
lemma (after readjusting the constant $c_E$). \qed\\
After this preparation we can prove the square-integrability of the range of the map $\thec$.
\bep\label{bh-square-integrability} In massless  scalar free field theory for $s\geq 3$ the map $\thec$, 
given by relation (\ref{thec-expansion2}), satisfies the bound
\beq
\sup_{A\in\mfa(\mco(r))_1}\|\thec(A)\|_{E,2}<\infty.
\eeq
\eep
\proof First, we recall that due to the bound~(\ref{ttau-bound}) and Lemma~(\ref{b-lemma}) there
holds
\beqa
\|\ctau_{\mup,\nup}^{(r)}\|&\leq& 2^{|\mup|+|\nup|}\sqrt{(|\mup|+|\nup|)!}\,\|b_{\ka,r}\|_{2}^{\mup+\nup}\non\\
&\leq& \sqrt{(|\mup|+|\nup|)!} \bigg(\fr{(c_{r})^{|\ka|+1}}{\ka!}\bigg)^{\mup+\nup} \label{ttau-bound-repeated}
\eeqa
for some constant $c_{r}$, independent of $\ka$.
Next, making use of the Minkowski inequality~(\ref{Minkowski}), we obtain for any $A\in\mfa(\mco(r))_1$
\beqa
\|\thec(A)\|_{E,2}&\leq&\!\!\!\!\!\!\!\!\!\!\sum_{\su{\mup,\nup \\ \exists_{j>1} \te{ s.t. } \mup(j)+\nup(j)\neq 0 \\ 
|\mup|+|\nup|\leq 3}}
\!\!\!\!\!\!\!\!\!\!\|\ctau_{\mup,\nup}^{(r)}\|\, \|\cS_{\mup,\nup}\|_{E,2}\non\\
&\leq& \sum_{\su{\mup,\nup \\ |\mup|+|\nup|\leq 3}} \sqrt{(|\mup|+|\nup|)!}\bigg(\fr{(c_{r}c_E)^{|\ka|+1}}{\ka!}\bigg)^{\mup+\nup}\non\\
&\leq& \sqrt{3!}\bigg(\sum_{\mup,|\mup|\leq 3}\bigg(\fr{(c_{r}c_E)^{|\ka|+1}}{\ka!}\bigg)^{\mup}\bigg)^2\non\\
&\leq& \sqrt{3!}\bigg(\sum_{k=0}^3 \bigg(\sum_{\ka\in\nat_0^s} \fr{(c_{r}c_E)^{|\ka|+1}}{\ka!}\bigg)^{k}\bigg)^2,
\eeqa
where in the second step we made use of the bounds (\ref{ttau-bound-repeated}), (\ref{cS-L2-bound})
and in the last step we used the fact that the multinomial coefficients are larger or equal to
one and we applied the multinomial formula~(\ref{multiform}). The last expression is clearly finite. \qed\\ 
\subsection{Square-Integrability of $\thet$}\label{square-integrability-2}

Now we proceed to the proof of the square-integrability of the range of the map $\thet$, defined by expansion~(\ref{lastexpansion}). Again, we will first  apply Lemma~\ref{master-harmonic} to the individual terms of this expansion and then use the Minkowski inequality (\ref{Minkowski}) to conclude the proof.
In order to control the terms appearing on the r.h.s. of estimate~(\ref{master-estimate}), we need two auxiliary 
lemmas proven below.

In Subsection \ref{Taylor} we introduced, for any $\tr>0$, a function $\chi(\mco_{\tr})\in\czn$ s.t. 
$\chi(\mco_{\tr})(\vx)=1$\label{symbol-chir-next} for $\vx\in\mco_{\tr}$. Here we demand in addition that $\chi(\mco_{\tr})(\vx)=0$
for $\vx\notin\mco_{\tr+\teps}$, for some fixed $\teps>0$, independent of $\tr$.
We denote the operator of multiplication by $\chi(\mco_{\tr})$ in configuration space by the same 
symbol.
\bel\label{F2} Suppose that $F\in S^\prime(\real^s)$ coincides with a bounded, measurable function in the 
region $\{\, \vy\in\real^s \,|\, |\vy|\geq \tr\,\}$ and its Fourier transform
$\tF$ is a positive, measurable function s.t. $\tF^{1/2}\in L^2(\real^s,d^sp)+L^{\infty}(\real^s,d^sp)$.
Then $\tF^{1/2}\chitr$ is a bounded operator and there holds
\beq
\|\chitr \tF\chitrz\|\leq c_{\tr,\teps}\sup_{|\vec{z}|\leq 2\tr+3\teps} |F(\vec{z}+\vx)| \ \textrm{ for }  
|\vx|\geq 3(\tr+\teps),\label{Fbound2}
\eeq
where $\chitrz(\vy)=\chitr(\vy-\vx)$, the constant $c_{\tr,\teps}$ is independent of $\vx$ and we denote the
operator of multiplication by $\tF$ in momentum space by the same symbol.
\eel
\proof In order to prove the first statement we make a decomposition $\tF^{1/2}=\tF^{1/2}_2+\tF^{1/2}_{\infty}$,
where $\tF^{1/2}_2\in L^2(\real^s,d^sp)$, $\tF^{1/2}_{\infty}\in L^{\infty}(\real^s,d^sp)$. Since 
$\tF^{1/2}_{\infty}$ is a bounded operator, it suffices to consider $\tF^{1/2}_2\chitr$. We pick
$f_1,f_2\in S(\real^s)$ and estimate
\beqa
|\lan f_1|\tF^{1/2}_2  \chitr f_2\ran|=(2\pi)^{-\fr{s}{2}}\big|\int d^sp\,d^sq \ \bar{\tf}_1(\vep) \tF^{1/2}_2(\vep)
\wt{\chi(\mco_{\tr})}(\vep-\veq)\tf_2(\veq)\big|\ph{4}& &\non\\
\leq c\|\bar{\tf}_1\tF^{1/2}_2\|_1\|\chi(\mco_{\tr}) \|_2\|f_2\|_2\leq c\|f_1\|_2
\|\tF^{1/2}_2\|_2\|\chi(\mco_{\tr}) \|_2\|f_2\|_2,& &
\eeqa
where in the second step we made use of the Young inequality\footnote 
{The Young inequality states that 
for any  positive functions $f\in L^{r_1}(\real^s,d^sp)$, $g\in L^{r_2}(\real^s,d^sp)$, $h\in L^{r_3}(\real^s,d^sp)$,
where 
$1\leq r_1,r_2,r_3\leq\infty$ s.t. $\fr{1}{r_1}+\fr{1}{r_2}+\fr{1}{r_3}=2$, there holds the
bound
\begin{eqnarray*}
\int d^sp\, d^sq \ f(\vep)g(\vep-\veq)h(\veq)\leq c_{r_1,r_2,r_3}\|f\|_{r_1}\|g\|_{r_2}\|h\|_{r_3}.
\end{eqnarray*}} \cite{RS2} and in the last estimate we applied H\"older's inequality.

Next, we  verify relation (\ref{Fbound2}). If $|\vx|\geq 3(\tr+\teps) $,
then $|\vy-\vx|\leq 2\tr+3\teps$ implies  $|\vy|\geq \tr$ and the  expression
\beq
\wFz(\vep):=(2\pi)^{-\fr{s}{2}}\int d^sy \, e^{i\vep\vy}F(\vy)\chi_{\vx}(\mco_{2(\tr+\teps)})(\vy) \label{Fx1}
\eeq
defines a bounded, continuous function.
The operator of multiplication by $\wFz$ in momentum space, denoted by the same symbol,
satisfies the identity
\beq
\chitr \wFz \chitrz =\chitr \tF\chitrz \label{Fequality1}.
\eeq
To justify this equality of two bounded operators,
 it suffices to compare their matrix elements between $f_1,f_2\in S(\real^s)$.
We introduce vectors $g_1=\chitr^*f_1$, $g_2=\chitrz f_2$ and compute
\beqa
\lan f_1 |\chitr \tF\chitrz |f_2 \ran=\int d^sp\, \bar{\tg}_1(\vep)\tF(\vep) \tg_2(\vep)=\tF(\bar{\tg}_1 \tg_2)
=F((2\pi)^{-\fr{s}{2}}\bar{g}_1^{\sswedge}*g_2)& &\non\\
=(\chi_{\vx}(\mco_{2\tr+2\teps})F)((2\pi)^{-\fr{s}{2}}\bar{g}_1^{\sswedge}*g_2)=
\lan f_1 |\chitr \wFz\chitrz |f_2 \ran,& &\qquad
\eeqa
where $g_2^{\sswedge}(\vx)=g_2(-\vx)$, in the fourth step we made use of the support properties of
$g_1$ and $g_2$ in configuration space and in the last step we reversed the first three steps
with $F$ replaced with $\chi_{\vx}(\mco_{2\tr+2\teps})F$. Finally, we obtain from  (\ref{Fx1})
\beqa
|\wFz(\vep)| 
\leq(2\pi)^{-\fr{s}{2}}\int d^sy\, |\chi(\mco_{2(\tr+\teps)})(\vy)| \sup_{|\vec{z}|\leq 2\tr+3\teps} |F(\vec{z}+\vx)|\ph{,}& &\non\\
=c_{\tr,\teps}\sup_{|\vec{z}|\leq 2\tr+3\teps} |F(\vec{z}+\vx)|. & &
\eeqa
From this bound and identity~(\ref{Fequality1}) there follows the estimate in the statement of the lemma. \qed\\
We recall that $h_r\in D(\mco_r)_{\real}$ has strictly positive Fourier transform and, by Lemma~\ref{TEpm-lemma}, 
the operators $\Thpm=\om^{-\ga}\tih^{1/2}\Lpm$ are bounded 
for any $\h\leq\ga<\fr{s-1}{2}$.
Thus we can set in Lemma~\ref{master-harmonic} $\hh=\tih^{-1}$, $g_{k,i}=\omm\tih\Lpm e_i$ 
for $k\in\{1,2\}$ and $i\in\nat$, where $\{e_i\}_1^\infty$ is the orthonormal basis of 
eigenvectors of the operator $T$,
given by (\ref{T-definition}). The resulting estimate, studied in Lemma~\ref{map2} below,
relies on the decay properties of the functions
\beq
\real^{s}\ni \vx\to\lan \omm\tih \Lpm e_i|U(\vx)\omm \tih \Lpm e_i\ran \label{function}
\eeq
which  appear on the r.h.s.  of  relation~(\ref{master-harmonic}). These properties are established
in the following lemma with the help of the relation
\beq
\Lpm=\om^{\mp\half}\chir\om^{\pm\half}\Lpm,  \label{chil1}
\eeq
which provides a link with  Lemma~\ref{F2}.
\bel \label{mbounds} Assume that $s\geq 3$ and let $e$ be a normalized eigenvector of the operator $T$,
 given by (\ref{T-definition}), corresponding to the eigenvalue $t$. Then there holds 
\begin{enumerate}
\item[(a)] $\lan \om^{-\half}\tih\Lm e | U(\vx)\om^{-\half}\tih\Lm e\ran=0$ for $|\vx|>4\rz$,
\item[(b)] $|\lan \om^{-\half}\tih\Lpm e | U(\vx)\om^{-\half}\tih\Lpm e\ran|
\leq \fr{c_{\rz}(m)e^{-\fr{m|\vx|}{2} } t^2}{(|\vx|+1)^{s-2} }$, 
\end{enumerate}
where the constant $c_{\rz}(m)$ is independent of $\vx$, $e$ and finite for any $m\geq 0$. (If $m>0$, the above relations hold for $s\geq 1$).
\eel
\proof
To prove part (a) we set again $\chi_{\vx}(\mco_{\rz})(\vy)=\chi(\mco_{\rz})(\vy-\vx)$ and note that
\beqa
& &\lan \om^{-\half}\tih\Lm e |U(\vx)\om^{-\half}\tih\Lm e\ran\non\\
& &\phantom{44444444444}=\lan\om^{-\half}\tih \Lm e |\chi(\mco_{2\rz})\chi_{\vx}(\mco_{2\rz}) U(\vx)
\om^{-\half}\tih \Lm e\ran=0,
\eeqa
for $|\vx|>4\rz$, since $h_r\in D(\mco_{\rz})$ and hence $\omm\tih\Lm e\in [\widetilde{D}(\mco_{2\rz})]$.
This latter statement follows from the fact that $\omm\tih\Lm e=(\omm\tih\Lm)\Lm e$ and $\omm\tih\Lm$
is a bounded operator due to Lemma~\ref{TEpm-lemma}.  In view of the uniform bound
\beq
|\lan \om^{-\half}\tih\Lpm e | U(\vx)\om^{-\half}\tih\Lpm e\ran|\leq 
\|\om^{\alr-\h}\tih^{1/2}\|^2_{\infty} \lan e |\Tbpm^2 e\ran\leq \|\om^{2\alr-1}\tih\|_{\infty} t^2,
\label{uniformbound}
\eeq
which involves the parameter $\alr\in [\fr{1}{2},\fr{s-1}{2}[$ from the
definition of the operator $T$, there also follows the ($-$) part of (b). To prove the (+) part we  estimate
\beqa
|\lan\om^{-\half}\tih\Lp e | U(\vx)\om^{-\half}\tih\Lp e\ran |
&=&|\lan\tih\om^\h \Lp e|\chi(\mco_{2\rz})\om^{-2}\chi_{\vx}(\mco_{2\rz})\tih\omp U(\vx)\Lp e\ran|\non\\
&\leq&t^2\| \om^{2\alr+1} \tih\|_{\infty}\,  \|\chi(\mco_{2\rz})\om^{-2}\chi_{\vx}(\mco_{2\rz})\|.
\eeqa
Now we are in position to apply  Lemma~\ref{F2}: We set $\tF(\vep)=(|\vep|^{2}+m^2)^{-1}$, $m\geq 0$ and obtain
\beq
\tF(\vep)^{1/2}=(|\vep|^2+m^2)^{-\h}\theta(-|\vep|+1)+(|\vep|^2+m^2)^{-\h}\theta(|\vep|-1)\in 
L^2(\real^s,d^sp)+L^{\infty}(\real^s,d^sp).
\eeq
For $s\geq 2$,  $F(\vx)=\fr{1}{|\vx|^{s-2}}(m|\vx|)^{\fr{s-2}{2}}K_{\fr{s-2}{2}}(m|\vx|)$, where $K_{\fr{s-2}{2}}$ is the modified
Bessel function of the second kind. It satisfies the following bound for $s\geq 3$ and $z\geq 0$ 
\beq
z^{\fr{s-2}{2}} K_{\fr{s-2}{2}}(z)\leq c e^{-\fr{z}{2}},
\eeq
where the constant $c$ is independent of $z$. (See Section 7.2 of \cite{GJ} for a proof).
Consequently, we obtain for $|\vx|\geq 6\rz+3\teps$
\beq
\|\chi(\mco_{2\rz})\om^{-2}\chi_{\vx}(\mco_{2\rz})\|
\leq\fr{c_{\rz}e^{-\fr{m}{2}(|\vx|-4r-3\teps) }} {(|\vx|-4\rz-3\teps)^{s-2}}.
\eeq
Making use of the uniform bound (\ref{uniformbound}),
we get the estimate from the statement of the lemma for a suitable constant $c_{\rz}(m)$ and $s\geq 3$.

It remains to consider the massive case in low dimensions. For $s=2$ there holds $F(\vx)=K_0(m|\vx|)$,
what implies $|F(\vx)|\leq c_{r,m}e^{-m|\vx|}$ for $|\vx|\geq 2r$. (See \cite{GJ} Proposition~7.2.1.~(c)).
For $s=1$ we have the explicit formula $F(\vx)=((2\pi)^{\h}m)^{-1}e^{-m|\vx|}$. In both cases we
easily verify, by an analogous argument as above, that the estimate in part~(b) of the lemma 
holds for $m>0$.  \qed\\
With our choice of the set $\Set$ (see Section~\ref{Part1}), the map $\thet$, given by 
expansion~(\ref{lastexpansion}), has the form
\beq
\thet(A)=\sum_{\su{\mub,\nub \\ |\mup|+|\nup|>3 \te{ or } |\mum|+|\num|>0}  }\tau_{\mub,\nub}(A)\,\Smn.
\label{lastestimate1}
\eeq
We recall from formula~(\ref{Smn-def}) and the bound (\ref{Sestimate2}) that 
\beq
\Smn=P_E a^*(\LL e)^{\mub}a(\LL e)^{\nub} P_E
\eeq
are bounded operators. In the following lemma we show that they are square-integrable under the restrictions
on $\mub,\nub$ appearing in the sum (\ref{lastestimate1}).
\bel\label{map2} Let $(\mub,\nub)$ be a pair of 2-multiindices $(\mub,\nub)$ s.t. $|\mup|+|\nup|>3$  or  $|\mum|+|\num|>0$.
Then, in massless scalar free field theory for $s\geq 3$ there holds the bound
\beq
\|\Smn\|_{E,2}\leq C_{r,E} E^{\fr{|\mub|+|\nub|}{2}} t^{\mub}t^{\nub},
\eeq
where $\{t_j\}_1^\infty$ are the eigenvalues of the operator $T$ given by (\ref{T-definition}) and
the constant $C_{r,E}$ is independent of $\mub,\nub$. 
\eel
\proof 
Any  pair of 2-multiindices $(\mub,\nub)$ satisfying the restrictions from the statement
of the lemma can be decomposed as follows: 
\beqa
\mub&=&\mub_a+\mub_b,\\
\nub&=&\nub_a+\nub_b, 
\eeqa
where  $|\mub_a|+|\nub_a|=|\mup_a|+|\nup_a|=4$ or $|\mub_a|+|\nub_a|=|\mum_a|+|\num_a|=1$.
For any compact subset $K\subset\real^s$ and $\fun\in\traceEBP$ we obtain
\beqa
\int_K d^sx\, |\fun(\Smn(\vx))|^2&\leq&
\int_K d^sx\, \fun_{\vx}(a^*(\LL e)^{\mub}a(\LL e)^{\mub})\, \fun_{\vx}(a^*(\LL e)^{\nub}a(\LL e)^{\nub})\non\\
& &\phantom{}=\int_K d^sx\, \fun_{\vx}(a^*(\LL e)^{\mub_a} a^*(\LL e)^{\mub_b}  a(\LL e)^{\mub_b} a(\LL e)^{\mub_a})\non\\
& &\phantom{4}\phantom{4\int d^sx}\cdot \,\fun_{\vx}(a^*(\LL e)^{\nub_a} a^*(\LL e)^{\nub_b}  a(\LL e)^{\nub_b} a(\LL e)^{\nub_a})\non\\
& &\phantom{44} \leq E^{|\mub_b|+|\nub_b|} t^{2\mub_b}t^{2\nub_b}
\int_K d^sx\, \fun_{\vx}(a^*(\LL e)^{\mub_a} a(\LL e)^{\mub_a})\non\\
& &\phantom{444444444444444444444444}\cdot \fun_{\vx}(a^*(\LL e)^{\nub_a} a(\LL e)^{\nub_a}),\non\\
\eeqa
where in the first step we made use of the Cauchy-Schwarz inequality and in the third step
we used  estimate~(\ref{Sestimate2}). Now we set in Lemma~\ref{master-harmonic}
$\hh=\tih^{-1}$, $g_{k,i}=\omm\tih\Lpm e_i$ for $k\in\{1,2\}$, $i\in\nat$, where the function $\tih>0$ entered into
the definition~(\ref{T-definition}) of the operator $T$. We obtain
\beqa
\sup_{\fun\in\traceEBP}\int_K d^sx\, |\fun(\Smn(\vx))|^2 \leq E^{|\mub|+|\nub|} t^{2\mub_b}t^{2\nub_b}(\enorm)^{|\mub_a|+|\nub_a|}
\ph{44444444}& & \non\\ 
\cdot\int_{\De K} d^sy\, |(\vac| b(\tih\om^{-\h}\LL e)^{\mub_a}b^*(U(\vy)\tih\om^{-\h}\LL e)^{\mub_a}\vac)|& &\non\\
\cdot
|(\vac| b(\tih\om^{-\h}\LL e)^{\nub_a}b^*(U(\vy)\tih\om^{-\h}\LL e)^{\nub_a}\vac)|& &\non\\
& &\phantom{}\non\\
\leq E^{|\mub|+|\nub|} t^{2\mub_b}t^{2\nub_b}(\enorm)^{|\mub_a|+|\nub_a|}& &\non\\ 
\cdot\int_{\De K} d^sy\, |\lan \tih\omm\LL e| U(\vy)\omm\tih\LL e\ran^{\mub_a}|\cdot
 |\lan\tih\omm\LL e| U(\vy)\tih\omm\LL e\ran^{\nub_a}|& &\non\\
\leq C_{r,E}^2 E^{|\mub|+|\nub|} t^{2\mub}t^{2\nub}.& & \quad
\eeqa
We note that the integral, obtained in the second step, converges with $\De K\nearrow\real^s$ due to the properties of the 2-multiindices
$\nub_a$, $\mub_a$ and Lemma~\ref{mbounds}. Thus the constant $C_{r,E}$ can be chosen independently of $K$ and the above
bound still holds after replacing $K$ with $\real^s$. Now the estimate in the statement of the lemma follows from the 
bound~(\ref{decomp-appl}) after readjusting the constant $C_{r,E}$. \qed\\
Now we are ready to prove the main result of this subsection.
\bep\label{map21} In massless scalar free field theory for $s\geq 3$ the map $\thet$, given by relation~(\ref{lastestimate1}),
satisfies the bound
\beq
\sup_{A\in\mfa(\mco(r))_1}\|\thet(A)\|_{E,2}<\infty. 
\eeq
\eep
\proof First, we recall that the bound (\ref{tauestimate2}) gives
$\|\tau_{\mub,\nub}\|\leq 2^{\fr{5}{2}(|\mub|+|\nub|)} (\mub!\nub!)^{-\h}$.
Next, making use of the Minkowski inequality (\ref{Minkowski}), we obtain for any $A\in\mfa(\mco(r))_1$
\beqa
\|\thet(A)\|_{E,2}&\leq& \sum_{\su{\mub,\nub \\ |\mup|+|\nup|>3 \te{ or } |\mum|+|\num|>0}  }\|\tau_{\mub,\nub}\|\,\|\Smn\|_{E,2}
\leq C_r\sum_{\mub,\nub}\fr{(2^5E)^{\fr{|\mub|+|\nub|}{2}}}{(\mub!\nub!)^\h}t^{\mub} t^{\nub}\non\\
&=& C_r\bigg(\sum_{\mup}\fr{(2^5E)^{\fr{|\mup|}{2}} }{(\mup!)^\h}t^{\mup}\bigg)^4
\leq C_r\bigg(\sum_{k=0}^{\infty}\fr{(2^5E \|T\|_1^2)^{\fr{k}{2}}}{k!^\h}\bigg)^4,
\eeqa
where in the second step we applied Lemma~\ref{map2} and in the last step we made use of the fact that
the multinomial coefficients are larger or equal to one, and of the multinomial formula~(\ref{multiform}). Clearly, the
last expression on the r.h.s. of the above estimate is finite. \qed\\
From relation~(\ref{R2-relation}) and Propositions~\ref{bh-square-integrability}, \ref{map21},  we conclude
that for any $r>0$, $E\geq 0$ and $A\in\mfa(\mco(r))$ there holds
\beq
\|R^{(2)}(A)\|_{E,2}<\infty,
\eeq
what completes the proof of Theorem~\ref{HA-in-models}. Thus we have 
verified that Condition~$\A$ holds in massless scalar free field theory, its even part and its sub-theory
generated by derivatives of the field for $s\geq 3$.

%% file: PhD-L1_app.tex
\chapter{Verification of Condition $\Bs$ in Massive Scalar Free Field Theory} \label{Condition-L1} 

In this appendix we verify that Condition $\Bs$, stated in Section~\ref{triviality-of-Apc},
holds in massive free field theory. In Section~\ref{prel-L1} we recall the relevant 
background material from Appendix~\ref{Preliminaries}. In Section~\ref{b-L1} 
we verify Condition $\B$. Section~\ref{a-L1} is devoted to its strengthened variant.

\section{Preliminaries}\label{prel-L1}

We infer from Propositions~\ref{massive-condition-Ns} and \ref{N-sharp} that in massive free field theory the map 
$\whXi_E:\mfa(\mco(r))\to B(\hil)$, given by $\whXi_E(A)=P_EAP_E$, has
the following  expansions for $A\in\mfa_{\scc}(\mco(r))$,
\beqa
\whXi_E(A)&=&\sum_{\su{\mub,\nub \\ |\mub|+|\nub|\neq 0}}\ctau_{\mub,\nub}^{(r)}(A)\,\cS_{\mub,\nub},
\label{L1-expansion}\\
\whXi_E(A)&=&\sum_{\su{\mub,\nub \\ |\mub|+|\nub|\neq 0 } }\tau_{\mub,\nub}(A)\Smn, \label{L1-expansion1} 
\eeqa
which converge in the norm topology in $B(\hil)$. Here we made use of the fact that
$\ctau_{0,0}^{(r)}=\tau_{0,0}=\om_0$. By definitions~(\ref{cS}) and (\ref{Smn-def}), 
the operators $\cS_{\mub,\nub}$, $\Smn$ have the form
\beqa
\cS_{\mub,\nub}&=&P_E a^*(h_{\ka,E})^{\mub}a(h_{\ka,E})^{\nub}P_E, \label{L1-cS}\\
\Smn&=&P_E a^*(\LL e)^{\mub}a(\LL e)^{\nub} P_E, \label{L1-Smn}
\eeqa
where the vectors $\{\hti_{\ka_j,E}^{\pm}\}_1^\infty$ are defined by~(\ref{h}) and $\{e_j\}_1^\infty$
are the eigenvectors of the operator $T$, given by (\ref{T-definition}), whose eigenvalues are denoted by $\{t_j\}_1^\infty$. Due to the bounds~(\ref{tS-bounds}) and (\ref{Sestimate2}) there holds 
\beqa
\|\cS_{\mub,\nub}\| &\leq& E^{\fr{|\mub|+|\nub|}{2}} \|\omm \hti_{\ka,E}\|_2^{\mub+\nub}
\leq \M^{\fr{|\mub|+|\nub|}{2}}\|\hti_{\ka,E}\|_2^{\mub+\nub}, \label{L1-energy-bound}\\
\|\Smn\|&\leq& E^{\fr{|\mub|+|\nub|}{2}}t^{\mub} t^{\nub},\label{L1-energy-bound1}
\eeqa
where in the second step of the first estimate we made use of the fact that in the massive theory $\|\omm\|=m^{-\h}$
and we set $\M=\fr{E}{m}$. We note that $\cS_{\mub,\nub}=\Smn=0$
for $|\mub|>[\M]$ or $|\nub|>[\M]$. Thus, by Lemma~\ref{ttau-bound-lemma}, the normal functionals $\ctau_{\mub,\nub}^{(r)}$,
$\tau_{\mub,\nub}$, entering into the sums (\ref{L1-expansion}), (\ref{L1-expansion1}),  satisfy the following estimates 
for any $l\geq 0$
\beqa
\|R^{-l}\ctau_{\mub,\nub}^{(r)}R^{-l}\|&\leq& (2c_l)^{2\M}\sqrt{(2[\M])!}\|b_{\ka,r}\|_{2,l}^{\mub+\nub},
\label{L1-tauestimate}\\
\|\tau_{\mub,\nub}\|&\leq& \fr{2^{\fr{5}{2}(|\mub|+|\nub|)}}{(\mub!\nub!)^\h}\leq 2^{5\M},\label{L1-tauestimate1}
\eeqa
where $c_l=(12+2m^2)^{l/2}$, the vectors $\{ b_{\ka_j,r}^{\pm}\}_1^\infty$ are defined by~(\ref{b}) and
the Sobolev norms are given by (\ref{Sobolev-norm}). Finally, by 
Lemmas~\ref{b-lemma} and \ref{h-lemma}, we obtain the following bound for any combination of $\pm$-signs 
and any $\la, l \geq 0$
\beq
\sum_{\ka\in\nat^s_0} \|b_{\ka,r}^{\pm}\|_{2,l} \, \|\hti_{\ka,E}^{\pm}\|_{2,\la}
\leq \sum_{\ka\in\nat^s_0}\fr{(c_{l,r}\,c_{\la,0,E})^{(|\ka|+1) } }{\ka!}<\infty. \label{L1-sp-convergence}
\eeq
This concludes the list of auxiliary results which we need to verify Condition $\Bs$.

\section{Verification of Condition $\B$ }\label{b-L1}

Let us first briefly describe our strategy:  By Theorem~\ref{harmonic}, observables of the
form $B^*B$, where $B\in\mfa$ is almost local and energy-decreasing, are integrable i.e. 
they belong to $\CC$. The operators $\cS_{\mub,\nub}$ are of similar form for $|\mub|\neq 0$
and $|\nub|\neq 0$. In fact, we will show in Lemma~\ref{L1-Sestimate-lemma} below that under such restriction 
$\|\cS_{\mub,\nub}\|_{E,1}<\infty$. The role of the time-smearing function $g$, entering 
into Condition~$\Bs$, is to eliminate all other terms.
\bel\label{0-Sestimate-lemma}  Let $g\in S(\real)$ be s.t. $\supp\,\tg\subset ]-m, m[$,
where $m>0$ is the mass of  free field theory. Let $n\geq 1$ and $h_1,\ldots, h_n\in L^2(\real^s,d^sp)$. 
Then
\beq
P_E\big(a(h_1)\ldots a(h_n)\big)(g)P_E=0.
\eeq
In particular, if  $|\mub|=0$ or $|\nub|=0$ but $|\mub|+|\nub|\neq 0$ then 
$\cS_{\mub,\nub}(g)=\Smn(g)=0$.
\eel
\proof Let  $\Psi_1,\Psi_2\in P_E\hil\cap D_S$. Then we obtain
\beqa
& &(\Psi_1|P_E\big(a(h_1)\ldots a(h_n)\big)(g)P_E\Psi_2)\non\\
&=&\int dt\, g(t)\,\int d^sp_1\ldots d^sp_{n}\,\ov{h}(\vep_1)e^{-i\om(\vep_1)t}\ldots 
\ov{h}(\vep_n)e^{-i\om(\vep_n)t}(\Psi_1|a(\vep_1)\ldots a(\vep_n)\Psi_2)\non\\
&=&(2\pi)^\h \int d^sp_1\ldots d^sp_{n}\,\tg(-(\om(\vep_1)+\cdots+\om(\vep_n)))\non\\
& &\ph{44444444444444444444444444}\cdot\ov{h}(\vep_1)\ldots\ov{h}(\vep_n)(\Psi_1|a(\vep_1)\ldots a(\vep_n)\Psi_2)=0,
\eeqa
where in the last step we exploited the support properties of $\tg$. 
Making use of the energy bounds (\ref{energy-bounds}) and the fact that $\|\omm\|<\infty$ in the
massive theory, we conclude that 
\beq
P_E\big(a(h_1)\ldots a(h_n)\big)(g)P_E=0.
\eeq
Given the restrictions on $(\mub,\nub)$ in the statement of the lemma, either  $\cS_{\mub,\nub}(g)$
or $(\cS_{\mub,\nub}(\ov{g}))^*$ is of the form considered above. The same is true for $\Smn$. \qed.\\
Now we are ready to estimate the $\|\,\cdot\,\|_{E,1}$-seminorms of the operators $\cS_{\mub,\nub}$ which are not
covered by the above lemma. We will exploit the 
fact that for any $C\in B(\hil)$, which satisfies $\|C\|_{E,1}<\infty$, there holds
\beq
\|C\|_{E,1}\leq 4\sup_{\fun\in\traceEBP} \int d^sx\,|\fun(C(\vx))|. \label{L1-decomp-appl}
\eeq
This bound follows immediately from  decomposition (\ref{decomp}).
\bel\label{L1-Sestimate-lemma} Let $(\mub,\nub)$ be a pair of 2-multiindices s.t. $|\mub|\cdot|\nub|\neq 0$
and $\la>\fr{s}{2}$. Then there holds
\beq
\|\cS_{\mub,\nub}\|_{E,1}\leq c_{\la}\M^{\fr{|\mub|+|\nub|}{2}}\|\hti_{\ka,E}\|_{2,\la}^{\mub+\nub},
\eeq
where $\M=\fr{E}{m}$ and $c_{\la}$ is a constant independent of $\mub$, $\nub$.
\eel
\proof 
For any compact subset $K\subset\real^s$ we obtain from the Cauchy-Schwarz inequality
\beqa
& &\sup_{\fun\in\traceEBP}\int_K d^sx\, |\fun(\cS_{\mub,\nub}(\vx))|\non\\
&\leq&\sup_{\fun\in\traceEBP}\big(\int_K d^sx\, |\fun(\cS_{\mub,\mub}(\vx))|\big)^\h
\sup_{\fun\in\traceEBP}\big(\int_K d^sx\, |\fun(\cS_{\nub,\nub}(\vx))|\big)^\h.\label{first}
\eeqa
We decompose $\mub$ into two 2-multiindices $\mub=\muh+\muc$ in such a way that $|\muc|=1$.
\beqa
\int_K d^sx\, |\fun(\cS_{\mub,\mub}(\vx))|&=&\int_K d^sx\, 
|\fun\big((a^*(h_{\ka,E})^{\muc}P_E a^*(h_{\ka,E})^{\muh}a(h_{\ka,E})^{\muh}P_E a(h_{\ka,E})^{\muc})(\vx)\big)|\non\\
&\leq& \|a(h_{\ka,E})^{\muh}P_E\|^2\int_K d^sx\,
 \fun\big((a^*(h_{\ka,E})^{\muc}a(h_{\ka,E})^{\muc})(\vx)\big)\non\\
&\leq&  \|\hti_{\ka,E}\|_2^{2\muh} \M^{|\mub|}\int_{\De K} d^sx\,|\lan \hti_{\ka,E}^{\muc}|U(\vx) \hti_{\ka,E}^{\muc}\ran|\non\\
&\leq& c_{\la}\M^{|\mub|}\|\hti_{\ka,E}\|_{2,\la}^{2\mub}.\label{second}
\eeqa
Here in the third step we applied the bound (\ref{L1-energy-bound}) and Lemma~\ref{master-harmonic}
with $\hh=\omm$. In the last step we made use of Lemma~\ref{Sobolev-application} and of the fact
that $\|f\|_{2}\leq \|f\|_{2,\la}$ for any $f\in L^2(\real^s, d^sp)_{\la}$. The second factor
on the r.h.s. of relation~(\ref{first}) satisfies an analogous bound. Thus we obtain
\beq
\sup_{\fun\in\traceEBP}\int_K d^sx\, |\fun(\cS_{\mub,\nub}(\vx))|\leq
\M^{\fr{|\mub|+|\nub|}{2}}c_{\la}\|\hti_{\ka,E}\|_{2,\la}^{\mub+\nub}.
\eeq
Making use of the bound~(\ref{L1-decomp-appl}) and redefining the constant~$c_{\la}$, we obtain the
estimate in the statement of the lemma. \qed\\
Next, we note that if $C\in B(\hil)$ satisfies $\|C\|_{E,1}<\infty$ and $g\in S(\real)$ is a
time-smearing function then there holds
\beq
\|C(g)\|\leq \|g\|_1\|C\|_{E,1}. \label{C-g-bound}
\eeq
This bound is a simple consequence of the Fubini theorem. It concludes the list of auxiliary results
needed to verify Condition $\B$. Now we are ready to show  that for any $r>0$ there holds the bound
\beq
\|A(g)\|_{E,1}\leq c_{l,E,r}\|R^l A R^l\|, \quad\quad A\in\mfa_{\scc}(\mco(r)), \label{L1-g}
\eeq
where $g$ is defined as in the statement of Lemma~\ref{0-Sestimate-lemma} and the
constant $c_{l,E,r}$ is independent of $A$.
\bet Condition $\B$ holds in massive scalar free field theory for any dimension of
space $s\geq 1$. \label{L1-final-theorem}
\eet
\proof Let $g\in S(\real)$ be s.t. $\supp\,\tg\subset ]-m, m[$ and $l\geq 0$.
Making use of relation~(\ref{L1-expansion}) we obtain
\beqa
\|A(g)\|_{E,1}&\leq& \|R^lAR^l\|\,\sum_{\su{\mub,\nub \\ |\mub|+|\nub|\neq 0 }}\|R^{-l}\ctau_{\mub,\nub}^{(r)}R^{-l}\|\, \|\cS_{\mub,\nub}(g)\|_{E,1}\non\\
&\leq& \|R^lAR^l\|\, \|g\|_1 c_{\la} (4c_l^2\M)^{\M} \sqrt{(2[\M])!}  \sum_{\mub,\nub}\big(\|b_{\ka,r}\|_{2,l}\|\hti_{\ka,E}\|_{2,\la}\big)^{\mub+\nub}\non\\
&\leq& \|R^lAR^l\|\,\|g\|_1 c_{\la}(4c_l^2\M)^{\M}\sqrt{(2[\M])!} \non\\
&\cdot&\big(\sum_{n_1=0}^{[\M]}(\sum_{\ka\in\nat_0^s} \|b^+_{\ka,r}\|_{2,l}\,\|\hti^+_{\ka,E}\|_{2,\la})^{n_1}\big)^2\big(\sum_{n_2=0}^{[\M]}
(\sum_{\ka\in\nat_0^s} \|b^-_{\ka,r}\|_{2,l}\,\|\hti^-_{\ka,E}\|_{2,\la})^{n_2}\big)^2. \non\\
\eeqa
Here in the second step we exploited estimate~(\ref{L1-tauestimate}), 
Lemmas~\ref{0-Sestimate-lemma} and \ref{L1-Sestimate-lemma}, and relation~(\ref{C-g-bound}).
In the last step we made use of the fact that the multinomial coefficients are larger
or equal to one and of the multinomial formula~(\ref{multiform}). The last expression is finite
by estimate~(\ref{L1-sp-convergence}). \qed\\
We conclude this section with a brief comment on the even part of massive scalar free field theory
$(\eumfa,\eal,\ehil)$, introduced in Section~\ref{scalar-free-field-theory}. There clearly holds for any $\uA\in\eumfa_{\scc}(\mco)$ and $g\in S(\real)$ as specified in Condition~$\B$
\beq
\|\uA(g)\|_{E,1}\leq \|\epim(\uA)(g)\|_{E,1}\leq c_0\|\epim(\uA)\|=c_0\|\uA\|. \label{even-L1-computation}
\eeq
Here in the first step we made use of formula~(\ref{even-functoriality}) and relation~(\ref{even-norm-conservation}),
in the second step we exploited  Condition~$\B$ (for $l=0$), which is valid in (full) massive scalar free field
theory by Theorem~\ref{L1-final-theorem},
and in the last step we applied equality~(\ref{even-norm-equality}). We obtain:
\bec\label{Even-Corollary} Condition $\B$ (a) holds in the even part of massive scalar free field theory $(\eumfa,\eal,\ehil)$ 
for any dimension of space $s\geq 1$.
\eec 
\nin We conjecture that this model satisfies part (b) of Condition $\B$ as well, but we
do not have a complete argument at the moment.
\section{Verification of Condition $\Bs$ }\label{a-L1}
In this section we verify the strengthened variant of Condition $\B$. In addition to the 
bound~(\ref{L1-g}) above, we have to show that the operators $A(g)$, where $A\in\hmfa_{\scc}$
and $\tg$ is supported in $]-m,m[$, can be approximated by elements from 
$\mfc(g)=\{\, C(g)\, |\, C\in\mfc\,\}$ in the $T^{(1)}$-topology  introduced in Section~\ref{triviality-of-Apc}. In view of relation~(\ref{spectral-part(a)})
we have to find, for any $E\geq 0$ and $\eps>0$, an observable $C\in\mfc$ s.t.
\beq
\|A(g)-C(g)\|_{E,1}\leq \eps. \label{L1sharp-second-relation}
\eeq
We will accomplish this task in the last part of this section. (See relations (\ref{Settf-application}) and (\ref{Settf-application1}) below). The key observation is that in massive scalar free field theory, for $|\mub|\cdot|\nub|\neq 0$, there holds
\beq
\Smn=P_EB^*_{\mub}B_{\nub}P_E, \label{Smn=B*B}
\eeq
where $\Smn$ is given by (\ref{L1-Smn}) and $B_{\mub}$, $B_{\nub}\in \mfa$  are energy-decreasing and almost local
observables which depend on $E$. This
fact is established in the following  lemma.
\bel\label{Smn-B*B} Let $s\geq 1$ and $e\in L^2(\real^s,d^sp)$ be s.t. $Je=e$. Then, in massive 
scalar free field theory, for any
$E\geq 0$ there exist operators $B_E^\pm\in\mfa$ which are almost local, energy-decreasing and
s.t. $a(\Lpm e)P_E=B_E^\pm P_E$.
\eel
\proof We consider only the (+) case as the ($-$) case is analogous. Without loss of generality
we can assume that $[\M]>1$ and $\|\Lp e\|\neq 0$. We pick a function $h\in S(\real)$
which satisfies, for $n\in\{0,1,\ldots,2[\M]\}$,
\beq
\int du\, e^{-\h u^2\|\Lp e\|^2} u^n h(u)=-i\de_{n,1}. \label{delta-n-1}
\eeq
Such function can be constructed as follows: Let $f_n(u)=e^{-\h u^2\|\Lp e\|^2} u^n$. In the 
subspace $X=\Span\{\, f_0,f_2,\ldots, f_{2[\M]}\,\}$ of $L^2(\real,du)$ we can construct, with the
help of the Gram-Schmidt procedure, an orthonormal basis $\{\,g_0,g_2,\ldots,g_{2[\M]}\,\}$
in $X$, consisting of Schwartz-class functions. We introduce the projection on $X$
\beq
P_X=|g_0)(g_0|+|g_2)(g_2|+\cdots+|g_{2[\M]})(g_{2[\M]}|
\eeq
and note that $(I-P_X)f_1\neq 0$, since the functions $\{f_n\}_1^{2[\M]}$ are linearly independent. 
We set
\beq
h=-i\fr{(I-P_X)f_1}{\|(I-P_X)f_1\|^2}.
\eeq
It is manifestly a Schwartz-class function which satisfies (\ref{delta-n-1}).

Next, we note that $e^{ iu(a(\Lp e)+a^*(\Lp e)) }\in\mfa(\mco(r))$ for any $u\in\real$. (In fact,
there exists a sequence $f_n^+\in (1+J)\Lpring$ s.t. $\lim_{n\to\infty}\|f_n^+-\Lp e\|_2=0$. Thus, by
Theorem~X.41~(d) of \cite{RS2}, the corresponding sequence of Weyl operators $W(u f_n)$ converges
to $e^{ iu(a(\Lp e)+a^*(\Lp e)) }$ in the strong operator topology). Also the weak integral 
\beq
A=\int du\, e^{iu(a(\Lp e)+a^*(\Lp e))}h(u)
\eeq
defines an element of $\mfa(\mco(r))$ by the von Neumann bicommutant theorem. This operator
satisfies
\beq
P_E A P_E= P_E\big( a^*(\Lp e)+a(\Lp e) \big)P_E. \label{observable=field}
\eeq
In fact, let $\Psi_1,\Psi_2\in P_E\hil$. Then, making use of the fact that $\Psi_1$, $\Psi_2$ belong to
$\DF$ in a massive theory, we obtain
\beqa
(\Psi_1|A\Psi_2)&=&\int du\, e^{-\h u^2\|\Lp e\|^2}(\Psi_1| e^{iu(a^*(\Lp e)} e^{iu(a(\Lp e))} \Psi_2)\, h(u)\non\\
&=& \int du\, e^{-\h u^2\|\Lp e\|^2} \sum_{n_1,n_2\in\nat_{0}}\fr{(iu)^{n_1+n_2}}{n_1!n_2!}
(\Psi_1|a^*(\Lp e)^{n_1} a(\Lp e)^{n_2} \Psi_2)\, h(u)\non\\
&=& (\Psi_1|\big(a^*(\Lp e)+a(\Lp e)\big)\Psi_2),
\eeqa
where in the first step we proceeded to the normal ordered form of $A$, in the second step we expanded the
resulting exponentials into (finite) power series and in the last step we made use of relation~(\ref{delta-n-1}). 
Finally, we pick a function $f\in S(\real)$ s.t. $\tf(p_0)=(2\pi)^{-\h}$ for $p^0\in [-m,-E]$ and $\tf(p^0)=0$
for $p^0\notin [-\h m,-2E]$. 
By smearing both sides of equality~(\ref{observable=field}) in time
with $f$ we obtain $A(f)P_E=a(\Lp e)P_E$ where we made use of the fact that $A(f)$ is energy-decreasing.
Since it is also almost local, the proof is complete.  \qed\\
Thus we have verified relation~(\ref{Smn=B*B}). It remains to establish suitable convergence properties
of expansion~(\ref{L1-expansion1}). For this purpose we need the following lemma which is similar to
Lemma~\ref{L1-Sestimate-lemma} above.
\bel\label{L1-Sestimate-lemma1} Let $(\mub,\nub)$ be a pair of 2-multiindices s.t. $|\mub|\cdot |\nub|\neq 0$.
Then there holds in massive scalar free field theory for any $s\geq 1$
\beq
\|\Smn\|_{E,1}\leq  c_{r,E} E^{\fr{|\mub|+|\nub|}{2}}t^{\mub+\nub},
\eeq
where the constant $c_{r,E}$ is independent of $\mub$, $\nub$.
\eel
\proof 
For any compact subset $K\subset\real^s$ we obtain from the Cauchy-Schwarz inequality
\beqa
& &\sup_{\fun\in\traceEBP}\int_K d^sx\, |\fun(\Smn(\vx))|\non\\
&\leq&\sup_{\fun\in\traceEBP}\big(\int_K d^sx\, |\fun(S_{\mub,\mub}(\vx))|\big)^\h
\sup_{\fun\in\traceEBP}\big(\int_K d^sx\, |\fun(S_{\nub,\nub}(\vx))|\big)^\h.\label{first1}
\eeqa
We decompose $\mub$ into two 2-multiindices $\mub=\muh+\muc$ in such a way that $|\muc|=1$.
\beqa
\int_K d^sx\, |\fun(S_{\mub,\mub}(\vx))|
=\int_K d^sx\,
|\fun((a^*(\LL e)^{\muc}P_E a^*(\LL e)^{\muh}a(\LL e)^{\muh}P_E a(\LL e)^{\muc})(\vx))| & &\non\\
\leq \|a(\LL e)^{\muh}P_E\|^2\int_K d^sx\,
 \fun\big((a^*(\LL e)^{\muc}a(\LL e)^{\muc})(\vx)\big)& &\non\\
\leq  E^{|\mub|} t^{2\muh} \sup_{\om(\vep)\leq E}|\tih(\vep)|^{-2}
\int_{\De K} d^sx\,|\lan \omp \tih (\LL e)^{\muc}| U(\vx) \omp \tih (\LL e)^{\muc} \ran|& &\non\\
\leq c_{r,E}^2 E^{|\mub|} t^{2\mub}.\quad
\eeqa
Here in the third step we applied the bound (\ref{L1-energy-bound1}) and Lemma~\ref{master-harmonic}
with $\hh=\tih^{-1}$, where $\tih$ entered into the definition~(\ref{T-definition}) of the operator $T$.
In the last step we made use of Lemma~\ref{mbounds} which guarantees the convergence of the integral in the
massive case. 
The second factor on the r.h.s. of relation~(\ref{first1}) satisfies an analogous bound. Thus we obtain
\beq
\sup_{\fun\in\traceEBP}\int_K d^sx\, |\fun(\cS_{\mub,\nub}(\vx))|\leq
c_{r,E}^2 E^{\fr{|\mub|+|\nub|}{2}} t^{\mub+\nub}.
\eeq
Since the r.h.s. is independent of $K$, we can take the limit $K\nearrow \real^s$. Now the statement
of the lemma follows from the bound~(\ref{L1-decomp-appl}) after readjusting the constant $c_{r,E}$. \qed\\
The last auxiliary result which we need to establish Condition~$\Bs$ is the following
summability property.
\bel~\label{L1-convergence-Smn} Let $g\in S(\real)$ be s.t. $\supp\,\tg\subset ]-m, m[$.
Then, for arbitrary dimension of space $s\geq 1$, in massive scalar free field theory of mass $m$ 
there holds the bound
\beq
\sum_{\su{\mub,\nub \\ |\mub|+|\nub|\neq 0 } }\|\tau_{\mub,\nub}\|\, \|\Smn(g)\|_{E,1}<\infty.
\eeq
\eel
\proof Making use of Lemma~\ref{0-Sestimate-lemma}, we get 
\beqa
\sum_{\su{\mub,\nub \\ |\mub|+|\nub|\neq 0 } }\|\tau_{\mub,\nub}\|\, \|\Smn(g)\|_{E,1}&=&
\sum_{\su{\mub,\nub \\ |\mub|\cdot|\nub|\neq 0 } }\|\tau_{\mub,\nub}\|\, \|\Smn(g)\|_{E,1}\non\\
&\leq& 2^{5\M}c_{r,E}\|g\|_1\sum_{\su{\mub,\nub \\ |\mub|\cdot|\nub|\neq 0 } }E^{\fr{|\mub|+|\nub|}{2}}t^{\mub+\nub}\non\\
&\leq& 2^{5\M}c_{r,E}\|g\|_1\big(\sum_{\su{\mup \\ |\mup|\leq\M }} E^{\fr{|\mup|}{2}}t^{\mup}\big)^4.
\eeqa
Here in the second step we made use of Lemma~\ref{L1-Sestimate-lemma1}, estimate~(\ref{L1-tauestimate1})
and the bound~(\ref{L1-g}). The last expression is finite due to the following relation
\beqa
\sum_{\su{\mup \\ |\mup|\leq\M }} E^{\fr{|\mup|}{2}}t^{\mup}
=\sum_{k=0}^{\M} E^{\fr{k}{2}} \sum_{\su{\mup \\ |\mup|=k}} t^{\mup}
\leq \sum_{k=0}^{\M} E^{\fr{k}{2}}\|T\|_1^k,
\eeqa
where in the last step we made use of the fact that the multinomial coefficients are
larger or equal to one and of the multinomial formula~(\ref{multiform}). \qed\\
According to Lemma~\ref{L1-convergence-Smn}, for any $\eps>0$ there exists a finite set $\Sett_{\te{f}}$ of
pairs of 2-multiindices s.t.
\beq
\|A(g)-\sum_{ (\mub,\nub)\in\Sett_{\te{f}} }\tau_{\mub,\nub}(A)\Smn(g)\|_{E,1}\leq\eps. \label{Settf-application}
\eeq 
In view of relation~(\ref{Smn=B*B}), proved in Lemma~\ref{Smn-B*B}, there exists $C\in\mfc$ s.t.
\beq
\sum_{ (\mub,\nub)\in\Sett_{\te{f}} }\tau_{\mub,\nub}(A)\Smn(g)=P_EC(g)P_E.\label{Settf-application1}
\eeq
Substituting this formula to relation~(\ref{Settf-application}) we establish 
property~(\ref{L1sharp-second-relation}). We summarize:
\bet Condition $\Bs$ holds in massive scalar free field theory for any dimension of
space $s\geq 1$.
\eet

%% file: PhD-Nnat_app1.tex
\chapter{Verification of Condition $\Nnat$ in Scalar Free Field Theory} \label{Condition-Nnat}

It is the goal of the present appendix to verify that Condition~$\Nnat$, stated in 
Section~\ref{additivity-of-energy}, holds both in massive and massless scalar free field theory.
The argument relies on Lemma~\ref{harmonic-1}, stated below, which is a variant of Theorem~\ref{harmonic}
adopted to the problem at hand. This result is combined with Lemma~\ref{mbounds} which describes the decay
of correlations between certain operators under translations in space. 

The necessary background material from Appendix~\ref{Preliminaries} is summarized in Section~\ref{prel}. 
The verification argument, based on the publication \cite{Dy08.2} of the author, is given in Section~\ref{verification-argument}.

\section{Preliminaries}\label{prel}
The main object of our interest is the map $\PiEc: \traceE\to\mfa_{\scc}(\mco(r))^*$, given by 
\beq
\PiEc(\fun)=\fun|_{\mfa_{\scc}(\mco(r))}, \quad\quad\ \fun\in\traceE
\eeq
for some $E\geq 0$, $r>0$. In order to find a suitable expansion of this map into rank-one mappings, we note that
there holds 
\beq
\PiEc(\fun)(A)=\fun(\whXi_E(A)),\quad\quad A\in\mfa_{\scc}(\mco(r)). \label{PiEc-whXi}
\eeq
The map $\whXi_E: \mfa(\mco(r))\to B(\hil)$, given by $\whXi_E(A)=P_EAP_E$, was thoroughly studied
in the literature \cite{BP90,Bo00} and we summarized the relevant results
in Appendix~\ref{Preliminaries}. In particular, we recall from Proposition~\ref{N-sharp}
the following decomposition into rank-one maps
\beq
\whXi_E(A)=\sum_{\mub,\nub}\tau_{\mub,\nub}(A)\Smn, \quad\quad A\in\mfa(\mco(r)),\label{whXi-expansion-repeated}
\eeq
where the sum extends over all pairs of 2-multiindices and converges in the norm topology of $B(\hil)$.
According to definition~(\ref{Smn-def}) and estimate~(\ref{Sestimate2}), the bounded operators $\Smn$ are given by
\beq
\Smn=P_E a^*(\LL e)^{\mub}a(\LL e)^{\nub} P_E
\eeq
and satisfy the bound 
\beq
\|\Smn\|\leq E^{\fr{|\mub|+|\nub|}{2}}t^{\mub} t^{\nub}. \label{Sestimate}
\eeq
Here $\{e_j\}_1^\infty$ are the eigenvectors of the operator $T$, defined by (\ref{T-definition}),
and $\{t_j\}_1^\infty$ are the corresponding eigenvalues. The normal functionals 
$\tau_{\mub,\nub}$ can be estimated as follows 
\beq
\|\tau_{\mub,\nub}\|
\leq 
\fr{2^{\fr{5}{2}(|\mub|+|\nub|)}}{(\mub!\nub!)^\h}. \label{new-tauestimate}
\eeq
(See relation~(\ref{tauestimate2})). In view of equality~(\ref{PiEc-whXi}) we obtain from 
(\ref{whXi-expansion-repeated}) the following expansion for any $\fun\in\traceE$,  $A\in\mfa_{\scc}(\mco(r))$
\beq
\PiEc(\fun)(A)=\sum_{\su{\mub,\nub \\ (|\mub|,|\nub|)\neq (0,0)} }\tau_{\mub,\nub}(A)\fun(\Smn),
\label{PiEc-expansion}
\eeq
where we made use of the fact that $\tau_{0,0}=\om_0$.
We associate with operators $\Smn$  elements of $\traceE^*$, denoted by the same symbol,
given by $\traceE\ni\fun\to\fun(\Smn)$. Their norms are clearly equal to the operator
norms of $\Smn$. Thus we obtain from Proposition~\ref{N-sharp} and definition of the $p$-norms
stated in Section~\ref{additivity-of-energy}
\beq
\|\PiEc\|_{p}^p\leq \sum_{\su{\mub,\nub \\ (|\mub|,|\nub|)\neq (0,0)} }
\|\tau_{\mub,\nub}\|^p\|\Smn\|^p<\infty.
\eeq
That is, the maps $\PiEc$ are $p$-nuclear w.r.t. the standard norms on $\lin(\traceE,\mfa_{\scc}(\mco(r)))$.
We will use the same expansion (\ref{PiEc-expansion}) to estimate the $p$-norms of these maps
w.r.t. the norms $\|\,\cdot\,\|_{\x}$, given by (\ref{Nnorm}). There clearly holds the bound
\beq
\|\PiEc\|_{p,\x}\leq\bigg(\sum_{\su{\mub,\nub\\ (|\mub|,|\nub|)\neq (0,0)} }
\|\tau_{\mub,\nub}\|^p \|\Smn\|^p_{\x}\bigg)^{\pin}.\label{start}
\eeq
To verify Condition $\Nnat$, we have to find estimates on
the norms $\|\Smn\|_{\x}$ whose growth with $N$ can be controlled at large spacelike distances $x_i-x_j$ 
for $i\neq j$. We undertake this task in the next section.

\section{Verification of Condition $\Nnat$}\label{verification-argument}

The crucial ingredient of the argument is following lemma which is inspired
by Lemma 2.2 from \cite{Bu90}. (See also Theorem~\ref{harmonic}). Similarly as in Lemma~\ref{master-harmonic},
the present estimate is uniform in the particle number and depends only on the energy of 
the state in question. 
This result substantiates the underlying physical idea of additivity 
of energy over isolated subregions.
\bel\label{harmonic-1} Let $E\geq 0$ and  $\hh$ be a Borel function on $\real^s$ which is
bounded on $\{\,\vep\in\real^s\,|\, \om(\vep)\leq E\,\}$. We denote the operator of
multiplication by $\hh$ on $L^2(\real^s,d^sp)$ by the same symbol.
Suppose that $g \in L^2(\real^s, d^sp)$ is in the domain of $\hh\omp$.
Then, for any  $x_1,\ldots, x_N\in\real^{s+1}$,  there holds the bound
\beqa
& &\|P_E\sum_{k=1}^N(a^*(\hh\omp g)a(\hh\omp g))(x_k)P_E\|\non\\
& &\ph{4444444444} \leq E\sup_{\om(\vep)\leq E}|\hh(\vep)|^{2}
\big\{ \| g \|_2^2+(N-1)\sup_{i\neq j}|\lan g|U(x_{i}-x_{j}) g\ran| \big\}.
\label{harmonice}
\eeqa
\eel
\proof 
We pick  single-particle vectors $\Psi_1, g_1\in L^2(\real^s,d^sp)$ and define  $Q=$\\ $\sum_{k=1}^N(a^*(g_1)a(g_1))(x_k)$. Then there holds
\beqa
(\Psi_1|QQ\Psi_1)\leq\sum_{l=1}^N(\Psi_1|(a^*(g_1)a(g_1))(x_l)\Psi_1)\sum_{k=1}^N|\lan U(x_k)g_1|U(x_l)g_1\ran|
\ph{4}& &\non\\
\leq (\Psi_1|Q\Psi_1)\big\{\|g_1\|_2^2+(N-1)\sup_{i\neq j}|\lan U(x_{j})g_1|U(x_{i})g_1\ran|\big\},& &
\eeqa
where we made use of the fact that $a(U(x_k)g_1)a(U(x_l)g_1)\Psi_1=0$ and of the Cauchy-Schwarz inequality.
Since $(\Psi_1|Q\Psi_1)^2\leq(\Psi_1| QQ \Psi_1)\|\Psi_1\|^2$, we obtain
\beqa
& &\sum_{k=1}^N(\Psi_1|(a^*(g_1)a(g_1))(x_k)\Psi_1)\non\\
& &\phantom{4444444}\leq \|\Psi_1\|^2\big\{\|g_1\|_2^2+(N-1)\sup_{i\neq j}|\lan U(x_{j})g_1|U(x_{i})g_1\ran|\big\}. \label{single}
\eeqa
Next, let $n\geq 1$ and $\Psi_n\in P_E\hil$ be an $n$-particle vector s.t. the corresponding symmetric wave-function 
$\Psi_n(\vep_1,\ldots,\vep_n)$ belongs to $S(\real^{s\times n})$. We also introduce a single-particle wave-function associated with $\Psi_n$ given by  $\Psi_1(\vep_1)_{\pp}=\om(\vep_1)^\half\bar{\hh}(\vep_1)\Psi_n(\vep_1,\ldots, \vep_n)$, where we treat $\pp$ as parameters. With the help of relation (\ref{single}) we get
\beqa
\sum_{k=1}^N(\Psi_n|(a^*(\hh\omp g)a(\hh\omp g))(x_k)\Psi_n)\ph{444444444444444444444444444444444444444444}& &\non\\
=n\int\ds \sum_{k=1}^N(\Psi_{1,\pp}|( a^*(g)a(g) )(x_k)\Psi_{1,\pp})\ph{44444444444}& &\non\\
 \leq n\int\dsp |\hh(\vep_1)|^2\om(\vep_1)|\Psi_n(p_1,\ldots, p_n)|^2\ph{444444444444444444444.}& &\non\\
\cdot\big\{ \| g \|_2^2+
(N-1)\sup_{i\neq j}|\lan g|U(x_{i}-x_{j}) g \ran|\big\}. \ph{444444}& &
\eeqa
Finally, we note that
\beqa
& &n\int\dsp\, |\hh(\vep_1)|^{2} \om(\vep_1)|\Psi_n(\vep_1,\ldots, \vep_n)|^2\non\\
& &\phantom{4444}\leq \sup_{\om(\vep)\leq E}|\hh(\vep)|^{2}
\int\dsp (\om(\vep_1)+\cdots+\om(\vep_n))|\Psi_n(\vep_1,\ldots, \vep_n)|^2\non\\
& &\phantom{4444}\leq \sup_{\om(\vep)\leq E}|\hh(\vep)|^{2}  E\|\Psi_n\|^2\!,
\eeqa
where we made use of the fact that the wave-function is symmetric. Since the operators $(a^*(g)a(g))(x_k)$
conserve the particle number and vectors of the form $\Psi=c_0\vac+\sum_{n=1}^{\infty} \Psi_n$, where  $\|\Psi\|^2=|c_0|^2+\sum_{n=1}^{\infty}\|\Psi_n\|^2<\infty$,
are dense in $P_E\hil$, we easily obtain the bound in the statement of the lemma. \qed\\
Our next task is to control the expressions appearing on the r.h.s.  of estimate~(\ref{harmonice}).
Similarly as in Subsection~\ref{square-integrability-2} 
we recall from Lemma~\ref{TEpm-lemma} that $\Thpm=\om^{-\ga}\tih^{1/2}\Lpm$ are bounded operators
for any $\h\leq\ga<\fr{s-1}{2}$ and $\tih>0$.
Thus we can set in Lemma~\ref{harmonic-1} $\hh=\tih^{-1}$ and $g=\omm\tih\Lpm e$,
where $e$  is an  eigenvector of the operator $T$ given by (\ref{T-definition}). 
The resulting estimate, studied in Proposition~\ref{semibound} below,
relies on the decay properties of the functions
\beq
\real^{s+1}\ni x\to\lan \omm\tih \Lpm e|U(x)\omm \tih \Lpm e\ran \label{function-new}
\eeq
which  appear on the right-hand side of  relation~(\ref{harmonice}). We studied these 
functions for $x=(0,\vx)$ in Lemma~\ref{mbounds}.
In order to obtain estimates  which are valid for arbitrary spacelike translations 
$x$, we recall, in a slightly generalized form, the following result from \cite{BDL87}.
\bel\label{damping-1} Let $\de>0$. Then there exists some continuous function $\f(\om)$ which decreases almost
exponentially, i.e. $\sup_{\om}|\f(\om)|e^{|\om|^\kappa}<\infty \textrm{ for any } 0<\kappa<1$,
and which has the property that for any pair of operators
$A$, $B$ s.t. $\vac$ belongs to their domains and to the domains of their adjoints,
satisfying 
\beq
(\vac| \, [A, e^{itH}Be^{-itH}] \, \vac)=0 \textrm{ for } |t|<\de,
\eeq
there holds the identity $(\vac|AB\vac)=\h\big\{(\vac|A\f(\de H)B\vac)+(\vac|B\f(\de H)A\vac)\big\}$.
\eel
\nin With the help of the above lemma we prove the desired bounds.
\bel\label{Cook} Assume that $s\geq 3$. Let $e\in L^2(\real^s, d^sp)$ satisfy $Te=te$, $Je=e$ and $\|e\|=1$.
Then, for any $\eps>0$, $0<\ka<1$ and $x\in\real^{s+1}$ s.t. $|\vx|\geq |x^0|$, there hold the estimates
\beq
|\lan\tih\omm\Lpm e|U(x)\tih\omm\Lpm e\ran |\leq \fr{c_{\rz,\eps,\ka}(m)e^{-\h(\fr{m}{5})^{\ka}(|\vx|-|x^0|)^{\ka}} t^2}{(|\vx|-|x^0|+1)^{s-2-\eps}},
\eeq
where the constant $c_{\rz,\eps,\ka}(m)\geq 1$ is independent of $x$ and $e$, and finite for any $m\geq 0$.
(If $m>0$, the bound holds for $s\geq 1$).
\eel
\proof
First, we define the operators $\phi^{+}(e)=\fr{1}{\sqrt{2}}\big(a^*(\tih\Lp e)+a(\tih\Lp e)\big)$, 
$\phi^{-}(e)= \fr{1}{\sqrt{2}}\big(a^*(i\tih\Lm e)+a(i\tih\Lm e)\big)$ and their
translates $\phi^{\pm}(e)(x)=U(x)\phi^{\pm}(e)U(x)^{-1}$. Since the projections $\Lpm$ and the multiplication
operators $\tih$ commute with
$J$, and $Je=e$, the operators $\phi^{\pm}(e)$ are
localized in the double cone of radius $2\rz$ centered at zero.
In fact, by definition of the projections
$\Lpm$, we can find, for any $\eps^\prime>0$,  functions $g_{\pm}\in D(\mco_r)_{\real}$ s.t. 
$\|\Lpm e-\ompm \tg_\pm\|_{2}\leq \eps^\prime$. Setting $F^\pm=(2\pi)^{-\fr{s}{2}}h_r*g_\pm$ we obtain from definitions~(\ref{canonical-field}), (\ref{canonical-momentum}) of canonical fields and momenta the following bound
\beq
\|\phi^{\pm}(e)\vac-\phi_\pm(F^\pm)\vac\|=\fr{1}{\sqrt{2}}\|\tih\Lpm e-\tih\ompm \tg_{\pm}\|_2\leq 
\fr{\|\tih\|_{\infty}\eps^\prime}{\sqrt{2}}. 
\eeq
Thus for $x\in\real^{s+1}$ s.t. $\mco_{2r}+x$ and $\mco_{2r}$ are spacelike separated and
for any combination of $\pm$ signs there holds
\beq
[\phi^{\pm}(e),\phi^{\pm}(e)(x)]=(\vac|[\phi^{\pm}(e),\phi^{\pm}(e)(x)]\vac)=0.\label{weak-locality}
\eeq
Now we are ready to prove the desired estimate: We assume without loss of generality that $x^0>0$, 
introduce functions $\Gpm(\tau)=\lan\tih\Lpm e|\om^{-1}U(\vx+\tau\e_0)\tih\Lpm e\ran$ for $0\leq \tau\leq x^0$, where $\e_0$ is the unit vector in the time direction, and consider the derivative
\beq
\bigg|\fr{d\Gpm(\tau)}{d\tau}\bigg|=2|(\vac|\phi^{\pm}(e)\phi^{\pm}(e)(\vx+\tau\e_0)\vac)|.
\eeq
We define $\de_{\tau}=|\vx|-\tau-4\rz$ and assume that $|\vx|-x^0 \geq 5\rz$ what guarantees that $\de_{\tau}>0$ for 
$0\leq \tau\leq x^0$. Then, by relation~(\ref{weak-locality}), $\phi^{\pm}(e)$ and
$\phi^{\pm}(e)(\vx+\tau \e_0)$ satisfy the assumptions of Lemma~\ref{damping-1} with $\de=\de_{\tau}$. Making 
use of this result, we obtain 
\beqa
\bigg|\fr{d\Gpm(\tau)}{d\tau}\bigg|&=&\h|\lan\omar\tih\Lpm e|\om^{2\alr}\f(\de_\tau\om)e^{\h(\de_\tau\om)^{\ka}} e^{-\h(\de_\tau\om)^{\ka}}U(\vx+\tau \e_0)\omar\tih\Lpm e\ran\non\\
&+&\lan\omar\tih\Lpm e|\om^{2\alr}\f(\de_\tau\om)e^{\h(\de_\tau\om)^{\ka}} e^{-\h(\de_\tau\om)^{\ka}}U(-\vx-\tau \e_0)\omar\tih\Lpm e\ran|\non\\
&\leq& \fr{1}{\de_\tau^{2\alr}} t^2 \| \tih\|_{\infty}\, \sup_{\om\geq 0}|\om^{2\alr}e^{\h\om^{\ka}}\f(\om)|
e^{-\h m^{\ka}(|\vx|-x^0-4r)^{\ka}}. \label{derivative-new}
\eeqa
Next, we set $\alr=\fr{s-1-\eps}{2}$ for $0<\eps<1$ and arrive at the following estimate
\beqa
|\lan \omm\tih\Lpm e|U(x)\omm\tih\Lpm e\ran |=|\Gpm(x^0)|&\leq& |\Gpm(0)|+\int_0^{x^0}d\tau \bigg|\fr{d\Gpm(\tau)}{d\tau}\bigg|\non\\
&\leq& \fr{c_{\rz,\eps,\ka}(m)e^{-\h (\fr{m}{5})^{\ka}(|\vx|-x^0)^{\ka}}  t^2}{(|\vx|-x^0+1)^{s-2-\eps} },\quad\label{Cookmethod}
\eeqa
where in the last step we applied Lemma~\ref{mbounds} and estimate~(\ref{derivative-new}).
Since the l.h.s. of relation~(\ref{Cookmethod}) satisfies a uniform bound analogous to (\ref{uniformbound}),
we obtain the estimate in the statement of the lemma. \qed\\
Now we are ready to prove the required estimates on the norms of the functionals~$\Smn$.
\bep\label{semibound} Given a family of points $\x\in\real^{s+1}$ we define 
$\de(\vxb)=\inf_{i\neq j}(|\vx_i-\vx_j|-|x_i^0-x_j^0|)$. For $s\geq 3$, $\de(\vxb)\geq 0$, $(|\mub|,|\nub|)\neq (0,0)$ 
and any $\eps>0$, $0<\ka<1$ the functionals $\Smn$ satisfy the bound
\beqa
\|\Smn\|_{\x}^2\!\!\!
&\leq&\!\!\! 16 c_{\rz,\eps,\ka}(m)\cEErs E^{|\mub|+|\nub|}t^{2(\mub+\nub)}
\bigg\{1+(N-1)\fr{e^{-\h(\fr{m}{5})^{\ka}\de(\vxb)^{\ka}}}{(\de(\vxb)+1)^{s-2-\eps}}\bigg\},\non\\
\eeqa
where the constant $c_{\rz,\eps,\ka}(m)\geq 1$ appeared in Lemma~\ref{Cook} and  is finite for any $m\geq 0$.
(If $m>0$, the bound holds for $s\geq 1$).
\eep
\proof
We denote by $\traceEBP$ the set of positive functionals from $\traceEB$. Making use 
of the definition of $\|\,\cdot\,\|_{\x}$, decomposition (\ref{decomp}) and the Cauchy-Schwarz inequality
we obtain
\beqa
\|\Smn\|^2_{\x}&=&\sup_{\fun\in\tracB}\sum_{k=1}^N|\fun_{x_k}(\Smn)|^2 
 \leq 16\sup_{\fun\in\traceEBP}\sum_{k=1}^N|\fun_{x_k}(\Bms\Bn)|^2 \non\\
& &\phantom{4444444}\leq 16\sup_{\fun\in\traceEBP}\sum_{k=1}^N\fun_{x_k}(\Bms\Bm)\,\,\fun_{x_k}(\Bns\Bn)\non\\
& &\phantom{4444444}\leq 16 E^{|\mub|}t^{2\mub}\| P_E\sum_{k=1}^N(\Bns\Bn)(x_k)P_E\|,\qquad \label{seminorm1}
\eeqa
where in the last step we
applied the bound (\ref{Sestimate}). We can assume, without loss of generality, that $\nub\neq 0$
and decompose it into two 2-multiindices $\nub=\nuh+\nuc$ in such a way that $|\nuc|=1$. Since $\Bn=\Bnh\Bnc$, we get
\beqa
& &P_E\sum_{k=1}^N(\Bns\Bn)(x_k)P_E\non\\
& &\ph{4444444444444}=P_E\sum_{k=1}^N(\Bncs P_E\Bnhs\Bnh P_E\Bnc)(x_k)P_E\non\\
& &\ph{4444444444444}\leq\|\Bnh P_E\|^2 P_E\sum_{k=1}^N(\Bncs\Bnc)(x_k)P_E\non\\
& &\ph{4444444444444}=E^{|\nuh|}t^{2\nuh}  P_E\sum_{k=1}^N\big(\Bncs\Bnc\big)(x_k)P_E,\qquad
\label{aux3}
\eeqa
where in the last step  we used again estimate (\ref{Sestimate}). From relations (\ref{seminorm1}) and (\ref{aux3})
we obtain the bound
\beq
\|\Smn\|^2_{\x}\leq 16E^{|\mub|+|\nuh|}t^{2(\mub+\nuh)}\|P_E\sum_{k=1}^N\big(a^*(\LL e)^{\nuc}a(\LL e)^{\nuc}\big)(x_k)P_E\|.\label{aux2}
\eeq
From Lemmas \ref{harmonic-1} and  \ref{Cook} we get 
\beqa
& &\|P_E\sum_{k=1}^N\big(a^*(\LL e)^{\nuc}a(\LL e)^{\nuc}\big)(x_k)P_E\|
\leq 
E\cEErs\big\{\|\tih\omm(\LL e)^{\nuc}\|^2 \non\\
& &\phantom{444444444}+(N-1)\sup_{i\neq j}|\lan\tih\omm(\LL e)^{\nuc}| U(x_i-x_j)\tih\omm(\LL e)^{\nuc}\ran| \big\}\non\\
& &\phantom{444444444}\leq c_{\rz,\eps,\ka}(m) \cEErs E t^{2\nuc}\bigg\{1+(N-1)\fr{e^{-\h(\fr{m}{5})^{\ka}\de(\vxb)^{\ka}}}{(\de(\vxb)+1)^{s-2-\eps}}\bigg\}.
\quad\quad\label{collect}
\eeqa
Substituting inequality
(\ref{collect}) to  formula (\ref{aux2}),
we obtain the estimate in the statement of the proposition. \qed\\
We note that the bound from Proposition \ref{semibound} has a similar structure to estimate~(\ref{Sestimate})
for the ordinary norms of $\Smn$. Therefore, making use of formulas (\ref{start}) and (\ref{new-tauestimate}), and proceeding as in the proof of Lemma~\ref{N-sharp}, we obtain
\beqa
\|\PiEc\|_{p,\x}\leq 4(c_{\rz,\eps,\ka}(m))^{1/2} \cEE
\bigg(\sum_{k=0}^\infty \fr{(2^5E)^{\half pk}\|T^p\|_1^k}{(k!)^{\half p}} \bigg)^{\fr{4}{p}}& &\non\\
\cdot\bigg\{1+(N-1)\fr{e^{-\h(\fr{m}{5})^{\ka}\de(\vxb)^{\ka}}}{(\de(\vxb)+1)^{s-2-\eps}}\bigg\}^\h.& &
\label{Nnat-final-bound}
\eeqa
It follows that  $\limsup_{\de(\vxb)\to\infty}\|\PiEc\|_{p,\x}$ satisfies a bound which is  
independent of $N$. Consequently, we get
\bet  Condition $\Nnat$ holds in  scalar free field theory in $s\geq 3$ dimensional space. 
(If $m>0$, the same is true for $s\geq 1$).
\eet
It is  obvious that Condition~$\Nnat$ holds also in the sub-theory $(\dmfa,\al,\hil)$, generated by
derivatives of the field, since all our bounds remain valid if one restricts attention to a smaller 
set of observables. It also holds in
the even part of scalar free field theory $(\eumfa,\eal,\ehil)$ as can be seen by the following
argument: We consider the map $\uPiEc:\traceE^{\ee}\to \eumfa_{\scc}(\mco)^*$ given by
\beq
\uPiEc(\efun)=\efun|_{\eumfa_{\scc}(\mco)}\,\,\,,\quad\quad \efun\in\traceE^{\ee}.
\eeq
Due to formula~(\ref{even-functoriality}) we obtain
\beq
\uPiEc(\efun)(\uA)=\PiEc(\ei \efun)(\epim(\uA))
\eeq
for any $\efun\in\traceE^{\ee}$ and $\uA\in\eumfa_{\scc}(\mco)$. Thus the expansion of the map 
$\PiEc$ into rank-one mappings, given by
(\ref{PiEc-expansion}), induces a corresponding expansion of the map $\uPiEc$
\beq
\uPiEc(\efun)(\uA)=\sum_{\su{\mub,\nub \\ (|\mub|,|\nub|)\neq (0,0)} }
\un{\tau}_{\mub,\nub}(\uA)\un{S}_{\mub,\nub}(\efun),
\label{uPiEc-expansion}
\eeq
where $\un{\tau}_{\mub,\nub}(\uA)=\tau_{\mub,\nub}(\epim(\uA))$ and $\un{S}_{\mub,\nub}(\efun)=S_{\mub,\nub}(\ei\efun)$.
Making use of the facts that $\|\uA\|=\|\epim(\uA)\|$ and $\|\efun\|=\|\ei\efun\|$, justified in Section~\ref{scalar-free-field-theory}, we obtain that $\|\un{\tau}_{\mub,\nub}\|\leq \|\tau_{\mub,\nub}\|$ and
$\|\un{S}_{\mub,\nub}\|_{\x}\leq \|\Smn\|_{\x}$. It follows that the $p$-norms $\|\uPiEc\|_{p,\x}$ satisfy
the bound~(\ref{Nnat-final-bound}). Thus we obtain
\bec\label{Nnat-even-derivatives} Let $s\geq 3$. Then Condition $\Nnat$ holds in the even part of scalar free field theory $(\eumfa,\eal,\ehil)$
and in the sub-theory $(\dmfa,\al,\hil)$ generated by
derivatives of the field. (If $m>0$, the same is true for $s\geq 1$).
\eec

%% file: PhD-Cnat_app.tex
\chapter{Verification of Condition $\Csq$ in Massive Scalar Free Field Theory} \label{Condition-Csq}

We showed in Theorem \ref{sharp-nat} that the qualitative part (a) of Condition~$\Csq$ holds in
all theories satisfying Condition~$\Cs$, in particular in (massive and massless) scalar free field theory 
in physical spacetime \cite{BP90}. Moreover, we argued in Section~\ref{vacuum-structure} that in physically meaningful,
massive theories there should also hold the strengthened, quantitative part~(b) of this condition. We 
demonstrated that this quantitative refinement
has a number of interesting consequences pertaining to the vacuum structure. It is the goal of the present
appendix to illustrate the mechanism which enforces Condition~$\Csq$~(b) by a direct computation in
the theory of massive, non-interacting particles.

This appendix is organized as follows: In Section~\ref{Main-Line} we present the proof relying on
some technical information stated in Lemma~\ref{technical} below. The remaining two sections
are devoted to the proof of this lemma.
\section{Main Line of Argument}  \label{Main-Line}
In Section~\ref{coincidence-measurement} we introduced,
for any $\bx\in\Gad$ and $\fun\in\traceE$,  the following  elements of $\BaR$
\beq
\fun_{\bx}(\AN)=\fun(A_1(\vx_1)\ldots A_N(\vx_N)) \label{forms-new}
\eeq
and considered the maps $\Pi_{E,N,\de}\in\lin(\traceE\times\Gad,\BaR)$  given by
\beq
\Pi_{E,N,\de}(\fun,\bx)=\fun_{\bx}.
\eeq
In Section~\ref{vacuum-structure} we introduced
the maps $\Pi_{E,N,\de}^{\scc}\in\lin(\traceE\times\Gad,\BacR)$ defined as
\beq
\Pi_{E,N,\de}^{\scc}(\fun,\vxb)=\Pi_{E,N,\de}(\fun,\vxb)|_{\mfa_{\scc}(\mcor)^{\times N}}.
\eeq
It is our goal to show that $\lim_{\de\to 0}\|\Pi_{E,N,\de}^{\scc}\|=0$.
Then, by Lemma~\ref{content}, there follows the statement of Condition~$\Csq$ (b).
\bet\label{verification} The massive scalar free field theory satisfies Condition $\Csq$
for any dimension of space $s\geq 1$.
\eet
\proof 
The main ingredient of the proof consists in  the following elementary evaluation of the $N$-linear form
$\Pi^{\scc}_{E,N,\de}(\fun,\uvx)$, where  $\fun\in\traceEB$, $\uvx\in\Gad$, on the generating elements of $\mfa_{\scc}(\mcor)$.
We pick $f_1,\ldots,f_N\in\LL$, abbreviate their translates  $U(\vx)f_j$ as $f_{j,\vx}$ and compute
\beqa
& &\ph{444}\Pi^{\scc}_{E,N,\de}(\fun,\uvx)\big(\,\,\{W(f_1)-\om_0(W(f_1))I\}\times\cdots\times \{W(f_{N})-\om_0(W(f_N))I\}\, \big)\non\\
& &\ph{444444444}=\fun\big(\,(W(f_{1,\xx_1})-\om_0(W(f_1))I)\ldots (W(f_{N,\xx_N})-\om_0(W(f_N))I)\, \big)\non\\
& &\ph{444444444}=\sum_{R_1,R_2}(-1)^{|R_2|}e^{-\fr{1}{2}\sum_{k=1}^{|R_2|}\|f_{j_k}\|_2^2}
   \fun(W(f_{i_1,\xx_{i_1}}+\cdots+f_{i_{|R_1|},\xx_{i_{|R_1|}}}))\non\\
& &\ph{444444444}= \sum_{R_1,R_2} (-1)^{|R_2|}e^{-\fr{1}{2}\sum_{k=1}^N\|f_k\|_2^2} e^{-\sum_{1\leq k<l\leq|R_1|}
\Re\lan  f_{i_k,\xx_{i_k}}|f_{i_l,\xx_{i_l}}\ran} 
\non\\
& &\ph{444444444444444444444444444444444}\cdot\fun(:W(f_{i_1,\xx_{i_1}})\ldots W(f_{i_{|R_1|},\xx_{i_{|R_1|}}}):)\non\\
& &\ph{444444444} =\sum_{R_1,R_2}(-1)^{|R_2|}e^{-\fr{1}{2}\sum_{k=1}^N\|f_k\|_2^2}(e^{-\sum_{1\leq k<l \leq |R_1|}\Re\lan f_{i_k,\xx_{i_k}}|f_{i_l,\xx_{i_l}}\ran}-1)\non\\
& &\ph{444444444444444444444444444444444}\cdot\fun(:W(f_{i_1,\xx_{i_1}})\ldots W(f_{i_{|R_1|},\xx_{i_{|R_1|}}}):)\non\\
& &\ph{4444444}+e^{-\fr{1}{2}\sum_{k=1}^N\|f_k\|_2^2}\sum_{R_1,R_2}(-1)^{|R_2|}
\fun(:W(f_{i_1,\xx_{i_1}}+\cdots+f_{i_{|R_1|},\xx_{i_{|R_1|}} }):), \label{crucial}
\eeqa
where the sum extends over all partitions $R_1=(i_1,\ldots,i_{|R_1|})$, $R_2=(j_1,\ldots,j_{|R_2|})$ of an
$N$-element set into two, possibly improper, ordered subsets. (If the condition $1\leq k<l\leq |R_1|$ is empty,
the corresponding sum is understood to be zero). In the second step we
made use of the fact that the Weyl operators are localized in spacelike
separated regions.
In the third step we applied the identity $W(f)=e^{-\h\|f\|_2^2}:W(f):$ and in the last step we 
divided the expression into two parts: The first part tends to zero for large spacelike separations, 
due to the decay of $\lan f_{1,x_1}|f_{2,x_2}\ran$ when $x_1-x_2$ tends to spacelike infinity.
In the next lemma we show that the last sum on the r.h.s. of (\ref{crucial}) vanishes for 
$N>2\fr{E}{m}$, so we can omit this last term in the subsequent discussion.
\bel Let $\fun\in\traceE$ and $N>2\fr{E}{m}$ be a natural number. Then there holds
\beq
S:=\sum_{R_1,R_2}(-1)^{|R_2|}\fun(:W(f_{i_1,\xx_{i_1}}+\cdots+f_{i_{|R_1|},\xx_{i_{|R_1|}} }):)=0,
\eeq
where the sum extends over all partitions of an $N$-element set.
\eel
\proof For any $f\in L^2(\real^s,d^sp)$ we introduce the map $M(f): B(\hil)\to B(\hil)$ given by
\beq
M(f)(C)=P_Ee^{ia^*(f)}Ce^{ia(f)}P_E,\quad\quad C\in B(\hil).
\eeq
The exponentials are defined by their Taylor expansions which are finite (in the massive theory) due
to the energy projections. The range of $M(f)$ belongs to $B(\hil)$ due to the energy
bounds~(\ref{energy-bounds}). We note that $M(f_1)M(f_2)=M(f_2)M(f_1)$  for any 
$f_1,f_2\in L^2(\real^s,d^sp)$ and that  $M(0)(C)=P_ECP_E$ for any $C\in B(\hil)$.
We denote by $\hat{I}$ the identity operator acting from $B(\hil)$ to $B(\hil)$. There
clearly holds
\beqa
S&=&\sum_{R_1,R_2}(-1)^{|R_2|}\fun\big(M(f_{i_1,\xx_{i_1}})\ldots M(f_{i_{|R_1|},\xx_{i_{|R_1|}} })(I)\big)\non\\
&=&\fun\big((M(f_{1,\xx_1})-\hat{I})\ldots (M(f_{N,\xx_N})-\hat{I})(I) \big)\non\\
&=&\fun\big((M(f_{1,\xx_1})-M(0))\ldots (M(f_{N,\xx_N})-M(0) )(I) \big), \label{positive-maps}
\eeqa
where the last equality holds due to the fact that $\fun\in\traceE$. Finally, we note that for any $C\in B(\hil)$
\beq
\big(M(f)-M(0)\big)(C)=\sum_{k+l\geq 1}P_E\fr{(ia^*(f))^k}{k!}C\fr{(ia(f))^l}{l!}P_E.
\eeq
Substituting this relation to (\ref{positive-maps}) we verify the claim. \qed\\
We will exploit formula  (\ref{crucial}) to show that for $N>2\fr{E}{m}$ the norms of the maps $\Pi^{\scc}_{E,N,\de}$
tend to zero with $\de\to\infty$. To this end, we introduce
the $*$-algebra $\mfar(\mcor)$,  generated by finite linear combinations of Weyl operators, 
and denote by $\BBB$  the space of (not necessarily bounded) $M$-linear forms on $\mfar(\mcor)$. 
We define the maps $\Piz_{E,M,\de}:\traceE\times\Gadd\to\BBB$, linear in the first argument, extending
by linearity the following expression
\beqa
& &\Piz_{E,M,\de}(\fun,\uvx)(W(f_1)\times\cdots\times W(f_M))\non\\
& &=e^{-\fr{1}{2}\sum_{k=1}^M \|f_{k}\|_2^2} (e^{-\sum_{1\leq i<j \leq M}\Re\lan f_{i,\xx_{i}}|f_{j,\xx_{j}}\ran}-1)
\fun(:W(f_{1,\xx_{1}}+\cdots+f_{M,\xx_{M}} ):).\,\,\,\,\,\,\,\,\,\,\,\, \label{Piz}
\eeqa
We obtain from (\ref{crucial}) the equality valid for $N>2\fr{E}{m}$
\beqa
\Pi^{\scc}_{E,N,\de}(\fun,\uvx)\big(\,\{W(f_1)-\om_0(W(f_1))I\}\times\cdots\times \{W(f_{N})-\om_0(W(f_N))I\}\, \big)& &\non\\
=\sum_{R_1,R_2}
(-1)^{|R_2|}\om_0(W(f_{j_1}))\ldots\om_0(W(f_{j_{|R_2|}}))\cdot& &\non\\
\cdot\Piz_{E,|R_1|,\de}(\fun,\uvx_{R_1})(W(f_{i_1})\times\cdots\times W(f_{i_{|R_1|} })),& & \label{crucial1}
\eeqa
where $\uvx_{R_1}:=(\vx_{i_1},\ldots,\vx_{i_{|R_1|}})$ is clearly an element of $\Ga_{|R_1|,\de}$.
To conclude the argument we need the following technical lemma.
\bel\label{technical} For any $M\in\nat$, $E\geq 0$, double cone $\mcor$ and sufficiently large $\de>0$ 
(depending on $M$, $E$ and $\mcor$) there exist maps $\Pizz_{E,M,\de}\in
\lin(\traceE\times\Gadd,(\mfa(\mcor)^{\otimes M})^*)$ which have the properties
\begin{enumerate}
\item[(a)] $\lim_{\de\to\infty}\|\Pizz_{E,M,\de}\|=0$,
\item[(b)] $\Pizz_{E,M,\de}(\fun,\vxb)(A_1\otimes\cdots\otimes A_M)=\Piz_{E,M,\de}(\fun,\vxb)(A_1\times\cdots\times A_M)$
for $A_1,\ldots, A_M\in\mfar(\mcor)$ and any $(\fun,\vxb)\in\traceE\times\Gadd$.
\end{enumerate}
\eel
\noindent In view of this lemma, whose proof is postponed to Section~\ref{long-proof},
equality~(\ref{crucial1}) can now be rewritten as follows, for sufficiently
large $N$, $\de$ and any $A_1,\ldots,A_N\in\mfa_{\cc}(\mcor)$ 
\beqa
\Pi^{\scc}_{E,N,\de}(\fun,\uvx)(A_1\times\cdots\times A_N)=
\Pizz_{E,N,\de}(\fun,\uvx)(A_{1}\otimes\cdots\otimes A_{N}), \label{equality}
\eeqa
where we made use of the facts that $\om_0(A_1)=\cdots=\om_0(A_N)=0$ and that $\mfar(\mcor)$ is
dense in $\mfa(\mcor)$ in the strong operator topology.
Consequently, for $N>2\fr{E}{m}$, the maps $\Pi^{\scc}_{E,N,\de}$ share the properties of 
$\Pizz_{E,N,\de}$ stated in  Lemma~\ref{technical}. In addition, we know from Theorem~\ref{sharp-nat} that 
the maps $\Pi^{\scc}_{E,N,\de}$ are compact for any $\de>0$. 
We conclude that Condition $\Csq$ is satisfied. \qed\\
Now it easily follows that Condition~$\Csq$ is also satisfied in the even part of massive
scalar free field theory $(\eumfa,\eal,\ehil)$, introduced in 
Section~\ref{scalar-free-field-theory}. There we have shown  that
for any $\uA\in\eumfa$ and $\efun\in\traceE^{\ee}$ there holds
$\|\uA\|=\|\epim(\uA)\|$ and $\|\efun\|=\|\ei\efun\|$. Making use of these
equalities and formula~(\ref{even-functoriality}), one easily obtains the bound
\beq
\|\un{\Pi}^{\scc}_{E,N,\de}\|\leq \|\Pi^{\scc}_{E,N,\de}\|,
\eeq
where the map on the l.h.s. is defined within $(\eumfa,\eal,\ehil)$ and the mapping on the
r.h.s. corresponds to the full massive scalar free field theory. Thus there follows the corollary:
\bet The even part of massive scalar free field theory $(\eumfa,\eal,\ehil)$ satisfies Condition $\Csq$
for any dimension of space $s\geq 1$.
\eet

 We remark that the assumption $m>0$ is used only in one (crucial) step in the proof of 
Theorem~\ref{verification}, namely to eliminate the last term in relation~(\ref{crucial}) and 
establish equality~(\ref{equality}). The properties of the maps $\Piz_{E,M,\de}$, stated in Lemma~\ref{technical},
hold in massless free field theory as well. However, we do not expect that the present
Condition~$\Csq$ holds in massless theories as it stands. There, due to the existence of states with
arbitrary number of particles in $\traceE$,  the maximal number of localization 
centers $N_0$, visible in an experiment, should depend not only on the energy $E$ of the state, but
also on the experimental accuracy $\eps$. We conjecture that an accordingly modified condition
has similar physical consequences to the present one and that it holds in massless free field theory.

\section{Proof of Lemma F.1.3}\label{long-proof} 

The goal of this section is to construct the maps $\Pizz_{E,M,\de}\in\lin(\traceE\times\Gadd,(\mfa(\mcor)^{\otimes M})^*)$
and verify that they have the properties (a) and (b) specified in Lemma \ref{technical}. We will define these maps
as norm-convergent sums of rank-one mappings, i.e.
\beq
\Pizz_{E,M,\de}=\sum_{i=1}^{\infty}\tau_i\,S_i, \label{Pizz-expansion}
\eeq
where $\tau_i\in(\mfa(\mcor)^{\otimes M})^*$ and $S_i\in\lin(\traceE\times\Gadd,\complex)$.

In order to construct a suitable family of functionals $\tau_i$, we recall the relevant results from \cite{Bo00}, 
which we reproduced in Proposition~\ref{Bfunctionals}:
For any 2-multiindex $\mub=(\mup,\mum)$ and an orthonormal basis  $\{\en_i\}_{1}^{\infty}$
of $J$-invariant eigenvectors in the single-particle space $L^2(\real^s, d^sp)$ there exists
a normal functional $\tau_{\mub}$ on $B(\hil)$ s.t. for any $f\in\LL$ there holds 
\beq
\tau_{\mub}(W(f))=e^{-\half\|f\|^2_2}\lan \en|\Fp\ran^{\mup}\lan \en|\Fm\ran^{\mum}, \label{state}
\eeq 
where  $f^+$, $f^-$ are the real and imaginary parts of $f$ in configuration space.
These functionals satisfy the bound
\beq
\|\tau_{\mub}\|\leq 4^{|\mup|+|\mum|}(\mup!\mum!)^\half. \label{stateestimate}
\eeq
Turning to the definition of suitable functionals on $B(\hil)^{\otimes M}$, we introduce
$M$-tuples of  multiindices $\umu^{\pm}=(\mu^{\pm},\ldots,\mu^{\pm})$ and
the corresponding $2M$-multiindices $\umu=(\umu^+,\umu^-)$. We extend 
the standard rules of the multiindex notation (see Section~\ref{multiindex-notation})  as follows
\beqa
|\umu|&=&\sum_{i=1}^M(|\mu^+_i|+|\mu^-_i|),\\
\umu!&=&\prod_{i=1}^M\mu_i^+!\mu_i^-!,\\
\lan \en | f \ran^{\umu}&=&\prod_{i=1}^{M}\lan \en  | f_i^+  \ran^{\mu^+_i}\, \lan \en | f_i^- \ran^{\mu^-_i},
\eeqa
where $f_1,\ldots, f_M\in \LL$. Now for any $2M$-multiindex $\umu$ we define a normal functional 
$\tau_{\umu}$ on $B(\hil)^{\otimes M}$ by the expression
\beq
\tau_{\umu}=\tau_{\mub_1}\otimes\cdots\otimes\tau_{\mub_M}. \label{Mtau}
\eeq
From relations (\ref{state}), (\ref{stateestimate}) and the polar decomposition of a normal functional
\cite{Sa} one immediately obtains:
\bel\label{bound-tau} Let $\{\en_i\}_1^{\infty}$ be an orthonormal basis in $L^2(\real^s)$ of $J$-invariant
eigenvectors. The functionals $\tau_{\umu}\in (B(\hil)^{\otimes M})^*$, given by 
(\ref{Mtau}), have the following properties
\begin{enumerate}
\item[(a)] $\tau_{\umu}(W(f_1)\otimes\cdots\otimes W(f_M))=e^{-\half\sum_{k=1}^M\|f_k\|_2^2}\lan \en|f\ran^{\umu}$,
\item[(b)] $\|\tau_{\umu}\|\leq 4^{|\umu|}(\umu!)^\half$,
\end{enumerate}
where $f_1,\ldots, f_M\in \LL$.
\eel
\nin In order to construct a basis $\{\en_i\}_1^{\infty}$ in $L^2(\real^s,d^sp)$ of $J$-invariant eigenvectors, which is suitable for our purposes, we modify slightly our discussion from Subsection~\ref{orthonormal} (based on \cite{BP90,Bo00}).
Let $Q_E$ be the projection on states of energy lower than $E$ in the single-particle space. We define the operators $\T_{E}^{\pm}=Q_E\Epm$ and $\T_{\ka}^{\pm}=e^{-\fr{|\om|^{\ka}}{2}}\Epm$, where  $0<\ka<1$. 
By a slight modification of  Lemma~\ref{TEpm-lemma} 
one finds that these operators satisfy
$\|\T_E^{\pm}\|_1<\infty$, $\|\T_{\ka}^{\pm}\|_1<\infty$, where
$\|\cdot\|_1$ denotes the trace norm. Next, we introduce the operator $\T$ given by 
\beq
\T=(|\T_{E}^{+}|^2+|\T_{E}^{-}|^2+|\T_{\ka}^{+}|^2+|\T_{\ka}^{-}|^2)^\h. \label{T}
\eeq
Making use of the estimate  $\|(A+B)^p\|_1\leq \|A^p\|_1+\|B^p\|_1$, valid for any $0<p\leq 1$ and any pair of positive operators $A$, $B$ s.t. $A^p$, $B^p$ are trace-class \cite{Ko84}, we obtain
\beq
\|\T\|_1\leq\|\T_E^+\|_1+\|\T_E^-\|_1 +\|\T_{\ka}^+\|_1+\|\T_{\ka}^-\|_1<\infty.  \label{lub3-new}
\eeq
Since $\T$ commutes with $J$, it has a $J$-invariant orthonormal basis of eigenvectors $\{\en_i\}_1^\infty$ 
and we denote the corresponding eigenvalues by $\{\tn_i\}_1^\infty$.

Now we proceed to the construction of the functionals $S_i\in\lin(\traceE\times\Gadd,\complex)$, to appear in the expansion
(\ref{Pizz-expansion}). Let $\oal^{\pm}=(\al_{1,2}^{\pm},\ldots,\al_{M-1,M}^{\pm})$ 
be  ${M \choose 2}$-tuples of multiindices and let $\oal=(\oal^+,\oal^-)$ be the corresponding $2{M \choose 2}$-multiindex.
First, we define the contribution to the functional which is responsible for the correlations between measurements:
\beqa
F_{\oal,\obe}(\uvx)
=\prod_{1\leq i<j\leq M}\fr{(-1)^{|\al_{i,j}^-|+|\al_{i,j}^+|}}{\sqrt{\al_{i,j}^+!\be_{i,j}^+!\al_{i,j}^-!\be_{i,j}^-!} }(\vac|a(\Lp \en_{\vx_i})^{\al_{i,j}^+} a^*(\Lp \en_{\vx_j})^{\be_{i,j}^+}\vac)\ph{4}& &\non\\
\ph{}\cdot(\vac|a(\Lm \en_{\vx_i})^{\al_{i,j}^-} a^*(\Lm \en_{\vx_j})^{\be_{i,j}^-}\vac),& & \label{F1}
\eeqa
where we use the short-hand notation $\Lpm \en_{i,\vx_j}=U(\vx_j)\Lpm \en_i$. The functionals in question are given by
\beqa
& &S_{\umu,\unu,\oal,\obe}(\fun,\uvx)=\fr{i^{|\umu^+|+|\unu^+|+2|\umu^-|} }{\umu!\unu!\sqrt{\oal!\obe!}}
F_{\oal,\obe}(\uvx)\fun(a^*(\LL \en_{\vx})^{\umu}a(\LL \en_{\vx})^{\unu}), \label{S1}
\eeqa
where $\fun\in\traceE$ and $\vxb\in\Gadd$.
The norms of these functionals satisfy the bound, stated in the following lemma, whose proof is postponed
to Section~\ref{technical-proofs}.
\bel\label{bound-S} The functionals $S_{\umu,\unu,\oal,\obe}\in\lin(\traceE\times\Gadd,\complex)$, given by (\ref{S1}),
satisfy the following estimates
\beq
\|S_{\umu,\unu,\oal,\obe}\|\leq\bigg(\fr{\M^{\fr{1}{2}(|\umu|+|\unu|)}}{\umu!\unu!}\tn^{\umu+\unu}\bigg)\bigg(\fr{1}{\sqrt{\oal!\obe!}}
\sqrt{\fr{|\oal^+|!|\oal^-|!|\obe^+|!|\obe^-|!}{\oal!\obe!}} g(\de)^{|\oal|+|\obe|}\tn^{\oal+\obe}\bigg), \label{Sestimate1}
\eeq
where $\M=\fr{E}{m}$, $\{\tn_i\}_1^{\infty}$ are the eigenvalues of the operator $\T$ given by (\ref{T}) and the function
$g$, which is independent of $\oal$ and $\obe$, satisfies $\lim_{\de\to\infty}g(\de)=0$.
\eel
\nin Given the estimates from Lemmas \ref{bound-tau} (b) and \ref{bound-S}, we can proceed to the study of
convergence properties of the expansion (\ref{Pizz-expansion}). For this purpose we need  some notation: For any
pair of ${M \choose 2}$-tuples of multiindices
$\oal^{\pm}=(\al_{1,2}^{\pm},\ldots,\al_{M-1,M}^{\pm})$ we define the associated $M$-tuples of multiindices $\ual^{\pm}$, $\uall^{\pm}$ as follows
\beqa
\uali^{\pm}&=&\sum_{\su{1<j\leq M \\ i<j}}\al_{i,j}^{\pm},\label{arrow1}\\
\ualli^{\pm}&=&\sum_{\su{1\leq j<M,\\ j<i}}\al_{j,i}^{\pm},\label{arrow2}
\eeqa
where $i\in\{1,\ldots, M\}$. The corresponding $2M$-multiindices are denoted by $\ual=(\ual^+,\ual^-)$, $\uall=(\uall^+,\uall^-)$.
The relevant estimate is stated in the following lemma, whose proof is given in Section~\ref{technical-proofs}.
\bel\label{convergence} The functionals $\tau_{\umu}\in(\mfa(\mcor)^{\otimes M})^*$ and 
$S_{\umu,\unu,\oal,\obe}\in\lin(\traceE\times\Gadd,\complex)$ satisfy
\beq
\sum_{\su{\umu,\unu \\ \oal,\obe \\  (|\oal|,|\obe|)\neq (0,0) }} \|\tau_{\umu+\unu+\ual+\ube}\| \, \|S_{\umu,\unu,\oal,\obe}\|<\infty
\eeq
for sufficiently large $\de>0$, depending on $M$, $E$ and the double cone $\mcor$. Moreover, the above
sum tends to zero with $\de\to\infty$.
\eel
\nin After this preparation we  proceed to the main part of this section.\\
\nin\bf Proof of Lemma \ref{technical}: \rm   We define  $\Pizz_{E,M,\de}\in\lin(\traceE\times\Gadd,(\mfa(\mcor)^{\otimes M})^*)$ as follows
\beq
\Pizz_{E,M,\de}(\fun,\vxb)=\sum_{\su{\umu,\unu \\ \oal,\obe \\  (|\oal|,|\obe|)\neq (0,0) }} \tau_{\umu+\unu+\ual+\ube}\, S_{\umu,\unu,\oal,\obe}(\fun,\vxb).
\eeq
In view of Lemma \ref{convergence} this map is well defined for sufficiently large $\de>0$ and satisfies
$\lim_{\de\to\infty}\|\Pizz_{E,M,\de}\|=0$ as required in part (a) of Lemma~\ref{technical}. In order
to verify part (b), it suffices to show that
\beqa
\Pizz_{E,M,\de}(\fun,\vxb)(W(f_1)\otimes\cdots\otimes W(f_M))=\Piz_{E,M,\de}(\fun,\vxb)(W(f_1)\times\cdots\times W(f_M))& &\non\\
=e^{-\fr{1}{2}\sum_{k=1}^M \|f_{k}\|_2^2} 
(e^{-\sum_{1\leq i<j \leq M}
(\lan f_{i,\xx_{i}}^+|f_{j,\xx_{j}}^+\ran+\lan f_{i,\xx_{i}}^-|f_{j,\xx_{j}}^-\ran )}  -1) & &\non\\
\cdot\fun(:W(f_{1,\xx_{1}}+\cdots+f_{M,\xx_{M}} ):),& &\label{Piz1}
\eeqa
where the second equality restates the definition of the map $\Piz_{E,M,\de}$, given by formula~(\ref{Piz}).
The l.h.s. can be evaluated making use of  Lemma~\ref{bound-tau} (a) and definition~(\ref{S1}) 
\beqa
& &\Pizz_{E,N,\de}(\fun,\uvx)(W(f_1)\times\cdots\times W(f_N))\non\\
&=&\!\!\!\!e^{-\fr{1}{2}\sum_{k=1}^M \|f_{k}\|_2^2}\!\!\!\!\!\!\!\!\! \sum_{\su{ \umu,\unu \\ \oal, \obe \\ (|\oal|,|\obe|)\neq (0,0) }}\!\!\!\!\!\!\!\!\!\!\fr{i^{|\umu^+|+|\unu^+|+2|\umu^-|} }{\umu!\unu!\sqrt{\oal!\obe!}}
\lan \en | f \ran^{\umu+\unu+\ual+\ube}F_{\oal,\obe}(\uvx)\fun(a^*(\LL \en_{\vx})^{\umu}a(\LL \en_{\vx})^{\unu}).\,\,\,\,\,\,\,\,\,
\quad \ \ \ \label{Piz21}
\eeqa
First, we consider the sum w.r.t.  $\umu,\unu$. There holds 
\beqa
 \sum_{\su{ \umu,\unu }}\fr{i^{|\umu^+|+|\unu^+|+2|\umu^-|} }{\umu!\unu!}
\lan \en | f \ran^{\umu+\unu}\fun(a^*(\LL \en_{\vx})^{\umu}a(\LL \en_{\vx})^{\unu})=\fun(:W(f_{1,\xx_{1}}+\cdots+f_{M,\xx_{M}} ):),
\quad\,\,\,\,\label{summunu}
\eeqa
as one can verify by expanding the normal ordered Weyl operator on the r.h.s. into the power series of
creation and annihilation operators of the functions $f_{j,\xx_{j}}^{\pm}$, expanding each function $f^{\pm}_j$
in the orthonormal basis $\{\en_i\}_1^{\infty}$ and making use of the multinomial formula~(\ref{multiform})
\beq
a^{(*)}(f^{\pm}_{j,\xx_j})^{m^{\pm}_j}=\sum_{\mu_j^{\pm},|\mu_j^{\pm}|=m^{\pm}_j}\fr{m_j^{\pm}!}{\mu_j^{\pm}!}
\lan \en | f_j \ran^{\mu_j^{\pm}}
a^{(*)}(\Lp \en_{\xx_j})^{\mu_j^{\pm}}.  \label{creation1}
\eeq
The sum w.r.t. $\oal,\obe$ in (\ref{Piz21}) gives
\beqa
\sum_{\su{\oal,\obe \\  (|\oal|,|\obe|)\neq (0,0) }} \fr{1}{\sqrt{\oal!\obe!}}\lan\, \en | f \ran^{\ual+\ube}F_{\oal,\obe}(\uvx)
=\big(\prod_{1\leq i<j \leq M}e^{-\lan f_{i,\xx_{i}}^+|f_{j,\xx_{j}}^+\ran} e^{-\lan f_{i,\xx_{i}}^-|f_{j,\xx_{j}}^-\ran}\big) - 1.
\quad\label{expo11} 
\eeqa 
This relation can be verified by expanding the exponential functions on the r.h.s. into the Taylor series, making
use of the identity
\beq
\lan f_{i,\xx_{i}}^{\pm}|f_{j,\xx_{j}}^{\pm}\ran^{k^\pm_{i,j}}=
\fr{(\vac|a(f_{i,\xx_{i}}^{\pm})^{k^+_{i,j}}a^*(f_{j,\xx_{j}}^{\pm})^{k^+_{i,j}}\vac)}{k^+_{i,j}!}
\eeq
and applying to the resulting expressions expansions (\ref{creation1}). Comparing (\ref{expo11}) and (\ref{summunu}) with (\ref{Piz1}) we conclude the proof of Lemma \ref{technical}. \qed
\section{Some Technical Proofs}\label{technical-proofs}
In this section we provide  proofs of Lemmas~\ref{bound-S} and \ref{convergence} which we
used to prove Lemma~\ref{technical} in the previous section.

The key ingredient of our proof of Lemma~\ref{bound-S} is
the observation that when the spatial distance between two local operators 
is large, then the energy transfer between them is heavily damped. We exploited
this idea in Section~\ref{coincidence-measurement}, where it was encoded in Lemma~\ref{mollifiers}.
In the present context it is more convenient to use a
 variant of Lemma~2.3 of \cite{BDL87} which is reproduced as Lemma~\ref{damping-1} in the present work.
 With the help of this result we prove the following  lemma which will help us to control 
the correlation terms $F_{\oal,\obe}$.
\bel\label{damping1} Let $\de>0$, $(\xx,\y)\in\Ga_{2,\de}$, $\{\en_i\}_{1}^{\infty}$ be the basis of the $J$-invariant eigenvectors of the operator $\T$ given by (\ref{T}), let $\{\tn_i\}_1^\infty$ be the corresponding eigenvalues  and let $\al,\be$ be multiindices. Then there holds, for any combination of $\pm$ signs,
\beq
|(\vac|a(\Lpm \en_{\xx})^{\al}a^*(\Lpm \en_{\y})^{\be}\vac)|\leq \sqrt{|\al|!|\be|!} g(\de)^{|\al|+|\be|}
\tn^{\al+\be},\label{correlations}
\eeq
where $\Lpm \en_{i,\xx}:=U(\xx)\Lpm \en_i$, the function $g$  is independent of $\al$, $\be$ and satisfies $\lim_{\de\to\infty}g(\de)=0$.
\eel
\proof We consider here only the (++) case, as the remaining cases are treated analogously.
We define the operators $\phi^{+}(\en_i)=\fr{1}{\sqrt{2}}\big(a^*(\Lp \en_i)+a(\Lp \en_i)\big)$ and their
translates $\phi^{+}(\en_i)(\xx)=U(\xx)\phi^{+}(\en_i)U(\xx)^{-1}$. Since the projections $\Lpm$ commute with
$J$, $J\en_i=\en_i$ and $\de>0$, these operators satisfy the assumptions of Lemma~\ref{damping-1}. (See the proof of 
Lemma~\ref{Cook} for a detailed justification).
Therefore, we obtain
\beqa
& &\lan \Lp \en_{i,\xx}| \Lp \en_{j,\y}\ran=2(\vac|\phi^{+}(\en_i)(\xx)\phi^{+}(\en_j)(\y)\vac)\non\\
& &\ph{444444}=(\vac|\phi^{+}(\en_i)(\xx)\f(\de H)\phi^{+}(\en_j)(\y)\vac)
+(\vac|\phi^{+}(\en_j)(\y)\f(\de H)\phi^{+}(\en_i)(\xx)\vac)\non\\
& &\ph{444444}=\fr{1}{2}\big(\lan \Lp \en_{i,\xx}|\f(\de\om)\Lp \en_{j,\y}\ran+\lan \Lp \en_{j,\y}|\f(\de\om)\Lp \en_{i,\xx}\ran\big),
\eeqa
where the function $\f$ was defined in Lemma~\ref{damping-1}.
Making use of this result, exploiting  the fact that the l.h.s. of (\ref{correlations}) vanishes for $|\al|\neq |\be|$
and setting $|\al|=|\be|=k$, we get
\beqa
(\vac|a(\Lp \en_{\xx})^{\al}a^*(\Lp \en_{\y})^{\be}\vac)\ph{444444444444444444444444444444444444444}\non\\
=(\vac|a(\Lp \en_{i_1,\xx})\ldots a(\Lp \en_{i_k,\xx})a^*(\Lp \en_{j_1,\y})\ldots a^*(\Lp \en_{j_k,\y})\vac)& &\non\\
=\sum_{\si\in S_k}\lan \Lp \en_{i_1,\xx}| \Lp \en_{j_{\si_1},\y}\ran \ldots \lan \Lp \en_{i_k,\xx}| \Lp \en_{j_{\si_k},\y}\ran& &\non\\
=\sum_{\si\in S_k}\fr{1}{2}\big(\lan \Lp \en_{i_1,\xx}|\f(\de\om)\Lp \en_{j_{\si_1},\y}\ran+\lan \Lp \en_{j_{\si_1},\y}|\f(\de\om)\Lp \en_{i_1,\xx}\ran\big)& &\non\\   
\ldots \fr{1}{2}\big(\lan \Lp \en_{i_k,\xx}|\f(\de\om)\Lp \en_{j_{\si_k},\y}\ran+\lan \Lp \en_{j_{\si_k},\y}|\f(\de\om)\Lp \en_{i_k,\xx}\ran \big), & &\quad\quad
\eeqa
where the sum extends over all permutations of a $k$-element set. For any $0<\ka<1$ there holds
$c_{\f}^2:=\sup_{\om}|\f(\om)e^{-|\om|^{\ka}}|<\infty$. Consequently, we get
\beqa
|\lan \Lp \en_{i,\xx}|\f(\de\om)\Lp \en_{j,\y}\ran|&=&
|\lan \Lp \en_{i,\xx}|\f(\de\om) e^{(\de|\om|)^{\ka}} e^{-(\de^{\ka}-1)|\om|^{\ka}}e^{-|\om|^{\ka}}\Lp \en_{i,\y}\ran|\non\\
&\leq& c_{\f}^2e^{-(\de^{\ka}-1) m^{\ka}} \|e^{-\fr{|\om|^\ka}{2}}\Lp \en_i\|\,\|e^{-\fr{|\om|^\ka}{2}}\Lp \en_j\|.
\eeqa
Finally, we note that 
$\|e^{-\fr{|\om|^\ka}{2}}\Lp \en_i\|=\|\T_{\ka}^+\en_i\|\leq \|\T \en_i\|=\tn_i $
and the claim follows. \qed\\
After this preparation we proceed to the proof of Lemma~\ref{bound-S}.\\
\bf Proof of Lemma~\ref{bound-S}: \rm Exploiting the energy bounds~(\ref{energy-bounds}) and the fact that 
$\|\omm\|=m^{-\h}$ in the massive theory, we obtain
\beqa
|\fun(a^*(\LL \en_{\vx})^{\umu}a(\LL \en_{\vx})^{\unu})|&\leq& \M^{\fr{1}{2}(|\umu|+|\unu|)}\|Q_E\LL \en\|^{\umu}\,
\|Q_E\LL \en\|^{\unu}\non\\
&\leq& \M^{\fr{1}{2}(|\umu|+|\unu|)}\tn^{\umu+\unu}.\label{mubound1}
\eeqa
Next, with the help of  Lemma~\ref{damping1} we analyze the expressions $F_{\oal,\obe}$, given by~(\ref{F1}),
\beqa
|F_{\oal,\obe}(\uvx)|\leq \prod_{1\leq i<j\leq M}\sqrt{\fr{|\al_{i,j}^+|!|\be_{i,j}^+|!|\al_{i,j}^-|!|\be_{i,j}^-|! }
{\al_{i,j}^+!\be_{i,j}^+!\al_{i,j}^-!\be_{i,j}^-!}}(g(\de)\tn)^{\al_{i,j}^++\be_{i,j}^++\al_{i,j}^-+\be_{i,j}^-}& &\non\\
\leq\sqrt{\fr{|\oal^+|!|\oal^-|!|\obe^+|!|\obe^-|!}{\oal!\obe!}} g(\de)^{|\oal|+|\obe|}\tn^{\oal+\obe},& &\label{Fbound1}
\eeqa
where we made use of the estimate $\prod_{1\leq i<j\leq M}|\al_{i,j}^+|!\leq (\sum_{1\leq i<j\leq M}|\al_{i,j}^+|)!=|\oal^+|!$. Altogether, combining~(\ref{mubound1}) and (\ref{Fbound1}), we obtain from (\ref{S1}) the bound~(\ref{Sestimate1}). \qed\\
We conclude this appendix with the proof of Lemma~\ref{convergence}.\\ 
\bf Proof of Lemma \ref{convergence}: \rm  First, we estimate the norms of the functionals $\tau_{\umu+\unu+\ual+\ube}$.
Making use of the bound stated in Lemma~\ref{bound-tau}~(b),  and of the fact 
that $(a+b+c)!\leq 3^{a+b+c}a!b!c!$ for any $a,b,c\in\nat_0$, which follows from the properties of the multinomial 
coefficients, we get
\beqa
\|\tau_{\umu+\unu+\ual+\ube}\|&\leq& 4^{|\umu|+|\unu|+|\ual|+|\ube|}\sqrt{(\umu+\unu+\ual+\ube)!}\non\\
&\leq& \bigg( (4\sqrt{3})^{|\umu|+|\unu|}\sqrt{(\umu+\unu)!}\bigg)\bigg((4\sqrt{3})^{|\oal|+|\obe|} \sqrt{\ual!\ube!}\bigg), \label{prodstateestimate1}
\eeqa
where we noted that $|\ual|=|\oal|$ and $|\ube|=|\obe|$. (See definitions (\ref{arrow1}) and (\ref{arrow2})). 
The factor $\sqrt{ {\ual!\ube!} }$  in this bound will be controlled by the factor $\sqrt{\oal!\obe!}$
appearing in the denominator in (\ref{Sestimate1}). We note the relevant estimate
\beqa
\fr{\ual!}{\oal!}=\fr{\ual^+!}{\oal^+!}\fr{\ual^-!}{\oal^-!}
&=&\prod_{i=1}^M\fr{(\sum_{\su{1<j\leq M,\, j>i}}\al_{i,j}^+)!}
{(\prod_{1<j\leq M,\, j>i}\al_{i,j}^+)!}\fr{(\sum_{1<j\leq M,\, j>i}\al_{i,j}^-)!}{(\prod_{1<j\leq M,\, j>i}\al_{i,j}^-)!}
\non\\
&\leq& M^{\sum_{1\leq i<j\leq M}(|\al_{i,j}^+|+|\al_{i,j}^-|) }\leq M^{|\oal|}, \label{multiindexest1}
\eeqa
where we made use of the properties of the multinomial coefficients. Similarly, the factor $\sqrt{(\umu+\unu)!}$
appearing in (\ref{prodstateestimate1}) will be counterbalanced by $\sqrt{\umu!\unu!}$ extracted from the denominator of (\ref{Sestimate1}). The relevant estimate relies on the property of the binomial coefficients
\beq
\fr{(\umu+\unu)!}{\umu!\unu!}\leq 2^{|\umu|+|\unu|}.
\eeq
With the help of the last two bounds and relations (\ref{Sestimate1}), (\ref{prodstateestimate1}) we obtain
\beqa
 \sum_{\su{\umu,\unu \\ \oal,\obe \\  (|\oal|,|\obe|)\neq (0,0) }} \|\tau_{\umu+\unu+\ual+\ube}\|\, \|S_{\umu,\unu,\oal,\obe}\|
\leq 
\bigg(\sum_{\umu,\unu}\fr{(4\sqrt{6\M})^{|\umu|+|\unu|} }{\sqrt{\umu!\unu!}}\tn^{\umu+\unu}\bigg)& & \non\\
\cdot\bigg(\!\!\!\!\!\!\!\!\sum_{\su{\oal,\obe \\  (|\oal|,|\obe|)\neq (0,0) } }\!\!\!\!\!\!\!\!\sqrt{\fr{|\oal^+|!|\oal^-|!|\obe^+|!|\obe^-|! }{\oal!\obe!}} (4\sqrt{3M}g(\de))^{|\oal|+|\obe|}\tn^{\oal+\obe}\bigg),& & \label{pnorm1}
\eeqa
where  we made use of the bound (\ref{multiindexest1}). The sum w.r.t. $\umu,\unu$ can
be easily estimated as it factorizes into $4M$ independent sums: Let $\mu$ be an ordinary multiindex, then
\beqa
\bigg(\sum_{\umu,\unu}\fr{(4\sqrt{6\M})^{|\umu|+|\unu|} }{\sqrt{\umu!\unu!}}\tn^{\umu+\unu}\bigg)
&=&\bigg(\sum_{\mu}\fr{(4\sqrt{6\M})^{|\mu|} }{\sqrt{\mu!}}\tn^{\mu}\bigg)^{4M}\non\\ 
&\leq&\bigg(\sum_{k=0}^{\infty}\fr{(4\sqrt{6\M})^{k} }{\sqrt{k!}}\sum_{\mu,|\mu|=k}\fr{|\mu|!}{\mu!}\tn^{\mu}\bigg)^{4M}\non\\
&\leq&\bigg(\sum_{k=0}^{\infty}\fr{(4\sqrt{6\M}\|\T\|_1)^{k}}{\sqrt{k!}}\bigg)^{4M},
\eeqa
where in the second step we made use of the fact that the multinomial coefficients are greater than or equal to one
and in the last step we used  the multinomial formula~(\ref{multiform}). Clearly, the last sum is convergent. (As
a matter of fact it would suffice to consider $k\leq \M$ since $S_{\umu,\unu,\oal,\obe}$, given by formula
(\ref{S1}), vanishes for $|\umu|>\M$ or $|\unu|>\M$). As for the sum w.r.t. $\oal$, $\obe$ on the r.h.s. of
(\ref{pnorm1}), it suffices to study the case $|\oal^+|\neq 0$. Then it factorizes into four independent
sums  and we discuss here one of the factors
\beqa
& &\sum_{\oal^+,|\oal^+|\neq 0}\sqrt{\fr{|\oal^+|!}{\oal^+!}}  (4\sqrt{3M}g(\de))^{|\oal^+|}\tn^{\oal^+}
\non\\
& &\ph{44444444444444}=\!\!\!\!\!\! \sum_{\su{\oal^+,|\oal^+|\neq 0 }}\!\!\!\!\!\!
(4\sqrt{3M}g(\de))^{|\oal^+|}\sqrt{\fr{(|\al^+_{1,2}|+\cdots+|\al^+_{M-1,M}|)!}{\al^+_{1,2}!\ldots \al^+_{M-1,M}!}}
\tn^{\oal^+}\non\\
& &\ph{44444444444444}\leq \sum_{\su{\oal^+, |\oal^+|\neq 0}}
(4\sqrt{3M^3}g(\de))^{|\oal^+|}\fr{|\al^+_{1,2}|!}{\al^+_{1,2}!}\ldots\fr{|\al^+_{M-1,M}|!}{\al^+_{M-1,M}!}\tn^{\oal^+}\non\\
& &\ph{44444444444444}\leq \sum_{\su{k^+_{1,2},\ldots,k^+_{M-1,M} \\ \sum_{1\leq i<j\leq M} k^+_{i,j}\neq 0}}
\prod_{1\leq i<j\leq M}(4\sqrt{3M^3}g(\de))^{k^+_{i,j}}\sum_{\al^+_{i,j},|\al^+_{i,j}|=k^+_{i,j}}\fr{|\al^+_{i,j}|!}{\al^+_{i,j}!}\tn^{\al^+_{i,j}}\non\\
& &\ph{44444444444444}\leq \sum_{\su{k^+_{1,2},\ldots,k^+_{M-1,M} \\ \sum_{1\leq i<j\leq M} k^+_{i,j}\neq 0}}(4\sqrt{3M^3}g(\de)\|\T\|_1)^{(k^+_{1,2}+\cdots+k^+_{M-1,M})}. \label{last1}
\eeqa
In the second step we made use of the fact that 
\beq
\fr{(|\al^+_{1,2}|+\cdots+|\al^+_{M-1,M}|)!}{|\al^+_{1,2}|!\ldots |\al^+_{M-1,M}|!}\leq M^{2(|\al^+_{1,2}|+\cdots+|\al^+_{M-1,M}|)}
\eeq
and in the last step we exploited the multinomial formula~(\ref{multiform}). The last expression on the r.h.s. of (\ref{last1}) 
is a convergent geometric series for sufficiently large $\de$ and it tends to zero with $\de\to\infty$,  since $\lim_{\de\to\infty}g(\de)=0$. \qed\\

%% file: PhD-Conventions.tex
\chapter*{Notational Conventions} \label{Conventions} 
\thispagestyle{empty}
\addcontentsline{toc}{chapter}{{Notational Conventions}}

\bf Minkowski space notation: \rm We work in Minkowski spacetime $\real^{s+1}$. We use
 the Minkowski metric $\eta_{\mu\nu}$ with signature $(+,\underbrace{-,\ldots,-}_{s})$. The
Minkowski scalar product of two four-vectors is given by $p\cdot x=p^0x^0-\vep\vx$, where
$\vep\vx=\sum_{i=1}^s p_ix_i$.\\

\nin\bf Fourier transform: \rm Consider the Schwartz-class functions
$f\in S(\real^{s+1})$, $g\in S(\real^{s})$ and $h\in S(\real)$. ($h$ is understood as a function of time).
We define their Fourier transforms as follows
\beqa
\tilde{f}(p)&=&(2\pi)^{-\fr{s+1}{2}}\int d^{s+1}x\, e^{ip^0x^0-i\vep\vx} f(x^0,\vx),\non\\
\tilde{g}(\vep)&=&(2\pi)^{-\fr{s}{2}}\int d^sx\, e^{-i\vep\vx} g(\vx),\non\\
\tilde{h}(p^0)&=&(2\pi)^{-\h}\int dx^0\, e^{ip^0x^0} h(x^0).\non
\eeqa 
The Fourier transforms of  distributions $F\in S^\prime(\real^{s+1})$, $G\in S^\prime(\real^{s})$
and $H\in S^\prime(\real)$ are defined according to the relations $\tilde{F}(\tilde{f})=F(f)$,
$\tilde{G}(\tilde{g})=G(g)$ and $\tilde{H}(\tilde{h})=H(h)$. In the formal notation
\beqa
\tilde{F}(p)&=&(2\pi)^{-\fr{s+1}{2}}\int d^{s+1}x\, e^{-ip^0x^0+i\vep\vx} F(x^0,\vx),\non\\
\tilde{G}(\vep)&=&(2\pi)^{-\fr{s}{2}}\int d^sx\, e^{i\vep\vx} G(\vx),\non\\
\tilde{H}(p^0)&=&(2\pi)^{-\h}\int dx^0\, e^{-ip^0x^0} H(x^0).\non
\eeqa

\newpage

\nin\bf Additional conditions: \rm\\

\begin{tabular}[]{ll}
Condition  &  Reference \\
\hline
\\
Condition~$A$    	&   Page   \pageref{cond-A} \\
Condition~$A^\prime$    &   Page   \pageref{cond-Ap} \\
Condition~$\Cs$    	&   Page   \pageref{cond-Cs} \\
Condition~$\Csq$    	&   Page   \pageref{cond-Csq} \\
Condition~$\Cnat$    	&   Page   \pageref{cond-Cnat} \\
Condition~$\B$    	&   Page   \pageref{cond-L1} \\
Condition~$\Bs$    	&   Page   \pageref{cond-L1s} \\
Condition~$\A$    	&   Page   \pageref{cond-L2} \\
Condition~$M$    	&   Page   \pageref{cond-M} \\
Condition~$\Ns$    	&   Page   \pageref{cond-Ns} \\
Condition~$\Nnat$       &   Page   \pageref{cond-Nnat} \\
Condition~$R$    	&   Page   \pageref{cond-R} \\
Condition~$S$    	&   Page   \pageref{cond-S} \\
Condition~$T$    	&   Page   \pageref{cond-T} \\
Condition~$\V$    	&   Page   \pageref{cond-V} \\
\\
\end{tabular}	

\nin\bf Frequently used symbols: \rm
 \thispagestyle{empty}
\begin{center}
\begin{longtable}[]{lp{11cm}l}
Symbol		&	Description				& Reference							\\	
\hline																		\\
$\|\,\cdot\,\|_{E,2}$ & Square-integrability seminorm on $\hmfa_{\ac}$  	& Page \pageref{square-integrable}\\
$\|\,\cdot\,\|_{E,1}$ & Integrability seminorm on $\CC$  	        & Page \pageref{symbol-integrable-semi}\\
$\nor\,\cdot\,\nor_{p}$ & $p$-norm w.r.t. the norm $\nor\,\cdot\,\nor$	& Page \pageref{symbol-pnorm}\\
$\|\,\cdot\,\|_p$ & Norm on $L^{p}(\real^s)$ / $p$-norm of an operator on $L^{2}(\real^s,d^sp)$ & Page  \pageref{Sobolev-norm}/\pageref{symbol-Hilbert-pnorm}\\
$\|\,\cdot\,\|_{\x}$ & Special norms on $\lin(\traceE,X)$		& Page \pageref{Nnorm}\\
$\|\,\cdot\,\|_{p,\x}$ & $p$-norm w.r.t. the norm $\|\,\cdot\,\|_{\x}$  & Page \pageref{symbol-pnormx}\\
$\|\,\cdot\,\|_{2,l}$ & Sobolev norms     & Page \pageref{Sobolev-norm}\\
$[\,\,\,\cdot\,\,\,]$ & Closure in $L^2(\real^s,d^sp)$ & Page \pageref{symbol-Lpm}\\
$a^*(\vep),a(\vep)$ & Creation and annihilation operators of a mode $\vep$	& Page \pageref{a(p)}\\
$a^*(f), a(f)$ & Creation and annihilation operators of a function $f$		& Page \pageref{symbol-a(f)}\\
$\uA$ & Observable $A$ restricted to $\ehil$ 					& Page \pageref{symbol-epi}\\
$\hmfa$	     &  Algebra of local observables	                               & Page   \pageref{symbol-hmfa} \\
$\dmfa$ & Algebra of observables generated by derivatives of the free field                      	& Page \pageref{symbol-dmfa}\\ 
$\eumfa$ & Algebra of observables of the even part of free field theory acting on $\ehil$        	& Page \pageref{symbol-eumfa}\\
$\emfa$ & Algebra of observables of the even part of free field theory acting on the (full) Fock space  & Page \pageref{symbol-eumfa}\\
$\mfa$	     &	Global algebra of observables                                  & Page   \pageref{symbol-mfa} \\	
$\mfar(\mco)$ & $*$-algebra of finite, linear combinations of Weyl operators & Page \pageref{symbol-mfar}\\
$\mfa(\mco)$ &	Local algebra attached to $\mco$                               & Page   \pageref{symbol-loc} \\	
$\al_x$ & Translation automorphisms                                            & Page	\pageref{symbol-al} \\
$\eal_x$ & Translation automorphisms of free field theory restricted to $\ehil$	& Page \pageref{symbol-eal}\\
$\wt{A}(\vep)$ & Fourier transform of a local observable			& Page \pageref{Fourier1}\\
$\hmfa_{\spp} $ & Pure-point subspace of  $\hmfa$ 				& Page \pageref{symbol-pure-point}\\
$\hmfa_{\scc} $ & Continuous subspace of $\hmfa$ 				& Page \pageref{symbol-continuous}\\
$\mfa_{\scc}(\mco) $ & Continuous subspace of $\mfa(\mco)$ 			& Page \pageref{symbol-local-continuous}\\
$\hmfa_{\ac} $ & Absolutely continuous subspace of $\hmfa$ 			& Page \pageref{def-ac}\\
$\hmfa_{\pc} $ & Point-continuous subspace of  $\hmfa$ 				& Page \pageref{symbol-point-continuous}\\
$\CC$                 & Space of integrable observables 	        & Page \pageref{symbol-integrable-space}\\
$\Ba$  & $N$-linear forms on $\mfa(\mco)$				& Page \pageref{symbol-Ba}\\
$B(\hil)$  & Bounded operators on $\hil$                                       & Page	\pageref{symbol-Bhil} \\
$\bti_{\ka,r}^\pm$ & Functions appearing in the Taylor expansion of $\tF^{\pm}$  & Page \pageref{b}\\
$\mfc$ & Algebra of particle detectors					& Page \pageref{symbol-particle-detectors}\\
$\compsup$ & $\complex^N$ equipped with the supremum norm			& Page \pageref{symbol-compsup}\\
$C_0^{\infty}(\real^s)$ & Smooth, compactly supported functions on $\real^s$ & \\ 
$\BE$  & Vectors of bounded energy		                               & Page	\pageref{symbol-BE} \\
$\DF$  & Subspace of finite particle vectors in Fock space				& Page \pageref{symbol-DF}\\
$D_S$ & Subspace of vectors with Schwartz-class wavefunctions in $\DF$ 		& Page \pageref{symbol-DS}\\
$D(\mco_r)$  & Smooth functions supported in $\mco_r$			& Page \pageref{symbol-DOr}\\
$\{e_j\}_1^\infty$ & Basis of $J$-invariant eigenvectors of $T$ & Page \pageref{symbol-J-basis}\\
$\Phi_{\FH}$ & Field content						& Page \pageref{symbol-field-content}\\
$\Phi^{\te{(d)}}_{\FH} $ & Field content of $(\dmfa,\al,\hil)$  		& Page \pageref{symbol-Phid}\\
$\Phi^{\te{(e)}}_{\FH} $ & Field content of $(\eumfa,\eal,\ehil)$   		 & Page \pageref{symbol-Phie}\\
$\fin(V,W)$ & Finite rank maps from Banach space $V$ to $W$		& Page \pageref{symbol-fin}\\
$\fin(V\times\Ga,W)$ & Finite rank maps form $V\times\Ga$ to $W$       & Page \pageref{symbol-lin-times}\\
$\fun_{\bx}$  & Special elements of $\Ba$				& Page \pageref{forms}\\
$\phipm$  &  Canonical field and momentum 				& Page \pageref{canonical-field}\\
$\wil\phipm^n\wir$ & Wick powers of the fields $\phipm$				& Page \pageref{symbol-Wick}\\
$\Gad$  & Set of admissible configurations of bounded regions		& Page \pageref{symbol-Gad}\\
$H$     & Hamiltonian			                                       & Page	\pageref{symbol-H} \\
$\hil$  & Hilbert space			                                       & Page	\pageref{symbol-hil} \\
$\ehil$ & Even part of the Fock space 						& Page \pageref{symbol-ehil}\\
$\chi_E$ & Approximate characteristic function of $\{\, \vep\in\real^s\, |\, \om(\vep)\leq E\}$ & Page \pageref{symbol-chiE}\\
$\chir$ & Approximate characteristic function of $\mco_r$ & Page \pageref{symbol-chir}/\pageref{symbol-chir-next}\\
$\hti_{\ka,E}^\pm$ & Functions appearing in the Taylor expansion of $\tF^{\pm}$  & Page \pageref{h}\\
$h_{\rzero}$ & Special functions from $D(\mco_{\rzero})$ & Page \pageref{symbol-hrzero}\\
$\ei$ & Isometric embedding $\trace^{\te{(e)}}\hookrightarrow\trace$ 		& Page \pageref{symbol-ei}\\
$J$  & Complex conjugation in $L^2(\real^s,d^sx)$   & Page \pageref{symbol-J}\\
$\ka$ & $s$-index & Page \pageref{symbol-ka}\\
$\lin(V,W)$ & Linear maps from Banach space $V$ to $W$			& Page \pageref{symbol-lin}\\
$\lin(V\times\Ga,W)$ & Maps form $V\times\Ga$ to $W$, linear in the first argument & Page \pageref{symbol-lin-times}\\
$\lin(V\times\Ga,W)$ & Maps form $V\times\Ga$ to $W$, linear in the first argument & Page \pageref{symbol-lin-times}\\
$\Lring,\Lpmring$  & Special subspaces of $L^2(\real^s,d^sp)$		& Page \pageref{symbol-Lpmring}\\
$L^2(\real^s,d^sx)$ & Square-integrable functions on $\real^s$ & Page \pageref{symbol-Sobolev-space} \\
$L^2(\real^s,d^sx)_l$ & Sobolev space & Page \pageref{symbol-Sobolev-space}\\
$\Lpm$ & Special subspaces/projections on $L^2(\real^s,d^sp)$ & Page \pageref{symbol-Lpm}\\
$\mub$  & Pair of multiindices  					& Page \pageref{symbol-2multi}\\
$\Sett$ & Finite subset of pairs of multiindices & Page \pageref{symbol-sett}\\
$\N(\eps)$  & $\eps$-content of a map/set				& Page \pageref{epsilon-content}/\pageref{symbol-eps-content-set}\\
$\mco$  & Open bounded region of $\real^{s+1}$                                 & Page	\pageref{symbol-mco} \\
$\mco(r)$  & Double cone of radius $r$	                                       & Page	\pageref{symbol-mco(r)} \\
$\mco_r$  & Ball in $\real^s$ of radius $r$                                    & Page	\pageref{symbol-mcor} \\
$\om_0$   & Vacuum state (or translationally invariant state)                  & Page	\pageref{symbol-om0} \\
$\ord(A)$  & Infrared order of $A\in\hmfa_{\scc}$	\thispagestyle{empty}   & Page \pageref{infrared-order}\\
$\Ord(A)$  & $\{\, \ord(A)\, |\, A\in\hmfa_{\scc}\,\}$			        & Page \pageref{infrared-order}\\
$\om^{\vxb}$  & Special elements of $\mfa(\mco)^*\ot\compsup$		& Page \pageref{symbol-omvxb}\\
$\om(\vep)$ & $\sqrt{|\vep|^2+m^2}$ & Page \pageref{symbol-om(p)}\\
$\cpo$  & Poincar\'e group 			                               & Page	\pageref{symbol-cpo} \\
$\veP$  & Momentum operators		                                       & Page	\pageref{symbol-veP} \\
$P_E$  & Spectral projection of $H$ on $\{\,\om\in\real_+\,|\,\om\leq E\,\}$   & Page	\pageref{symbol-PE} \\
$\PC$ & Particle content						& Page \pageref{symbol-particle-content}\\
$\Pi_E, \whXi_E, \Xi_{\be}$ & Maps appearing in Condition $\Cs$  	& Page \pageref{symbol-PiE}\\
$\Pi_{E,N,\de}$ & Maps appearing in Lemma~\ref{equivalence}		& Page \pageref{symbol-Gad}\\
$\Pi_{E,N,\de}^{\scc}$ & Map appearing in Condition~$\Csq$		& Page \pageref{symbol-PiENdc}\\
$\Ppr$ & Spectral projection on the ball of radius $r$ centered at $r$  & Page \pageref{symbol-Ppr}\\
$\Pi_E^{\scc}$ & Maps appearing in Condition~$\Nnat$			& Page \pageref{symbol-PiEc}\\
$\epi$ & Representation of  $\emfa$ in $B(\ehil)$  							& Page \pageref{symbol-epi}\\
$Q_E$ & Projection on vectors of energy below $E$ in  $L^2(\real^s,d^sp)$   & Page \pageref{symbol-QE} \\
$R$ & $(1+H)^{-1}$ / Parameter appearing in Lemma~\ref{AHRB-lemma}      & Page \pageref{symbol-R}/\pageref{symbol-new-R}\\
$R^{(2)}$ & Map appearing in Theorem~\ref{HA-in-models} & Page \pageref{R2-in-models}\\
$S(\real^s)$ & Schwartz-class functions on $\real^s$ & \\
$S_E$  & States from $\traceE$		                                       & Page	\pageref{symbol-SE} \\
$S_{E}(\mco)$ & Special sets from the ranges of $\Pi_E$  		& Page \pageref{set-SEm}\\
$S_{E,N.\de}(\mco)$ & Special sets from the ranges of $\The_{E,N,\de}$	& Page \pageref{symbol-SENde}\\
$\Sp_{B}(\al_{\real^{s+1}})$  &  Arveson spectrum of $B\in\mfa$ w.r.t. $\al_x$ & Page	\pageref{symbol-Arveson} \\
$\si^{(t)}_{\fun}$ & Asymptotic functional approximants			& Page \pageref{symbol-asymptotic-functional-appr}\\
$\si^{(+)}_{\fun}$ & Asymptotic functionals  				& Page \pageref{symbol-asymptotic-functional}\\
$\si_{\mup,\mum}$ & Special functionals on $\mfa(\mco)$ in free field theory & Page \pageref{funcbound}\\
$\bsi^{(r)}_{\mupi,\mumi}$ &  Special functionals on $\mfa(\mco)$ in free field theory  & Page \pageref{symbol-very-spacial-functionals}\\
$\The_{E,N,\de}$  & Maps appearing in Condition~$\Cnat$			& Page \pageref{theta}\\
$s$ & Dimension of space & \thispagestyle{empty}\\
$\Set$ & Suitable subset of $\nat_{0}^{2}$ & Page \pageref{symbol-set}\\
$\cS_{\mub,\nub}$ & Operators appearing in expansion (\ref{thec-expansion1}) of the map $\thec$ & Page \pageref{cS}\\
$\Smn$ &  Operators appearing in expansion (\ref{lastexpansion}) of the map $\thet$ & Page \pageref{Smn-def}\\
$\trace$  & Trace-class operators on $\hil$                                    & Page	\pageref{symbol-trace} \\
$\traceE$  & $P_E\trace P_E$		                                       & Page	\pageref{symbol-traceE} \\
$\traceEP$  & Positive elements from $\traceE$                                   & Page	\pageref{symbol-traceE+} \\
$\trace_{\infty}$ & Functionals of polynomially damped energy		& Page \pageref{symbol-trace-infty}\\
$\trace_{(p,r)}$ &  $\Ppr\trace\Ppr$					& Page \pageref{symbol-traceppr}\\
$\trace^{\te{(e)}}$ & Predual of $B(\ehil)$					& Page \pageref{symbol-tracee}\\
$\tau_{\mub,\nub}$ &  Functionals  appearing in expansion (\ref{lastexpansion}) of the map $\thet$ & Page \pageref{functionals1}\\
$\ctau_{\mub,\nub}^{(r)}$  & Functionals  appearing in expansion (\ref{thec-expansion1}) of the map $\thec$ & Page \pageref{ctau}\\ 
$\htau_{\mupi,\mumi}^{(r)}$ & Functionals appearing in expansion (\ref{finiterank0}) of the map $\theh$  & Page \pageref{htau}\\
$T^{00}$ & Stress-energy tensor						& Page \pageref{symbol-T00}\\
$\TEpm,\Thpm$ & Suitable nuclear operators on $L^2(\real^s,d^sp)$ & Page \pageref{symbol-Toper}\\
$T$ & $(|\TEp|^2+|\TEm|^2+|\Thp|^2+|\Thm|^2)^\h$ & Page \pageref{T-definition}\\
$\{t_j\}_1^\infty$ & Eigenvalues of $T$ & Page \pageref{symbol-T-eigenvectors}\\
$\theo$ & Map appearing in decomposition (\ref{symbol-first-decomposition}) of $\whXi_E$ & Page \pageref{onesum}\\
$\thet$ & Map appearing in decomposition (\ref{symbol-first-decomposition}) of $\whXi_E$ & Page \pageref{twosums}\\
$\theh$ & Map appearing in decomposition (\ref{symbol-second-decomposition}) of $\whXi_E$ & Page \pageref{thehdef}\\
$\thec$ & Map appearing in decomposition (\ref{symbol-second-decomposition}) of $\whXi_E$ & Page \pageref{thecdef}\\
$U$   & Unitary representation of translations in $\hil$               		& Page	\pageref{symbol-U} \\
$W(f)$	& Weyl operator  						& Page \pageref{Weyl-operator}\\

\end{longtable}	

\end{center}

\thispagestyle{empty}

%% file: PhD-Acknowledgements.tex
\addcontentsline{toc}{chapter}{Acknowledgements}
\thispagestyle{plain}
\vspace*{6em}
\noindent 
{\Huge \bf Acknowledgements}
\vspace{3.5em}\\

\noindent I would like to thank Prof.~D.~Buchholz for formulating the interesting problem
in the field of particle structures and automorphism groups, and giving me the opportunity to 
work on it.
Without his advice and encouragement this work would not have been possible. In particular,
I am indebted to him for suggesting to me the approach to the study of the vacuum structure pursued 
in Section~\ref{vacuum-structure} and for pointing out to me his observations summarized 
in Section~\ref{QM2}.


I am also grateful to Prof.~K.~Fredenhagen for his immediate consent to write the 
additional report.

Thanks are  due to my MSc Thesis adviser, Prof.~J.~Derezinski, for introducing me to the
subject of scattering in quantum field theory and to  Prof.~J.~Bros for an outline of its
early developments. I am also grateful to Prof.~S.~Gierowski for pointing out to me the value of
abstract thinking.




In the course of this work I benefited from discussions with many past or present members
and visitors  of the Quantum Field Theory Group in G\"ottingen. 
In particular, I would like to mention 
Jan Schlemmer, Ansgar Schneider, Antonia Miteva, Helmut H\"olzler, Gandalf Lechner, Nikolay Nikolov,
Katarzyna Rejzner, Julian Pook and Daniela Cadamuro.

Many thanks to my family: My parents for their numerous visits to G\"ottingen and holidays we spent together.
My grandparents for encouraging telephone conversations.  My brother 
\L ukasz for endless discussions about art and science.

Finally, I would like to thank great friends who supported me at various stages of this work:
Jasmin Kernspecht, Kostas Daoulas, Aneta Kieba\l a, Anna Podlewska and Christian Podlewski.

Financial support from EC Research Training Network 'Quantum Spaces - Non-commutative Geometry',
Deutsche Forschungsgemeinschaft and Graduiertenkolleg "Mathematische Strukturen in der modernen Quantenphysik"
is gratefully acknowledged. I also acknowledge travel grants from
the Research Institute in Oberwolfach, Max Planck Institute for Mathematics in Bonn and Universities of Mainz, 
Hamburg and Leipzig, as well as from Wilhelm und Else Heraeus-Stiftung. A fellowship supported by the Austrian 
Federal Ministry of Science and Research, the High Energy Physics Institute of the Austrian Academy of Sciences and
the Erwin Schr\"odinger International Institute of Mathematical Physics supported my
stay at the 4th Vienna Central European Seminar on Particle Physics and Quantum Field Theory.
